\documentclass{aa}  
\usepackage{graphicx}
\usepackage[subrefformat=parens,labelformat=parens]{subfig}

\usepackage{txfonts}
\usepackage{natbib}
\usepackage{breqn}
\usepackage{lscape}
\usepackage{hyperref}

\usepackage{siunitx}
\usepackage{lscape}

\newcommand{\HeI}[1]{\mbox{He\,{\sc i}~$\lambda${#1}}}
\newcommand{\HeII}[1]{\mbox{He\,{\sc ii}~$\lambda${#1}}}

\newcommand{\CIV}[1]{\mbox{C\,{\sc iv}~$\lambda${#1}}}
\newcommand{\OIII}[1]{\mbox{O\,{\sc iii}~$\lambda${#1}}}

\newcommand{\NIII}[1]{\mbox{N\,{\sc iii}~$\lambda${#1}}}

\newcommand{\SiIII}[1]{\mbox{Si\,{\sc iii}~$\lambda${#1}}}

\newcommand{\kms}[1]{\mbox{km\,s$^{-1}$~{#1}}}

\newcommand\rurl[1]{\href{http://#1}{\nolinkurl{#1}}}

\begin{document}

   \title{MONOS: Multiplicity Of Northern O-type Spectroscopic systems.\thanks{Table~\ref{t-orbsol-def} is available in electronic form, and the Appendix~\ref{app:rv} tables are only available in electronic form
at the CDS via anonymous ftp to \rurl{cdsarc.u-strasbg.fr} (130.79.128.5)
or via \url{http://cdsweb.u-strasbg.fr/cgi-bin/qcat?J/A+A/}}} 
   \subtitle{II. Orbit review and analysis for 35 single-lined spectroscopic binary systems and candidates}

   \author{ E. Trigueros P\'aez \inst{1, 2}
      \and R. H. Barb\'a \inst{3}
      \and I. Negueruela \inst{1}
      \and J. Ma\'{\i}z Apell\'aniz \inst{2}
      \and S. Sim\'on-D\'{\i}az \inst{4,5}
      \and G.Holgado \inst{2}
        }

   \institute{Departamento de F\'{\i}sica Aplicada. Universidad de Alicante.
              Carretera San Vicente del Raspeig s/n.
              E-\num{03690} San Vicente del Raspeig, Spain.
              \email{emilio.trigueros@gcloud.ua.es}
        \and
             {Centro de Astrobiolog\'{\i}a, CSIC-INTA. Campus ESAC. 
             Camino bajo del castillo s/n. 
             E-\num{28692} Villanueva de la Cañada, Madrid, Spain.}
        \and
             {Departamento de Astronom\'{\i}a. Universidad de La Serena.
             Av. Cisternas 1200 Norte. 
             La Serena, Chile.}        
        \and
             {Instituto de Astrof\'{\i}sica de Canarias. 
             E-\num{38200} La Laguna, Tenerife, Spain.}
        \and
             {Departamento de Astrof\'{\i}sica. Universidad de La Laguna.
             E-\num{38205} La Laguna, Tenerife, Spain}
        }

   \date{Received xxx xxx, 2021; accepted xxx xxx, 2021}

 
  \abstract
   {Massive stars are a key element for understanding the chemical and dynamical evolution of galaxies. Stellar evolution is conditioned by many factors: Rotation, mass loss, and interaction with other objects are the most important ones for massive stars. During the first evolutionary stages of stars with initial masses (i.e., $M_{\rm ZAMS}$) in the $M_{\rm ZAMS}\sim$18-70~M$_\odot$ range, they are of spectral type O. Given that stars in this mass range spend roughly 90\% of their lifetime as O-type stars, establishing the multiplicity frequency and binary properties of O-type stars is crucial for many fields of modern astrophysics.}
   {The aim of the MONOS project is to collect information to study northern Galactic O-type spectroscopic binaries. In this second paper, we tackle the study of the 35 single-line spectroscopic binary (SB1) systems identified in the previous paper of the series, analyze our data, and review the literature on the orbits of the systems.}
   {We have measured $\sim$ \,$4500$ radial velocities for a selection of diagnostic lines for the $\sim$\,$700$ spectra of the studied systems in our database, for which we have used two different methods: a Gaussian fit for several lines per object and cross-correlation with synthetic spectra computed with the {\sc FASTWIND} stellar atmospheric code. We have also explored the photometric data delivered by the TESS mission to analyze the light curve (LC) of the systems, extracting 31 of them. We have explored the possible periods with the Lomb-Scargle method and, whenever possible, calculated the orbital solutions using the {\sc SBOP} and {\sc GBART} codes. For those systems in which an improved solution was possible, we merged our radial velocities with those in the literature and calculated a combined solution.}
   {As a result of this work, of the 35 SB1 systems identified in our first paper we have confirmed 21 systems as SB1 with good orbits, discarded the binary nature of six stars (9~Sge, HD~\num{192281}, HDE~\num{229232}~AB, 68~Cyg, HD~108, and $\alpha$~Cam), and left six stars as inconclusive due to a lack of data. The remaining two stars are 15~Mon~Aa, which has been classified as SB2, and Cyg~OB2-22~C, for which we find evidence that it is most likely a triple system where the O star is orbiting an eclipsing SB1. We have also recalculated 20 new orbital solutions, including the first spectroscopic orbital solution for V747~Cep. For Cyg~OB2-22~C, we have obtained new ephemerides but no new orbit.}
   {}
\maketitle

\section{Introduction} \label{intro}
$\,\!$\indent One of the key pillars in our understanding of the chemical and dynamical evolution of galaxies is our knowledge about massive stars, and O-type stars in particular.
These stars play a crucial role in this regard due to their short life span, during which they greatly affect their surroundings (UV radiation and strong stellar winds), and violent death (supernova explosions). 
However, since a significant fraction of massive stars are found in multiple systems, a large percentage of them being short-period systems\footnote{The exact value to classify binaries as "short" or "long" period systems is a subject of debate. \citet{SanaEvan11} adopted $P_{\break} = 10$ d to separate short- and long-period systems to empirically describe the cumulative distribution function of binary periods. \cite{Sanaetal12a} adopted a $P_{\break} = 6$ d as a limit for systems that interact during the main sequence, merging at a higher rate.}, any study of massive stars will be incomplete without a thorough understanding of their multiplicity and the role of that characteristic in the formation, evolution, and death of massive stars \citep{ZinnYork07, Masoetal09, Chinetal12, Sanaetal13b, Sotaetal14, Barbaetal17, Apellanizetal19}. 

\cite{Sanaetal12a} proposed that nearly $70\%$ are expected to exchange mass with a companion during their lifetimes and that almost a third will do so while both components are still on the main sequence. Therefore, it is crucial to obtain accurate knowledge of their orbital and stellar properties in order to understand the role of massive stars as a population in the evolution of the galaxies.

O-type stars are the initial evolutionary phases of  massive stars with initial masses  (i.e., $M_{\rm ZAMS}$) in the range $M_{\rm ZAMS}\sim\,18-70\; \mathrm{M_{\sun}}$, and, nowadays, even some O and B supergiants are interpreted as H-burning objects \citep{Bouretal12, Higginsetal19}.
Given that stars in this mass range spend roughly 90\% of their lifetime as O-type stars, establishing the multiplicity frequency and binary properties of O-type stars is crucial for many fields of modern astrophysics, including for calibrating contemporary binary evolution and stellar population synthesis models.
Furthermore, for a given mass, the O-type phase is likely to be the one with the lower mass-loss rates, and hence it is a particularly appropriate evolutionary point to study the multiplicity properties; their spectra tend to be emission-free, and hence it is easier to measure their radial velocities (RVs). 
Nevertheless, it is important to note that we expect to find, and indeed do find, binaries with members in all evolutionary stages (i.e., main-sequence stars, Wolf-Rayet stars, or collapsed companions).

One of the difficulties we have encountered is the lack of homogeneity in the quality of information regarding the O-type star multiplicity, especially in the absence of an updated catalog with revised information about published spectroscopic orbits.
Although some efforts have been made, such as the $\rm{S_{B^9}}$ Catalog \citep{Pouretal04} or \citet{Masoetal98}, those catalogs are out of date and leave out orbits for new systems or revised orbits for previously known ones, and a critical reanalysis is needed.

As we stated in the first paper of this series, \citet{Apellanizetal19}, hereafter MONOS~I we have started an ambitious project that aims to bring homogeneity to the extensive but diverse literature and data on Galactic O-type spectroscopic binaries.
The overall project involves both hemispheres. 
The MONOS (Multiplicity Of Northern O-type Spectroscopic systems) project analyzes targets with $\delta > -20^{\rm o}$, and the rest of the southern hemisphere is being studied with the {\em OWN Survey} project \citep{Barbaetal17,barbetal10} and with MOSOS  (Multiplicity Of Southern O-type Stars), which has the same goals and will be structured in the same way as the MONOS series.

The Galactic O-Star Catalog (GOSC; \citealt{Maizetal04b, Sotaetal08, Maizetal17c}) is the main resource for the sample selection for the MONOS project. 
In MONOS~I we selected the spectroscopic and$/$or eclipsing O+OBcc binaries (i.e., systems composed of an O star plus an OB or a compact object companion) with previously published orbits and $\delta > -20^{\circ}$.
At that point, we expressly excluded systems that had been tentatively identified as spectroscopic binaries but that have no published orbits (eclipsing and/or spectroscopic). Therefore, the MONOS catalog is still a heterogeneous collection of objects selected according to the criterion of having been previously studied; in the future, we plan to extend it to be a magnitude complete catalog.
MONOS~I was focused on the spectral classification and multiplicity status of the systems, while in this second paper of the series we study in detail the sample of the 35 systems classified as single-lined spectroscopic binaries (SB1) in MONOS~I. 
Here, we review the information available in the literature about their orbital parameters, updating this information whenever possible by using high-resolution spectroscopy from LiLiMaRlin (Library of Libraries of Massive-Star High-Resolution Spectra)\footnote{LiLiMaRlin is a library of libraries of massive-star high-resolution optical spectra built by collecting data from our spectroscopic surveys: CAF\'E-BEANS \citep{Neguetal15a} OWN, IACOB \citep{SimDetal15b, SimonDetal20} and NoMaDS  \\ \citep{Maizetal12}  and programs and searches in public archives (CARMENES, FIES, Mercator, OHP, HARPS, FEROS, and UVES).}  \citep{Maizetal19a}. 
For most of the systems we present new RV measurements, and we determine 20 new orbital solutions. 
In forthcoming papers of the MONOS series we will study the double-lined spectroscopic binaries (SB2) and more complex systems, and we will present new spectroscopic orbits for O-type binary systems without known published orbits. 

This paper is structured as follows. 
In Sect.~\ref{methods} we describe the data analyzed in this work and the RV measurement methods used. 
In Sect.~\ref{data-analysis} we present the procedure followed for the analysis of each object. 
In Sect.~\ref{systems} we analyze each object in the sample, combining information available in the literature and new information derived from our analysis. 
In Sect.~\ref{summary} we summarize the findings of this paper. 
Last, we have included three appendixes: In Appendix~\ref{app:tables} we present the tables with the orbital solutions for each system and details about the spectra available in our database. 
In Appendix~\ref{app:curves} we show the figures of RV curves for the studied systems with an orbital solution and, additionally, the light curves (LCs) discussed in the text. 
The RV measurements determined in this work appear in the last appendix,~\ref{app:rv}.

\section{Observations and methodology} \label{methods}

\subsection{Spectroscopic observations}

\subsubsection{LiLiMaRlin database and sample description} \label{sample}

$\,\!$\indent As we mentioned in MONOS~I, our spectroscopic data are extracted from the spectral library LiLiMaRlin, built by collecting data obtained with different instruments and telescopes during the last two decades. We recently added to LiLiMaRlin more than 5000 spectra, gathered with two more instruments: the CARMENES (Calar Alto high-Resolution search for M dwarfs with Exoearths with Near-infrared and optical Échelle Spectrographs) spectrograph attached to the 3.5-m telescope at Calar Alto Observatory, and the HARPS (High Accuracy Radial Velocity Planet Searcher) spectrograph installed on ESO's 3.6~m telescope at La Silla Observatory in Chile, pushing the number of spectra available in the library close to \num{32000}. We are currently in the process of adding spectra from the Ultraviolet and Visual Echelle Spectrograph (UVES) at the Very Large Telescope (VLT), and that will increase the number to a total of close to \num{55000} spectra.
The total sample analyzed in this work corresponds to $\sim 700$ spectra for 32 (of the 35) systems classified as SB1 in the previous article of the series, MONOS~I; the distribution of the spectra available in the database can be found in Table~\ref{t-numspc}.
Spectra provided by LiLiMaRlin are normalized, telluric-line subtracted and corrected to the solar system barycentric frame of reference.

The SB1 sample investigated can be divided into three different groups, depending on the spectral characteristics of the stars and the number and quality of spectra available (see Table~\ref{t-obj}):
(a) well-behaved objects (21), with well-defined absorption lines for which we have enough good quality spectra to make an accurate RV analysis; (b) systems (6) with spectra of variable quality or without enough data to make a full analysis as described in Sect.~\ref{data-analysis}; and (c) systems (8) with a very small number of (or no) spectra, or without enough quality to allow a meaningful analysis of their orbital properties and/or definite confirmation of their SB1 nature.

\begin{table}[!htp]
    \small
    \centering
    \renewcommand{\arraystretch}{1.2}
    \caption{SB1 systems and candidates studied in this work. 
    Here we summarize the data available in the LiLiMaRlin database.
    }
    \label{t-obj}
    \begin{tabular}{l | l | l}
        \hline \hline
        Well behaved /      & Variable quality /    &  Low quality /  \\
        Good quality        & Not enough data       &  few or no spectra \\
        \hline
        \hyperlink{hd164438}{HD~\num{164438}}      & \hyperlink{2-A11}{Cyg~OB2-A11}        & \hyperlink{als15133}{ALS~\num{15133}} \\
        \hyperlink{V479}{V479~Sct}                 & \hyperlink{als15148}{ALS~\num{15148}} & \hyperlink{2-22C}{Cyg~OB2-22~C}       \\
        \hyperlink{9Sge}{9~Sge}                    & \hyperlink{2-1}{Cyg~OB2-1}            & \hyperlink{2-22B}{Cyg~OB2-22~B}       \\  
        \hyperlink{X1}{Cyg~X-1}                    & \hyperlink{2-20}{Cyg~OB2-20}          & \hyperlink{2-41}{Cyg~OB2-41}          \\
        \hyperlink{bd36}{BD~+36\,4063}             & \hyperlink{2-15}{Cyg~OB2-15}          & \hyperlink{als15131}{ALS~\num{15131}} \\
        \hyperlink{hde229234}{HDE~\num{229234}}    & \hyperlink{2-11}{Cyg~OB2-11}          & \hyperlink{2-70}{Cyg OB2-70}          \\
        \hyperlink{hd192281}{HD~\num{192281}}      &                                       & \hyperlink{als15115}{ALS~\num{15115}} \\
        \hyperlink{hde229232}{HDE~\num{229232}~AB} &                                       & \hyperlink{2-29}{Cyg OB2-29}          \\
        \hyperlink{68cyg}{68~Cyg}                  & & \\
        \hyperlink{hd108}{HD~108}                  & & \\
        \hyperlink{V747}{V747~Cep}                 & & \\
        \hyperlink{hd12323}{HD~\num{12323}}        & & \\
        \hyperlink{hd16429}{HD~\num{16429}~A}      & & \\
        \hyperlink{hd14633}{HD~\num{14633}~AaAb}   & & \\
        \hyperlink{hd15137}{HD~\num{15137}}        & & \\
        \hyperlink{aCam}{$\alpha$~Cam}             & & \\
        \hyperlink{hd37737}{HD~\num{37737}}        & & \\
        \hyperlink{15mon}{15~Mon~AaAb}             & & \\
        \hyperlink{hd46573}{HD~\num{46573}}        & & \\            
        \hyperlink{tet1ori}{$\theta^{1}$~Ori~CaCb} & & \\
        \hyperlink{hd52533}{HD~\num{52533}~A}      & & \\
        \hline
    \end{tabular}
\end{table}

We consider good quality spectra those with a signal-to-noise ratio (S/N) around 150, and low quality those around 50 or with clear normalization issues.  It is important to note that this division is simply based on the quality and quantity of our data and has no relevance to the binary status of the objects. This assessment of the data is relevant since it determines the methods that can be used to measure the RV of the object. However, several factors affect our ability to detect the companion in an SB1 system (e.g., spectral type of both components, rotation and relative brightness among others).
We can assume that we detect companions five times fainter than the primary and up to ten times fainter for the most favorable systems.

\subsubsection{Radial velocity measurements} \label{rv_measure}

$\,\!$\indent Accurate RV measurements are the key to a proper study of spectroscopic binaries.
Different methods of RV measurements could yield different results, depending on the observations (e.g., resolving power), data quality, and the nature of the binary system itself.
We distinguish between methods based on the fitting of a function (e.g., Gaussian) to one or several lines profiles and more complex methods like cross-correlation (x-corr), using different techniques (e.g., Fourier analysis) or data processing. Each one has its pros and cons. Function fitting has the advantages of simplicity, the possibility of easy comparison between results for different lines, and the possibility of looking at the residuals to find weak components or other reasons for a bad fit to the data. On the other hand, line profiles can be distorted by wind infilling, magnetic and$/$or pulsational effects, or line blending, so using one or just a few lines can lead to biased results when fitting a simple function. 
The x-corr method has the advantage of using the information of many lines at the same time but the inconveniences of the dependence on the choice of template and of the possibility of contamination of the results by emission or interstellar lines and normalization or noise issues. 
Given those pros and cons, we implemented two different methods to determine RVs for the systems: spectral x-corr using synthetic spectra and Gaussian profile fitting for individual lines.

For the first set of systems (first column in Table~\ref{t-obj}), we used the x-corr method to obtain the RV measurements and compared our results with the published orbits for the systems. 
On top of that we measured several different lines individually with the Gaussian fitting, including some metallic ones that are less prone to be affected by stellar winds (see Table~\ref{t-rest_wav}).
For the second and third set of systems, we could not obtain reliable synthetic spectra to be used as templates for the x-corr method (see Sect.~\ref{xcorr} for an explanation of the process) for all the stars due to the quality of the spectra (i.e., available lines to carry out the procedure described below) or because we did not have any suitable spectra; thus, we could only use the Gaussian fit for those systems. 
The last set was also measured if possible, but the lack of enough quality data from our end did not let us evaluate the published orbital solutions properly. 

In Appendix~\ref{app:rv} we provide the RVs that we obtained for each object. 
Table~\ref{ex-rvt:HD192281} shows the first rows of the table in which we present such measurements for HD~\num{192281} as an example. The structure of the table is as follows: the first two columns identify the spectrum with the code used in the LiLiMaRlin database and the RJD (HJD-\num{2400000}) of the observation. 
If we used the cross-correlation method for the object the measured RV will be in the "XCorr" column, followed by the Gaussian measurements for different lines selected.
For each column, we present the mean and standard deviation after applying a $3\sigma$ clipping and the number of dropped spectra during the clipping. 

An important consideration has to be made regarding the error in the measurements. Although we obtain the formal errors of the fit for each measurement method, we take a slightly more conservative approach.
We consider a good upper bound for the error of our measurements to be 5~\kms,  which corresponds to the $\sigma$ of the RV measurement of well-behaved lines for the single stars after applying the $3\sigma$ clipping (see Sect. \ref{single}). 
The formal fitting error associated with each measurement is also provided in the RV tables in Appendix~\ref{app:rv}.

Almost all the systems presented in this work have previously
published RVs. 
In many cases, we derived a new orbital solution by combining published RV data with ours. 
For those systems, given that different authors adopted different rest wavelengths for the spectral lines, we firstly applied an appropriate velocity correction to bring the measurements to the same rest frame.
When we have spectra in common, we can also measure the RV shift empirically. Finally, in cases where a systematic shift in RV not correlated with the adopted rest wavelengths was found (i.e., when measurements retrieved from the literature came from an average of different lines or when we did not know the rest wavelengths used), the RV shift was determined from the orbital $\gamma$ value obtained from the orbital fit for each RV data set separately. In the sections devoted to each object, we detail the process followed and the shift adopted for each system.

\subsubsection{Cross-correlation} \label{xcorr}

$\,\!$\indent In order to obtain RV measurements, we implemented the x-corr technique in a Python code. 
The spectral templates used in our program are based on synthetic spectra obtained with the stellar atmosphere code {\sc FASTWIND} \citep{SantolayaRetal97, Pulsetal05, RivGetal12}. 
In particular, for each star, we use the {\sc FASTWIND} spectrum corresponding to the best-fitted model resulting from the quantitative spectroscopic analysis performed by \cite{Holgadoetal18, Holgadoetal20}, by means of the {\sc IACOB-GBAT/FASTWIND}\footnote{The version used in this work is the v10.1} tool \citep{SimDetal11d}.

The original {\sc FASTWIND} synthetic spectrum considered for the x-corr analysis includes five \ion{H}{i} lines, eleven \ion{He}{i} lines, and six \ion{He}{ii} lines present in the spectral range $4000-7000$\,\AA; however, as described below, in the end not all lines were used for the RV determination. 
Each synthetic spectrum includes the convolution with the corresponding projected rotational velocity ($v \sin i$) and macroturbulent velocity ($V_{\mathrm{mac}}$), as determined by \cite{Holgadoetal18} by using the IACOB-BROAD tool \citep{SimDHerr14}.

For the x-corr analysis, we only considered a $20$\AA\ spectral window centered on each line. 
Depending on the signal-to-noise and spectral coverage of each spectrum, we tried to include as many lines as possible, typically \ion{He}{i} 4144, 4388, 4471, 4713, 5016 and 5876 \AA, \ion{He}{ii} 4200, 4542, 4686 and 5412 \AA, and the blends $H\gamma$+\HeII{4338}, and $H\beta$+\HeII{4859}. 
In practice, the actual set of lines selected was different for each object due to the varying wavelength coverage and S/N.

The RV is determined from the x-corr function's peak, which is fitted with a parabolic function. 
The RV errors are derived from the errors determined for the fitted parabola's parameters.

\subsubsection{Gaussian profile fitting} \label{gauss}

$\,\!$\indent A single Gaussian profile fitting routine was implemented in Python to determine RVs from absorption line profiles. 
The aim of this routine is to perform a simple check to the RVs derived by using the cross-correlation method, and also to extend the RV determination to other metal lines available in the spectra, such as \OIII{5592}, \CIV{5812} or \SiIII{4553}, which are not included in the set of synthetic spectra associated with the grid of {\sc FASTWIND} models build to be used with {\sc IACOB-GBAT}.
For the profile fitting, an adjacent continuum is defined around the line, and a nonlinear least-squares minimization method is applied to both the profile and the adjacent continuum.

For the fitting procedure, we create a Gaussian model with four initial parameters, the continuum level, the center, $\sigma$, and amplitude. 
The initial set of parameters for the automatized process are obtained from a manual fit of one spectrum.
The fitting method can freely adjust such parameters within a window for each spectrum fitted (typically, we allow a $20\%$ variation) to enhance performance, and any of them could be fixed if necessary. 
The errors are obtained from the fit itself.
The rest wavelengths for the used spectral lines are listed in Table~ \ref{t-rest_wav}.

\begin{table}
    \centering
    \small
    \renewcommand{\arraystretch}{1.2}
    \caption{Rest (air) wavelengths for the lines used for RV measurement (Sect. \ref{gauss}). The number of significant digits for each value reflects the accuracy of the rest wavelengths, which depends on, among other things, their singlet or multiplet nature.}
    \label{t-rest_wav}
    \begin{tabular}{c|r@{.}l}
    \hline \hline
        Line & \multicolumn{2}{c}{ \AA\ } \\
    \hline
        \NIII{4379}  &  4379&11  \\
        \HeI{4471}   &  4471&48  \\
        \HeII{4542}  &  4541&591 \\
        \SiIII{4553} &  4552&622 \\
        \HeII{4686}  &  4685&682 \\
        \HeI{4713}   &  4713&2   \\
        H$\beta$     &  4861&33  \\
        \HeI{4922}   &  4921&9   \\
        \HeI{5015}   &  5015&7   \\ 
        \HeII{5412}  &  5411&53  \\
        \OIII{5592}  &  5592&252 \\
        \CIV{5801}   &  5801&33  \\
        \CIV{5812}   &  5811&97  \\
        \HeI{5876}   &  5875&621 \\
        H$\alpha$    &  6562&8   \\
        \HeI{6678}   &  6678&152 \\
        \HeI{7065}   &  7065&19  \\
        \HeII{8237}  &  8236&8   \\
    \hline
    \end{tabular}
\end{table}

\subsection{Photometric time series} \label{photometry}

$\,\!$\indent Eclipsing systems and ellipsoidal variables are an important subset of binaries, a characteristic that is recognized as E in the spectroscopic binary status (SBS) nomenclature (for a detailed explanation of such a classification, see MONOS~I). We explored the possibility of detecting extrinsic variability in the systems studied by exploring the exquisite Transiting Exoplanet Survey Satellite ({\em TESS})  database.

We retrieved time series from the TESS database, obtaining 31 LCs.
Although some stars were observed at 2-min cadence (and they are accessible through MAST archives\footnote{\url{https://mast.stsci.edu/portal/Mashup/Clients/Mast/Portal.html}}), we decided to extract LCs directly from the TESS full frame images (FFIs) with 30 min cadence using the Python package {\sc lightkurve}\footnote{\url{https://docs.lightkurve.org/index.html}} \citep{Lightkurve18} version 1 in order to control the LC extraction process. 
The images are cut by using the {\sc astrocut} package \footnote{\url{https://github.com/spacetelescope/astrocut}, \citep{Brasseur19}}.
Aperture photometry was performed on images cutouts of $15\times15$ pixels (about $315''\times315''$). 
The source mask was defined for each object, depending on the brightness of the close neighbors, and it was tuned interactively in order to minimize the contamination. 
Given that the TESS pixel size is large (21''), the extraction done over several pixels can potentially include neighbors. 
In the {\sc lightkurve} package, we have the possibility to examine the cutout image using the Gaia Data Release 2 (DR2) catalog. 
Therefore, the size of the mask varied from 2 pixels for crowded fields to 16 pixels for bright isolated stars. 
The sky background mask (including scattered light) was selected using the remaining lowest brightness pixels in the cutout, for the majority of the cases, over one hundred pixels.
The background was modeled using principal component analysis (PCA), following the package recommendations.

The main objective of the analysis of TESS data is to search for variability on the order of a few days or less. 
The low-frequency signals or slopes present in the extracted TESS time series have been removed by a Savitzky-Golay filter implemented in the flatten method on the {\sc lightkurve} package.

We extracted TESS time series for 31 stars. 
Two stars have not been yet observed (HD 164438 and V479 Sct), while two stars are located in crowded fields (Cyg~OB2-22~B and $\theta^1$~Ori~CaCb).

Furthermore, we explore different public photometric databases, such as the Hipparcos Epoch Photometric (HIP) and the Kamogata-Kiso-Kyoto Wide-field Survey\footnote{We extracted four HIP and two KWS LCs; we are only presenting here one HIP LC.}  \citep[KWS;][]{Maehara14} in order to obtain additional clues about variability associated with the orbital cycle, or also to determine intrinsic variability that can lead RV variations.

\section{Data analysis and results} \label{data-analysis}

$\,\!$\indent Each star with available spectroscopic data was analyzed following the subsequent procedure.
Firstly, we measured the RVs of the available spectra for such an object in our database LiLiMaRlin, a sample of such measurements can be seen in Table~\ref{ex-rvt:HD192281}. 
For all the systems with an available synthetic spectrum, we measured the RV via the x-corr method, where we select the suitable lines for each object, adding or removing more lines to the initial set depending on the spectral characteristics of the object and the quality of our data. 
Then we inspected the \HeI{5876} and \HeII{5412} lines and measured them with the Gaussian fitting method. 
We also checked the availability of metallic lines, such as \OIII{5592} and \CIV{5812}. 
Finally, we explored other lines of \ion{He}{i} and \ion{He}{ii}. All the measurements were visually inspected to ensure that the fitting procedures were correct. 

The Lomb-Scargle (LS) method (\citealt{Lomb76, Scargle82}, see also \citeauthor{VanderPlas18} \citeyear{VanderPlas18} for details of the method) was used to search for periodicities in the photometric and RV time series.
Orbital elements were determined employing two codes {\sc GBART} \citep{GBART} and {\sc SBOP} \citep{SBOP}.
Results obtained with both codes are compatible, although in some systems with limited data, convergence issues favored the use of one code or another due to the differences in the implemented algorithms. 
As initial parameters, we used LS periods, the amplitude of RVs, and if available, previously published orbital parameters. 
If the orbital solution obtained is coherent with the previous one, we explored combining RVs to obtain an improved solution. 
For some systems, our RVs were insufficient for an orbital analysis; therefore, the orbital solutions are derived from combined data.

We summarize the orbital parameters determined for each system in Table~\ref{t-orbsol-def}. 
Table~\ref{t-orbsol-all} (in Appendix~\ref{app:tables}) lists the orbital solutions found in the literature and our orbital solutions obtained with different methods. 
LCs and RV curves are plotted in Appendix~\ref{app:curves}.

\begin{table*}[!htp]
\small
\centering
\renewcommand{\arraystretch}{1.3}\tabcolsep=0.12cm
\caption{Some RV measurements for HD~\num{192281} using the x-corr method and the selected lines with the Gaussian method. For each column we calculated the mean RV and the standard deviation after applying a $3\sigma$ clipping; the number of measurements dropped by the clipping procedure is also noted. \label{ex-rvt:HD192281}}
\begin{tabular}{l r@{.}l  r@{.}l  r@{.}l  r@{.}l  r@{.}l  r@{.}l  r@{.}l }
\hline \hline 
\multicolumn{1}{c}{Spectra} & \multicolumn{2}{c}{RJD} & \multicolumn{2}{c}{XCorr} & \multicolumn{2}{c}{\HeI{5876}} & \multicolumn{2}{c}{\HeII{4542}} & \multicolumn{2}{c}{\HeII{4686}} & \multicolumn{2}{c}{\HeII{5412}} & \multicolumn{2}{c}{\OIII{5592}} \\
\multicolumn{1}{c}{} & \multicolumn{2}{c}{d} & \multicolumn{2}{c}{km\,s$^{-1}$} & \multicolumn{2}{c}{km\,s$^{-1}$} & \multicolumn{2}{c}{km\,s$^{-1}$} & \multicolumn{2}{c}{km\,s$^{-1}$} & \multicolumn{2}{c}{km\,s$^{-1}$} & \multicolumn{2}{c}{km\,s$^{-1}$} \\
\hline
000707\_P & \num{51733}&604 & -28&8 & -37&5 & -26&1 & -49&2 & -27&0 & .&. \\
100807\_I & \num{55416}&728 & -27&6 & -38&1 & -27&9 & -69&0 & -22&4 & -8&8 \\
111108\_M & \num{55874}&389 & -31&0 & -36&6 & -27&4 & -57&6 & -28&7 & 1&5 \\
131208\_C & \num{56635}&382 & -34&8 & -48&6 & -42&8 & -30&5 & -44&8 & .&. \\
180327\_C\_V & \num{58205}&626 & -33&0 & -41&4 & .&. & .&. & -27&0 & .&. \\
\ldots    & .&.         & .&.   & .&.   &  .&.  &  .&.  & .&.   & .&.\\ 
\hline
RV mean & .&. & -28&8 & -40&6 & -28&2 & -46&7 & -26&0 & -6&0 \\
$\sigma$ RV & .&. & 4&9 & 7&9 & 5&6 & 13&2 & 4&2 & 5&7 \\
Drop & .&. & \multicolumn{2}{c}{0} & \multicolumn{2}{c}{1} & \multicolumn{2}{c}{1} & \multicolumn{2}{c}{0} & \multicolumn{2}{c}{3} & \multicolumn{2}{c}{2} \\
\hline
\end{tabular}
\renewcommand{\arraystretch}{1.0}
\end{table*}

\section{Individual systems} \label{systems}

$\,\!$\indent In this section, we review and discuss the published orbits and binary status for each of the 35 objects, including the RV measurements determined from our analysis of the collected spectroscopic data and also photometric time series.
The stars are grouped by constellations in the sky, and sorted by Galactic longitude, as it was performed in the previous MONOS~I paper. 
The SBS classification for each star derived is stated with the name.
Those stars where we were unable to validate the SBS are marked as unconfirmed (unc.).
Those systems that present ellipsoidal variations are marked as El.

\subsection{Sagittarius-Sagitta} \label{Sgr-Sge}
 
\paragraph{\textbf{HD~\num[detect-all]{164438}} =  BD~$-$19~4800 = ALS~4567} \label{hd164438}
\textbf{SB1}

\begin{figure}[!htp]
    \centering
    \includegraphics[width=.45\textwidth]{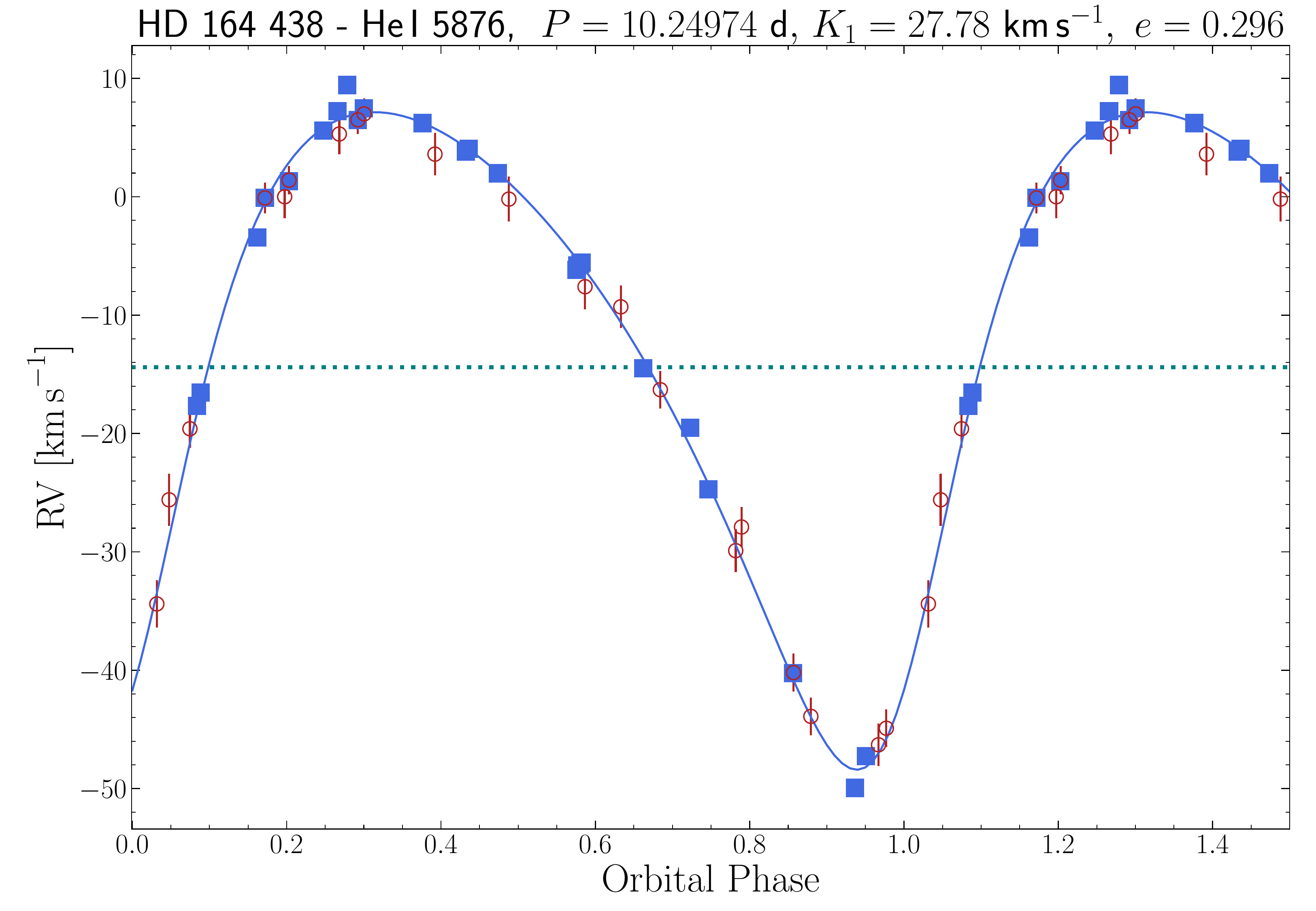}
    \includegraphics[width=.45\textwidth]{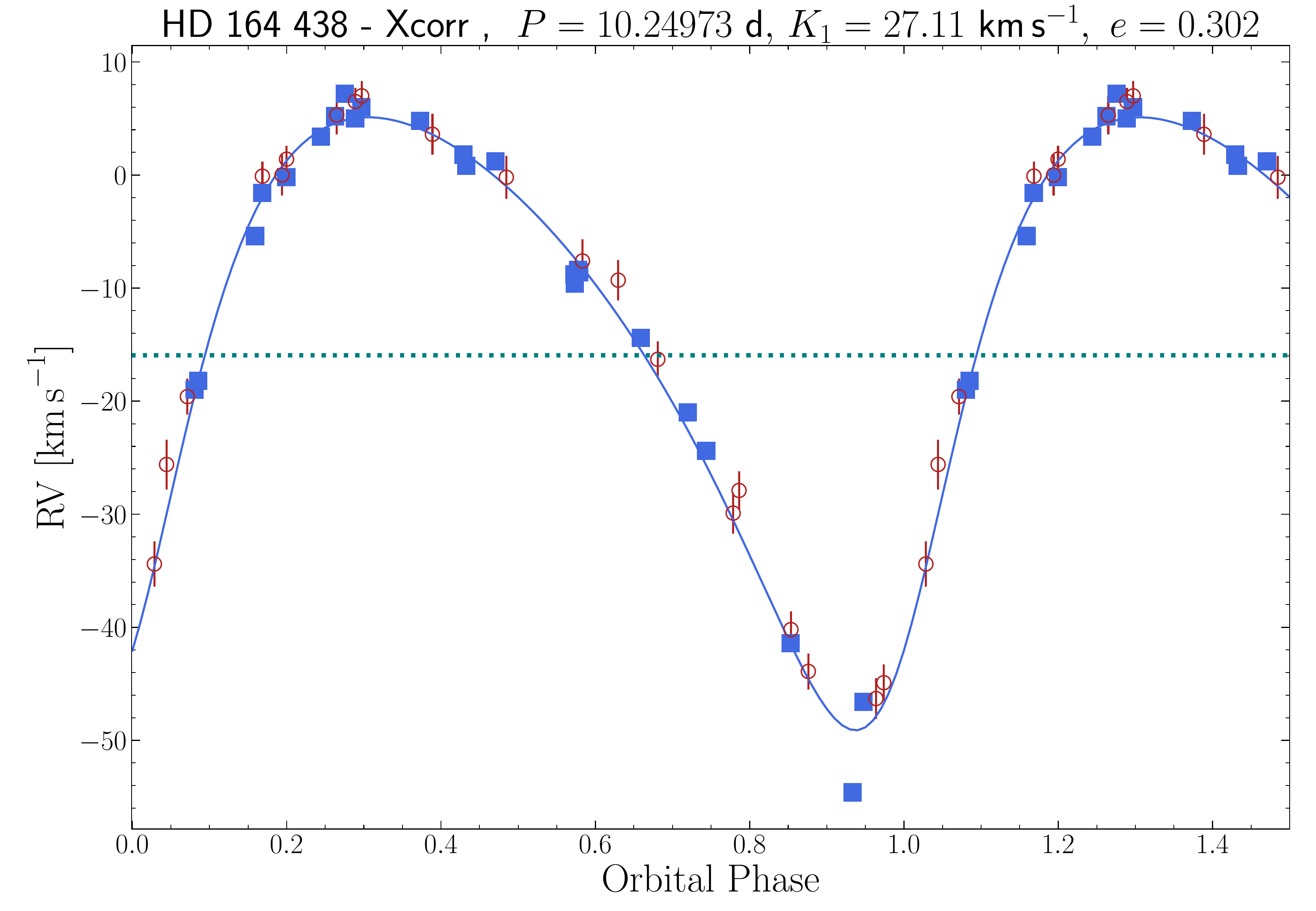}
    \caption{
    Orbital solutions for HD~\num{164438} obtained only with MONOS data.
    Upper panel: New orbital solution for HD~\num{164438} obtained from our RV measurements of the \HeI{5876} line using a Gaussian fit. The orbital solution was obtained using the SBOP code. 
    Lower panel: Orbital solution using our x-corr RV measurements. The RVs determined by \cite{Mayeretal17} (red circle) are shifted by $6.5$~\kms to bring them to our velocity rest frame. 
    MONOS RVs are shown with the navy square symbols. 
    The dotted blue line shows the $\gamma$ of the system.
    }
    \label{HD164-monos-gsol-fig}
\end{figure}

\hypertarget{hd164438}{} This SB1 system, classified as O9.2~IV, has an orbital solution presented by \cite{Mayeretal17}. 
The period of the system is $10.25$~d, with a small semi-amplitude $K_1=26.9$~\kms, and a fairly eccentric orbit ($\sim 0.3$). 
We collected 27 spectra spanning 4300~days (about 12 years); five FEROS\footnote{Fiber-fed Extended Range Optical Spectrograph installed at the MPG/ESO 2.2-meter telescope located at ESO’s La Silla Observatory.} spectra are in common with \cite{Mayeretal17}.
Both of our orbital solutions (\HeI{5876} and x-corr RVs) are compatible with that derived by \cite{Mayeretal17}. 
For \HeI{5876}, we applied a correction of $+4$~\kms to the \cite{Mayeretal17} RVs to bring them to our rest frame (Fig.~\ref{HD164-monos-gsol-fig}). 
An additional systematic difference of $+2.5$~\kms was also applied based on the RVs obtained from the five FEROS epochs in common.

\paragraph{\textbf{V479~Sct} = ALS~5039} 
\textbf{SB1}

\hypertarget{V479}{} This star is the counterpart of the X-ray source RX J$1826.2-1450$ \citep[LS 5039;][]{Motcetal97}, a $\gamma$-ray binary that has been extensively studied at all wavelengths from the radio to the TeV regime, as it is one of the few confirmed massive X-ray binaries associated with radio emission \citep{McSwainetal04}.
There are only a handful of high-mass binaries with significant emission at energies above 100~MeV: PSR~B$1259-63$ and LS~I~$+61^\circ303$ are two well-studied examples, both containing a Be star \citep{Fermi09}, while V479~Sct was the first found with an O-type star and has the shortest orbital period of the whole set.
This star has also been proposed to be a runaway \citep{Riboetal02, Maizetal18b}, probably ejected in a violent episode from either Ser~OB2 or Sct~OB3. 

V479~Sct is composed of a massive ON6~V((f))z star \citep{Maizetal16} and a compact object companion. 
The nature of the compact object is still a matter of debate: options that have been explored are a micro-quasar, a black hole (BH) companion, and a young non-accreting neutron star (NS) interacting with the wind of the O-type star \citep[see][and references therein]{Dubus13}. 

The first optical orbital solutions published (see Table~\ref{t-orbsol-all}) correspond to a short period, $P\sim 4.4$~d, and a high eccentricity, $e\sim 0.4$ \citep{McSwainetal01, McSwainetal04}.
More recent orbital solutions point to a different period, $P=3.906$~d, and a smaller eccentricity, $e \sim 0.24-0.35$, \citep{Casaresetal05, Aragona09, Sartetal11}. 
It is worth noticing that $P=3.9$~d was also found independently at higher energies \citep{Hadaetal12, Fermi09, HESS06b}.

We collected 18 spectra spanning about 19 years, although some of them have a low S/N (due to the faintness of the source, $B=12.2$).
We determined a new orbital solution using RVs derived from the \HeII{5412} line, obtaining a period of $P=\num{3.9061}$~d and an eccentricity of $e=0.28$, confirming the solution derived by \cite{Sartetal11}. 
We also explore the RVs of \HeI{5876} line, obtaining a similar orbital solution albeit with a higher eccentricity of $e=0.34$.
Overall, our orbital solutions are compatible with that obtained by \cite{Sartetal11}, with the exception of the $\gamma$ value of the system. Ours is $+6$~\kms higher, probably due to differences in the lines used and the rest wavelengths assumed. \cite{Sartetal11} used the average of RVs determined from \HeII{4200}, 4686 and 5412 lines.

A systematic blue-shift in the RVs for H and \ion{He}{i} respect to \ion{He}{ii} has been detected in previous studies \citep{Casaresetal05,Aragona09,Sartetal11}, and we confirm this finding.
Finally, we also explored the orbital solution derived from a metallic ion, \OIII{5592}, finding a comparable solution, except for the $\gamma$ value (Fig.~\ref{orb-fig:1} upper left panel), which is shifted by $+22.6$~\kms with respect to the value of \cite{Sartetal11} (or $+16$~\kms with respect to our \HeI{5876} solution). 
This is an interesting result for two reasons: 
first, the \OIII{5592} line is expected to be much less affected by the wind interaction with the compact object than the \ion{He}{II} lines; and second, the $\gamma$ value difference of about $16-23$~\kms is relevant, given that this object is a runaway star.
The systemic $\gamma$ value combined with {\it Gaia} data can be used to trace back the system's trajectory accurately and then help to determine from which cluster or association the system was ejected.
The visual inspection of the O-type spectrum reveals a strong nitrogen enrichment and a carbon depletion, a characteristic also noted by \cite{Aragona09}, which may be a sign that the system underwent mass transfer.

\paragraph{\textbf{9~Sge} = HD~\num[detect-all]{188001} = QZ~Sge = BD~$+$18~4276 = ALS~\num[detect-all]{10596}} 
\textbf{Single}

\hypertarget{9Sge}{} This O7.5~Iabf runaway star (\citealt{Mdzi04, SchiRose08}) was proposed as an SB1 system by \cite{Aslanovetal84}, who determined an orbital period of $78.3$~d, a small RV semi-amplitude, and a moderate eccentricity of $0.38$.
Adopting a similar period ($P=78.74$~d), \cite{Underhilletal95} recalculated the probable orbital solution, obtaining a larger eccentricity ($e=0.556$), but a different orbital orientation. 
Both published orbital solutions show large scatter, throwing doubt into the binary status of the star.
\cite{McSwetal07} investigated the consistency of those orbital solutions through the analysis of 97 historic RVs and eight new ones, gathered during almost 80 years, finding a noisy period of $P=29.83$~d, which they considered as spurious, concluding that the star is probably single. 
It should be taken into account that the collected RVs by those authors are derived from measurements of lines produced by different ions, which as we will see, show different behavior.

\begin{figure*}[!htp]
    \includegraphics[width=.5\textwidth]{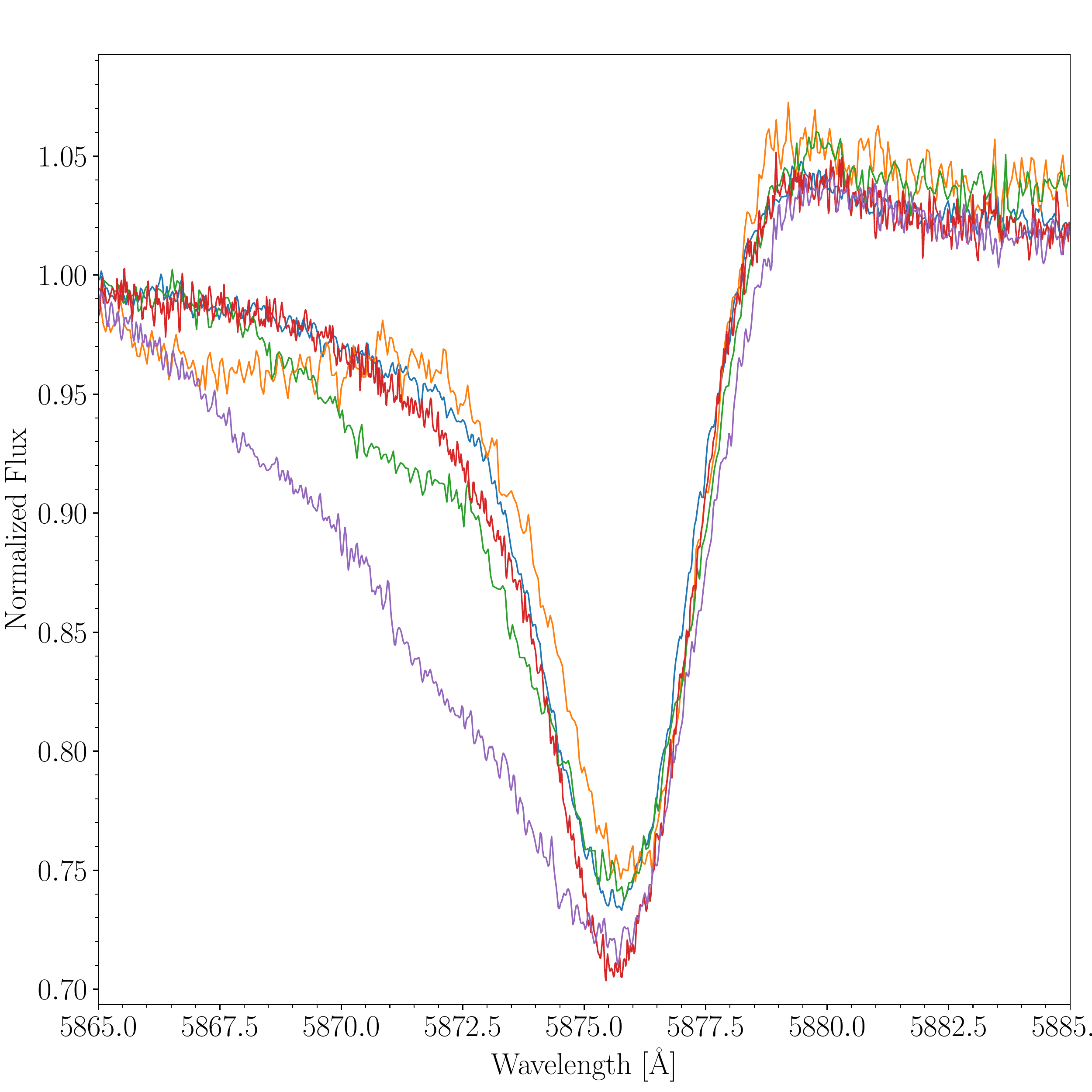}
    \includegraphics[width=.5\textwidth]{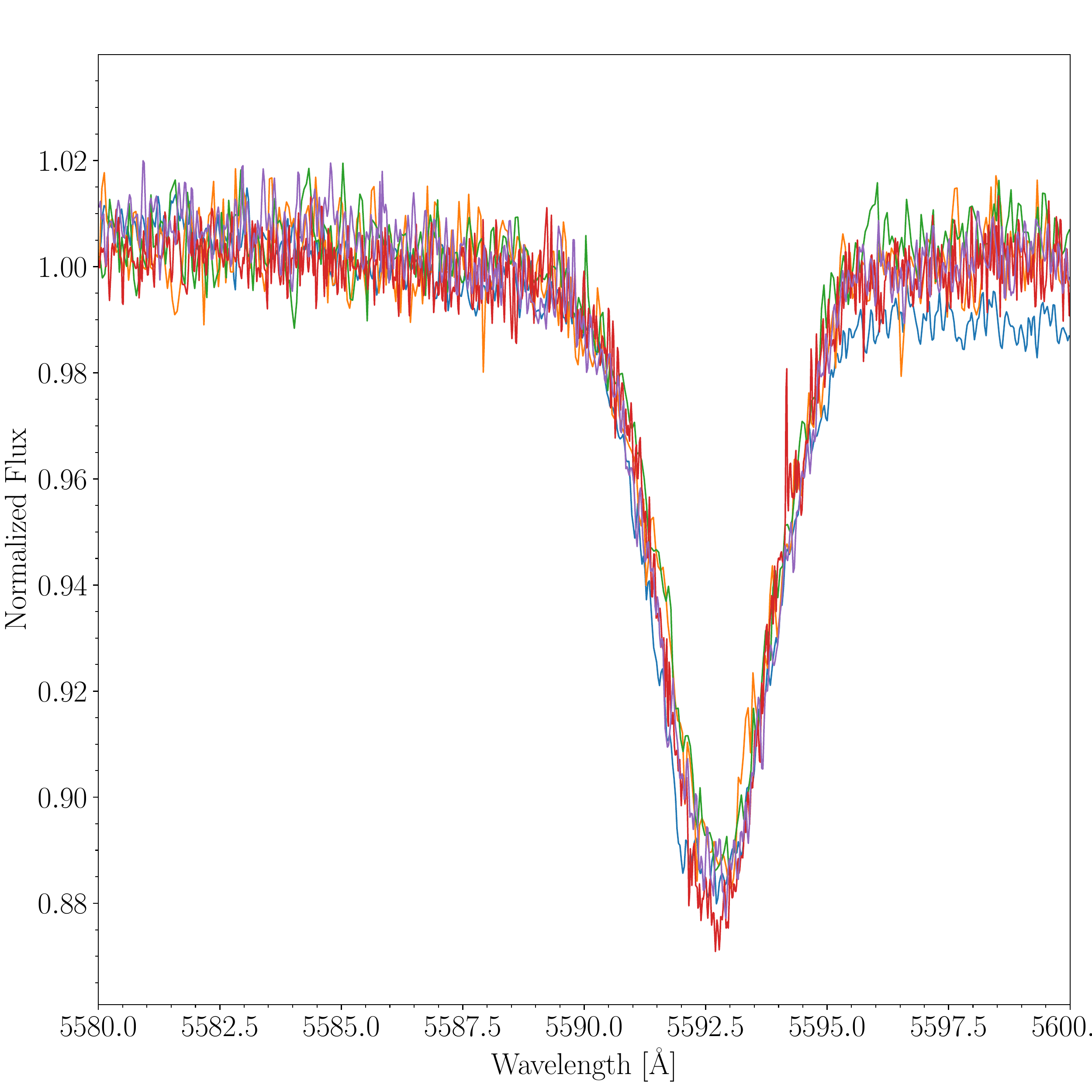}    
    \caption{
    Diagnostic lines of 9~Sge that show the spectral variability of the star.
    Left panel: \HeI{5876} line of 9~Sge at different epochs. It shows a P-Cygni profile that varies due to the effect of the stellar wind. 
    Right panel: \OIII{5592} line. It shows a more symmetric profile, indicating that the line is less affected by winds. 
    Plotted spectra are: 040530\_F (blue), 040827\_P (green), 051110\_P (orange), 110615\_M\_3 (red), and 110910\_I\_2 (purple). 
    See Table~\ref{t-numspc} for a description of naming convention used to identify each spectrum.
    }
    \label{9_Sge-spectra}
\end{figure*}

We collected 108 spectra spanning 15 years of monitoring, most of them obtained in two observing runs (around the years 2011 and 2013), specifically devoted by the IACOB project to investigate variability due to stellar oscillations in this star.
Some spectral lines show significant profile variations, especially the Balmer lines and \ion{He}{i} lines, while metallic ions show smaller variations. 
Figure~\ref{9_Sge-spectra} shows representative profiles at different epochs for the \HeI{5876} and \OIII{5592} lines, as representative for stronger and weaker profile variability, respectively.

\begin{figure*}
    \includegraphics[width=.5\textwidth]{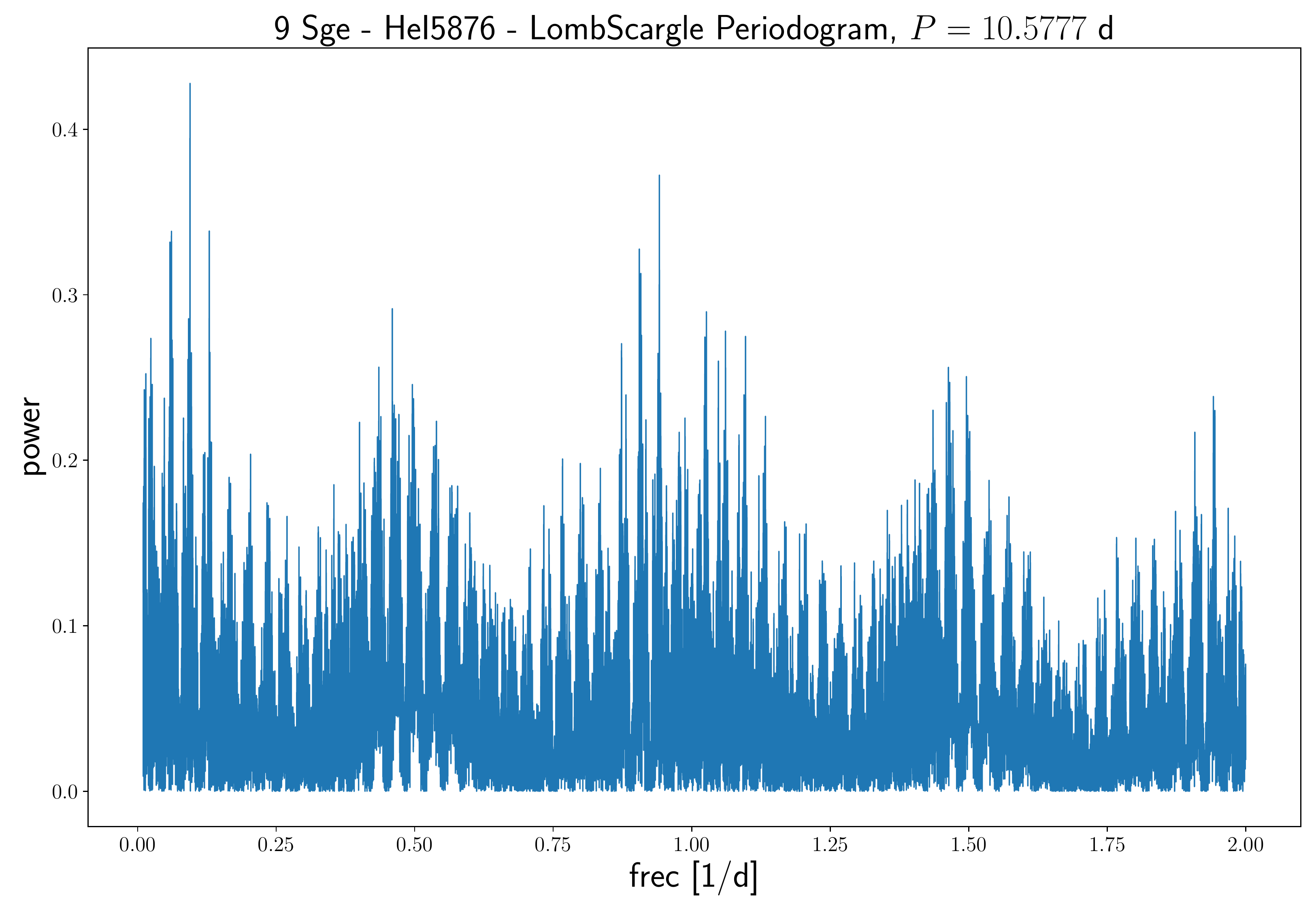}
    \includegraphics[width=.5\textwidth]{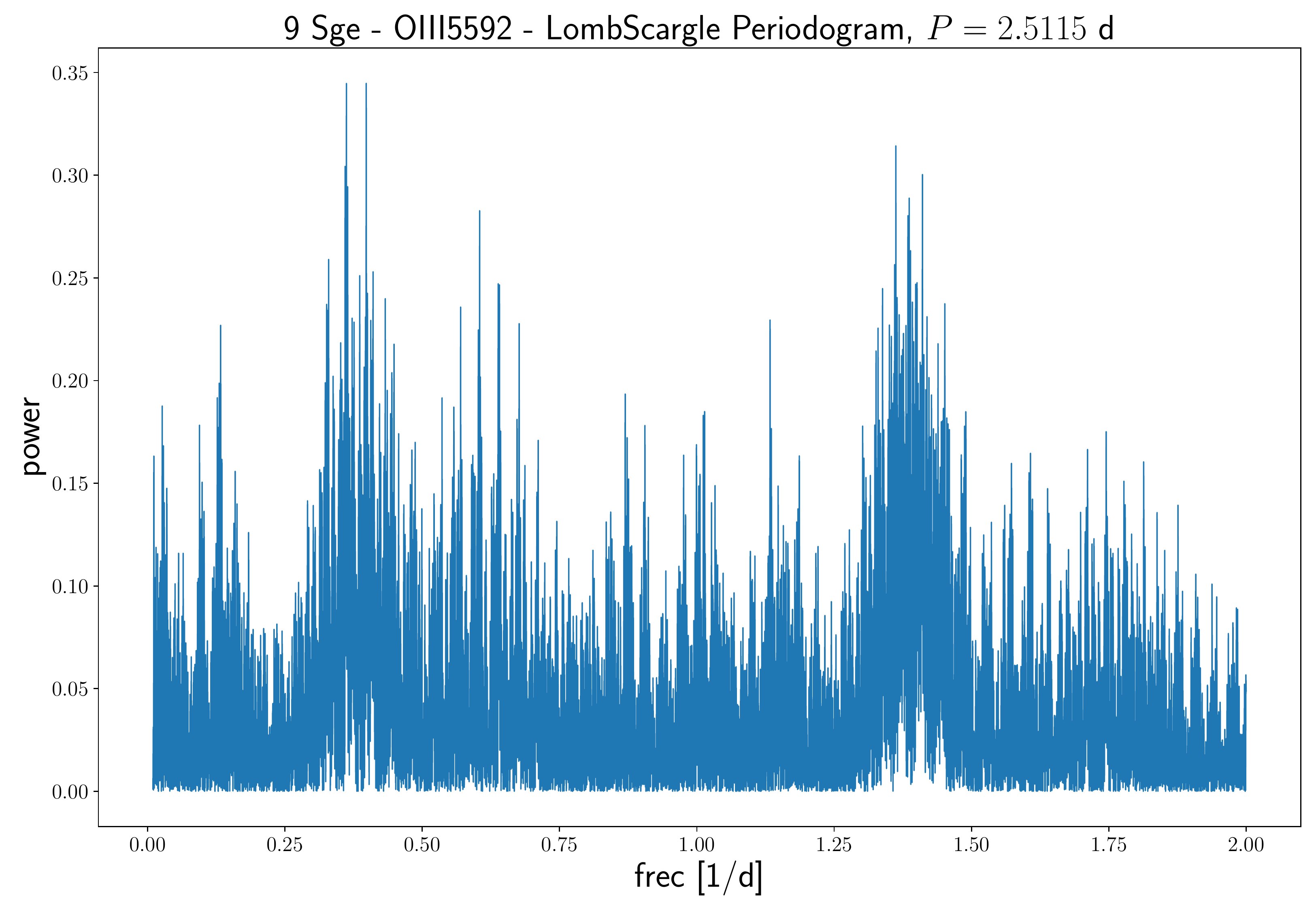}
    \caption{
    LS power spectrum derived from the RVs of \HeI{5876} (left panel) and the \OIII{5592} (right panel) absorption lines of 9~Sge.
    }
    \label{9_Sge_LS}
\end{figure*}

We performed a periodic signal search with the LS method by using the RVs determined for different ions. 
Periodograms do not present any conclusive common periodicity. Different ions display structured periodograms with low amplitude peaks at different frequencies, most of them centered around 0.1 and 0.4~d$^{-1}$.
As an example, Fig.~\ref{9_Sge_LS} shows the periodogram, again, for the \HeI{5876} and \OIII{5592} lines.

\begin{figure}
    \centering
    \includegraphics[width=.45\textwidth]{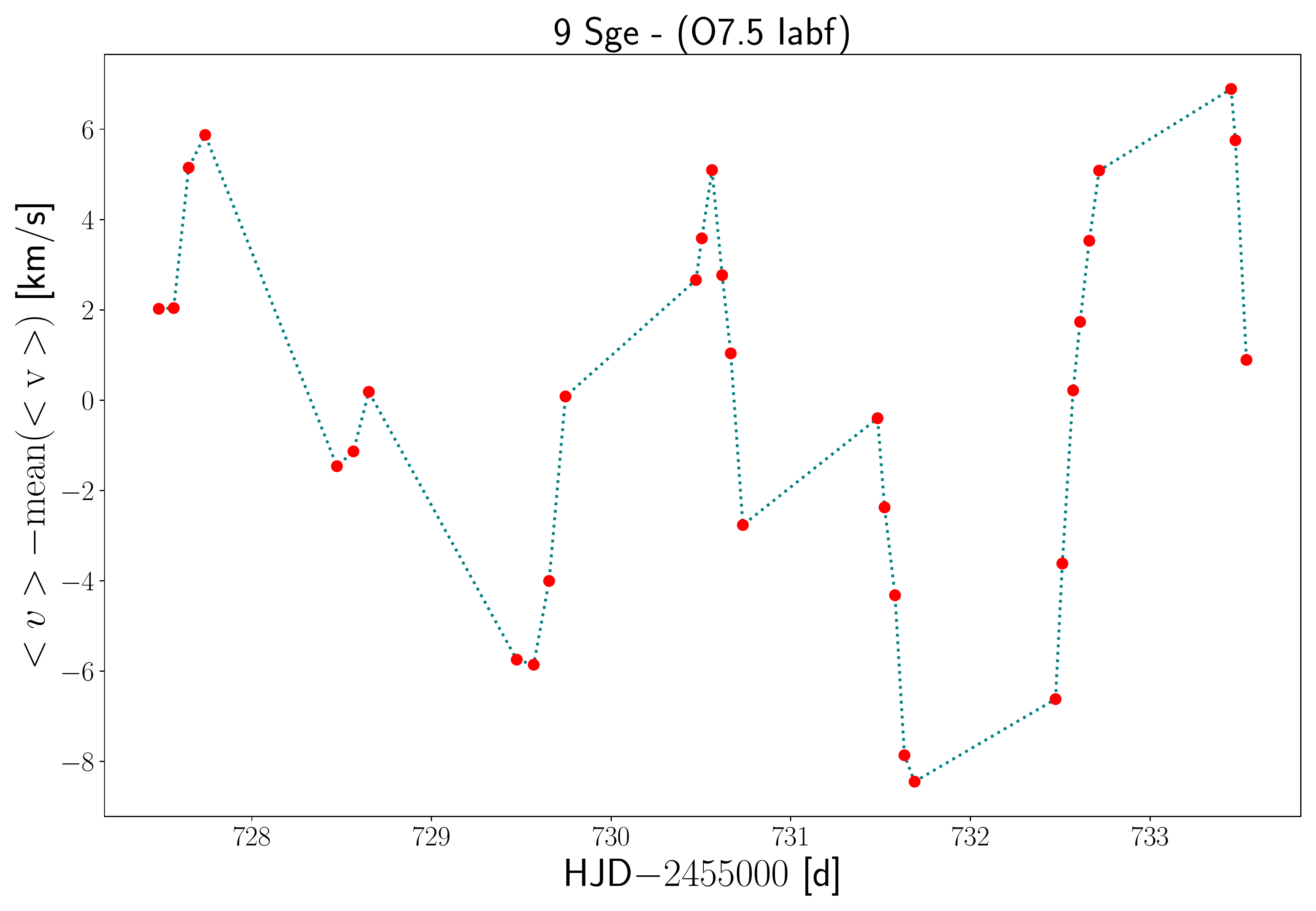}
    \caption{
    RVs of the \OIII{5592} line measured on a timescale of seven days in 9 Sge. 
    These RV variations are not compatible with binary motion.}
    \label{9sge_RVpp}
\end{figure}

Peak-to-peak radial velocity variations (RV$_{\rm pp}$) for different ions are about 25~\kms, on a timescale of a few days (Fig.~\ref{9sge_RVpp}).
Visual inspection of RV subsets obtained during consecutive nights does not find a coherent pattern, as expected for a spectroscopic orbit. 
This picture suggests that the star is pulsating and not an SB1 system. 
Indeed, as shown by \citet{SimonDetal20a}, the effect of stellar oscillation on the measured RV in late O and early B supergiants can be as high as 20\,--\,25~\kms. 

\begin{figure}
    \centering
    \includegraphics[width=.45\textwidth]{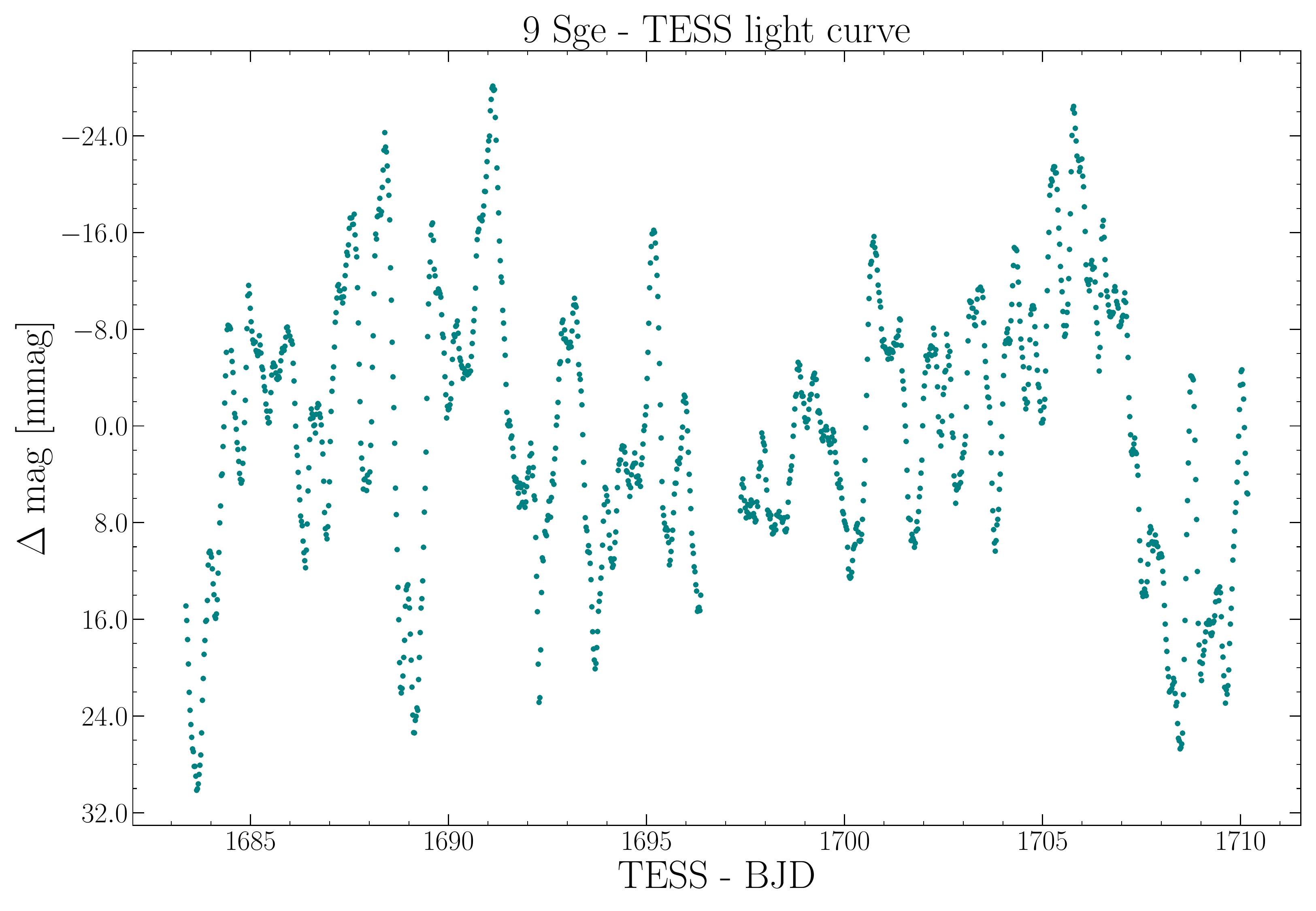}
    \caption{Normalized TESS LC of 9~Sge.}
    \label{TESS-9_Sge}
\end{figure}

Low-frequency photometric variability has been detected in O-type supergiants linked to stellar oscillations \\ \citep[cf.][]{Burssens20}. 
The TESS LC (sector 14) is plotted in Fig.~\ref{TESS-9_Sge}. 
The stochastic low-frequency variability of about $60$~mmag is easily identified. 
In this sense, the case of 9~Sge seems to be similar to HD~\num{151804} \citep[V973~Sco, O8~Iaf;][]{Ramiaetal18}, which presents a similar type of variability. 
Given the RV and photometric variability, we conclude that 9~Sge is not a spectroscopic binary system at a detection level of $\sim20$~\kms.

For a kinematic analysis, the RV of a runaway star is a key value, but in this case it is difficult to decide which value is representative for the star. 
The x-corr RV value is determined from the cross-correlation of the spectrum of the star with a {\sc FASTWIND} synthetic spectrum.
As mentioned previously, the available grid of {\sc FASTWIND} synthetic spectra only includes \ion{H}{}, \ion{He}{i}, and \ion{He}{ii} lines. 
In the case of O-type supergiants, the shape of some of those lines could be affected by the wind, and so x-corr RV measurements could be subject to a systematic shift. 
In Table~\ref{t-single}, we present the average RV measurement derived from x-corr and Gaussian fit for different lines in 9~Sge (and other single stars, as will be mentioned later).
The difference in the RV values is apparent. 
For 9~Sge in particular, we assume that the RV value corresponding to the \OIII{5592} line is likely to be more meaningful because this line is less affected by winds. This RV$=23.7~$\kms could be compatible with the runaway nature of the star, although a review of the proper motions of the star using \textit{Gaia} data is important.

\subsection{Cygnus} \label{Cyg}

\paragraph{\textbf{Cyg~X-1} = V1357~Cyg = HDE~\num[detect-all]{226868} = BD~$+$34~3815 = ALS~\num[detect-all]{10678}}  
\textbf{SB1E}

\hypertarget{X1}{} Cyg~X-1 was the first Galactic source suggested and then confirmed to host a stellar mass BH.
It is one of the brightest X-ray sources and one of the most observed objects in the sky at all the wavelengths, from radio to GeV~$\gamma$ rays. 
A spectroscopic monitoring of this famous O-supergiant carried out by \cite{WebsterMurdin72} revealed the SB1 nature of the system, with a period of $5.6$~d.
The orbital elements were revised by \cite{GiesBolton82}, finding a refined period of $P=\num{5.59974}$~d and a small eccentricity that is not statistically significant (see Table~\ref{t-orbsol-all}). 
The first orbital solution derived by using digital reticon data was produced by \cite{Ninkovetal87}, who derived essentially the same elements as \cite{GiesBolton82}. 
\cite{Brocksoppetal99a} collected 421 RVs from 14 different sources and determined a period $P=\num{5.599829}$~d, consistent with the photometric one.
Through a specific five-year spectroscopic monitoring program, \cite{giesetal03} determined new orbital elements using the RVs of the \HeI{6678} absorption line, adopting the period from \cite{Brocksoppetal99a} and assuming $e=0$. 
Following that study, \cite{gies08} refined the orbital solution with more RV data, but again adopting a fixed period.
A new dynamical model for the system was developed by \cite{Orosetal11}, who, by combining all published RVs and photometric data, found a small but significant eccentricity ($e=0.018$) after adopting the period given by \cite{Brocksoppetal99a}.

In LiLiMaRlin, we collected 31 spectrograms (between 2008 and 2019). 
Although our data set is much smaller than those of previous studies, it has the advantage of being the newest since the measurements presented by \cite{giesetal03} and \cite{gies08}, which cover the year intervals 1998-2002 and 2002-2003, respectively. 

The spectrum of Cyg~X-1 is variable during the orbital cycle and hardness state, with the variability especially noticeable in H$\alpha$ (cf. \citealt{Yanetal08,gies08}). 
This spectral variability has an impact on the line profile of different ions, and thus RVs determined could depart from the expected values for a spectroscopic orbit. 
For example, the orbital solutions calculated from our Gaussian RV measurements of \HeI{5876} and \HeII{5412} show small eccentricities ($e=0.075$, and $e=0.017$, respectively), but they are within the errors and can be considered circular.
In the case of \OIII{5592}, the orbital solution converges to a quasi-circular orbit ($e=0.008 \pm 0.007$) (these preliminary orbits are not included in Table~\ref{t-orbsol-all}, see below).

Consequently, we explored the circular orbit solution and obtained one for each line. 
The best period determined from our orbital solutions is $P = \num{5.59975} \pm \num{0.00004}$ d, in good agreement with previous values.
The semi-amplitude also has different values depending on the ion measured. 
For example, $K_1 = 83.8$~\kms is obtained for \HeII{5412}, while for \OIII{5592}, we find $K_1 = 77.5$~\kms. 
Figure~\ref{orb-fig:1} (upper right panel) shows the orbital solution for \HeII{5412}, and Table~\ref{t-orbsol-all} lists the orbital elements.
In the case of \HeI{5876}, the profile presents an incipient red wing emission, resulting in a blue shift of the barycentric velocity of the orbit to $\gamma = -14.7$ \kms, and a lower semi-amplitude, $K_1 = 74.2$~\kms. 
It is interesting to note that the higher value of $K_1 = 83.8$~\kms obtained for \HeII{5412} leads to a change in the value of the mass function, increasing it by about 35\%, from $f(m) = 0.23 \pm 0.007$~M$_\odot$ \citep{giesetal03} to $f(m) = 0.341 \pm 0.016$~M$_\odot$.

The TESS LC obtained in sector 14 shows a complex structure (Fig.~\ref{tess-fig:1} upper panels).
The amplitude of the ellipsoidal variation grows through the observing cycle, from about 45~mmag at the beginning and reaching 75~mmag at the end of the cycle.
These changes in the amplitude could be associated with stochastic variations observed during the orbital cycle.
The folded LC illustrates these variations: the primary minimum (around $\phi = 0.25$), immediately after the first quadrature, is more stable, but the secondary minimum, around $\phi=0.75$, presents noticeable changes, perhaps related to the activity of the system. 

In the future, we plan to do a more complete analysis of the Cyg~X-1 orbit incorporating new data.

\paragraph{\textbf{BD~$+$36~4063} = ALS~\num[detect-all]{11334}} 
\textbf{SB1E (El.)}

\hypertarget{bd36}{} This ON9.7~Ib active mass-transfer binary system \citep{willetal09b} belongs to the special group of ON stars \citep{2016aj....151...91w} and shows substantial variability.
\cite{willetal09b} presented the first spectroscopic orbit based on the analysis of seven intermediate resolution ($R=2400$) blue spectra obtained during six consecutive nights. 
Radial velocities were determined by using a cross-correlation method against a TLUSTY synthetic spectrum \citep{LanzHube07}, and assuming a circular orbit (Table~\ref{t-orbsol-all}).
They adopted the photometric period, $P=\num{4.8126}$~d, determined from ellipsoidal variations in a photometric time series obtained in 1999 and 2003-2007.
The star is classified as eclipsing in the Variable Star Index \citep[VSX;][]{Watson06}. 

In LiLiMaRlin, we have nine spectra obtained during 2018 and 2019, covering the complete orbital cycle.
The spectral behavior shows the back-and-forth movement of the absorption lines during the orbital cycle. 
During quadratures, emission lines are noticeable opposite to some absorptions, for example H$\alpha$, H$\beta$, \HeI{5876,} and \HeI{6678} lines.
Given the complexity of the line profiles in the spectrum, we chose to measure the line that shows the most symmetric profile using Gaussian fitting, namely, \HeII{5412}~\AA.
The new orbital solutions determined using this line can be qualified as very good. 
The period, $P = \num{4.81202}$~d, and semi-amplitude, $K_1=158.3$~\kms, are very consistent with the solution determined by \cite{willetal09b}, but an eccentricity is detected, $e=0.09$ (this preliminary orbit is not in Table~\ref{t-orbsol-all}, see below). 

In order to improve the orbital elements, especially the period, we combined our RV measurements with those determined by \cite{willetal09b}, after a velocity shift of $+11.8$~\kms, to put them in our rest frame.
Figure~\ref{orb-fig:1} (middle left panel) shows the orbital solution with the combined sets of RVs. 
The period found is very robust, considering that the time interval between observations is more than ten years. 
In addition, the eccentricity derived, $e=0.01\pm0.02$ is consistent with a circular orbit. 
Therefore, we adopt the combined solution as the final one. 
Table~\ref{t-orbsol-all} shows all orbital elements for this binary.

It is interesting to note that although the spectrum of the system shows a complex behavior along the orbital cycle, the orbital solution is very reliable despite being determined by combining different data sets with different resolving power and measured with different methodologies. 
This stability gives the possibility to unravel the invisible companion star in a dedicated spectroscopic study.

The TESS LC (Fig.~\ref{tess-fig:1} middle left panel) shows ellipsoidal variations and confirms the VSX classification.\ Therefore, its SBS changes to SB1E.

\paragraph{\textbf{HDE~\num[detect-all]{229234}} =  BD~$+$38~4069 = ALS~\num[detect-all]{11297}} 
\textbf{SB1E (El.)}

\hypertarget{hde229234}{} This O9~III system in NGC~6913 was proposed as an SB1 by \cite{Liuetal89}.
It has two orbital solutions published by \cite{Boecheetal04}, and more recently, by \cite{Mahyetal13}. 
Both solutions are compatible with a circular orbit, with a period of $3.511$~d, and a small semi-amplitude, $K_1=48$~\kms\ (Table~\ref{t-orbsol-all}).
These orbital solutions were derived from RVs determined by the average of values obtained from Gaussian fitting to \ion{He}{i} and \ion{He}{ii} absorption lines. 

In the framework of the MONOS project, we collected ten spectra strategically placed at both quadratures, in two groups separated by 1000 and 2100 days after the last observation by \cite{Mahyetal13}.
To obtain a combined orbital solution, we need to bring the published RVs to our rest frame. 
Firstly, we obtained an orbital solution by using only our RVs, determined from the line \HeI{4471}, and subtracted the appropriated $\gamma$ to each of the published data sets, namely, $14.3$~\kms for the RVs from \cite{Boecheetal04}, and $13$~\kms for the values from \cite{Mahyetal13}. 
Then, we combined our RV measurements with those obtained in previous works shifted by these appropriate amounts.
The LS periodogram of the combined data yields a period of $P = \num{3.51039}$~d, confirming previous findings.
We also confirm the parameters of the previously published orbital solutions, but an apparent scatter up to 10~\kms is seen in the RVs. 
We suggest that it can be related to stellar pulsations or wind variability.
Figure~\ref{orb-fig:1} (middle right panel) illustrates the mean orbital solution obtained by combining the RVs measured by \cite{Boecheetal04}, \cite{Mahyetal13}, and ours. 
As previously mentioned, RV variations have been observed in O-type giants and supergiants related to pulsations, an effect that can be confirmed through the inspection of the TESS photometric time series.
Figure~\ref{tess-fig:1} (lower panels) shows the TESS LC obtained in sector 14 and 15.
It shows periodic ellipsoidal variations with an amplitude of about 35~mmag and superimposed stochastic variations, which can introduce a small scatter in the RVs. Therefore, its SBS changes to SB1E.

\paragraph{\textbf{HD~\num[detect-all]{192281}} = V2011~Cyg = BD~$+$39~4082 = ALS~\num[detect-all]{10943} = SBC9 2383} 
\textbf{Single}

\hypertarget{hd192281}{} This O4.5~IV(n)(f) runaway star \citep{Maizetal18b} and fast rotator \citep[$v \sin i = 292$~\kms,][]{SimDHerr14} was identified as an SB1 system by \cite{Bara93}, who determined an orbital period of $5.48$~d, a small semi-amplitude $K_1=16.8$~\kms, and an eccentricity of $0.19$. 
This orbital solution was derived using photographic spectra obtained at a reciprocal dispersion of $44$~\AA/mm (equivalent to intermediate resolution), which raises doubts about the reliability of that solution, taking into account the broad absorption profile of the star. 
\cite{DeBeRauw04} studied the star in detail and did not find any evidence of RV variability due to binarity on timescales of several days to a year.
Moreover, they detected a periodic modulation of the emission wings of \HeII{4686} line with a period of about $1.5$~d, which is probably rotationally modulated, although non-radial pulsations could also be considered.

\begin{figure}
    \centering
    \includegraphics[width=.5\textwidth]{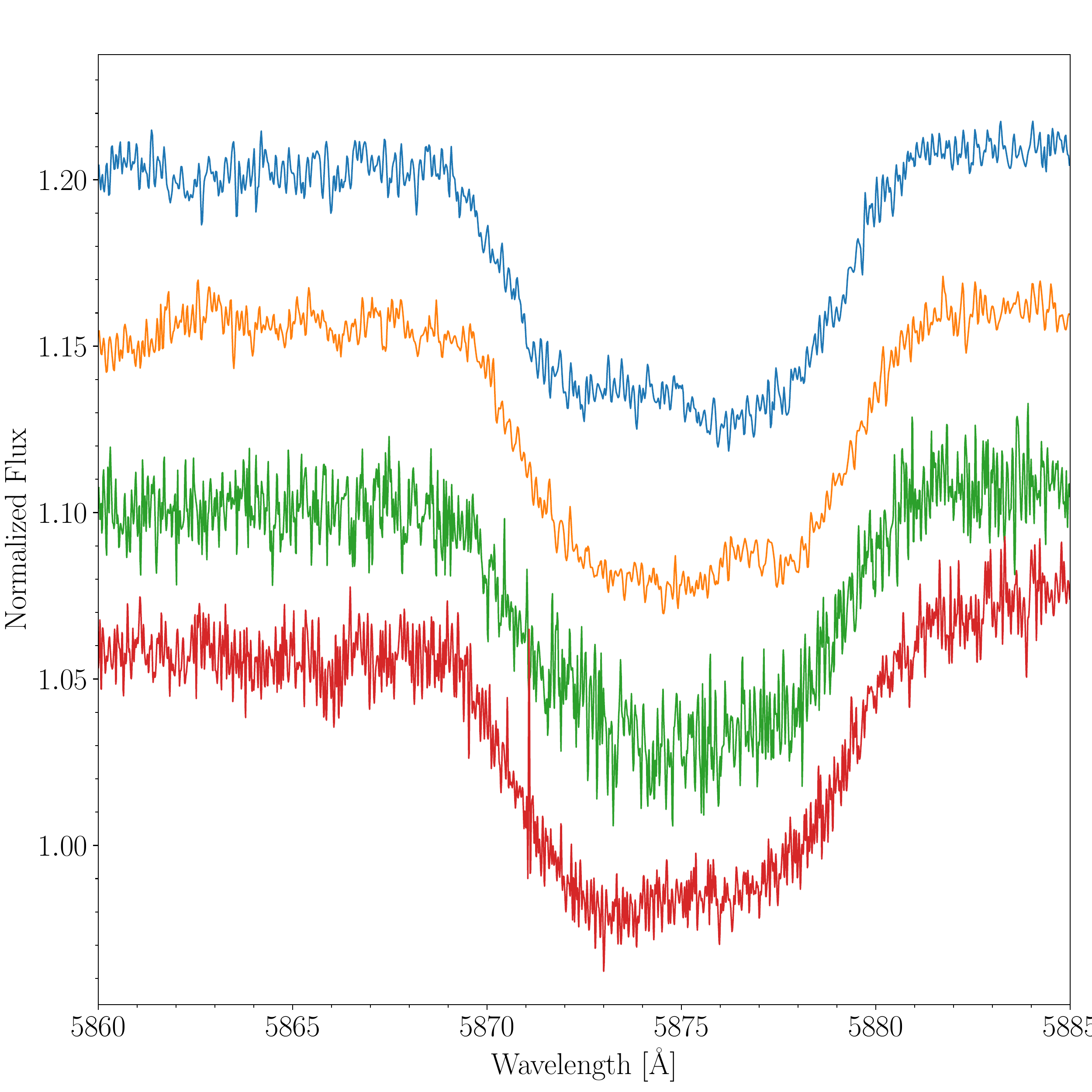}
    \includegraphics[width=.5\textwidth]{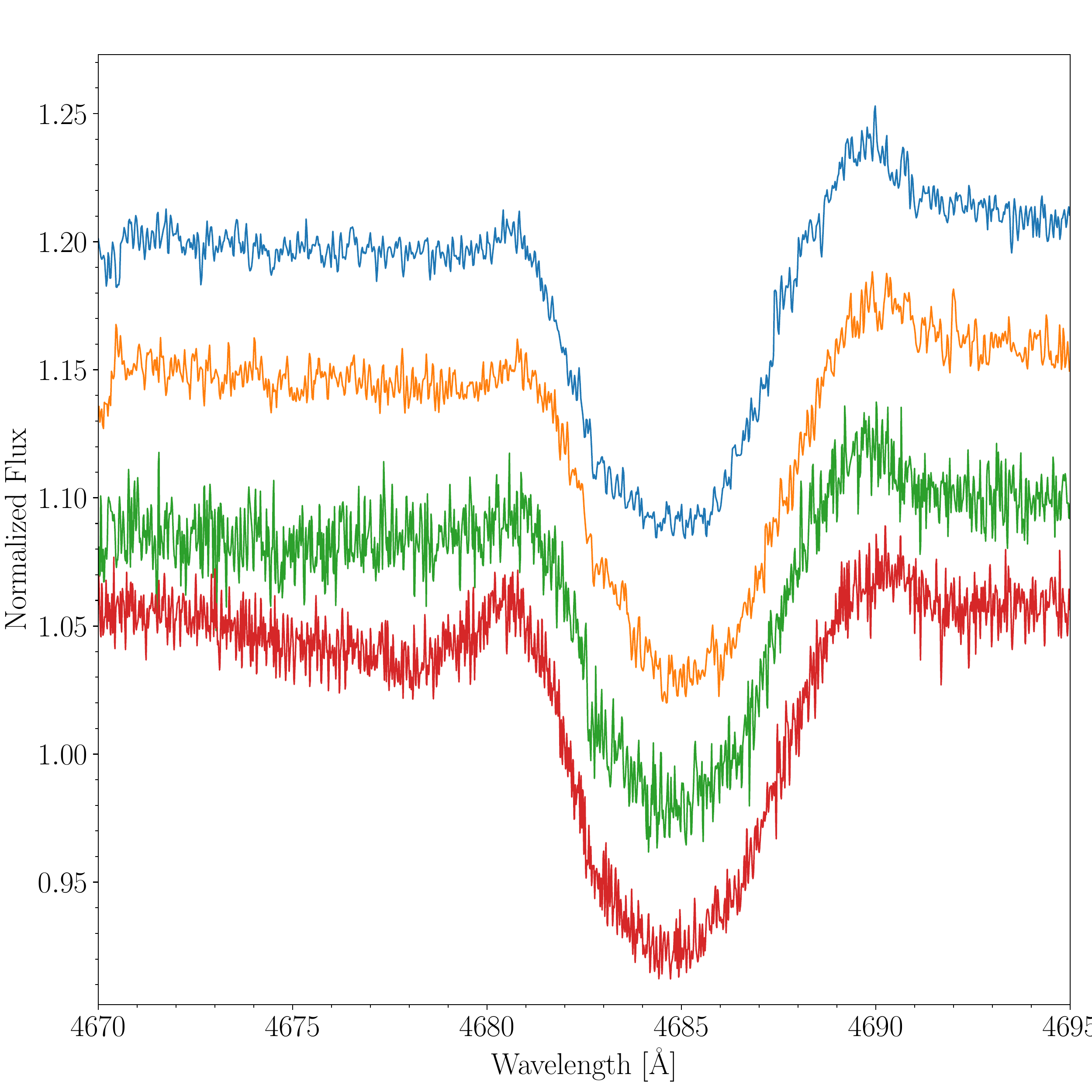}
    \caption{\HeI{5876} (upper) and \HeII{4686} (lower) absorption lines of HD~192\,281 at different epochs: 100807\_I (blue), 101024\_I  (orange), 111108\_M (green), and 150901\_P (red).}
    \label{f-HD_192_281}
\end{figure}

We analyzed 23 LiLiMaRlin spectra spanning 18 years. 
The main feature is the spectral variability, as it was highlighted by \cite{DeBeRauw04}. 
Noticeable changes are seen in the \HeII{4686} emission wings, and also in other absorption profiles (e.g., \HeI{5876}; see Fig.~\ref{f-HD_192_281}). 
Our RVs determined by using the x-corr method show a mean value of $-28.8\pm 4.9$~\kms, with an RV$_{\rm pp} \sim 16$~\kms. 
In addition, the RVs determined by Gaussian fitting of prominent absorption lines \HeII{4542}, \HeII{5412}, and \HeI{5876} show scattered values with mean values $-28.2 \pm 5.6$, $-26.0 \pm 4.2$, and $-40.6 \pm 7.9$~\kms, respectively.

\cite{Bara93} reported light variations with an amplitude of $0.04$~mag and a probable period of $9.59$~d. 
This possible periodic photometric variability is not confirmed by \textit{HIPPARCOS} data (see S. Otero's remark in the VSX database entry for V2011~Cyg).
We also checked for possible photometric variability by using the KWS \citep{Maehara14} $V-$band observations between 2011 and 2019.
These data, with a mean value of $V=7.572\pm0.032$~mag, do not show any periodicity. 
Likewise, the TESS LC obtained in sectors 14 and 15 shows stochastic light variation with a timescale of 0.9~d and an amplitude of about 25~mmag (Fig.~\ref{tess-fig:1} middle right panel).
The photometric and RV variability indicates that this object is therefore not an SB1 system, and thus we modify its status to single.

\paragraph{\textbf{HDE~\num[detect-all]{229232}~AB} = BD~$+$38~4070~AB = ALS~\num[detect-all]{11296}~AB} %
\textbf{Single}

\hypertarget{hde229232}{} This early O-type star was identified as an SB1 system by \cite{Willetal13}, who determined a preliminary period $P=6.2$~d, in a low semi-amplitude ($K_1=15.6$~\kms) circular orbit. 
\cite{Aldoretal15} identified the B companion, with a $\Delta m = 1.2$, and a minimum separation of $12.8$~mas. 
As noted in MONOS~I, the maximum separation is about $40$~mas. 

\begin{figure}
    \centering
    \includegraphics[width=.5\textwidth]{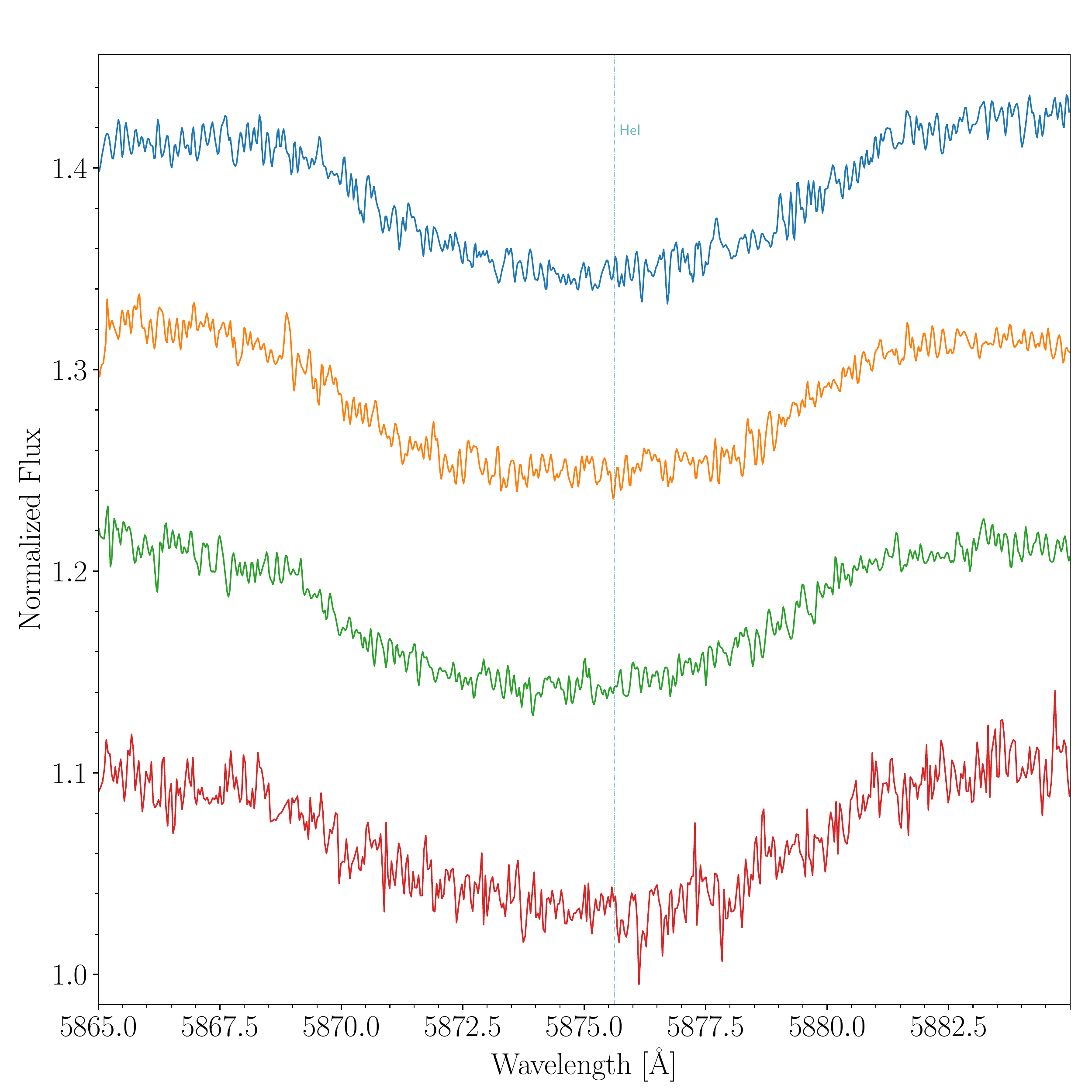}
    \caption{\HeI{5876} line of HD~\num{229232}~AB at different epochs: 131208\_C (red), 161028\_I (orange), 181125\_I (blue), and 190713\_I (green).}
    \label{fig:HDE229232}
\end{figure}

For this star, we only have 9 spectra in LiLiMaRlin. 
We measured several \ion{He}{i} and \ion{He}{ii} lines, and were unable to find any periodicity (Fig.~\ref{fig:HDE229232}). 
Using {\sc IACOB-BROAD} \citep{SimDHerr14}, we found that this star is a fast rotator with $v\sin i = 280$~\kms, very similar to the value of $273 \pm 19$~\kms found by \cite{Willetal13}. 
RVs present an RV$_{\rm pp} \sim 30$~\kms, about $11\%$ of the $v \sin i$, which together with the lack of a clear periodic signal in the LS periodogram, even combining our measurements with those of \cite{Willetal13}, leads us to think that this star is likely a pulsating variable instead of an SB1. 
Interestingly, the LS periodogram of TESS data obtained in sector 14 show a clear peak at frequency $f_1 = \num{1.65761}$~d$^{-1}$ ($P = \num{0.60328}$~d).
Figure~\ref{tess-fig:2} (upper left panel) shows the folded LC using that period. 
The rotational or ellipsoidal cycle is apparent with an amplitude of about $7$~mmag.
The analysis of x-corr RVs using such a period offers an interesting result: it is possible to get orbital solutions with a pretty similar compatible period ($P = \num{0.60513}$~d;
Fig.~\ref{orb-fig:1} lower left panel).
In that solution, eccentricity is assumed as $e=0$.
Given the large $v\sin i$, we suggest the possibility that this period could be related to a modulation by stellar rotation; the expected rotational periods are around $ P = 0.6 - 0.85$~d.
If the orbital solution is validated should implicate a very low-mass companion due to the very small mass-function, $f(m) = \num{0.0011}~\mathrm{M}_\odot$. 
A low orbital inclination is not compatible with the fast rotator nature of the O star, unless there is a strong misalignment of the stellar spin with respect to the orbit. 
Therefore, we conclude that HDE~\num{229232}~AB is not a binary, and the status changed to single.

\paragraph{\textbf{ALS~\num[detect-all]{15133}} = RLP\,1592 = [MT91]~70} 
\textbf{SB1 unc.}

\hypertarget{als15133}{} \cite{Kobuetal12} identified this star as a very long-period ($P=2259$~d) SB1 system. 
The RVs were determined from spectra obtained during a period of time spanning 4400 days at three different observatories. 
The orbital solution was obtained after the combination of RVs but applying a systematic shift of $-10$~\kms to Wisconsin Infrared Observatory data. 
Given the long period of the system and the small semi-amplitude ($K_1=9\pm1$~\kms), the proposed SB1 status should be checked. 
\cite{Aldoretal15} did not detect any close astrometric companion at a scale of a tenth of an arcsecond, although a faint companion is placed $4.4$~arcsec away (cf. MONOS I).

There is only one spectrum of this star in the LiLiMaRlin database, and thus we cannot give additional information about the quality of the orbital solution.
As a comparative example, the RV determined by Gaussian fitting to the \HeI{5876} absorption line is $-6.6\pm 0.6$~\kms, corresponding to an orbital phase $0.256$, using the ephemeris proposed by \cite{Kobuetal12}. 
This value is shifted $+6.9$~\kms with respect to the RV expected from that solution. 
Part of this shift ($+3.5$~\kms) is explained by the different choice of the rest wavelength for this line compared with those authors. 
Therefore, we leave this star in the SB1 category, but we were unable to confirm the results; thus, as we stated at the beginning of this section, the "unc." in the SBS classification.

The TESS LC (sectors 14 and 15) shows stochastic light variations with a dispersion $\sigma = 1.5$~mmag, a significant value compared with the median error of measurements, $\epsilon = 0.35$~mmag). The LS periodogram reveals a somewhat significant signal composed of two main periods, $P_1= 2.600$~d and $P_2 = 4.585$~d ($f_1 = 0.385$~d$^{-1}$ and $f_2 = 0.218$~d$^{-1}$, respectively). As proposed by \citet{Burssens20}, the main peak could be related to the rotational frequency. Thus, for the O9.5 IV component, the rotational frequency $\nu_{\rm rot} \sim 0.385$ d$^{-1}$ brings expected rotational velocities $v \sim 145-260$~\kms, adopting stellar radii from 7.4 to 13.3~$R_\odot$, for a O9.5~V or O9.5~III star \citep{Martetal05a}.

\paragraph{\textbf{Cyg~OB2-A11} = ALS~\num[detect-all]{21079} = [CPR2002]~A11 = [MT91]~267} 
\textbf{SB1}

\hypertarget{2-A11}{} \cite{Kobuetal12} determined the first spectroscopic orbit of this O7~Ib(f) member of Cyg OB2 association, with a period $P=15.511$~d, a small semi-amplitude ($K_1=24$~\kms), and a moderate eccentricity of $0.21$. 
They also suggest that the large residual in the RV-curve solution could be produced by photospheric line variations (a common feature in O-type supergiant stars; see, e.g., \citet{Simonetal21}), or due to the presence of an unresolved third body. 
We note that the ephemeris published by \cite{Kobuetal12} might not be precise because listed dates are given in fractions of $0.25$~d. 

We obtained 13 spectra over seven years, distributed along the orbit. 
The orbital solution we found using only our data is consistent with that obtained by \cite{Kobuetal12}, although the period is somewhat shorter, $P=15.446$~d and a marginally higher eccentricity was found (this preliminary orbit is not shown in Table~\ref{t-orbsol-all}, see below).

To complete the analysis, we explored a combined orbital solution, obtaining a similar period $P=15.443$~d, and congruent, albeit slightly lower, eccentricity $e = 0.136$ (Table~\ref{t-orbsol-all}).
In this solution, we used the RV determined by Gaussian fitting to the \HeI{5876} absorption line in order to compare with the results obtained by \cite{Kobuetal12} (applying a $+3.5$~\kms correction to bring them to our rest frame, Fig.~\ref{orb-fig:1} lower right panel).
Both the combined and individual orbital solutions show a scatter in the RV residuals, for example, in the combined solution the probable error is $4.2$~\kms, that is,  about 15\% of the semi-amplitude. 
This phenomenon is not unexpected in O-type supergiants affected by variable stellar wind and pulsations. 
They are manifested through the H$\alpha$ profile changes and also in the morphology of the TESS LC. 
Again, the TESS LC obtained in sectors 14 and 15 displays stochastic variability with a timescale of about 1.0 d and an amplitude of about 30 mmag (Fig.~\ref{tess-fig:2} upper right panel).

\paragraph{\textbf{Cyg~OB2-22~C} = V2185~Cyg = ALS~\num[detect-all]{15127} = [MT91]~421}%
\textbf{SB1 unc. + E}

\hypertarget{2-22C}{} \cite{Pigulskietal98} discovered that Cyg~OB2-22~C is a detached eclipsing system with a period $P=4.161$~d \citep[improved by][to $P=4.1621$]{Salaetal15}, in a circular orbit (or $e \cos \omega \sim 0$). 
The LC of this O9.5~III\,n star shows a flat minimum, indicating total or annular eclipses, diluted by the presence of an important contribution from a third light. 
Despite many studies about the massive stellar content and binaries in Cyg~OB2, this star has never been observed systematically to obtain an RV orbital solution. 
In fact, the assumption about the characteristics of the secondary as a star of spectral type B9-A0~V was inferred from a noisy and contaminated LC \citep{Pigulskietal98}.

We only have seven spectra of this object, covering only four nights (three in 2012 and one in 2019), and with a poor S/N, not enough for RV orbital analysis.
Nevertheless, we explored the behavior of several absorption lines of \ion{He}{i} and \ion{He}{ii}.
These lines are very broad ($v \sin i = 266$ \kms). 
The best stellar line in our spectra is \HeI{5876}, which shows a well-defined profile without a clear signal of spectroscopic splitting, although small changes in the core and wings are apparent.
We measured RVs using the Gaussian fitting method.
These RVs show relatively small changes of about 20 to 30~\kms, a value unexpectedly low for a close binary.
The question that arises is about the orbital phases that these RVs correspond to.
Since the ephemeris determined by \cite{Pigulskietal98} are from about 25 years ago, we need more recent ephemeris to avoid propagating errors in the determination of the orbital phases.

We undertake the task of calculating a new ephemeris by using the time of minima provided by \cite{Pigulskietal98}, and calculating new ones with the data used by \cite{Salaetal15}, and based on the TESS time series.
Although the extracted TESS data (sectors 14 and 15) includes the entire Cyg~OB2-22 cluster, the eclipses can be noticed, even if affected by the heavy dilution and stochastic variability produced by the other stars within the aperture (Fig.~\ref{tess-fig:2} Middle left panel). 
The LS periodogram shows a sharp maximum at $P=4.1621$~d, close to the value found by \cite{Pigulskietal98} and \cite{Salaetal15}.
The folded LC is also shown in Fig.~\ref{tess-fig:2} (middle right panel).

\begin{figure*}
    \includegraphics[width=.5\textwidth]{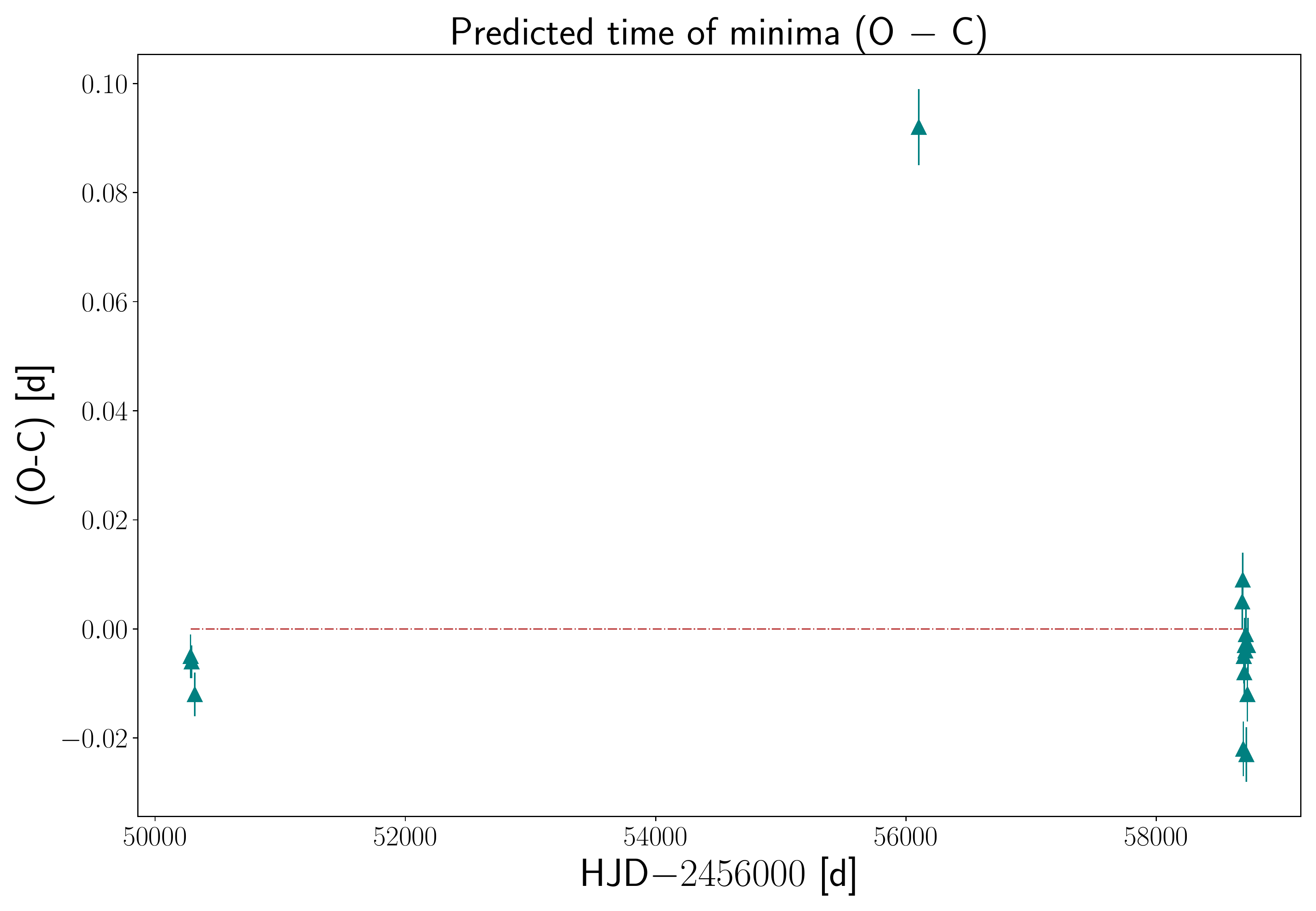}
    \includegraphics[width=.5\textwidth]{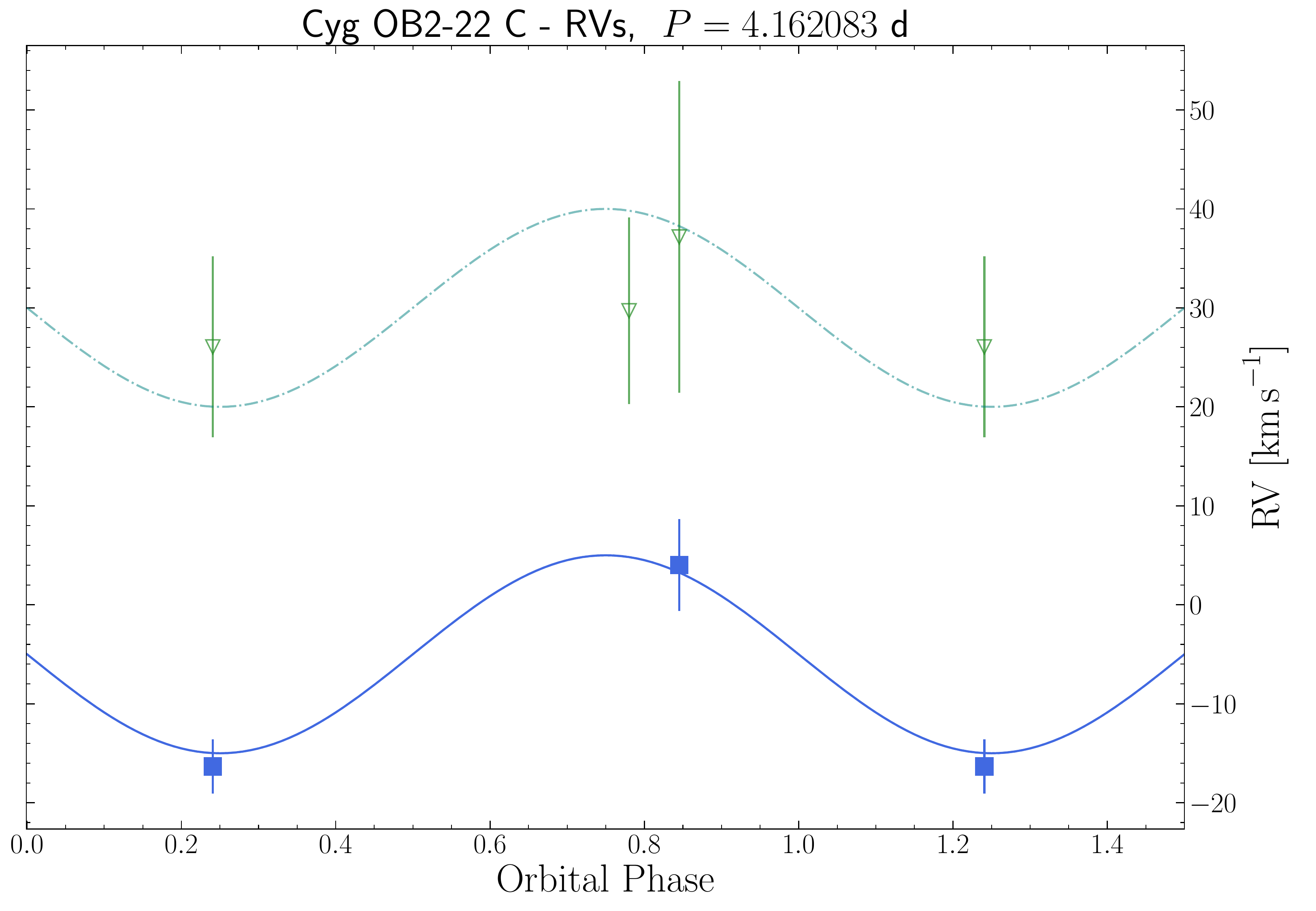}
    \caption{
    Results of the analysis for Cyg~OB2-22~C.
    Left panel: Heliocentric time of minima determined for Cyg~OB2-22~C. 
    Right panel: \HeI{5876} (blue squares) and \HeII{5412} (green triangles) RVs of Cyg~OB2-22~C folded with the adopted ephemeris. It is important to note that they do not represent the orbit of the O star.}
    \label{f-cygob2-22c}
\end{figure*}

Table~\ref{tab:minima-Cyg_OB2-22_C} lists the time of minima recompiled and calculated, while Fig.~\ref{f-cygob2-22c} (left panel) displays the observed minus predicted time of minima ($O-C$) adopting a linear ephemeris.
As it is noted in that figure, time of minima are deviated from that linear ephemeris, suggesting two alternatives: 
a) the period is variable; or 
b) there is a travel-time effect due to the presence of a third body.
Obviously, these sparse observations are not enough to solve the puzzle.
Thus, the adopted ephemeris is:

$$\rm{Min\,I} = {\rm HJD}\ \num{2456102.860}  + \num{4.162083} \times \it{E}\, ,$$

\noindent  $E$ being the orbital cycle.
Figure~\ref{f-cygob2-22c} (right panel) shows the RV values for the \HeII{5412} and \HeI{5876} lines folded with this ephemeris.
We note that RVs near the expected upper quadrature ($\phi=0.73$) do not show a special behavior compared with those near $\phi=0.21$, halfway to the lower quadrature.
This scenario suggests that the O-type star is not part of the eclipsing system.
In that case, we would probably be in the presence of a triple system, made up of a close binary (probable early or mid B-type components), and an O star in a more distant orbit.
A system with these characteristics is known: 29~CMa \citep[HD\,57\,060;][]{Sotaetal14}, composed of an O-type supergiant star that shows clear eclipses with $P = 4.39$~d, but no spectroscopic features in the O spectrum moving with that period have yet been found. Another similar (but even more complex) system is $\tau$~CMa~Aa,Ab \citep{ApellanizBarba20}.

Therefore, given the RV variations detected in the O-type component, we propose a new SBS as SB1 unc. + E.

\begin{table}
    \centering
    \caption{Heliocentric time of minima determined for Cyg~OB2-22~C.}
    \label{tab:minima-Cyg_OB2-22_C}
    \begin{tabular}{c c c l }
    \hline
    \hline
        $T_{\rm min}$     &  error & (O-C) & Source \\
        HJD$-\num{2400000}$ &  (d)   & (d)   & \\
    \hline
\num{50284.263} & 0.004 & -0.005 & P\&K98\\
\num{50292.586} & 0.003 & -0.006 & P\&K98\\
\num{50317.553} & 0.004 & -0.012 & P\&K98\\
\num{56102.952} & 0.007 &  0.092 & Sa15\\
\num{58687.519} & 0.005 &  0.005 & TESS \\
\num{58691.685} & 0.005 &  0.009 & TESS \\
\num{58695.816} & 0.005 & -0.022 & TESS \\
\num{58699.995} & 0.005 & -0.005 & TESS \\
\num{58704.154} & 0.005 & -0.008 & TESS \\
\num{58708.321} & 0.005 & -0.003 & TESS \\
\num{58712.482} & 0.005 & -0.004 & TESS \\
\num{58716.647} & 0.005 & -0.001 & TESS \\
\num{58720.787} & 0.005 & -0.023 & TESS \\
\num{58729.122} & 0.005 & -0.012 & TESS \\
\num{58733.293} & 0.005 & -0.003 & TESS \\
    \hline
    \end{tabular}
    \tablebib{
    (P\&K98) \citet{Pigulskietal98};  (Sa15) \citet{Salaetal15}.
    }
\end{table}

\paragraph{\textbf{Cyg~OB2-22~B} = ALS~\num[detect-all]{19499}~A = Schulte~22~B = [MT91]~417~B} 
\textbf{SB1 unc.}

\hypertarget{2-22B}{} This star, classified as O6~V((f)) in GOSSS I \citep[Galactic O-Star Spectroscopic Survey;][]{Maizetal11}, is the astrometric companion to the O3~If* star Cyg OB2-22 A, separated $1.5$~arcsec.
It was identified as an SB1 system by \cite{Kobuetal14}. 
They obtained a period $P=38\pm0.2$~d and an eccentricity $e=0.21$, but they warned that the very small semi-amplitude ($K_1=9.5$~\kms) RV curve is not well sampled at all orbital phases, making the derived parameters possibly uncertain.

In LiLiMaRlin, we obtained four spectra over two nights, but the spectra show evident light contamination from the earlier close companion Cyg~OB2-22~A. 
Given the difficulty to isolate the two close components A and B with \'echelle spectrographs, we will revisit this object in the future using data from long-slit observations.
Thus, we classify the status as SB1 unc.

\paragraph{\textbf{Cyg~OB2-41} = ALS~\num[detect-all]{15144} = [MT91] 378} 
\textbf{SB1}

\hypertarget{2-41}{} \cite{Kobuetal14} published the first spectroscopic solution for this O9.7~III(n) system. 
The orbit is characterized by a period $P=29.41$~d, a semi-amplitude $K_1=36.3$~\kms, and an eccentricity $e=0.23$.
 
In LiLiMaRlin, we collected two spectra separated by 93 days. 
Both data points are located near the quadrature.
Combining our RVs obtained for the \HeI{5876} absorption line, with the RVs values determined by \cite{Kobuetal14} (applied the $+3.5$~\kms correction), we redetermined the orbital solution, confirming its overall shape but finding a slightly shorter period $P=29.37$~d and a larger semi-amplitude of $K=41.8$~\kms, which makes the mass function of the system a bit larger (Fig.~\ref{orb-fig:2} upper left panel).

TESS time series obtained in sectors 14 and 16 show a pattern of irregular stochastic variations with a small amplitude of about $6$~mmag.
The periodogram reveals a main periodic signal at $3.317$~d, with a secondary peak at $4.483$~d.
Following \citet{Burssens20}, if the main periodic feature corresponds to the rotational period of the primary star, the rotational frequency $\nu_{\rm rot} \sim 0.3$~d$^{-1}$ could correspond to a rotational velocity $v \sim 200$~\kms, adopting a stellar radius of 13.3~$R_\odot$ for a O9.5~III star \citep{Martetal05a}.
Therefore, we conclude that the photometric period is not related to the orbital motion of the star (middle left panel of Fig.~\ref{tess-fig:3}).

\paragraph{\textbf{ALS~\num[detect-all]{15148}} = RLP 853 = [MT91] 448} 
\textbf{SB1E (El.)}

\hypertarget{als15148}{} \cite{Kobuetal14} discovered that this early O-type star is a short-period ($P=3.17$~d), small semi-amplitude ($K_1=27.7$~\kms), and eccentric ($e=0.1\pm0.06$) SB1 system. 
The period was also confirmed by \cite{Salaetal15}, who detected photometric ellipsoidal variations.

For this object, we have ten spectra covering five different nights over seven years.
We calculated a new orbital solution by combining our \HeI{5876} RVs with those obtained by \cite{Kobuetal14}, taking into account the $+3.5$~\kms shift due to the different rest wavelength adopted for the \HeI{5876} line, and found a very similar solution. 
It is worth noticing that the small eccentricity found by both \citet{Kobuetal14} and us ($0.1$, and $0.06$, respectively) is compatible with a circular orbit. 
The new orbital solution is listed in Table~\ref{t-orbsol-all}, and plotted in Fig.~\ref{orb-fig:2} upper right panel. 

The TESS LC shows clearly the ellipsoidal variation detected by \cite{Salaetal15}, confirming the orbital period (Fig.~\ref{tess-fig:2} lower panels). Therefore, its SBS changes to SB1E.

\paragraph{\textbf{Cyg~OB2-1} = ALS~\num[detect-all]{11401} = Schulte~1 = [MT91]~59} %
\textbf{SB1E+Ca}

\hypertarget{2-1}{} In MONOS~I, we classified this star as O8~IV(n)((f)), and identified an astrometric companion at $1\farcs2$ ($\Delta z=2.66$~mag).
\cite{CabNetal14} detected that companion using FGS/HST\footnote{Fine Guidance Sensor (FGS) on board the Hubble Space Telescope (HST)} ($\Delta m_{\rm F583}=2.58$~mag), although they only provided a lower limit separation (751.3~mas). 
The system is cataloged as WDS 20312+4132 in the Washington Double Star Catalog.

The system was revealed as an SB1 by \citet{Kimietal08}, who presented a preliminary orbit, which was improved by \citet{Kobuetal14}. 
\citet{Lauretal15} discovered that Cyg OB2-1 is also an eclipsing system. 
Those authors confirm a significant eccentricity ($e=0.14$), for the given short period determined, $P=\num{4.8523}$~d. 
Thus, the SBS status of the system changes from SB1+Ca to SB1E+Ca.

Since we have only two LiLiMaRlin spectra (obtained in 2018 and 2019), we combined our RV measures with those determined by \citet{Kobuetal14} to obtain a refined solution, in a similar procedure as performed for other systems in the Cyg~OB2 association.
In this case, we combined RVs determined for the absorption line \HeI{4471}, and the derived orbital solution is compatible with that calculated by \citet{Kobuetal14} (Fig.~\ref{orb-fig:2} middle left panel).

The TESS LC obtained in the observation of sectors 14 and 15 clearly displays two narrow eclipses of different depths (Fig.~\ref{tess-fig:3} upper panels). 
The LC also shows superimposed stochastic variations of about 4~mmag.
The orbital eccentricity is confirmed based on the phase separation of both eclipses.
The depth of the secondary eclipse suggests an important contribution of the secondary component to the total light of the system, but the visual inspection of our LiLiMaRlin spectra located near the upper quadrature does not show any evidence of spectral lines belonging to that component.
The pulsational variability superimposed on the LC could be explained in terms of the heartbeat variables \citep{Thompsonetal12}, meaning short-period highly eccentric binary systems that have dynamical tidal distortions and tidally induced pulsations.

\paragraph{\textbf{ALS~\num[detect-all]{15131}} = RLP 666 = [MT91] 390} 
\textbf{SB1 unc.}

\hypertarget{als15131}{} This SB1 system discovered by \cite{Kobuetal14} has a very small semi-amplitude ($K_1=5.4$~\kms), with a period $P=4.63$~d. 
It is worth noticing that this semi-amplitude is similar to the expected RV$_{\rm pp}$ due to pulsations for an O-type dwarf star such as this O7.5~V((f)).

The orbital solution is not well constrained, and the system must be monitored to improve it, but the faintness of the system imposes a challenge. 
Considering that we only have one spectrum, we cannot analyze the validity of the published orbit, and more data are therefore needed.

TESS time series obtained in sectors 14 and 15 shows stochastic variations with a maximum amplitude of about 10 mmag. 
The LS periodogram does not show prominent frequencies, the two dominants being \num{0.35973}~d$^{-1}$ and \num{0.87193}~d$^{-1}$ (\num{2.7799}~d and \num{1.14688}~d, respectively), which are not related to the proposed orbital period.

\paragraph{\textbf{Cyg~OB2-20} = ALS~\num[detect-all]{11404} = [MT91]~145} 
\textbf{SB1}

\hypertarget{2-20}{} \cite{Kimietal09} calculated the first orbital solution for this SB1 system, later revised by \cite{Kobuetal14}. 
The system was reclassified as O9.7~IV in MONOS~I.

For this system, we collected three IACOB spectra, unintentionally near the same orbital phase, but with the advantage of increasing the time line by more than 4000 days since the  \cite{Kobuetal14} observations.
Again, combining our RVs determined for the \HeI{4471} line, with those by \cite{Kobuetal14} (shifted by $+4.7$~\kms to bring their RVs to our rest frame), we slightly refined the period to $P=25.125$~d, certifying its validity (Fig.~\ref{orb-fig:2} middle right panel).
The TESS LC in sectors 14 and 15 shows a noise level of $1$~mmag without any clear periodicity.

\paragraph{\textbf{Cyg~OB2-70} = ALS~\num[detect-all]{15119} = RLP 706 = [MT91] 588}   
\textbf{SB1 unc.}

\hypertarget{2-70}{} The orbital solution for this O9.5~IV(n) system determined by \cite{Kobuetal14} is roughly constrained. 
Although data were collected over 14 years, the small semi-amplitude ($K_1=14.5$~\kms), large eccentricity ($e=0.51$), and long period ($P=245.1$~d) are factors that complicate the task of refining the orbit. 
It should be mentioned that those authors highlighted that there are other possible shorter periods ranging from $34.9$~d to $174$~d. 
Although no data are yet available in our database for this object, we plan to collect them in a future observational campaign.

The TESS time series obtained in sectors 14 and 15 shows stochastic irregular variations with an amplitude of about $5$~mmag. The periodogram displays a weak periodic signal around $P=2.718$~d.

\paragraph{\textbf{Cyg~OB2-15} = ALS~\num[detect-all]{15102} = Schulte~15 = [MT91]~258} 
\textbf{SB1}

\hypertarget{2-15}{} This eccentric SB1 system, classified as O8~III in MONOS I, has a period $P = 14.66$~d \citep{Kimietal08, Kobuetal14}.

Based on four LiLiMaRlin spectra obtained near the quadratures, we determined a new orbital solution by combining our \HeI{5876} RV measurements with those by \cite{Kobuetal14} (Fig.~\ref{orb-fig:2} lower left panel) shifted $+3.5$ \kms, as we did for other Cyg~OB2 systems.
We confirm the previous finding, with a period $P=\num{14.6575}$~d and an eccentricity of $e=0.138$.
The TESS data obtained in sectors 14 and 15 shows stochastic irregular variations with a $\sigma = 0.9$~mmag.

\paragraph{\textbf{ALS~\num[detect-all]{15115}} = RLP 680 = [MT91] 485}  
\textbf{SB1 unc.}

\hypertarget{als15115}{} This very long-period ($4\,066$~d~$= 11.1$ a) O8~V SB1 system discovered by \cite{Kobuetal14} has a not so well constrained orbital solution due to the large eccentricity $e = 0.75$ and very small semi-amplitude, $K_1 = 15.0$~\kms. 
At the moment, only one orbital cycle has been covered. 
We do not have any spectra in the LiLiMaRlin database, and so we cannot evaluate the validity of the orbit. 
We will visit this object in a future observational campaign.

Again, the TESS time series obtained in sectors 14 and 15 shows stochastic variations with a $\sigma = 1.3$~mmag. The LS periodogram shows a weak signal at $P=3.306$~d ($f=0.302$~d$^{-1}$), which could represent a rotational velocity $v_{\rm rot} \sim 130$~\kms, for a $R=8.5$\,R$_\odot$ corresponding to an O8 V star \citep[cf.][]{Martetal05a}.

\paragraph{\textbf{Cyg~OB2-29} = Schulte 29 = ALS~\num[detect-all]{15110} = [MT91] 745}   
\textbf{SB1 unc.}

\hypertarget{2-29}{} \cite{Kobuetal14} derived an orbital solution for this fast rotator O7.5~V(n)((f))z SB1 system.
As the orbital solution shows a high eccentricity ($e=0.6$) and a long period ($P=151.2$~d), but a small semi-amplitude ($K_1=17.4$~\kms), it will be essential to observe the system during the periastron passage in order to confirm the solution. Although no data are yet available within the LiLiMaRlin database, the system is integrated into our future observing campaigns.
The TESS LC extracted from sectors 14 and 15 shows stochastic variability with $\sigma = 1.2$~mmag. The periodogram displays a barely detected peak at $P=1.518$~d ($f=0.659$~d$^{-1}$), which could represent a rotational velocity $v_{\rm rot} \sim 300$~\kms, for a $R=8.9$\,R$_\odot$ corresponding to an O7.5~V star \citep[cf.][]{Martetal05a}. This probable high rotation is congruent with the "(n)" qualifier in the spectral classification.

\paragraph{\textbf{Cyg~OB2-11} = BD~$+$41~3807 = ALS~\num[detect-all]{11438} = [MT91]~734 = Schulte~11} 
\textbf{SB1}

\hypertarget{2-11}{} This rare O5.5~Ifc star, GOSSS spectroscopic classification standard \citep{Maizetal16}, was identified as an SB1 by \cite{Kobuetal12}. 
These authors presented an orbital solution with a period $P=72.4$~d, large eccentricity ($e=0.5$), and small semi-amplitude ($K_1=26$~\kms). 

We obtained six spectra for this system in three different epochs: four of them in 2011, one in 2016 and the last one in 2019. 
Unfortunately, the three data sets are about the same orbital phase ($\sim 0.65-0.75$). 
Following the same procedure as in previous analyses, to obtain a new orbital solution, we combine our RVs determined from the \HeI{5876} line with the values provided by \cite{Kobuetal12}.
Their RVs were shifted by $+3.5$~\kms to bring them to our wavelength rest frame.
Thus, the orbit is confirmed, with a similar period ($P=72.488\pm0.05$~d) but slightly different eccentricity ($e=0.37\pm0.06$) and semi-amplitude ($K_1=23.5\pm1.1$~\kms; Fig.~\ref{orb-fig:2} lower right panel).
The RV residuals derived from the orbital solution are relatively large given the sharp profile of \HeI{5876} line.
It is worth noticing that this line shows an asymmetric profile, which may be indicating that it is affected by stellar winds. 
Our spectra show H$\alpha$ as a P~Cyg emission profile, with small variations between 2011 and 2019.
Thus, RV changes induced by profile distortions in lines affected by winds are not discarded. 
We determined the RVs for \OIII{5592} line, which presents a narrow symmetric profile, with the RV$_{\rm pp} \simeq 3$~\kms in our three epochs. 
This RV variation is also consistent for other \ion{He}{i} and \ion{He}{ii} absorption lines. 

We inspected the TESS LC obtained in sectors 14 and 15, detecting a stochastic variability of about 35~mmag and a typical timescale for variations of about $0.5-1.1$~d (Fig.~\ref{tess-fig:3} middle right panel). 
The periodogram analysis using the LS method gives five main frequencies/periods, which are listed in Table~\ref{tab:period-Cyg_OB2-11}.
As we mentioned before, this multifrequency variability has been detected in O-type supergiants, and it can be linked to gravity (g) modes (cf. \citealt{Burssens20}), which are associated with the slowly pulsating OB-type stars.

\begin{table}
    \centering
    \caption{Significant frequencies in the TESS photometric time series of Cyg OB2-11.}
    \label{tab:period-Cyg_OB2-11}
    \begin{tabular}{c | c | c }
    \hline
    \hline
        Frequency  &  Period & Power \\
          d$^{-1}$ & d & \\
    \hline
        0.9341 & 1.0705 & 47.7 \\
        1.5911 & 0.6285 & 37.0 \\
        1.6825 & 0.5944 & 31.9 \\
        1.3895 & 0.7197 & 31.6 \\
        1.8765 & 0.5329 & 30.4 \\
    \hline
    \end{tabular}
\end{table}

\paragraph{\textbf{68~Cyg} = HD~\num[detect-all]{203064} = V1809~Cyg = BD~$+$43~3877 = ALS~\num[detect-all]{11807} = SBC9~1295} 
\textbf{Single}

\hypertarget{68cyg}{} Classified as a runaway \citep{Maizetal18b, 1974rmxaa...1..211c}, this  O7.5 IIIn((f)) star is a famous fast rotator ($v \sin i=299$~\kms, \citealt{SimDHerr14}), which has been previously often studied.
In MONOS~I, a binary status SB1? was set due to the quality of the orbital solutions determined by \citet[see also \citealt{1984Ap+SS.102...97C}]{Aldusevaetal82}, and by \cite{Giesetal86}.
The former authors proposed a poorly constrained circular orbit with $P=5.1$~d and a small semi-amplitude of $K_1\sim30$~\kms from the RVs measurements of H$\delta$ using micro-photometer in plates with a reciprocal dispersion of $44$~\AA/mm. 
The latter authors also determined RVs from scanned plate spectrograms at more significant reciprocal dispersion ($12$~\AA/mm), deriving a different period for the RVs variations ($P=3.1781$~d) and semi-amplitude ($K_1=6.9$~\kms) without any evidence about the periodicity found by \cite{Aldusevaetal82}. 
Noteworthy, both studies warned that RV variations in this fast rotator could be related to inhomogeneities in the stellar wind, which induces profile variations, or even pulsations.
Additionally, different studies have also found random velocity variability within a range of about $25$~\kms \, \citep[e.g.,][]{Contietal77, Bohannanetal78, Garmanyetal80, Stoneetal82}.
The star is known to show substantial variability in the discrete absorption components (DACs) present in resonance P Cygni profiles in the FUV \citep{Kaperetal96}. 
The timescale for the most persistent DACs variations is about $1.3$~d. 
\cite{Lefevreetal09} classified the star as a long-term "unsolved" variable using Hipparcos photometry, with an amplitude of about 0.03 mag.

\begin{figure}
    \centering
    \includegraphics[width=7.5cm]{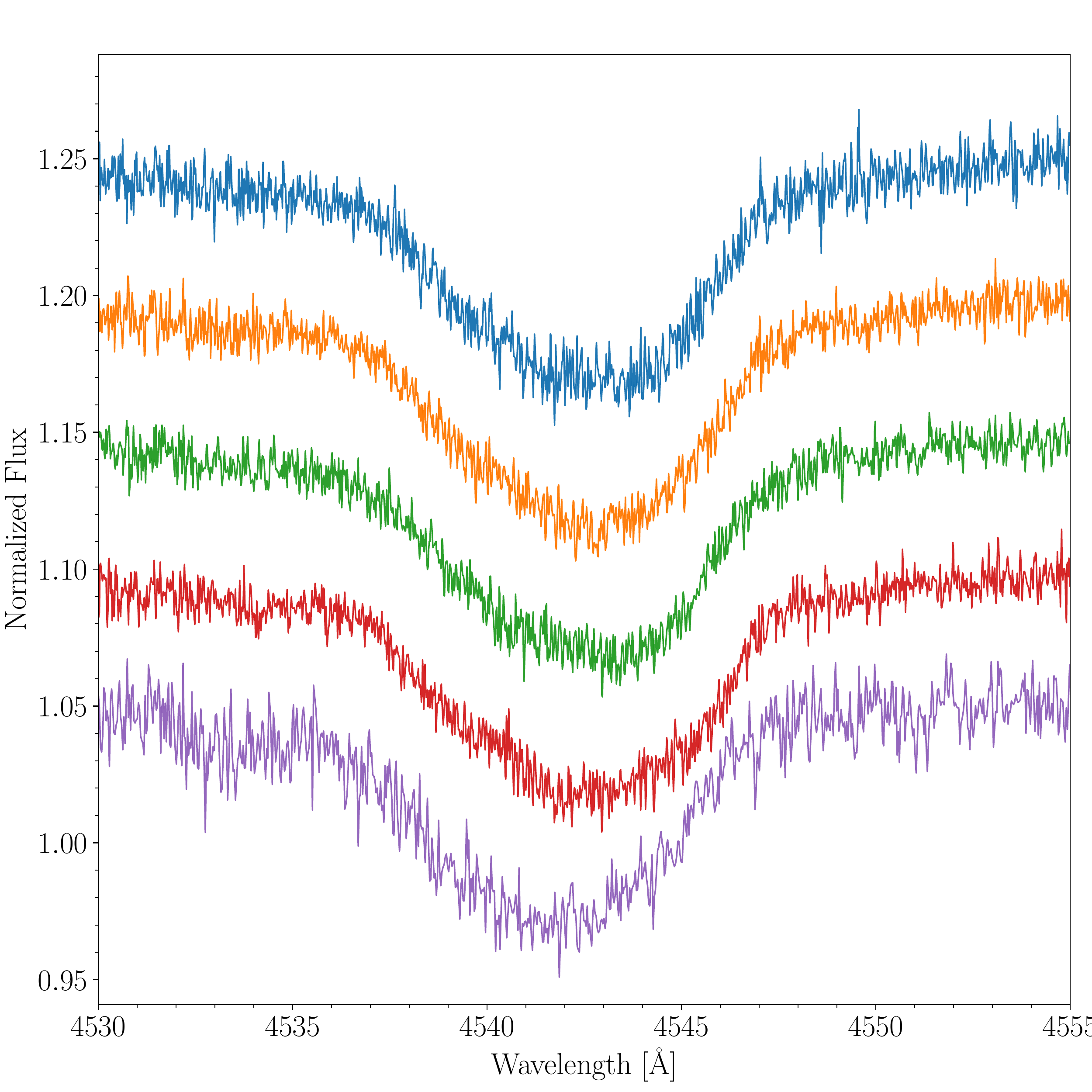}
    \includegraphics[width=7.5cm]{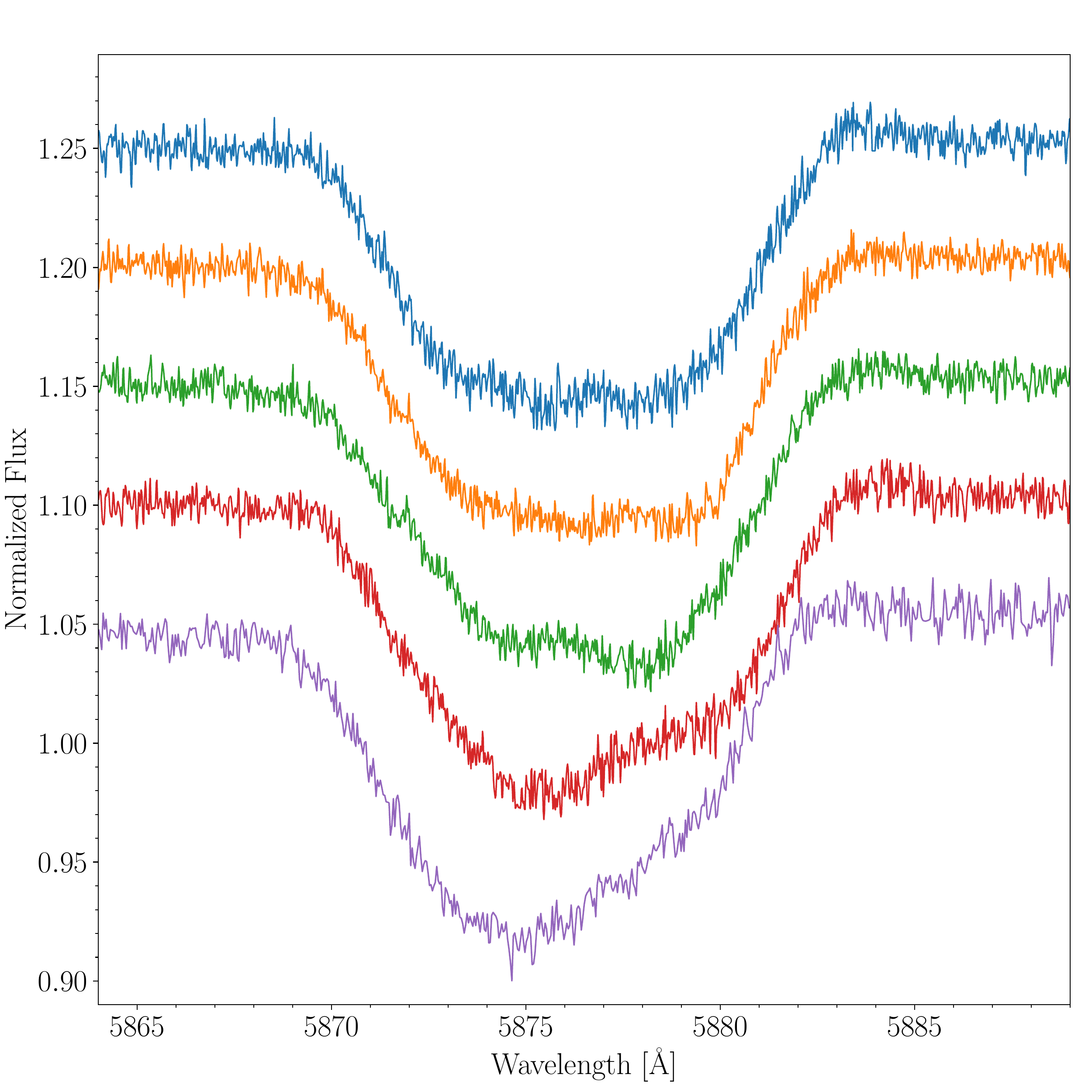}
    \includegraphics[width=7.5cm]{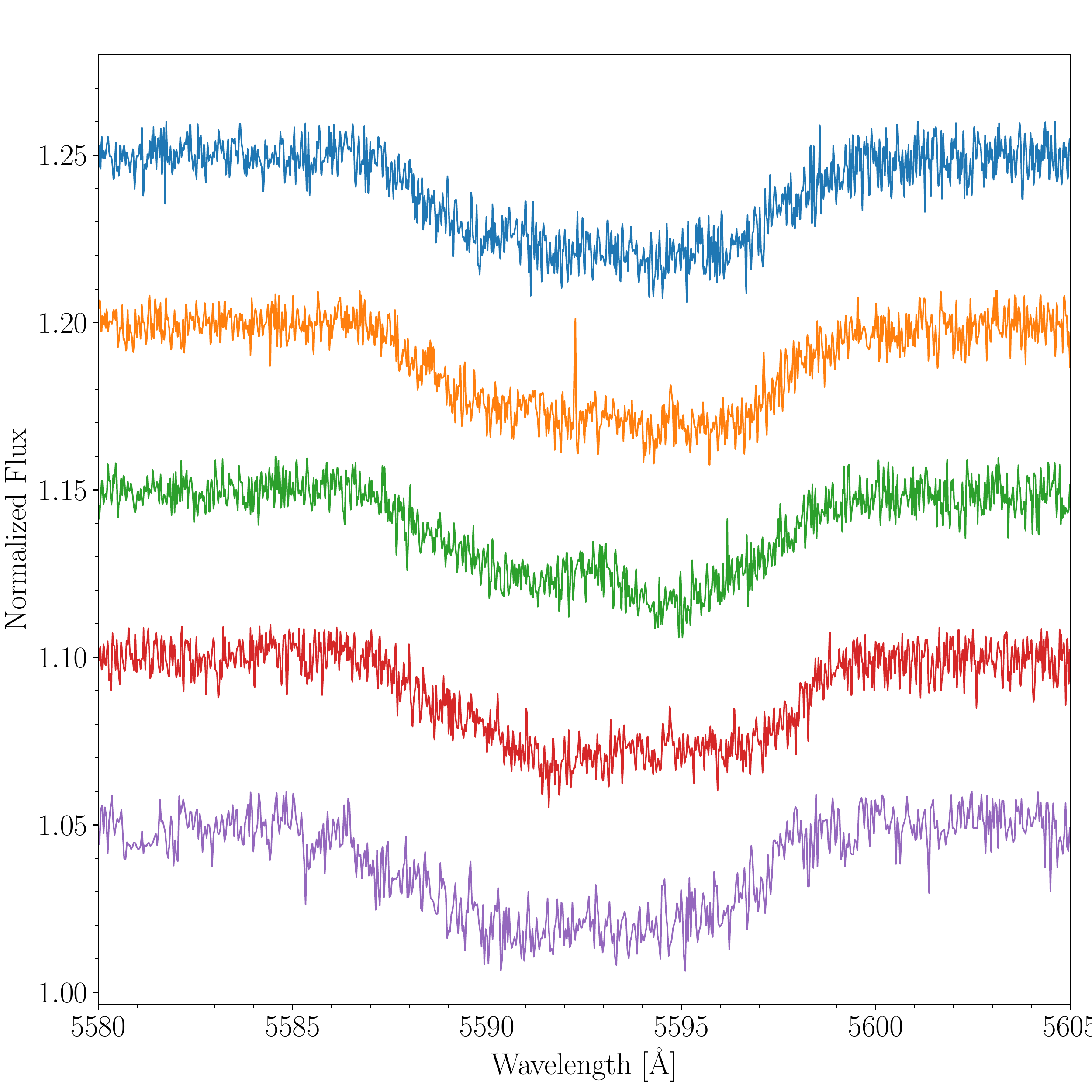}
    \caption{
    \HeII{4542} (upper), \HeI{5876} (mid), and \OIII{5592} (lower) lines of 68 Cyg for several epochs to show its variability. 
    The spectra shown are: 110613\_M (blue), 110617\_M\_4 (red), 110619\_M (green), 110619\_M\_2 (orange), and 140607\_C (purple).}
    \label{f-68_Cyg-spectra}
\end{figure}

The LiLiMaRlin sample consists of 33 spectra collected during eleven years (2008-2019), plus one additional spectrum obtained in 1996.
Figure~\ref{f-68_Cyg-spectra} shows representative profiles of the \HeI{5876}, \HeII{4542}, and \OIII{5592} lines obtained in 2011 and 2014 in order to illustrate the complexity of the line variations.
Profiles show the typical broad shape for a fast rotator, with structured "features" moving in the core of the line.
Although it seems that there is a correlation of these features between the different ions, they also have different strengths.
Another interesting behavior is the fast change in RV on a timescale of few hours.
RV measurements determined in 15 spectra obtained during three consecutive days show variations up to $12$~\kms\ in six hours. 
Considering all of our RV measurements, the range for variations is about RV$_{\rm pp} \sim 50$~\kms, that is, 14\% of the rotational velocity, without any clear periodic signal, which places the system on the edge of what is expected to find in a Giant O-type pulsating star (see \citet{Simonetal21}, and Britavskiy et al., in prep).

TESS data obtained in sectors 15 and 16 show a clear pattern of irregular stochastic variations with an amplitude of about 10~mmag (Fig.~\ref{tess-fig:3} lower panels).
Although the periodogram obtained through the LS method reveals a clear main periodic signal, Table~\ref{tab:period-68_Cyg} shows the five most important frequencies/periods.
It is a hard task to disentangle the contribution of low-frequency pulsational modes and rotational modulation. 
\cite{Burssens20} proposed a procedure to classify potential rotational variables based on the contribution of a single frequency and its harmonics and subharmonics.
Therefore, given the significant contribution of the first frequency detected, $f_1 = 0.711634$ d$^{-1}$ ($P=1.405216$ d), can be understood as rotational modulation. 
As mentioned, 68 Cyg is a fast rotator, and thus, following \cite{Burssens20}, the expected rotation frequency will be $\nu_{\rm rot} \sim 0.42 - 0.75$ d$^{-1}$ (adopting a radius $R_\odot ~ 8 - 14 R_\odot$, corresponding to an O8 V-III star from \citealt{Martetal05a}), which is in the range of the observed frequency.

Therefore, based on the lack of a coherent RV signal and the variability pattern, we suggest that the RV orbital solutions found for this system are spurious and that RV variation is probably associated with rotational activity (spots, stellar winds or both). 

\begin{table}
    \centering
    \caption{Relevant frequencies in the TESS photometric time series of 68 Cyg.}
    \label{tab:period-68_Cyg}
    \begin{tabular}{c | c | c }
    \hline
    \hline
        Frequency & Period & Power \\
         d$^{-1}$ & d & \\
    \hline
        0.7116  & 1.4053 & 84.0 \\
        0.2545  & 3.9293 & 72.5 \\
        0.1791  & 5.5835 & 45.7 \\
        0.5594  & 1.7876 & 32.8 \\
        3.2010  & 0.3124 & 29.2 \\
    \hline
    \end{tabular}
    
\end{table}

\subsection{Cepheus-Camelopardalis} \label{Cep-Cam}

\paragraph{\textbf{HD~108} = BD~$+$62~2363 = ALS~6036 = SBC9~2} %
\textbf{Single}

\hypertarget{hd108}{} This magnetic O6.5-8.5~f?p var star shows complex spectral variations \citep{Nazeetal04b,Nazeetal08d}. 
HD~108 has been proposed as a runaway by some authors, \citep{BekenBower74, Underhill94b, Nazeetal08d, Tetzetal11} but different peculiar velocities have been given, so this feature is still a matter of debate.

It has been extensively monitored during extended observing campaigns, which demonstrate that the notable line-profile variations have a timescale of decades \citep{Nazeetal01b}. 
\cite{Hutchings75} proposed that HD~108 is an SB1 system in a circular orbit of period $P=4.61$~d, and very low semi-amplitude ($K_1 < 12$ \kms, depending on the spectral line chosen). 
Following this study, \cite{Vreuxetal79}, based on high-resolution spectra obtained during two consecutive weeks, could not detect such aperiodicity and proposed a shorter one, $P=1.02$~d. 
Subsequently, \cite{Aslanovetal89} combined their observations (obtained in 1982-3) with those RVs determined by \cite{Hutchings75} and \cite{Vreuxetal79} and derived a new short period, $P=5.79$~d. 
\cite{Underhilletal94} presented a different scenario to explain the spectral variations. 
She proposed that spectral variations are produced by a polar jet moving almost perpendicular to the line of sight, and that the star could be related to the luminous blue variables.
Then, the binary saga continued with a new orbital solution found by \cite{Barannikov99}, but a different orbital configuration.
Based on the analysis of fifteen-year-long spectroscopic and photometric monitoring, he presented a very long-period ($P=1\,628$~d) and highly eccentric ($e=0.43$) orbital solution.
Since the binary scenario was not convincing, \cite{Nazeetal01b} analyzed observations collected at the Observatoire de Haute-Provence (France) over 15 years, concluding that this binary scenario with the aforementioned periods is not supported, and the star shares several characteristics of Oe- or Be-type stars.
Later, \cite{Nazeetal08d} proposed that the spectroscopic variations were caused by the magnetic effects on the star, which were modeled by \cite{ShultzWade17} as an oblique rotator in an extremely slowly rotating star, confirming the rotational period of about $55$~a.

\begin{figure}
    \centering
    \includegraphics[width=.45\textwidth]{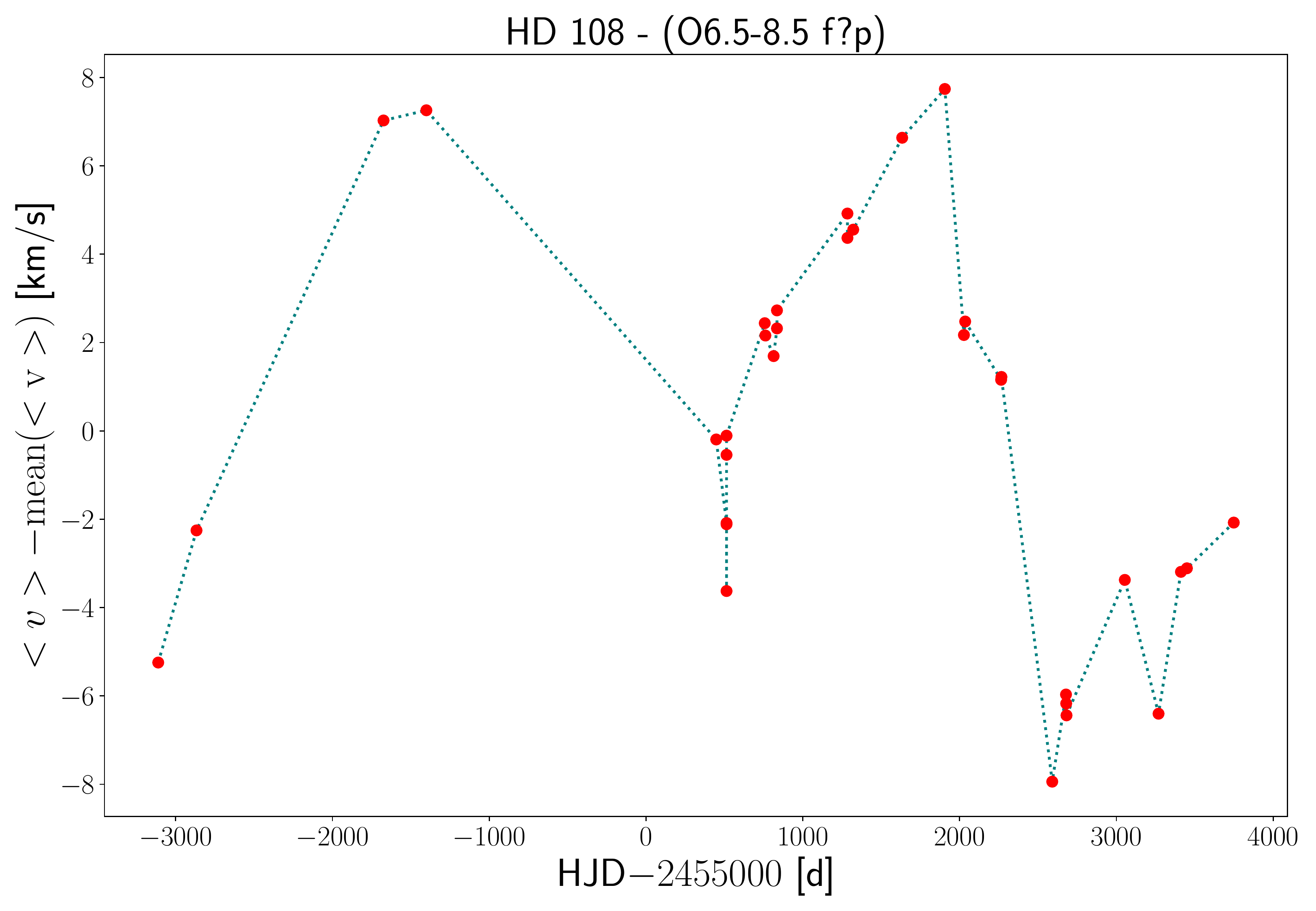}
    \includegraphics[width=.45\textwidth]{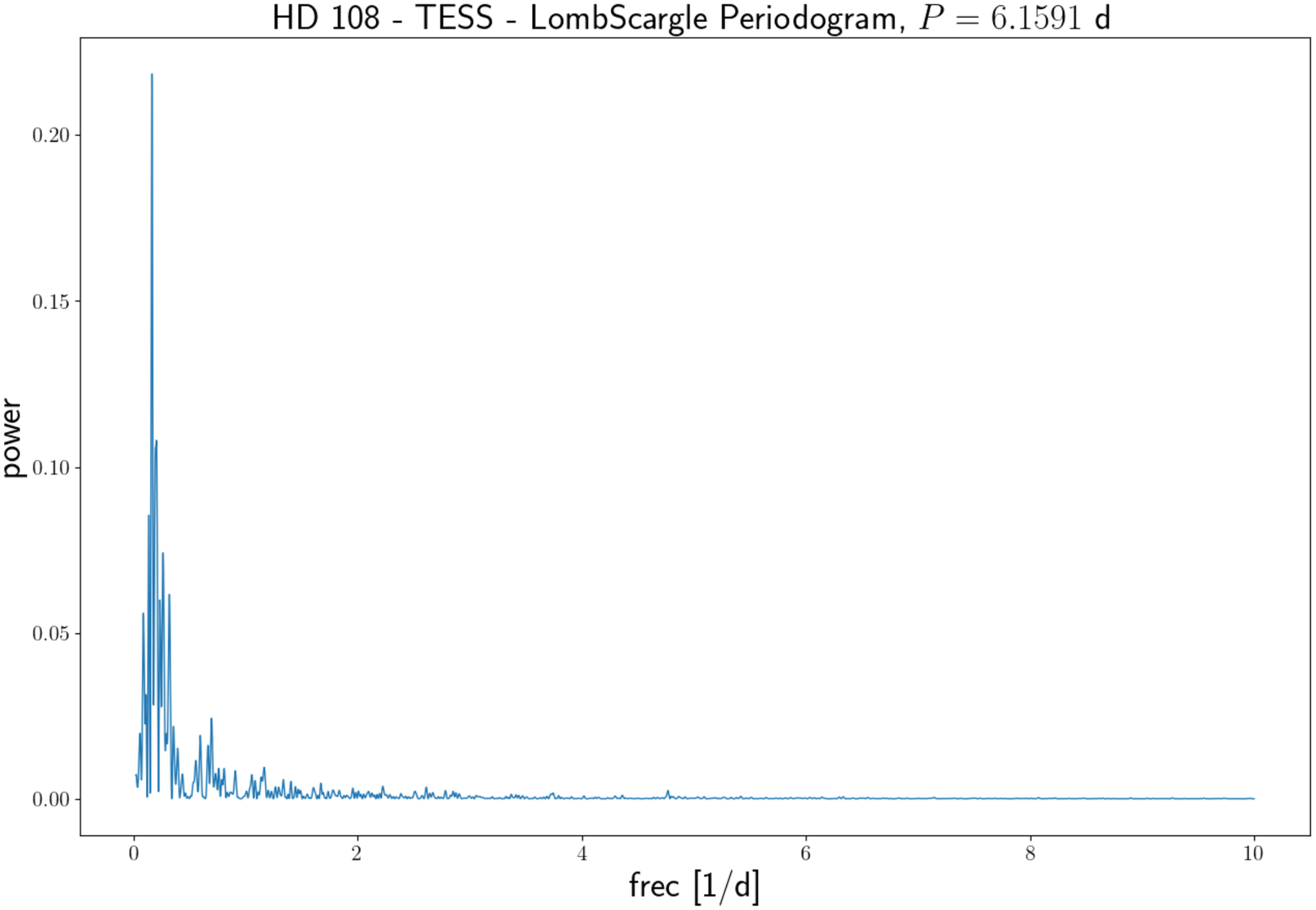}
    \caption{Variability of HD~108.
    Upper panel: RV variations in the line \OIII{5592} for HD~108.
    Lower panel: LS periodogram of TESS data in sectors 10 and 11 for HD~108.}
    \label{hd108_RVpp}
\end{figure}

The analysis of our sample of 41 epochs spaced over 18 years ultimately discards both orbital solutions and confirms the more recent scenario proposed by \cite{Nazeetal08d}. 
We measured the two \HeII{4542} and \HeII{5412} lines and three metallic ones, namely \SiIII{4553}, \OIII{5592} and \CIV{5812} employing the Gaussian method. 
We searched for periodicity using the LS method with each line but no clear peaks were found. 
We also tried to obtain an orbital solution adopting both historical solutions as initial parameters but, again, we were unable to find a convergent solution. 
Then, we measured the rotational velocity obtaining a $v \sin i = 46.2 \pm 2.2$~\kms, in line with the results of \citet{Nazeetal01b,Nazeetal08d,Martetal10a}. 
The RV$_{\rm pp}$ of the \OIII{5592} line is $\sim 15$~\kms (Fig.~\ref{hd108_RVpp} upper panel); accordingly to  \citet{Simonetal21} a star-like HD~108 could have RV$_{\rm pp}$ variations up to $\sim 25$~\kms and not being associated with binarity.
Consequently, the accumulated evidence points to HD~108 being a single star.

Interestingly, TESS data obtained in sectors 10 and 11 show a clear periodic signal with $P=6.1591$\,d (amplitude 5~mmag) with stochastic variability superimposed (amplitude up to 10~mmag). 
Figure~\ref{hd108_RVpp} (lower panel) displays the LS periodogram obtained with the time series spanning 51 days. 
A noticeable maximum is apparent at the mentioned period.
Figure~\ref{tess-fig:4} (upper panels) in the appendix shows the time series and folded LC.
This new period is not related to the proposed previous periods.
The folded LC resembles those observed in ellipsoidal or rotational variables. 
The former possibility should be discussed. 
\cite{Martetal10a} proposed that the rotational profile of the star is dominated by atmospheric macroturbulence.
Subsequently, \cite{ShultzWade17} interpreted the spectral variability with their oblique rotator model, reinforcing \cite{Nazeetal10a} proposal that the star is an extremely slow rotator.
The question then arises whether the coherent modulation shown by TESS data is representative of a permanent periodic phenomenon or is it circumstantial to the observing time window.
It should be noted that \cite{AaertsRogers15} and \cite{Grassitellietal15} point out that internal gravitational waves (responsible for stochastic photometric variability) are connected to the macroturbulence, and therefore they can be important \citep[evidence supported spectroscopically][]{SimDetal17, Bowmanetal20}. 
This opens the possibility of observational exploration to determine if the low amplitude photometric modulation is produced by a certain coherence in the pulsations in this extremely slow rotating star, mimicking rotational modulation.
We consider that in order to solve this question, simultaneous high-resolution spectroscopic and photometric observations with large signal-to-noise must be carried out.

\paragraph{\textbf{V747~Cep}, BD~$+$66~1673, ALS~\num[detect-all]{13375}} 
\textbf{SB1E}

\hypertarget{V747}{} \cite{Majaetal08} identified V747~Cep as a short-period ($P=\num{5.33146}$~d) Algol-type eclipsing binary in an eccentric orbit ($e \simeq 0.3$). 
These authors found that the combined spectrum is O5~Vn((f)), making it the earliest star in the open cluster Berkeley 59.
This spectral classification was slightly modified to O5.5~V(n)((f)) by \citet{Maizetal16}. 
No spectroscopic orbit has been published yet, but the target was included in MONOS~I due to the photometric (eclipsing) period derived by \citet{Majaetal08}.

Using the 20 spectra collected in LiLiMaRlin, we can confirm the binary nature of V747~Cep. 
The orbital period derived from RVs determined for all the strongest lines is consistent with the photometric one, $P=\num{5.3324}$~d, with a large eccentricity ($e=0.37$; Fig.~\ref{orb-fig:3} upper left panel), although different lines bring slightly different semi-amplitudes.

Figure~\ref{tess-fig:4} (middle left panel) displays the TESS LC of V747~Cep obtained in sectors 18 and 24, folded with the spectroscopic period and setting the primary eclipse as the origin of the orbital phase, $T_0 = \num{2458810.33}$~HJD.
The LC of V747 Cep resembles that of Cyg~OB2-1. 
It shows two sharp eclipses, with a noticeable increase in flux after the primary eclipse, when the secondary star is receding. 
The second part of the LC displays a slow decline with stochastic oscillations, which are apparent after the secondary eclipse egress. 
Thus, the LC can be described as a very eccentric system with sharp eclipses, reflection effect noticeable during the periastron passage, and mixed with pulsations.
As it was mentioned in the case of Cyg~OB2-1, the pulsational variability could be explained in terms of the heartbeat phenomenon \citep{Thompsonetal12}: short-period, highly eccentric binary systems that have dynamical tidal distortions and tidally induced pulsations.

\begin{figure}
    \centering
    \includegraphics[width=.5\textwidth]{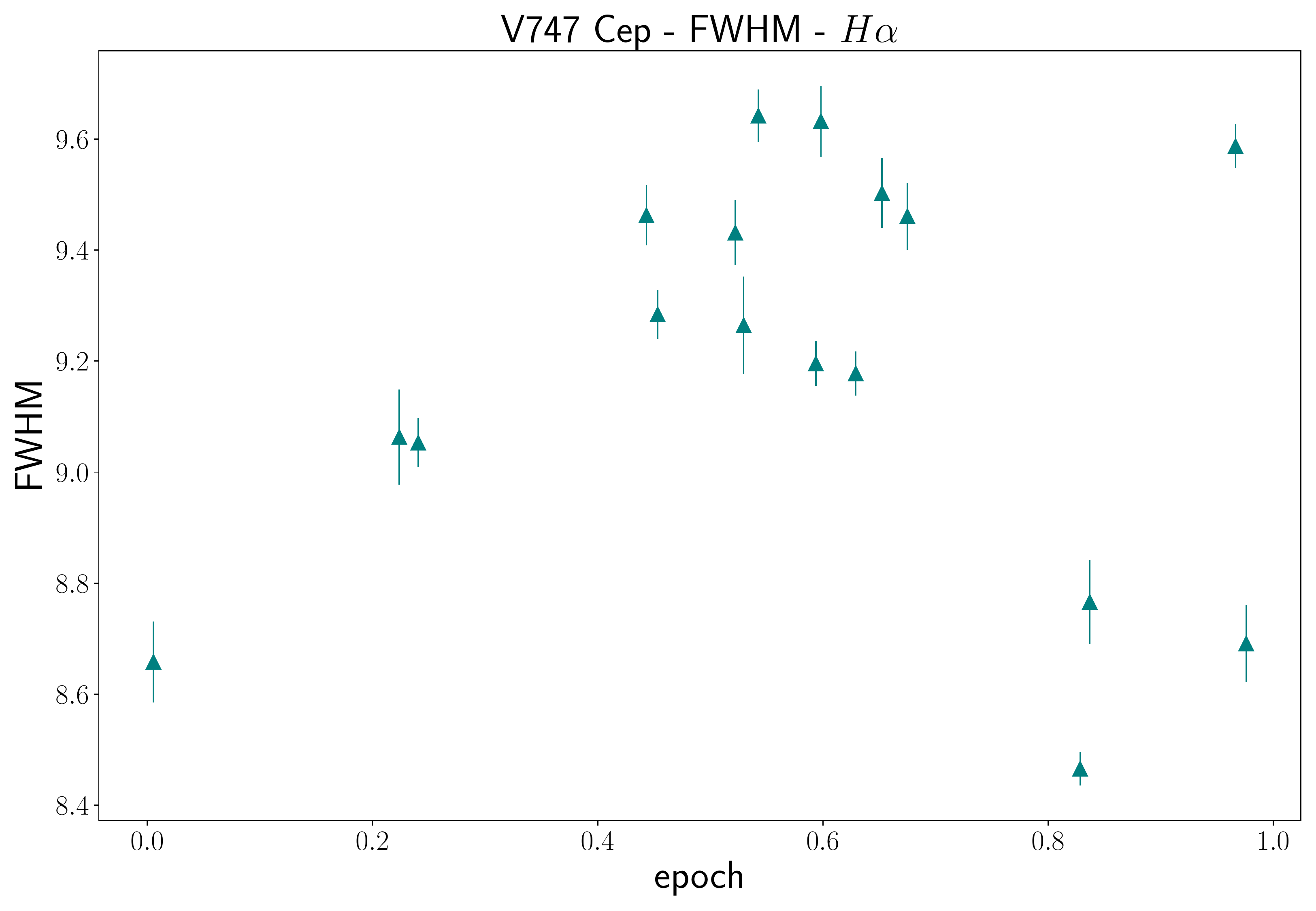}
    \includegraphics[width=.5\textwidth]{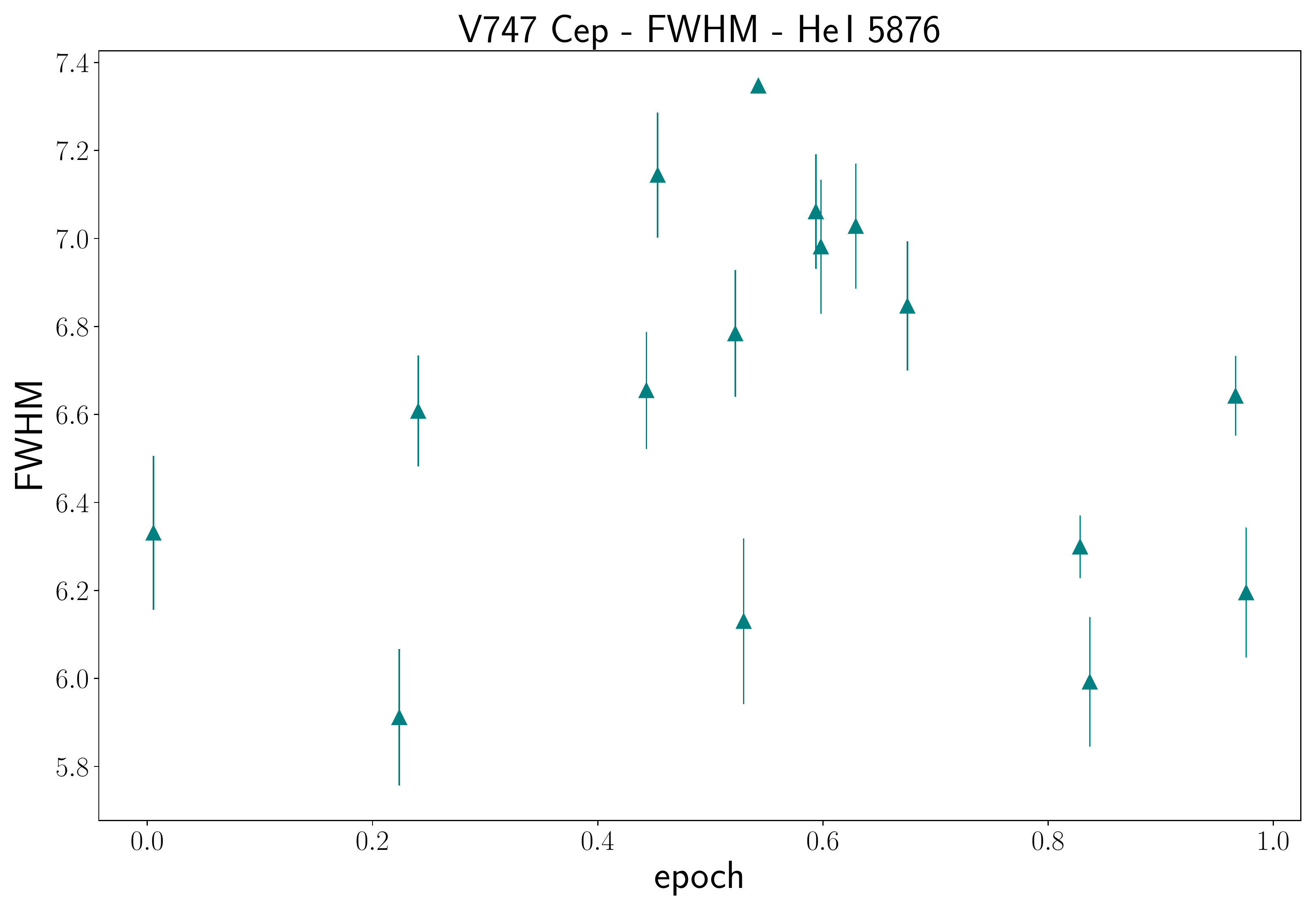}
    \caption{\HeI{5876} (upper) and $H\alpha$ (lower) FWHM for V747~Cep to show its variability.}
    \label{FWHM-V747_Cep}
\end{figure}

Given the presence of two minima in the LC, we explore the SB2 status for the system.
A variation in full width at half maximum (FWHM) with the orbital cycle is noticeable in several absorption lines. 
In eccentric SB2 systems, the spectroscopic lines of the two components are expected to present the largest separation in RV during quadrature, especially close to the periastron passage. 
Figure~\ref{FWHM-V747_Cep} shows the FWHM for $H\alpha$ and \HeI{5876} during the orbital cycle. 
The FWHM of those lines show larger values in the quadrature corresponding to the apastron, but strangely, during the periastron FWHM changes are less noticeable. 
The behavior in orbital phase of other lines in the blue portion of the spectrum is limited by the low S/N because the star experiences a heavy reddening.

For our orbital solution, we use the RVs determined from the \HeII{5412} line. 
The period converges to $P=\num{5.33237}$~d, with a moderate semi-amplitude of $K_1=99.6$~\kms, an eccentricity $e=0.37$, and a longitude of periastron $\omega = 180.2 ^\circ$. 
It is important to note that our $\omega$ is consistent with the value proposed by \cite{Majaetal08}, about $180^\circ$. 
The orbital configuration puts the periastron passage just before the primary eclipse, this scenario being consistent with the proposed heartbeat variability, as \cite{Thompsonetal12}
demonstrated for 17 systems using Kepler data.

V747~Cep has become an extremely interesting system that deserves special analysis and, therefore, will be the subject of a forthcoming paper.

\paragraph{\textbf{HD~\num[detect-all]{12323}} = BD~$+$54~441 = ALS~6886 = LS~I~+55 22 = SBC9~106}
\textbf{SB1E (El.)}

\hypertarget{hd12323}{} This is an interesting late ON star discovered by \citet[see also \citealt{Walb76}]{ContAlsc71}, now classified as ON9.2~V \citep{Sotaetal14}.
The star was recognized as an SB1 system by \cite{BoltonRogers78}, who found a preliminary period $P=3.07$~d and an eccentricity $e=0.21$, quite high for such a short period. 
Their orbital solution was obtained from RV measurements of 20 absorption lines of \ion{H}{i}, \ion{He}{i}, \ion{He}{ii}, and \ion{Si}{iv} in high-resolution spectrograms secured on plates. 
Relying on that orbital solution, \cite{SticLloy01} added two new RVs obtained by cross-correlation of {\em IUE} spectra to find a new orbital solution with an even shorter period ($P=2.07$~d). 
Interestingly, these authors discard four RVs from the literature, arguing that they were more negative than expected for the ephemeris calculated.
Those RVs correspond to a mean value of  $-91.2\pm7.1$~\kms published by \cite{Contietal77} and three more published by \cite{Musaevetal89}.
Furthermore, the system has been recognized as a blue straggler of the Per~OB1\footnote{Although its membership is not certain, with Cas~OB8 the other possible origin. A preliminary review of the {\it Gaia} data seems to support Per~OB1 as the parent association.} association by \cite{Kendetal95} and classified as a runaway by \cite{Maizetal18b}. 

For this SB1 system, we collected 20 spectra spanning nine years. 
Spectra along this series show the same morphology and apparent back and forward changes in RVs, which resemble the movement of a primary O-type star, without evidence of any spectral feature associated with a companion.
To perform the comparison with previous studies, we chose at first instance RVs determined from the x-corr method and then individual lines, because earlier RVs determinations were derived from the average values from different absorption lines.
Our RVs determined by x-corr show a variability range of about $67$~\kms, in concordance with the amplitude of the orbital solution calculated by \cite{SticLloy01}.
Short timescale variations in RVs are apparent. 
For example, three spectra obtained during $3.6$~h in 2004 show an RV variation of about $12$~\kms. 
This sharp change in RV in a few hours suggests that the star could indeed have a short orbital period of about $1-2$~days, or otherwise, observations were obtained during the periastron passage of an eccentric orbit with a somewhat longer period.
We searched for periods between 0.2 and 40 days by using the LS method on our RVs determined through x-corr and Gaussian fitting for several lines, resulting in three significant periods: $1.9251$~d, $9.5550$~d, and $1.7516$~d, in order of power from the strongest to weakest.
The next step is to combine our RVs with those obtained in previous studies. 
Thus, we included all data in the literature plus our RV determinations using x-corr for a total of 30 RVs spanning more than 18,000 days. 
The periodogram obtained by the LS method shows that the highest peak corresponds to a $P = \num{1.92536}$~d. 
Using {\sc SBOP}, we explored the orbital solution and the solution converges to $P=\num{1.925140}$~d while fixing $e=0$ (Fig.~\ref{orb-fig:3} upper right panel). 
This is a very noticeable difference in the value of eccentricity compared with the previous orbital solutions.

Given the excellent quality of our spectra and RVs measured, it is interesting to explore photometric data.
The TESS time series observed in sector 18 shows a noticeable coherent signal with $P = \num{0.96265}$~d, which resembles an ellipsoidal modulation with an amplitude of about 10~mmag.
If we double the photometric period, we obtain a value $P = \num{1.9253}$~d that is the same as the spectroscopic period found by using all RVs available (Fig.~\ref{tess-fig:4} lower panels). 
Therefore, the spectroscopic and photometric data point to the presence of ellipsoidal variations in an SB1 system, and thus, the SBS classification changes to SB1E.

Furthermore, this system is remarkably highlighted as a runaway \citep{Maizetal18b}. Given the small mass function $f(m)=\num{0.0053}$~M$_\odot$, the companion may be a compact object. 
A specific study will follow in a separate paper.

\paragraph{\textbf{HD~\num[detect-all]{16429}~AaAb} = BD~$+$60~541~AaAb = ALS~7374~AB = SBC9~2474 = V482~Cas~AB} 
\textbf{SB1E+Cas}

\hypertarget{hd16429}{} \cite{McSw03} suggested that this member of the Cas~OB6 association is a triple system composed of two components separated by 0.28\arcsec. 
The primary, labeled as Aa, would be an O9.5~II star, while the secondary, labeled as Ab, would be an O8~III/V + B0~V? system. 
However, in a more recent analysis, \cite{ApellanizBarba20} used STIS/HST\footnote{Space Telescope Imaging Spectrograph on board the HST} to separate both components and found the system to be composed of an O9.2~III star and an O9.5~IV star for components Aa and Ab, respectively.

\cite{McSw03} assumed that the Aa component was single and only computed the RV curve for the O8~III/V component in the spectroscopic binary, which yielded a period $P=3.0544$~d and a moderate eccentricity, $e=0.167$. 
The orbital semi-amplitude $K_{\rm Ab1}=135$~\kms is relatively high. 

The system was classified as a variable star of $\beta$~Cep type by \cite{Hilletal67}, with a period of \num{0.37822}~d. 
The analysis of the collected photometric data from Hipparcos, KWS, and TESS does not show any clear evidence of a $P=0.378$~d signal. Folding the TESS (sector 18) data with the spectroscopic period, we see shallow eclipses with pulsational variations superimposed, as for the previous cases of V747~Cep and Cyg~OB2-1. 
(Fig.~\ref{tess-fig:5} upper panels).

Due to the small spatial separation between the Aa and Ab components, our 29 spectra collected in LiLiMaRlin are composite due to the contribution of both stars. 
Therefore, all measured lines present some amount of blending between the spectroscopic binary and the stationary component.
From the RVs of \ion{He}{ii} and \ion{C}{iv}, we were able to recover the period $P=3.054$~d, while the periodogram for the \ion{He}{i} lines gives a shorter period of around $P=2.5$~d. 
This discrepancy could be due to the blended nature of the lines.
The presence of the stationary stellar component affects each ion differently. 
Also, we note that given that the O9.2 III star has the strongest \ion{He}{ii} lines, it indicates that the binary is the Aa component instead of Ab.

Because of the complexity of the system, we will analyze it in detail in a future paper of this series, which will include further spatially resolved STIS/HST observations. 
Moreover, the SBS status of HD~16\,429~AaAb changes from SB1+C to SB1E+Cas. 
It is worth noticing that for this object, the spatial separation was made by using STIS/HST instead of GOSSS data (see MONOS I). Nevertheless, we kept the \textbf{s} in the SBS classification in order to maintain a consistent nomenclature.

\paragraph{\textbf{HD~\num[detect-all]{15137}} = BD~$+$51~579 = ALS~7218 = SBC9~2473} %
\textbf{SB1}

\hypertarget{hd15137}{} This O9.5~II-IIIn runaway star \citep{Mdzi04} was identified as an SB1 by \citet{Boyajianetal05}, who suggested a possible supernova ejection from NGC~654 and that is probably hosting a compact companion. 
These authors determined the first preliminary orbital solution, proposing a period $P=28.61$~d and large eccentricity $e=0.52$.
\citet{McSwetal07} improved the orbital solution by combining new spectroscopic observations and published data, with $P=30.35$~d, similar $e=0.48$, and small semi-amplitude $K_1=13$~\kms. 
\cite{DeBeetal08} detected significant variability in the line profiles of \ion{He}{i} and \ion{H}{$\beta$}, on a timescale of a few hours, with a frequency of about 2~d$^{-1}$. 
They suggested that these spectral variations could be related to non-radial pulsations or some circumstellar rotating structures.

\cite{McSwetal10} reanalyzed the orbital solution, using 91 new spectra, determining a completely new one, with an even longer period, $P=\num{55.3957}$~d and an eccentricity $e=0.62$ (although an orbital period of about 65 days is also considered). 
They also detected low amplitude profile variations of about $10$~\kms (roughly $60\%$ of the proposed orbital semi-amplitude) with a timescale of a few hours. 
Given the noticeable profile variations reported by \citet{DeBeetal08} and \citet{McSwetal10} plus the large scatter in the orbital solution, pulsations are likely to be significant. 

In our spectra, we found that peak-to-peak variations for \OIII{5592} of RV$_{\rm pp} = 44 \pm 6$~\kms, about 16\% of its rotational velocity ($v\sin i = 270 \pm 14$~\kms). 
This is slightly larger than the expected pulsational variability in an O-type giant (\citet{Simonetal21}). 
We combined the RVs determined for \HeI{5876} and \HeII{5412} by using Gaussian fitting in 18 LiliMaRlin spectra (spanning ten years) with additional data collected from the literature. 
RV measurements determined by \cite{McSwetal10} were shifted appropriately in order to bring them to our wavelength rest frame.
The first interesting result is that the three of our RVs with lower values are placed on dates expected for the periastron passage, according to the ephemerides published by \cite{McSwetal10}, seemingly confirming a probable orbital period of $55.4$~d.

To verify the periodicity, we search for periods using the LS method, the most significant signal is at $55.389$~d, and the second power peak is about $65.27$~d.
Using the first period as input, we recalculated a new orbital solution that converges to very similar values to that in \cite{McSwetal10}, $P=55.399$~d, $K_1=13.9$~\kms, and $e=0.59$ (Fig.~\ref{orb-fig:3} middle left panel). 
The second period is most likely an alias due to the observation window, since the difference in the peak frequencies corresponds to a difference of a year \footnote{It is worth noticing that one can also find an orbital solution with this period, but the scatter is significantly larger; thus, we consider the shorter period to be the correct one.}.

\begin{figure}
    \centering
    \includegraphics[width=.5\textwidth]{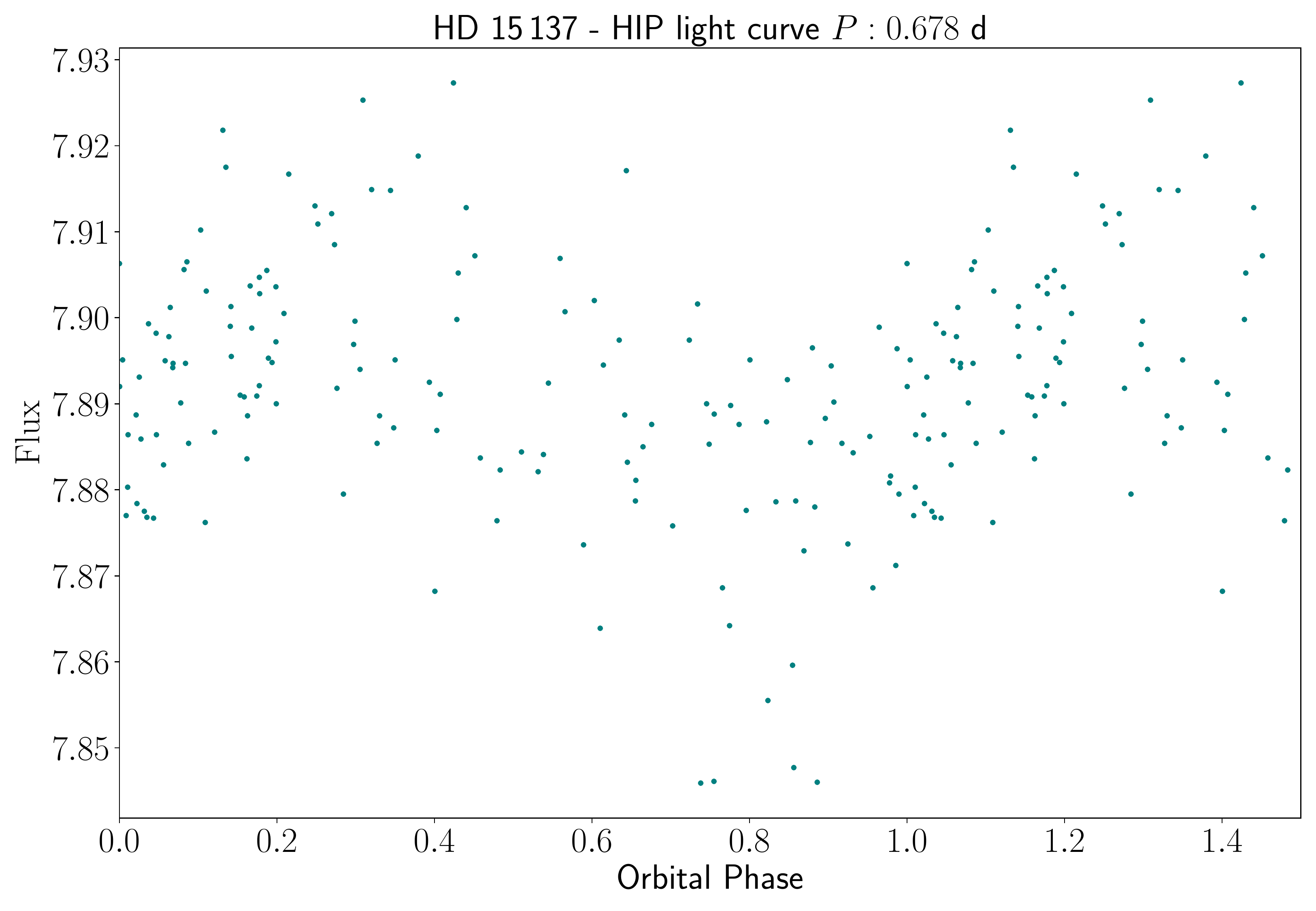}
    \caption{HIPPARCOS LC for HD~\num{15137} folded with period $0.678$~d.}
    \label{fig:hd15137_hip_fold}
\end{figure}

\begin{figure}
    \centering
    \includegraphics[width=.5\textwidth]{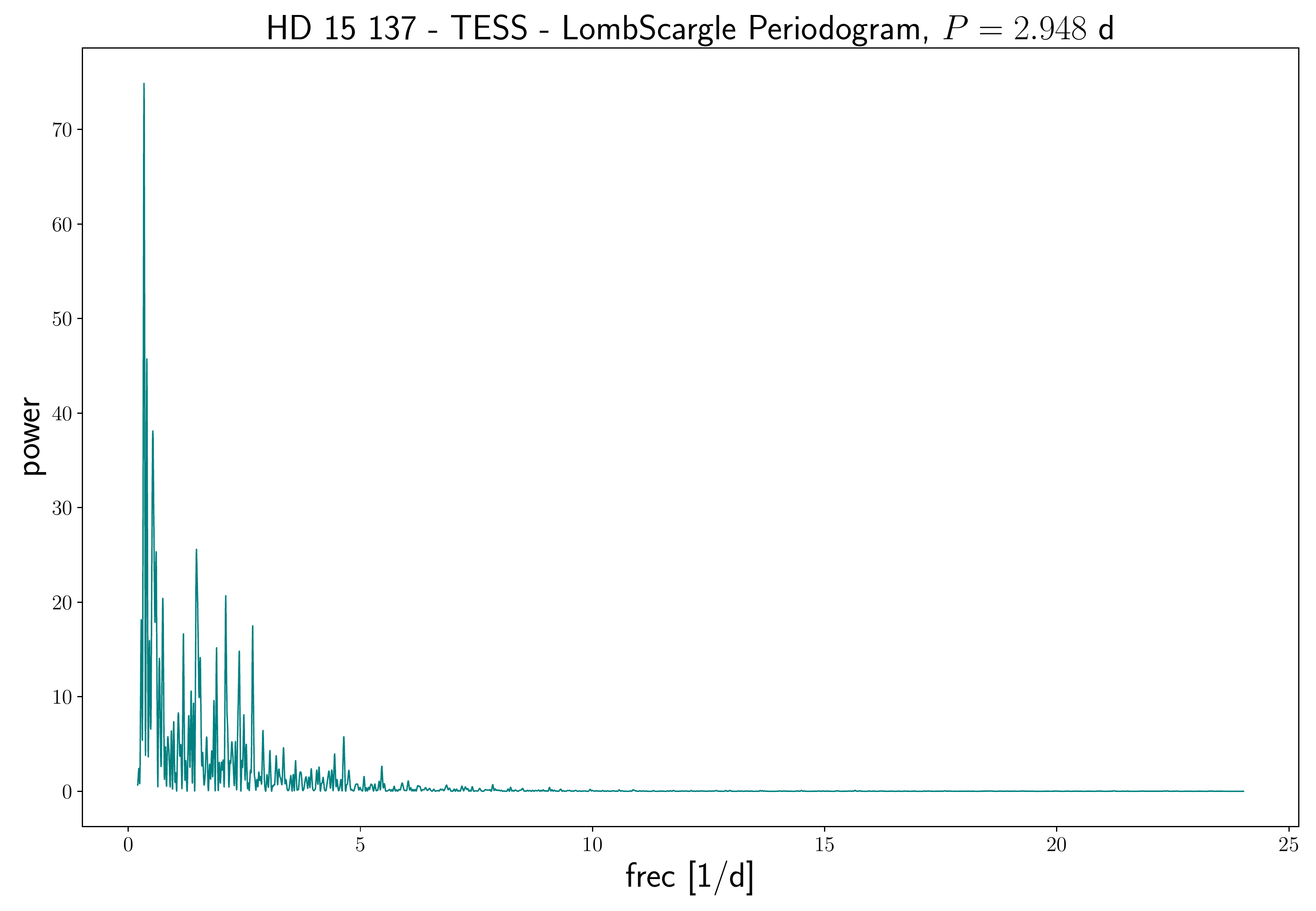}
    \caption{LS periodogram of TESS data in sector 18 for HD~\num{15137}.}
    \label{fig:hd15137_tess}
\end{figure}

Subsequently, we explore in the literature reports of photometric variability: the star is not cataloged in the VSX database. 
We retrieved the HIP \citep{ESA97a} data for the star (HIP \num{11473}) and searched for photometric variations.
The periodogram reveals a series of frequencies around $1.4748$~d$^{-1}$, ($0.678$~d, Table~\ref{tab:period-HD15137}). 
The periodic signal is remarkable when folding data with the period (Fig.~\ref{fig:hd15137_hip_fold}).
The TESS LC in sector 18 of the star reveals a noticeable apparent stochastic variation with an amplitude of about 25 mmag (Figs.~\ref{fig:hd15137_tess} and ~\ref{tess-fig:4} middle right panel). 
Again, we performed a periodogram analysis: the main frequencies are listed in Table~\ref{tab:period-HD15137}, standing out $f_1 = 0.33917$~d$^{-1}$ ($P = 2.94839$~d). 
This main frequency could be related to rotational modulation, because if we consider the expected radius of the star (about 13-23~$R_\odot$ for an O9.5~III-I star, \citet{Martetal05a}), the rotational frequency will be $\nu_{\rm rot} \sim 0.25 - 0.45$~d$^{-1}$.
This main feature is accompanied by a number of low frequencies, which combined give the "noisy" appearance of the LC.

\begin{table}
    \centering
    \small
    \caption{The five relevant frequencies in the Hipparcos and TESS photometric series of HD~\num{15137}.}
    \label{tab:period-HD15137}
    \begin{tabular}{c | c | c || c | c | c}
    \hline
    \hline
        Frequency &  Period & Power & Frequency &  Period & Power \\
         d$^{-1}$ &       d &       & d$^{-1}$  & d       & \\ 
    \hline
        1.4748 & 0.678\,05 & 19.75 & 0.3392 & 2.948\,11
        & 74.9 \\
        1.4862 & 0.672\,87 & 13.55 & 1.4699 & 0.680\,32 & 25.6 \\
        1.6034 & 0.623\,67 & 12.73 & 2.1005 & 0.476\,08 & 20.7 \\
        1.3090 & 0.763\,94 & 12.71 & 0.7436 & 1.344\,81 & 20.4 \\
        1.4632 & 0.683\,43 & 12.23 & 2.6811 & 0.372\,98 & 17.5 \\
    \hline
    \end{tabular}
\end{table}

\paragraph{\textbf{HD~\num[detect-all]{14633}~AaAb} = BD~$+$40~501~AaAb = ALS~\num[detect-all]{14760}~AaAb = SBC9~120} 
\textbf{SB1+Ca}

\hypertarget{hd14633}{} Four orbital solutions have been published (see  Table~\ref{t-orbsol-all}) for this system. 
All solutions are consistent with a period of about $P=15.4$~d, low semi-amplitude $K=17-31$~\kms, and a large eccentricity ($e\sim0.7$). 
The small mass function suggests the presence of a low-mass companion \citep{McSwainetal11}.

We collected 22 spectra for this object.
RVs were determined by using the Gaussian fitting for several \ion{He}{i}, and \ion{He}{ii} lines, and also through the x-corr method.
All RV data sets bring orbital solutions compatible with previous findings.
Combining our RVs with the available published ones, we derived an improved solution with a slightly longer period of $P=15.409$~d (Fig.~\ref{orb-fig:3} middle right panel). 
Due to the runaway nature of the star \citep{Maizetal18b}, with a still uncertain origin, it is worth noticing the slightly higher value of $\gamma \sim -42$~\kms of our solution.

\cite{Boyajianetal05} suggested that this system was ejected from the open cluster NGC~654 by a supernova explosion in a close binary, resulting in a system hosting an NS.
Later, \cite{McSwainetal11} supported this scenario, detecting a non-thermal X-ray flux component, showing variability during the orbital cycle, which is presumably associated with the NS companion. 
The only mention on the Ab component is that of \cite{Aldoretal15}, where it was resolved on the x-axis and unresolved on the y-axis. 
If that detection is confirmed, the NS companion scenario may lose strength.

This runaway star, now classified as ON8.5~V, was one of the first O-type stars recognized to present a ''nitrogen anomaly,'' displaying strong \ion{N}{iii} absorption lines, and weak \ion{C}{iii} lines, compared with stars with similar spectral type \citep{Walborn70, Walb71c}. 
\cite{Martetal15b} also found that HD~\num{14633} is a particularly enriched object, and as the case of V479~Sct, \cite{McSwainetal04} proposed that such an enrichment could be explained via mass transfer of CNO-processed gas prior to the supernova explosion.
The rejuvenation process due to the mass transfer makes this object an interesting system to study the formation of early-type blue stragglers. 
We need to take into account that the cluster origin is not yet proven, and a detailed analysis including {\it Gaia} parallaxes and proper motions is needed to confirm that scenario. 
The TESS data obtained in sector 18 shows stochastic irregular variations with a $\sigma = 1.8$~mmag  with a median error of the measurements $\epsilon = 0.08$~mmag.

\paragraph{\textbf{$\alpha$~Cam} = HD~\num[detect-all]{30614} = HR~1542 = BD~$+$66~358 = ALS~\num[detect-all]{14768} = SBC9~279} %
\textbf{Single}

\hypertarget{aCam}{} This bright O9~Ia runaway star has a long record of studies. 
The star was classified as an SB1 by \cite{Bohannanetal78}, who proposed a period of $48.6$~d, and a semi-amplitude of about $15$~\kms.
Using the {\em International Ultraviolet Explorer (IUE)}, \cite{de-Jageretal79} observed the star continuously during 72 hours, revealing gradual and erratic short-term changes in the velocity edge of the UV resonance lines, which were interpreted as changes in mass-loss and the ionization equilibrium in the envelope (see also \citealt{Kaperetal96}).
At the same time, \cite{Ebbetsetal80} monitored the star during four months using a coud\'e spectrograph and a digital detector, finding noticeable changes in the H$\alpha$ P~Cyg profile with a timescale of 6~hours to 1~day.
They suggested that the observed variability is due to the rotation of an inhomogeneous expanding envelope.
Another suggestion about the binary hypothesis was proposed by \cite{Stoneetal82}, who found no significant periods under $180$~d and that the semi-amplitude should be larger than $8$~\kms.

Using high-dispersion spectrogram plates obtained between 1976 and 1985 at the 6-m reflector of the Special Astrophysical Observatory of the USSR Academy of Sciences and the 2-m telescope at the Shemakha Astrophysical Observatory, \cite{Zeinalov86} proposed that $\alpha$~Cam is an SB1 system with a period $P = 3.6784$~d, a semi-amplitude $K_1=9.0$~\kms, and a high eccentricity ($e=0.45$).
The SB1 scenario proposed by those authors was put in doubt by \cite{McSwetal07}.
Photospheric fluctuations, non-radial pulsations, or variations in the stellar wind were proposed by a number of authors as the source of RV variability (amplitude $30$~\kms) \citep{Giesetal86, Markova02, Prinjaetal06, Kholtyginetal07}.
In particular, it is interesting to remark the results achieved by \cite{Prinjaetal06}, who used over 200 spectra to identify a $0.36$-day period in the profile changes in \HeI{5876}, probably produced by non-radial pulsations. 
This signal was persistent over two months, but it was not present two years later, highlighting the complexity of the analysis of these profile variations.

\begin{figure}
    \centering
    \includegraphics[width=.5\textwidth]{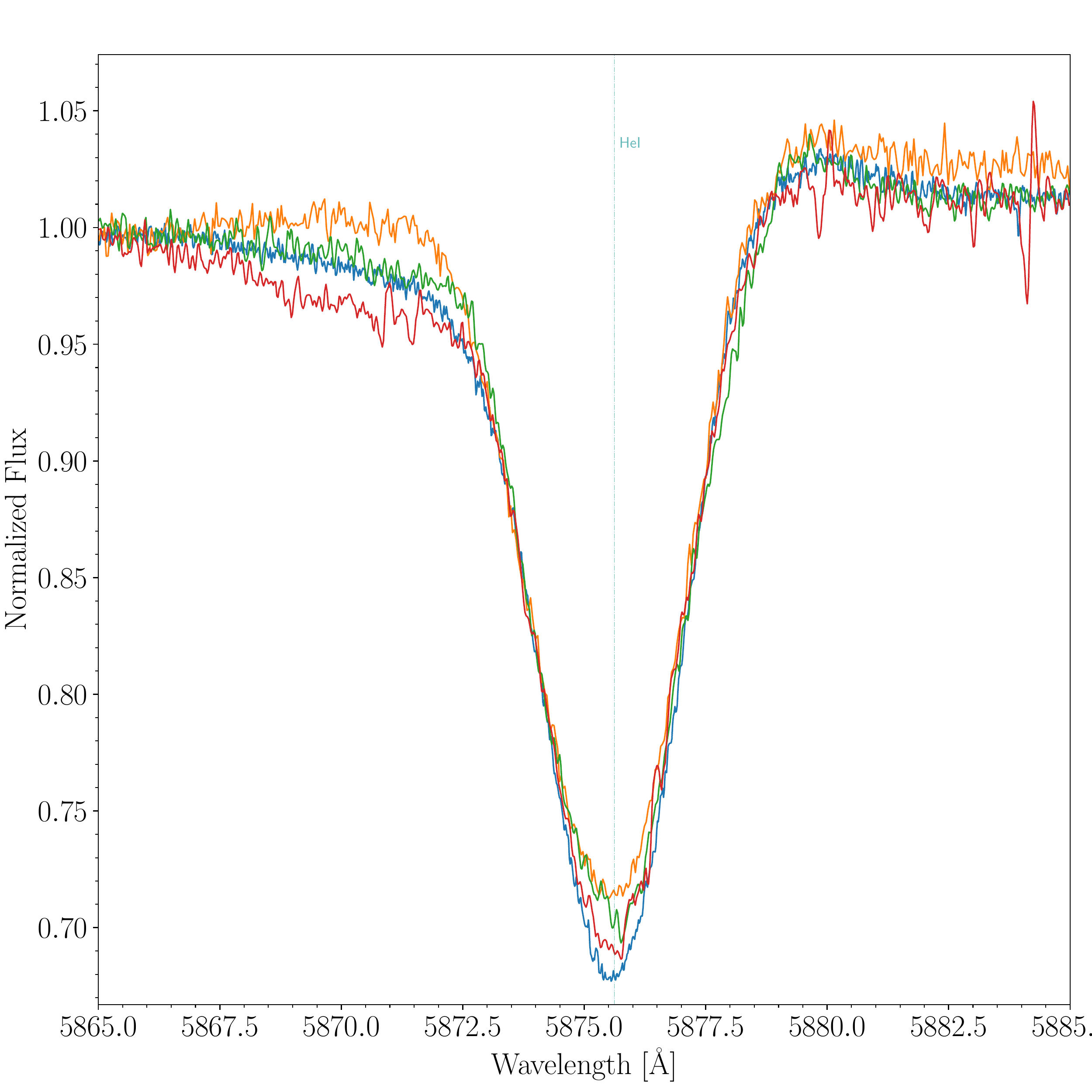}
    \caption{\HeI{5876} lines of $\alpha$~Cam for several epochs to show its variability. The spectra shown are: 171204\_M (blue), 121229\_C (orange), 110115\_I (green), and 081108\_I (red).}
    \label{fig:alphaCam_HeI}
\end{figure}

For the LiLiMaRlin database, we collected 55 spectra distributed from 1995 to 2018. 
A large fraction of them were obtained between 2013 and 2017.
As demonstrated by \cite{Prinjaetal06}, the profile of the line \HeI{5876} shows a complex behavior with a variable and extended blue wing and incipient emission in the red wing (Fig.~\ref{fig:alphaCam_HeI}).
We measured RVs in the strongest \ion{He}{i}, \ion{He}{ii} lines, and in two metallic lines (\OIII{5592}, \CIV{5812}), searching for periodicities, without identify any clear periodicity.
The amplitude of the variations is RV$_{\rm pp} \sim25$~\kms, which is in the range of values determined in other O-type supergiants with moderate rotation (we measured the rotational velocity of $\alpha$~Cam to be $v\sin i=113 \pm 6$~\kms) that display pulsations (\citet{Simonetal21}).

The TESS LC obtained in sector 19 reinforces the scenario of pulsations (Fig.~\ref{tess-fig:5} lower left panel). 
It shows clear variations with an amplitude of about 7~mmag, but it should be taken into account that such an amplitude is with respect to the mean flux, because long-term variability through the 27-day TESS observing cycle was smoothed during the normalization process. 
The LS periodogram brings as main period $P = 3.77786$~d ($f_1 = 0.2647$~d$^{-1}$), which is similar to the period found by \cite{Zeinalov86}. 
This kind of incoherent low-frequency modes are commonly observed in OB-type supergiants, for example HD~\num{152424} or HD~\num{37128}  \citep[cf.][]{Burssens20}. 
Therefore, we classify this system as a single star instead of an SB1.

\subsection{Auriga} \label{Aur}

\paragraph{\textbf{HD~\num[detect-all]{37737}} = BD~$+$36~1233 = ALS~8496 = SBC9~352}  %
\textbf{SB1E}

\hypertarget{hd37737}{} \cite{PetriePearce61} presented the first report about RV variability in this O9.5~II-III(n) system. 
\cite{Giesetal86} classified the star as an SB1 system, establishing a first orbital solution with a period $P=2.49$~d based on ten RV measurements spanning about 700~d. 
Their orbital solutions contemplated two alternatives: a circular orbit and an eccentric one ($e=0.132$). 
Later, \cite{SticLloy01} supported the circular orbit solution, adding a single {\em IUE} RV to the values determined by \cite{Giesetal86}.
Interestingly, \citet{McSwetal07} added 19 RVs measurements to the first data set and derived a very different solution, with $P=7.84$~d, $e=0.43$, and a semi-amplitude of $72$~\kms. This result was confirmed by \cite{Alexeeva13}, who found a very similar solution for this system.

In LiLiMaRlin we collected 17 spectra for this system. 
In the light of the two different periods proposed, we analyzed the LS periodogram of RVs determined for six absorption lines (\HeI{4471}, 5876, 7065; \HeII{4542} and 5412, and \OIII{5592}) and those obtained with the x-corr method.
All periodograms show a clear maximum at frequency $f=\num{0.1274}$~d$^{-1}$ ($P\simeq7.84$~d). 
In view of the congruence of our result with previous findings, we combined all the available published RVs with our measurements using the x-corr method to determine a new orbital solution, which is plotted in the lower-left panel of Fig.~\ref{orb-fig:3}, and listed in Table~\ref{t-orbsol-all}.
The visual inspection of the spectra does not reveal a contribution of the companion, although clear line profile variations are present. 
This star is also a fast rotator, and we measured a projected rotational velocity of $v \sin i=201\pm11$~\kms.

Recently \cite{Burggraaffetal18} using Multi-site All-Sky CAmeRA (MASCARA) data detected eclipses and found a period very similar to the spectroscopic one of $P=7.84673$~d. 
The TESS photometric time series obtained in sector 19 shows a striking eclipsing system (Fig.~\ref{tess-fig:5} middle panels).
The eclipses have depths of about 80~mmag and 40~mmag, for the primary and secondary minima, respectively, unevenly separated due to the high eccentricity of the system.
The LC presents two distinctive features. 
The first one is the sinusoidal variation synchronized with the orbital motion: the period of this pulsation is one-tenth ($0.784$~d) of the orbital one.
The second feature is the clear pulse-like maximum between eclipses, just after the periastron passage. 
As we mentioned previously in the case of V747~Cep,
this type of featured LC could be explained in terms of the heartbeat variability \citep{Thompsonetal12}. 
This system is particularly important to study the coupling between tidally induced pulsations and orbital motion.

Given the presence of eclipses, the system is now classified as an SB1E.

\subsection{Orion-Monoceros}\label{Ori-Mon}

\paragraph{\textbf{15~Mon~AaAb} = S~Mon~AaAb = HD~\num[detect-all]{47839}~AaAb = BD+10~1220~AaAb = ALS~9090~AaAb = SBC9~1725} %
\textbf{SB2a}

\hypertarget{15mon}{} Located at a distance of 719~pc, 15~Mon~AaAb is the main ionizing source in the area of NGC~2264 cluster \citep{Masoetal98, Apellaniz19a}. 
It has been known as an RV variable for over a century \citep{Frostetal04}. 
Recently, \cite{ApellanizBarba20} have used STIS/HST long-slit spectroscopy to spatially disentangle the blue-violet spectra of the Aa and Ab components ($\Delta B = 1.55$~mag and $\rho = 143$~mas at the time of the observation). 
Thus the SBS status of this system changes to SB2a.
The spectral classification determined for Aa is O7~V((f))z, while for Ab is B1:~Vn. 

Several orbital solutions for the inner system (there is also a B companion located farther away) have been published. The first one by \cite{Giesetal93}, which also happens to be the only purely spectroscopic one, gave a period of  $P=25.32$~a, improved later after including spectroscopic and astrometric data \cite[$P=23.62$~a, ][]{Giesetal97}.
\citet{Cvetetal09, Cvetetal10} reanalyzed the system and found a much longer period of $P=74$~a using interferometry. 
Recently \cite{Apellaniz19a} found an astrometric period of $P=108$~a for the inner pair. 
This trend is explained by the fact that we have not yet seen a full orbit; the last periastron passage was in 1996, so the orbit is not very well constrained.

Besides its binary nature, 15~Mon is also a known irregular variable star with an amplitude of about $70$~mmag. 
The TESS images (sectors 6 and 33) collect the light of all the stars in 15~Mon~AB. The time series shows some hints about the stochastic variations. The new feature is the presence of a $\beta$~Cep (BCEP) pulsator in the group, revealed by a strong peak in the LS periodogram at $P=0.1595$~d.
In LiLiMaRlin we gathered 65 spectra for this system and also HST/STIS observations (an additional STIS spectrum is scheduled to be observed at a slightly larger separation). 
The HST and LiLiMaRlin data will be combined together with other astrometric observations to derive a new spectro-astrometric orbit in the near future.

\paragraph{\textbf{HD~\num[detect-all]{46573}} = BD~$+$02~1295 = ALS~9029} %
\textbf{SB1}

\hypertarget{hd46573}{} This interesting O7~V((f))z object of the Monoceros OB2 association, has been classified as a runaway by \cite{Maizetal18b} and has a  bow shock clearly visible in the IR. 
It was proposed as a possible spectroscopic binary by \citet{Masoetal98}. 
\cite{Mahyetal09} have identified the star as an SB1 system with a preliminary period $P=10.67$~d, very small semi-amplitude $K_1=8.5$~\kms, and large eccentricity $e=0.47$ (although the circular orbit is not discarded). 
The orbital solution shows large residuals in RVs, and, therefore, it is not well constrained.

In MONOS, we have 16 spectra available spanning 15 years. 
In order to compare with the results of \cite{Mahyetal09}, we use RVs determined from the \HeI{5876} absorption line. 
The two FEROS spectra in common with their data show a difference of $0.5$ and $0.2$~\kms, respectively. 
As for other systems, we obtained a new orbital solution by combining both data sets.
The new solution is in moderate agreement with the previous one, finding a period of $P = 10.654$~d but a much higher eccentricity of $e=0.63$ and a larger semi-amplitude $K = 11.2$~\kms (Fig.~\ref{orb-fig:3} lower right panel). 
The possible circular solution suggested by \citet{Mahyetal09} is discarded. 

\citet{Burssens20} reported that the TESS time series (sector 6) shows stochastic low-frequency variability. The same behavior is seen in the data obtained in sector 33, with a $\sigma \sim 1.8$~mmag. The LS periodogram of sector 33 brings a dominant frequency ($f_1 = 0.4$~d$^{-1}$) different from sector 6 data ($f_1 = 0.861$~d$^{-1}$).
It is interesting to note that the time lapse between the observations in sectors 6 and 33 is two years, and during that period of time, the periodograms show very different dominant frequencies. The LS periodogram of both sectors combined points to the shortest frequency as dominant, $f_1 = 0.399$~d$^{-1}$. This result suggests that the dominant frequency determined with TESS observations from only one sector may not necessarily be associated with the rotational period of the star. Using observations with a longer time baseline can lead to a more reliable finding.

\paragraph{\textbf{$\theta^{1}$~Ori~CaCb} = HD~\num[detect-all]{37022}~AB =  BD~$-$05~1315 = ALS~\num[detect-all]{14788}~AB} 
\textbf{SB1+Sa}

\hypertarget{tet1ori}{} This O7~f?p~var (see MONOS~I) star is the main ionizing source of the Orion nebula and the brightest and most massive member of the Trapezium Cluster. 
$\theta^{1}$~Ori~C has been studied for roughly 90 years and about a hundred papers have been dedicated to it. Several periods have been ascribed to the system, ranging from a few days to longer than a century, each associated with a different cause.
$\theta^{1}$~Ori~C is, therefore, particularly complex to analyze.

The shortest period of $15.424\pm0.001$~d is rotational and magnetically induced \citep[see][and references therein]{Stahetal08, Stahetal96, SimDetal06}.
Several works have studied the system by using interferometric and spectroscopic data, finding a  period of $\sim 11$~a \citep[see][and references therein]{Krauetal09b, Lehmetal10, Balegaetal15, Gravity18} for the astrometric companion. 
\citet{Patietal08} and \citet{Stahetal08} found that a spectroscopic orbit with a period of $\sim22$~a and a much lower eccentricity (around $e\sim0.14$) was also compatible with the astrometric solutions.
Finally, \citet{Vitrichenko02} and \citet{Lehmetal10} found another spectroscopic companion with a period of $P = 61.5$~d, at an estimated separation of $\sim1$~mas, although they could not definitely assert that the interpretation of such a period as an orbital period was correct. 
This would be the less massive star in the system with about $\sim 1~\mathrm{M}_{\odot}$. 
In Table~\ref{t-orbsol-all}, we present the parameters for both the shorter- and longer-period orbital solutions.

\begin{figure}
    \centering
    \includegraphics[width=.5\textwidth]{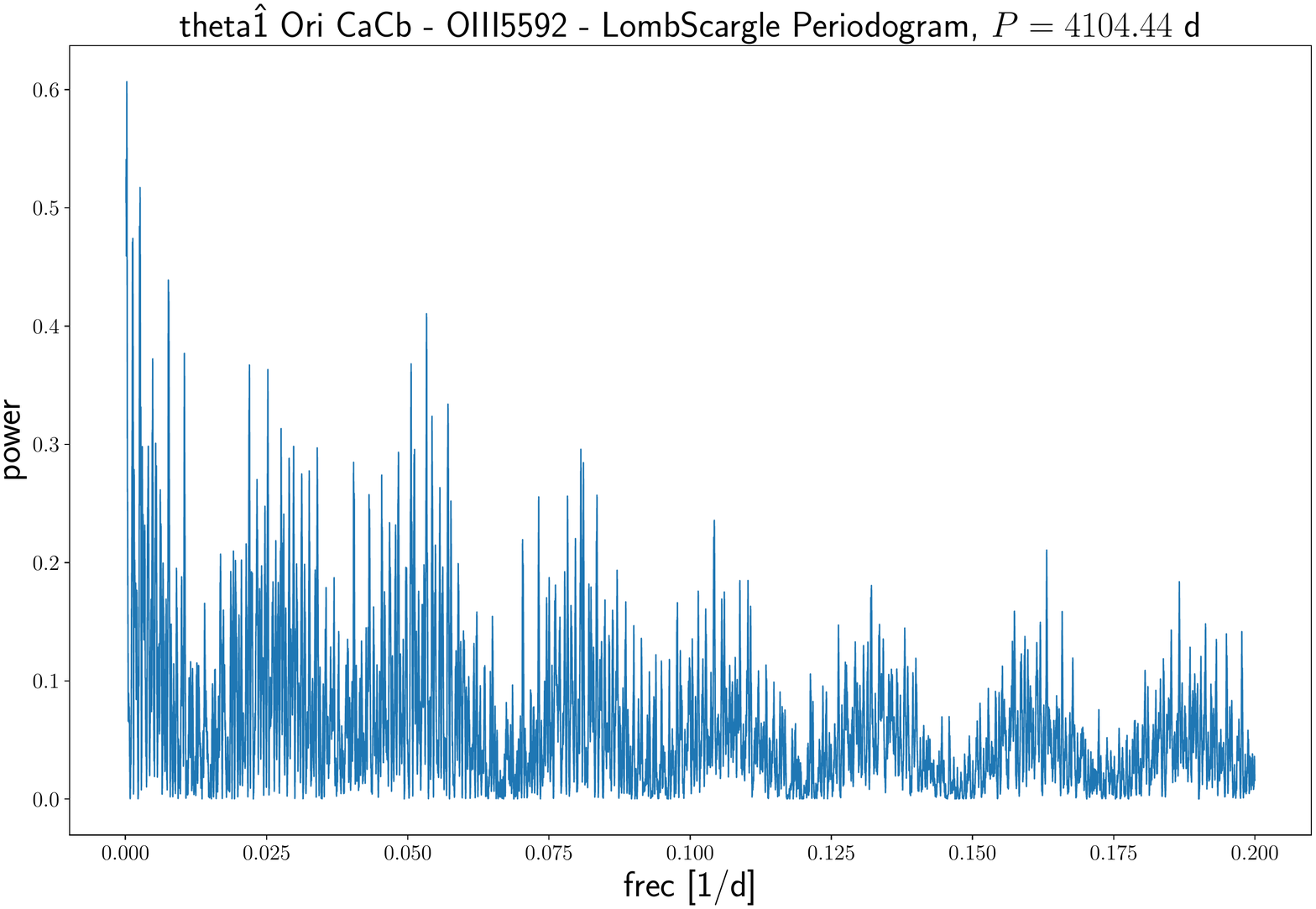}
    \caption{Periodogram of the \OIII{5592} line for $\theta^1$~Ori~CaCb.}
    \label{fig:theta1C_LS}
\end{figure}

In the LiLiMaRlin framework, we have collected 63 spectra spanning 25 years. 
The main result is that we can confirm the orbital period of about \num{4000}~days, as proposed by \citet{Krauetal09b}.
The preliminary orbit is shown in Table~\ref{t-orbsol-all} (Fig.~\ref{orb-fig:4}, left panel). 
This period corresponds to the higher peak in the LS periodogram calculated with our RVs.
Figure~\ref{fig:theta1C_LS} shows the LS periodogram for the \OIII{5592} line, but we can also see the same peak for other lines, such as \HeII{4542}. 
Nevertheless, this preliminary orbit should be treated with caution because the short, 15~d period, is also present.
In the case of the \OIII{5592} line, the RV$_{\rm pp} \sim 35$~\kms, due to the orbital motion, is only roughly double the variation we detect over periods of a few days, associated with the magnetic period; consequently, the RV curve presents a large scatter.

Given the importance of the system in the context of very young binaries, it will be analyzed in a separate paper, including all LiLiMaRlin spectroscopic data, more than 150 additional spectra obtained with the SONG telescope at the Teide Observatory\footnote{\url{https://phys.au.dk/song/research-and-facilities/}}, and combining astrometric information derived from HST observations.

\paragraph{\textbf{HD~\num[detect-all]{52533}~A} = BD~$-$02~1885~A = ALS~9251~A = SBC9~429} %
\textbf{SB1E}

\hypertarget{hd52533}{} This O8.5~IVn star was identified as an SB1 system by \cite{Giesetal86}, with a preliminary period $P=3.3$~d. 
The star was originally proposed as a runaway, but \cite{deWietal05} found that it is the brightest member of a small cluster.
\cite{McSwetal07} reanalyzed this system, proposing that it could be a triple system composed of a close binary with the hotter component and an early B-type component more separated, although high-resolution spatial observations have not confirmed this configuration. 
These authors determined a new orbital solution for the SB1 configuration, with a substantially longer period, $P=\num{22.1861}$~d, large semi-amplitude $K_1=105$~\kms, and moderate eccentricity $e=0.3$.

In the LiLiMaRlin database, we gathered 15 spectra obtained during a time spanning ten years. 
We search periods in the RVs determined from the x-corr method using only \ion{He}{ii} lines.
Firstly, the period search was performed using only our RV measurements, and then, combining them with the published values. 
In both cases, the best period found was $P=\num{21.966}$~d, slightly shorter than that determined by \cite{McSwetal07} (Fig.~\ref{orb-fig:4} right panel). 
Furthermore, we found a significant difference between the $\gamma$ value determined from our data ($+36.4$~\kms) and that determined by \cite{McSwetal07} ($+76.6$~\kms). 
Thus, to obtain a combined solution, we subtracted $40.2$~\kms from \cite{McSwetal07} RV velocities to bring them to our reference frame.
Finally, we calculated the combined solution presented in Table~\ref{t-orbsol-all}.
However, this orbital solution should be taken as preliminary because the large difference in $\gamma$ values between our orbital solution with respect to that determined by \cite{McSwetal07} could indicate that the SB1 is part of a higher-order multiple system. 
This difference of $40$~\kms is beyond the expected RV errors in both data sets.

A visual inspection of the spectra reveals clear line variations in the \ion{He}{i} lines, but no double lines. 
This effect may be caused by contamination of the secondary component of the SB1 or even a probable neighbor third body, since the dim Ab is located at 0.6\arcsec\ and thus is not seen in our spectra. 
This effect is much smaller, although still slightly noticeable in some spectra in the \ion{He}{ii} lines. 
We do not see clear double lines in any spectrum. 
This may be due to the lines of the secondary being broadened by rotation or simply being too dim. 
Thus we still consider HD~\num{52533}~A to be an SB1.

Interestingly, the star is cataloged as an eclipsing binary system in VSX by Sebasti\'an Otero, giving a period $P=\num{21.9675}$~d, which is entirely consistent with the spectroscopic one. 
Recently, \citet{PozoNetal19} confirmed this finding, deriving $P=\num{21.9652}$~d.
The TESS LC obtained in sector 7 is outstanding (Fig.~\ref{tess-fig:5} lower right panel): eclipses are deep and sharp; they are unevenly separated, a signature of high eccentricity. Given the relatively long period of the system, it becomes a very promising case for the determination of absolute parameters, because the orbital inclination must be very close to $90^\circ$ to allow the eclipses.

\section{Summary and discussion}\label{summary}

\begin{table*}[!htp]
\small
\centering
\renewcommand{\arraystretch}{1.3}
\caption{
Results of the MONOS~II analysis for systems where we do not find the O star to show a spectroscopic orbital motion. LPV (line-profile variable) indicates that we were unable to determine the origin of the line variation. 
The RVs of the objects are obtained as described in Sect.~\ref{rv_measure}. Errors are in parentheses, corresponding to the last digits.
\label{t-single}}
\begin{tabular}{l c c r@{.}l r@{.}l r@{.}l r@{(}l l}
\hline \hline
\multicolumn{1}{c}{Name} & \multicolumn{2}{c}{SBS classification} & \multicolumn{2}{c}{$\rm VR_{x-corr}$} & \multicolumn{2}{c}{$\rm VR_{\HeII{5412}}$} & \multicolumn{2}{c}{$\rm VR_{\OIII{5592}}$} & \multicolumn{2}{c}{$v \sin i$} & \multicolumn{1}{c}{Notes} \\
\multicolumn{1}{c}{} & \multicolumn{1}{c}{MONOS~I} & \multicolumn{1}{c}{MONOS~II} & \multicolumn{2}{c}{km\,s$^{-1}$} & \multicolumn{2}{c}{km\,s$^{-1}$} & \multicolumn{2}{c}{km\,s$^{-1}$} & \multicolumn{2}{c}{km\,s$^{-1}$} & \multicolumn{1}{c}{}\\
\hline
9~Sge           & SB1?    & Single     & 16&9(53)  & 20&3(52)   & 23&7(47)  &  69&3)  & LPV  \\
HD~192\,281     & SB1     & Single     & -28&8(49) & -26&0(42)  & -6&0(57)  & 277&14) & LPV  \\
HDE~229\,232 AB & SB1+Ca? & Single     & -20&7(59) & -16&6(113) & .&.       & 313&16) & LPV  \\
Cyg~OB2-22~C    & SB1E?   & SB1 unc.+E &  .&.      & 31&0(46)   & .&.       & 266&14) & Triple system \\     
68~Cyg          & SB1?    & Single     & 30&1(106) & 49&3(84)   & 43&5(208) & 312&16) & LPV  \\
HD~108          & SB1?    & Single     & .&.       & -75&5(47)  & -71&3(43) &  46&2)  & LPV  \\
$\alpha$~Cam    & SB1?    & Single     & 15&0(77)  & 20&7(46)   & 22&7(54)  & 113&6)  & LPV  \\
\hline
\end{tabular}
\renewcommand{\arraystretch}{1.0}
\end{table*}

$\,\!$\indent The aim of this paper is to review and update the multiplicity status of 35 O-type stars identified as SB1 systems in MONOS~I, of which 33 have previously published spectroscopic orbital solutions and two (Cyg~OB2-22~C and V747~Cep) only have previous photometric orbits based on their eclipsing nature. For the analysis, we used the LiLiMaRlin spectroscopic database, which contains more than 700 spectra for 32 objects. We determined RVs from several absorption lines by using Gaussian profile fitting as well as through a cross-correlation method against tailored templates generated with the {\sc FASTWIND} stellar atmosphere code. In this way, we obtained about $\num{4500}$ RV measurements, which are provided in Appendix~\ref{app:rv}. Moreover, we reviewed in detail the plentiful literature on all these objects, collecting all the information available on RVs and orbital solutions, which is listed in Table~\ref{t-orbsol-all}.

We analyzed in detail the RV behavior of each star and calculated orbital solutions by considering different strategies, which involve RVs obtained with different methodologies, in many cases combining our measurements with those collected from the literature. We point out that the vast majority of our RV measurements are obtained in later epochs than most of those collected from the literature. Significantly increasing the time span covered by observations is important, not only to improve the accuracy of the orbital periods but also to determine the possible existence of an additional long period due to a third object. Complementarily, in order to strengthen our conclusions about the multiplicity status for each star, we explored different photometric data sets, in most cases resorting to TESS data, and performed a photometric analysis of the relevant time series for many of the objects considered. 

We analyzed and revised the SBS qualifier for all the objects, updating 21 of them (60\% of the sample). After a careful review, we present spectroscopic orbital solutions for 21 SB1 systems, including V747~Cep, for which no such solution had previously been published (Table~\ref{t-stat}). Table~\ref{t-orbsol-def} shows the spectroscopic orbital solutions for 28 systems that we adopt as definitive, including the 20 new determinations and 8 solutions from previous works. The spectroscopic orbits of two of the systems listed, 15~Mon~AaAb (now considered an SB2) and HD~\num{16429}~Aa, will be analyzed in forthcoming papers.
Further details of these 28 orbital solutions are given in Table~\ref{t-orbsol-all}. For those systems where we have calculated more than one orbital solution, we favor those with smaller residuals.

There are six faint objects (all belonging to the Cyg~OB2 association) that remain as candidate SB1 because our current data sets (either spectroscopic or photometric) are insufficient to confirm their published solutions. These objects are difficult to observe due to their faintness, crowding, and/or the characteristics of the orbital solutions proposed, that is, very long periods (up to $4066$~d) and small semi-amplitudes ($K_1 \leq 17.5$~\kms).

\subsection{Single stars}\label{single}

$\,\!$\indent Our extensive RV monitoring cannot find significant orbital motion for six stars, leading to the conclusion that they are not true spectroscopic binaries. Therefore, we suggest that these six stars are likely single. This set of stars shares some remarkable characteristics. Firstly, all of them are classified as runaway stars (although HD~108 is not confirmed as such) and present photometric variability that can be related to stellar oscillations (9~Sge, HD~192\,281, 68~Cyg, HD~108, $\alpha$~Cam) or stellar rotation (HDE~\num{229232}~AB, 68~Cyg).
Three of them are fast rotators with luminosity class III to V: 68~Cyg, HD~\num{192281}, and HDE~\num{229232}~AB, with the last two displaying the earliest spectral types in the sample (O4).
Two others are late O-type supergiants, 9~Sge and $\alpha$~Cam, while the last one, HD~108, is a peculiar object with a strong magnetic field.
Table~\ref{t-single} summarizes the new status of these likely single stars, and in Table~\ref{t-orbsol-all} we include the list of previously proposed orbital solutions for completeness. A seventh system (Cyg~OB2-22~C) is listed in Table~\ref{t-single}, although its status is not yet confirmed. Based on current data, it is most likely a triple system where the O star is in a long orbit around a short-period eclipsing binary. The value listed for its RV corresponds to our measurement for the O star, as the outer period is probably too long to be measured any time soon.

We note that we cannot fully rule out binarity in the case of these stars due to pulsational activity. \citet{SimonDetal20a} have recently shown that the effects of stellar pulsations are an important factor to take into account in the study of spectroscopic multiplicity among O and B supergiants. These stars display different types of stellar pulsations, which result in RV variations \citep[a common feature in O- and B-type supergiant stars; see, e.g.,][]{Fullertonetal96, Aertsetal17, Simonetal18, SimonDetal20a}.
There is a correlation between the amplitude of the jitter in the RV variations (RV$_{\rm pp}$) and their rotational velocities.
For example, O-type supergiants with low and moderate rotational velocities ($v\sin i \leq 100$~\kms) can show RV variations with typical amplitudes of 
$\sim 20$~\kms.

For fast rotators, an RV$_{\rm pp}$ up to 10\% of the $v\sin i$ is probably due to the stellar pulsations instead of binarity (Britavskiy et al., in prep). 
As illustrated in \citet{SimonDetal20a} and \citet{Simonetal21}, these RV changes produced by pulsations display an incoherent variation pattern, different from that expected from organized orbital motion.
If any such star is part of a binary, its RV curve will display a combination of the intrinsic RV variations and the changes due to orbital motion.
Therefore, if the RV sampling is sparse, the orbital period can remain hidden, especially if the amplitude of the curve is similar to (or smaller than) that of intrinsic variability. Furthermore, a false period can be determined if the sampling is not appropriate, an issue that has been known for a long time \citep[see][]{Tanner48}.

It is important to note that the special characteristics, such as the runaway nature and the presence of magnetic fields, can be explained as a by-product of the binary evolution. Two scenarios have been proposed for the origin of OB runaway stars: (a) objects acquiring larger spatial velocities after a supernova kick or (b) an ejection resulting from a binary-binary encounter \citep{Hoogetal00}. In both cases, the runaway status could be acquired by the binary itself or the components of the binary in the case that it is disrupted. In fact, \citet{Boyajianetal05} proposed that two of the stars in our
sample, HD ~\num{15137} and HD~\num{14633}~AaAb, have been ejected from parental clusters as a result of supernova events. Several studies \citep[cf.][]{FerrarioWick05, Schneideretal19, Schneideretal20} have proposed binary mergers as a potential mechanism to form the strong magnetic fields observed in some O-type stars.

\subsection{Nature of companions and evolutionary stage}

$\,\!$\indent This work is focused on ascertaining the validity of the orbits of systems classified as SB1 in MONOS I and on determining new orbits. All systems whose orbits we have certified retain their SB1 status, therefore putting some constraints on the detection of secondary stars. These limits depend on observational factors (S/N, resolving power, etc., of the spectra) and natural factors (relative brightness of stellar components, stellar rotation of both stars, orbital inclination, mass ratio, etc.) and must be evaluated on a case-by-case basis, as follows.

A striking result of our analysis is the relatively large number of systems that display light variations due to orbital geometry or induced stellar changes (eclipsing systems and ellipsoidal variables): We count ten such systems from the 21 SB1 systems (27 including the unconfirmed ones) with spectroscopic orbital solutions (Table~\ref{t-stat}). In addition, we must mention Cyg OB2-22 C, which is indeed an eclipsing binary, although it is likely a triple system in which the O star is not part of the eclipsing pair.
The frequency of eclipsing and ellipsoidal variables is almost 50\% of SB1 systems, which is large compared with the number quoted in MONOS~I ($< 10\%$).
This much higher rate, due to the excellent quality of the LCs provided by TESS, substantially increases our capability of modeling the systems and thus determining their nature and evolutionary state and constraining the properties of the invisible component.

Four SB1 stars show characteristics typical of Algol eclipsing systems: Cyg~OB2-1, V747~Cep, HD~\num{37737}, and HD~\num{52533}~A. 
All of them are eccentric systems with a sharp and well-defined secondary eclipse. This latter feature should be taken into account when defining future strategies to search for spectral features of the secondary star. 

The first three systems also share an interesting feature in their LC: stellar oscillations.
These stars are all binaries with orbital periods between $4.5$ and $7.8$~d, although their primaries are diverse, with spectral types ranging from O5.5~V(n)((f)) to O9.5~II-III(n).

\begin{figure}
    \centering
    \includegraphics[width=.5 \textwidth]{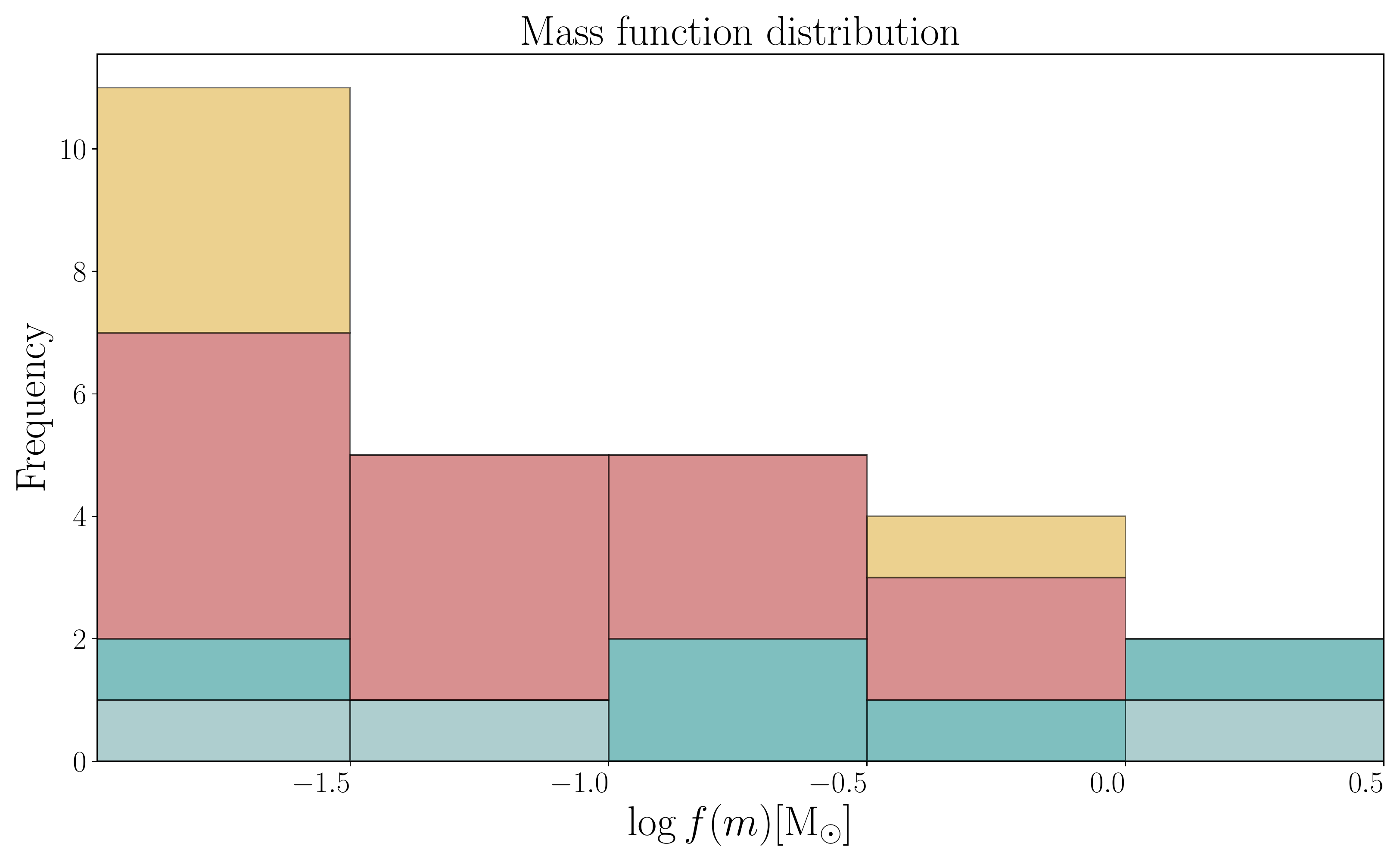}
    \caption{
    Histogram of the mass-function distribution of the SB1 systems in the MONOS~II sample. 
    We show the ellipsoidal systems in light gray and the eclipsing systems in teal. Systems with a normal star as a companion (but not eclipsing or ellipsoidal variables) are plotted in red. 
    Systems with a probable compact object as a companion are plotted in yellow (two of them are eclipsing systems).
    }
    \label{fig:fm}
\end{figure}

Among the SB1 systems analyzed, we find objects at very different evolutionary stages: near the ZAMS (Zero Age Main Sequence), still unevolved,  with ongoing mass-transfer, and  post-mass transfer, including systems with collapsed companions.
The mass function distribution shows preferentially small values (Fig.~\ref{fig:fm}), which points to two main possible scenarios: 
(a) unevolved low-mass companions or 
(b) collapsed BH or NS companions.
Of course, we may have low orbital inclinations in some systems, but the large number of eclipsing and ellipsoidal systems indicates that the preference for lower-mass companions is real.
\citet{Kobuetal14} have addressed this topic in detail for their sample of spectroscopic binaries in Cyg~OB2. We plan to come back to this issue for the entire spectroscopic binary sample in a future work.

\subsubsection{Unevolved systems}

$\,\!$\indent Possibly the youngest system in our sample, $\theta^1$~Ori~CaCb is one of best documented O-type stars near the ZAMS and is similar to the case of the triple O-type system Herschel~36~A \citep{Campillayetal19}.
$\theta^1$~Ori~CaCb is also a peculiar star: one of only six Of?p\,var stars in the Milky Way. 
Important questions are still open to debate regarding this system, such as the possible spectroscopic solar-like third component proposed by \citet{Lehmetal10}. 

Given the spectroscopic characteristics of the primaries and orbital parameters, we infer that 12 other systems are likely to be in an evolutionary stage prior to the first episode of mass transfer (Table~\ref{t:fm-sys}).
This group is composed of O stars on the main sequence or giants in wide orbits. 
Seven of them are eclipsing systems or present ellipsoidal variations. Of these seven, six are short-period binaries, but the seventh one, HD~52\,533~A, is an eclipsing system with $P=22$~d, pointing to an orbital inclination $i\sim90^\circ$.
In what follows, we estimate primary masses based on the calibration of \cite{Martetal05a}. 
Adopting $M_1 = 18-21$~M$_\odot$ for the O8.5\,IVn primary, the mass of the secondary is in the range $M_2 = 9-9.9$~M$_\odot$. This places the star in the spectral range B1\,--\,1.5\,V and suggests that it could be detected in high S/N spectra. For the other six systems, the inclination is not so well constrained, but we can still obtain meaningful estimates of the mass of the secondaries.

In the case of Cyg~OB2-1, an eclipsing system with sharp eclipses, the O8\,IV primary and high inclination combine to give an estimated mass for the secondary in the range $4-6$~M$_\sun$ (corresponding to a mid or late B-type star). 
Another system with sharp eclipses, V747~Cep, is earlier and more massive. Assuming a mass of about $27-30$~M$_\sun$ for the O5.5\,V primary results in a secondary of about $8-11$~M$_\sun$ ($i=60-90^\circ$), probably an early B-type star.

The eccentric eclipsing system HD~37\,737, O9.5~II-III(n) is a special case. 
Although the primary is already a giant, the sharpness of the eclipses strongly suggests that it is still inside its Roche lobe. 
Assuming a wide range of masses for the primary, $18-25$~M$_\sun$, the secondary would be in the range of $5-7$~M$_\sun$ (i.e., a mid-B-type star) for $i=60-90^\circ$.
Similarly, in the ellipsoidal system with a giant primary HDE~\num{229234}, we estimate a mass in the range $18-23$~M$_\sun$ for the O9\,III star, which leads to a secondary in the range $4-7$~M$_\sun$ (mid B-type) for $i=30-60^\circ$.

Two binaries with O supergiant primaries are intriguing systems: Cyg~OB2-11, O5.5~Ifc and $P=72$~d, and Cyg~OB2-A11, O7 Ib(f) and $P=15.5$~d.
Both stars have been detected as variable X-ray sources \citep{Rawetal15}.
In the case of Cyg~OB2-11, the wide orbit permits the possibility of a detached system, where components are evolving separately. 
\citet{Rawetal15} proposed that the X-ray variability is due to wind-wind interactions. 
Adopting a conservative range for the primary mass of $35-50$~M$_\sun$ , and orbital inclinations $i=20-70^\circ$, the secondary mass would be in the $7-19~$M$_\odot$, room for a mid B-type, early B-type, or late O-type star, as also suggested by \citet{Kobuetal14}. The colliding wind hypothesis renders the lower masses less likely.
The other O supergiant system, Cyg~OB2-A11, is a bright X-ray source\footnote{Using the flux values of \citet{Rawetal15} and the distances from \citet{Sara20} we calculated an L$_x$ of 3.05 and 2.05$\times 10^{32}$~ergs/s for Cyg~OB2-A11 and Cyg~OB2-11 respectively}, almost one order of magnitude brighter than expected for a single star. It presents strong variability in the X-ray flux of at least a factor of two through the orbital cycle \citep{Rawetal15}.
These characteristics led \citet{Rauw11} to propose that it is a colliding wind system.
Again, adopting a conservative range of values for the primary mass, $M_1=25-40$~M$_\sun$, we derive $M_2=2.5-12$~M$_\sun$ for orbital inclinations $i=20-90^\circ$. 
The spectral morphology and orbital parameters of Cyg OB2-A11 resemble those of the fast X-ray transient HD~\num{74194} \citep[LM~Vel;][]{Gameetal15}, and its X-ray luminosity is not incompatible with a quiescent X-ray binary. 
Therefore, a scenario with a compact companion cannot be ruled out. Contrarily, the colliding wind scenario requires a rather massive secondary and hence a high inclination.

\subsubsection{Systems undergoing interaction}

$\,\!$\indent Another supergiant system, the eclipsing binary BD~+36~4063, ON9.7~Ib, shows signatures of evolutionary processes in action: a nitrogen enhancement and an LC with indications of strong deformation of the stars, which suggests a near contact or semidetached configuration. 
\citet{willetal09b} proposed that the ON supergiant is undergoing mass transfer to an invisible and massive companion hidden by a thick disk.  

The TESS LC of the eclipsing system HD~\num{16429}~Aa is very likely typical of a semidetached configuration, although the presence of strong variability due to stellar pulsations renders the shape of the LC very complex. 
Adopting a primary mass of $18-23$~M$_\sun$ for this short-period system, we suggest that the secondary lies in the range $8-11$~M$_\sun$ (i.e.,\ an early B star).
Given the broadening of the eclipses, the secondary star seems oversized for the expected mass, which could be an indication that the system has undergone significant mass interchange, and now is probably in a similar configuration to AB~Cru \citep[O8~III + BN0.2:~Ib:][$P=3.41$~d]{Maizetal16, Lorenzetal94}.
As mentioned before, this complex multiple system deserves a dedicated analysis.

\begin{landscape}
\begin{table}
    \centering
    \small
    \renewcommand{\arraystretch}{1.2}
    \caption{Statistics of the 35 systems in this paper. Systems that we find not to be spectroscopic binaries are labeled as line-profile variable (LPV). It should be noted that we have included Cyg~OB2-22~C in the doubtful category (see text).}
    \label{t-stat}
    \begin{tabular}{*5c}
    \hline \hline
            & SB   & New Orb. & LPV  & Inconclusive/No data  \\
    \hline
            & 22   & 20       & 6    & 7              \\ 
        \%  & 63   & 57       & 17   & 20             \\
    \hline
    \end{tabular}
\end{table}

\begin{table}
\small
\centering
\renewcommand{\arraystretch}{1.2} \tabcolsep=0.14cm
\caption{
Adopted spectroscopic solutions of the SB1 systems in the MONOS sample. 
The $T_0$ values correspond to the periastron passage for the eccentric orbits. 
Errors are in parentheses, corresponding to the last digits. 
An "f" means that the parameter was fixed for that solution. 
An asterisk after the mass function indicates that the value was calculated by us with the parameters of the orbital solution.
We denote ellipsoidal systems with an ''$\dag$'' after the SBS classification.
\label{t-orbsol-def}}
\begin{tabular}{l c c r@{.}l r@{.}l r@{.}l r@{.}l r@{.}l r@{.}l r@{.}l l l}
\hline \hline
\multicolumn{1}{c}{Name} & \multicolumn{2}{c}{SBS Classification} &  \multicolumn{2}{c}{Period} & \multicolumn{2}{c}{$T_{0}$} & \multicolumn{2}{c}{$e$} & \multicolumn{2}{c}{$\omega$} & \multicolumn{2}{c}{$K_1$} & \multicolumn{2}{c}{$\gamma_{1}$} & \multicolumn{2}{c}{f(m)} & \multicolumn{1}{c}{$v \sin i$} & \multicolumn{1}{c}{Ref} \\
\multicolumn{1}{c}{} & \multicolumn{1}{c}{MONOS~I} & \multicolumn{1}{c}{MONOS~II} & \multicolumn{2}{c}{d} & \multicolumn{2}{c}{JD$-2\,400\,000$} & \multicolumn{2}{c}{} & \multicolumn{2}{c}{degrees} & \multicolumn{2}{c}{km\,s$^{-1}$} & \multicolumn{2}{c}{km\,s$^{-1}$} & \multicolumn{2}{c}{M$_{\odot}$} & \multicolumn{1}{c}{km\,s$^{-1}$} & \multicolumn{1}{c}{} \\
\hline
HD~\num{164438}       & SB1     & SB1        & 10&249\,74(17)       & \num{55501}&114(74)          & 0&296(13)     & 220&7(26)   & 27&78(40)  & -14&4(25)  & 0&0198(9)      & 55(3)   & TP. HeI \\
V479~Sct              & SB1     & SB1        & 3&906\,09(12)        & \num{55017}&468(240)         & 0&272(107)    & 263&5(270)  & 19&73(243) & 26&52(185) & 0&002\,77(105) & 157(8)  & TP. OIII \\
Cyg~X-1               & SB1     & SB1E       &  5&599\,74(f)        & \num{56792}&37(11)           & 0&0(f)        & .&.         & 83&81(128) & 2&66(11)   & 0&341          & 95(5)   & TP. HeII \\
BD~+36\,4063          & SB1     & SB1E$^\dag$& 4&812\,17(3)         & \num{58381}&04(2)            & 0&01(1)       & 22&0(83)    & 160&0(3)   & 0&5(18)    & 2&06(10)       & 96(5)   & TP. comb. HeII \\
HDE~\num{229234}      & SB1     & SB1E$^\dag$& 3&510\,395(12)       & \num{56527}&114(13)          & 0&04(1)       & 56&0(20)    & 49&9(6)    & -29&6(5)   & 0&0451(16)     & 95(5)   & TP. comb. HeI \\
ALS~\num{15133}       & SB1     & SB1 unc.   & 2\,259&0(46)         & \num{53355}&0(134)           & 0&34(11)      & 157&9(260)  & 9&0(1)     & -13&0(1)   & 0&1419 *       & \ldots  & Ko12 \\
Cyg~OB2-A11           & SB1     & SB1        & 15&442\,8(25)        & \num{55761}&47(99)           & 0&136(59)     & 47&0(24)    & 23&4(14)   & -18&58(99) & 0&0199(37)     & 100(7)  & TP. comb. HeI \\
Cyg~OB2-22~B          & SB1     & SB1 unc.   & 38&0(2)              & \num{56719}&9(16)            & 0&21(20)      & 244&0(31)   & 9&5(17)    & -20&4(12)  & 0&0032 *       & \ldots  & Ko14 \\
Cyg~OB2-41            & SB1     & SB1        & 29&370\,3(49)        & \num{56542}&66(91)           & 0&21(5)       & 255&0(12)   & 41&8(15)   & -2&4(12)   & 0&208(29)      & \ldots  & TP. comb. HeI \\
ALS~\num{15148}       & SB1     & SB1E$^\dag$& 3&170\,36(11)        & \num{58679}&22(9)            & 0&06(5)       & 71&0(49)    & 28&3(13)   & -6&9(9)    & 0&0074(11)     & 177(9)  & TP. comb. HeI \\
Cyg~OB2-1             & SB1+Ca  & SB1E+Ca    & 4&852\,24(21)        & \num{56338}&35(29)           & 0&17(7)       & 238&0(200)  & 69&9(45)   & -27&5(32)  & 0&164(33)      & 203(10) & TP. comb. HeI \\
ALS~\num{15131}       & SB1     & SB1 unc.   & 4&625\,2(12)         & \num{55971}&8(67)            & 0&15(15)      & 240&0(52)   & 5&4(9)     & -17&4(6)   & 0&0001 *       & \ldots  & Ko14 \\
Cyg~OB2-20            & SB1     & SB1        & 25&125\,2(11)        & \num{53922}&81(22)           & 0&375(22)     & 124&0(41)   & 43&0(12)   & -11&13(81) & 0&164(14)      & 14(1)   & TP. comb. HeI \\
Cyg~OB2-70            & SB1     & SB1 unc.   & 245&1(3)             & \num{56549}&1(84)            & 0&51(17)      & 242&0(16)   & 14&5(29)   & -17&8(14)  & 0&049 *        & \ldots  & Ko14 \\
Cyg~OB2-15            & SB1     & SB1        & 14&657\,47(38)       & \num{52162}&41(54)           & 0&138(36)     & 53&0(14)    & 40&0(13)   & -10&2(10)  & 0&095(11)      & 163(8)  & TP. comb. HeI \\
ALS~\num{15115}       & SB1     & SB1 unc.   & \num{4066}&0(45)     & \num{52229}&0(29)            & 0&75(f)       & 352&0(9)    & 15&0(23)   & -12&1(11)  & 0&412 *        & \ldots  & Ko14 \\
Cyg~OB2-29            & SB1     & SB1 unc.   & 151&2(8)             & \num{56543}&7(38)            & 0&60(20)      & 15&0(10)    & 17&5(23)   & -10&7(8)   & 0&043 *        & \ldots  & Ko14 \\
Cyg~OB2-11            & SB1     & SB1        & 72&488(50)           & \num{54925}&4(11)            & 0&374(58)     & 269&4(72)   & 23&5(11)   & -27&1(8)   & 0&078(16)      & 75(3)   & TP. comb. HeI \\
V747~Cep              & SB1E    & SB1E       & 5&332\,37(15)        & \num{54401}&172\,3(1367)     & 0&37(3)       & 180&2(65)   & 99&6(36)   & -10&7(25)  & 0&436(49)      & 220(11) & TP. HeII \\
HD~\num{12323}        & SB1     & SB1E$^\dag$& 1&925\,14(6)         & \num{56229}&34(2)            & 0&0(f)        & .&.         & 29&82(221) & -46&1(12)  & 0&0053(12)     & 121(6)  & TP. comb. CC \\
HD~\num{16429}~Aa     & SB1+Ca  & SB1E+Cas   & 3&054\,42(f)         & \num{51893}&22(10)           & 0&17(4)       & 169&0(12)   & 136&0(6)   & -50&0(3)   & 0&76(1)        & \ldots  & Mc03 \\
HD~\num{15137}        & SB1     & SB1        & 55&399\,4(30)        & \num{40017}&483(819)         & 0&585\,2(267) & 153&18(351) & 13&92(57)  & -41&67(32) & 0&0082(16)     & 270(14) & TP. HeI+HeII \\
HD~\num{14633}~AaAb   & SB1+Ca? & SB1+Ca     & 15&409\,24(8)        & \num{51792}&635(32)          & 0&695(13)     & 139&6(14)   & 19&1(3)    & -41&55(16) & 0&0041(3)      & 121(6)  & TP. comb. HeI \\
HD~\num{37737}        & SB1     & SB1E       & 7&847\,031(32)       & \num{56342}&54(63)           & 0&407(21)     & 161&0(32)   & 70&8(22)   & -9&4(10)   & 0&220(21)      & 201(11) & TP. comb. CC \\
15~Mon~Aa             & SB1a    & SB2a       & \num{27112}&2(14819) & \num{50105}&48(\num{151941}) & 0&716(98)     & 69&2(111)   & 13&0(1)    & 33&6(10)   & 2&12           & \ldots  & Cv10 \\
HD~\num{46573}        & SB1     & SB1        & 10&653\,9(f)         & \num{53732}&53(14)           & 0&63(3)       & 265&04(720) & 11&2(5)    & 50&1(4)    & 0&000\,7(1)    & 77(3)   & TP. comb. HeI \\
$\theta^{1}$~Ori~CaCb & SB1?+Sa & SB1+Sa     & \num{4201}&0(200)    & \num{56546}&0(107)           & 0&546(90)     & 85&9(146)   & 15&2(19)   & 29&3(9)    & 0&9(4)         & 23(2)   & TP. OIII \\
HD~\num{52533}~A      & SB1     & SB1E       & 21&965\,2(14)        & \num{44615}&34(76)           & 0&364(55)     & 337&8(82)   & 82&2(44)   & 39&25(324) & 1&02(17)       & 299(15) & TP. comb. CC \\
\hline     
\end{tabular}
\renewcommand{\arraystretch}{1.0}

\tablebib{ (Ko12) \citet{Kobuetal12}           ; (Ko14) \citet{Kobuetal14}      ; (Mc03) \citet{McSw03}          ; (Cv10) \citet{Cvetetal10}; (TP) This work.
}

\end{table}
\end{landscape}

\subsubsection{Likely post-interaction systems}

$\,\!$\indent Five SB1 systems could have a compact object as an invisible companion: V479~Sct, Cyg X-1, HD~\num{15137}, HD~\num{14633}~AaAb, and HD~\num{12323}. 
If so, they must have gone through a supernova explosion and most likely a previous episode of mass transfer.
Three of them, V479~Sct, Cyg~X-1, and HD~\num{14633}~AaAb, display strong X-ray emission. The first two are certain, while the last is likely to have a compact companion.

As for the other two, the situation is unclear. HD~15~137 has a very small mass function. Even though \citet{McSwetal10} and \citet{McSwainetal11} were unable to detect X-ray emission from the companion, they could not discard the possibility of a quiescent BH.
HD~\num{12323}, being a runaway system with a small mass function, has similar characteristics, which is suggestive of a possible compact object as a companion. We note, however, that a compact object in such a close orbit should in all likelihood produce easily detectable high-energy emission, for which, to our knowledge, only an upper bound has been reported \citep{Chlebowskietal89}, although this implies a failed detection according to the authors. 

Cyg X-1 is one of the best studied systems in our sample. The close orbit suggests that the system underwent substantial mass transfer from the progenitor of the BH onto the O star, likely the cause of its low hydrogen content and moderately high rotation (we measure a $v \sin i=95$~\kms) \citep{Ziolkowski14}. This was supported by the relatively low mass of the O9.7\,Iab companion. Typically quoted values were close to $\rm{M_{BH}} = 14.8\pm0.1~\rm{M_{\sun}}$ and $M_{*} = 19.1\pm~2\rm{M_{\sun}}$ \citep[and references therein]{Orosetal11}. In recent years, however, a higher distance for Cyg X-1 has been measured, leading to a much higher companion mass. The recent work of \citet{MillerJetal21} suggests a much more massive system, with $\rm{M_{BH}} = 21.4~\rm{M_{\sun}}$ and $M_{*}\sim 41.1~\rm{M_{\sun}}$, in line with \citet{Ziolkowski05} or the upper limits in \citet{Mastroserioetal19}. Although the \textit{Gaia} Early Data Release 3 (EDR3) astrometric data favor the higher distance, we note that our value for $K_{1}$ is about 10\% higher than that in \citet{MillerJetal21}, suggesting that more work is needed on this iconic system.

\begin{table*}[!htp]
    \centering
    \renewcommand{\arraystretch}{1.2}
    \caption{Estimated component masses based on the mass function and primary's spectral types and a range of orbital inclinations. The mass function for each system is from the calculated orbit (see Table~\ref{t-orbsol-def}). \label{t:fm-sys}}
    \begin{tabular}{l l r@{.}l r@{.}l r@{-}l r@{-}l c}
    \hline \hline
        Object & Spectral type & \multicolumn{2}{c}{P} & \multicolumn{2}{c}{$f(M)$} & \multicolumn{2}{c}{M$_1$} & \multicolumn{2}{c}{M$_2$} & $i$ \\
         &  & \multicolumn{2}{c}{d} & \multicolumn{2}{c}{M$_\sun$} & \multicolumn{2}{c}{M$_\sun$} & \multicolumn{2}{c}{M$_\sun$} & degrees \\
        \hline
        \multicolumn{11}{l}{Pre-mass transfer systems}\\
        \hline
        HD~\num{164438}  & O9.2~IV        & 10&2 & 0&0198 & 15&20 & 1.8&4.6 & $30-90$ \\
        HDE~\num{229234} & O9 III         &  3&5 & 0&0450 & 18&23 & 3.1&7   & $30-70$ \\ 
        Cyg~OB2-41       & O9.7~III(n)    & 29&4 & 0&2080 & 15&23 & 4.3&10  & $35-90$ \\
        ALS~\num{15148}  & O6.5~V         &  3&1 & 0&0074 & 24&27 & 1.7&2.1 & $60-90$ \\
        Cyg~OB2-1        & O8~IV(n)((f))  &  4&9 & 0&1640 & 18&21 & 4.3&6   & $60-90$ \\
        Cyg~OB2-20       & O9.7~IV        & 25&1 & 0&1640 & 15&18 & 3.9&9   & $35-90$ \\
        Cyg~OB2-15       & O8~III         & 14&7 & 0&0950 & 20&25 & 3.8&11  & $25-90$ \\
        Cyg~OB2-11       & O5.5~Ifc       & 72&5 & 0&0780 & 35&50 &   7&19  & $20-70$ \\ 
        V747\,Cep        & O5.5 V(n)((f)) &  5&5 & 0&4360 & 27&30 &   8&11  & $60-90$ \\
        HD~\num{37737}   & O9.5 II-III(n) &  7&8 & 0&2200 & 18&25 &   5&7   & $60-90$ \\
        HD~\num{46573}   & O7~V((f))z     & 10&7 & 0&0007 & 21&23 & 0.8&1.5 & $30-60$ \\
        HD~\num{52533}~A & O8.5~IVn       & 22&0 & 1&0200 & 18&21 &   9&12  & $60-90$ \\
        \\
        \hline
        \multicolumn{11}{l}{Ongoing mass-transfer systems}\\
        \hline
        BD~+36~4063         & ON9.7~Ib    &  4&8 & 2&06 & 15&23 & 11&18 & $60-90$ \\
        HD~\num{16429}~AaAb & O9.2 III    &  3&0 & 0&76 & 18&23 &  8&11 & $60-90$ \\
        \\
        \hline
        \multicolumn{11}{l}{Collapsed companions} \\
        \hline
        V479~Sct            & ON6~V((f))z   &  3&9 & 0&0028 & 24&27  & 1.4&12  & $10-60$ \\
        Cyg~X-1             & O9.7~Iabp var &  5&6 & 0&3410 & 20&40  &   6&16  & $35-60$ \\
        HD~\num{12323}      & ON9.2~V       &  1&9 & 0&0053 & 15&18  & 1.4&1.6 & $60-90$ \\
        HD~\num{15137}      & O9~II-IIIn    & 55&4 & 0&0082 & 15&20  & 1.5&9   & $10-60$ \\
        HD~\num{14633}~AaAb & ON8.5~V       & 15&4 & 0&0041 & 18&20  & 1.3&9   & $10-60$ \\
        \\
        \hline
        \multicolumn{11}{l}{Colliding wind binary or with compact companion} \\ 
        \hline
        Cyg~OB2-A11     & O7~Ib(f)    & 15&4 & 0&0199 & 25&40 & 2.5&7 & $30-60$ \\ 
        \hline
    \end{tabular}
\end{table*}
\begin{figure*}[!htp]
    \centering
    \includegraphics[width=.495\textwidth]{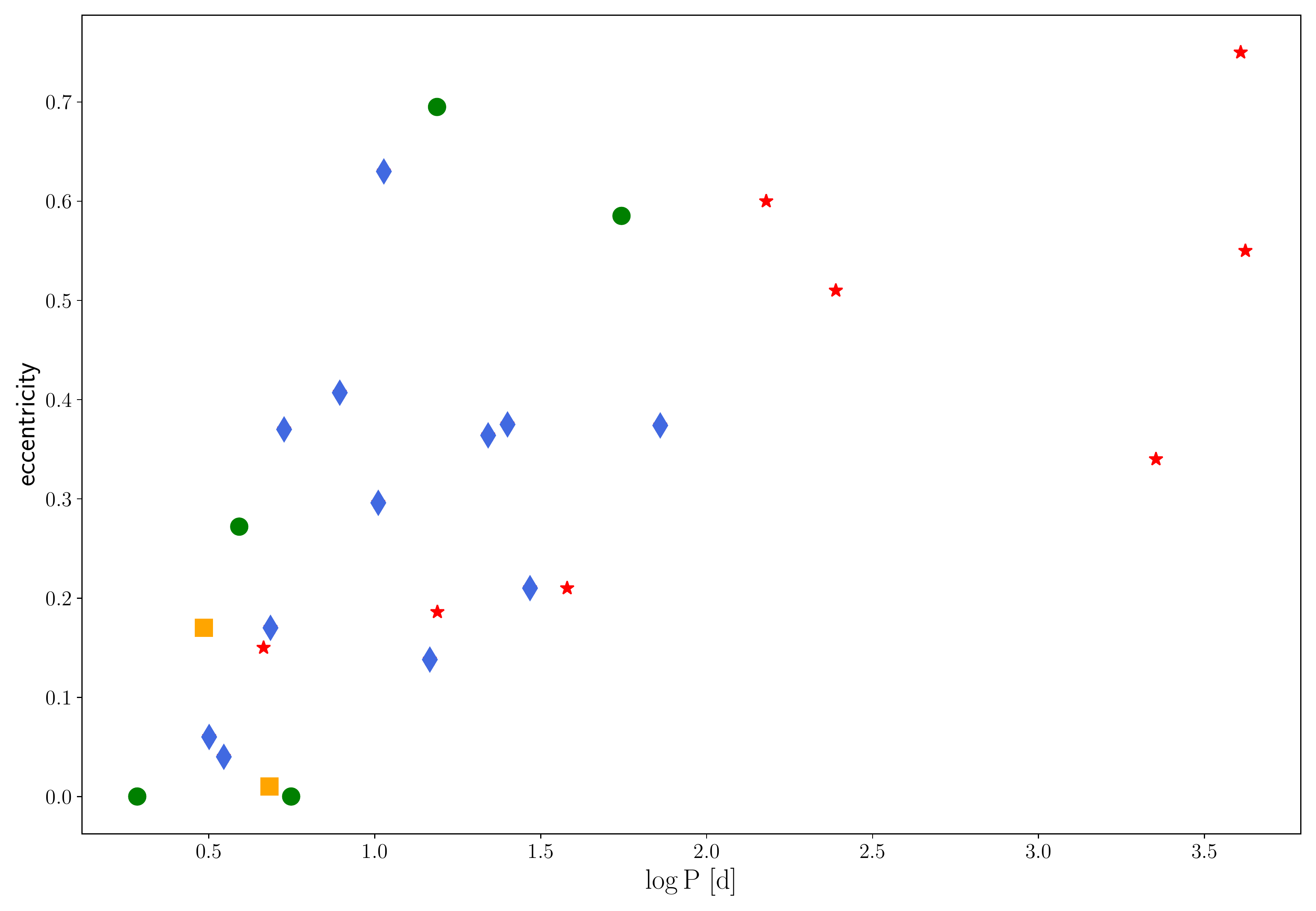} 
    \includegraphics[width=.495\textwidth]{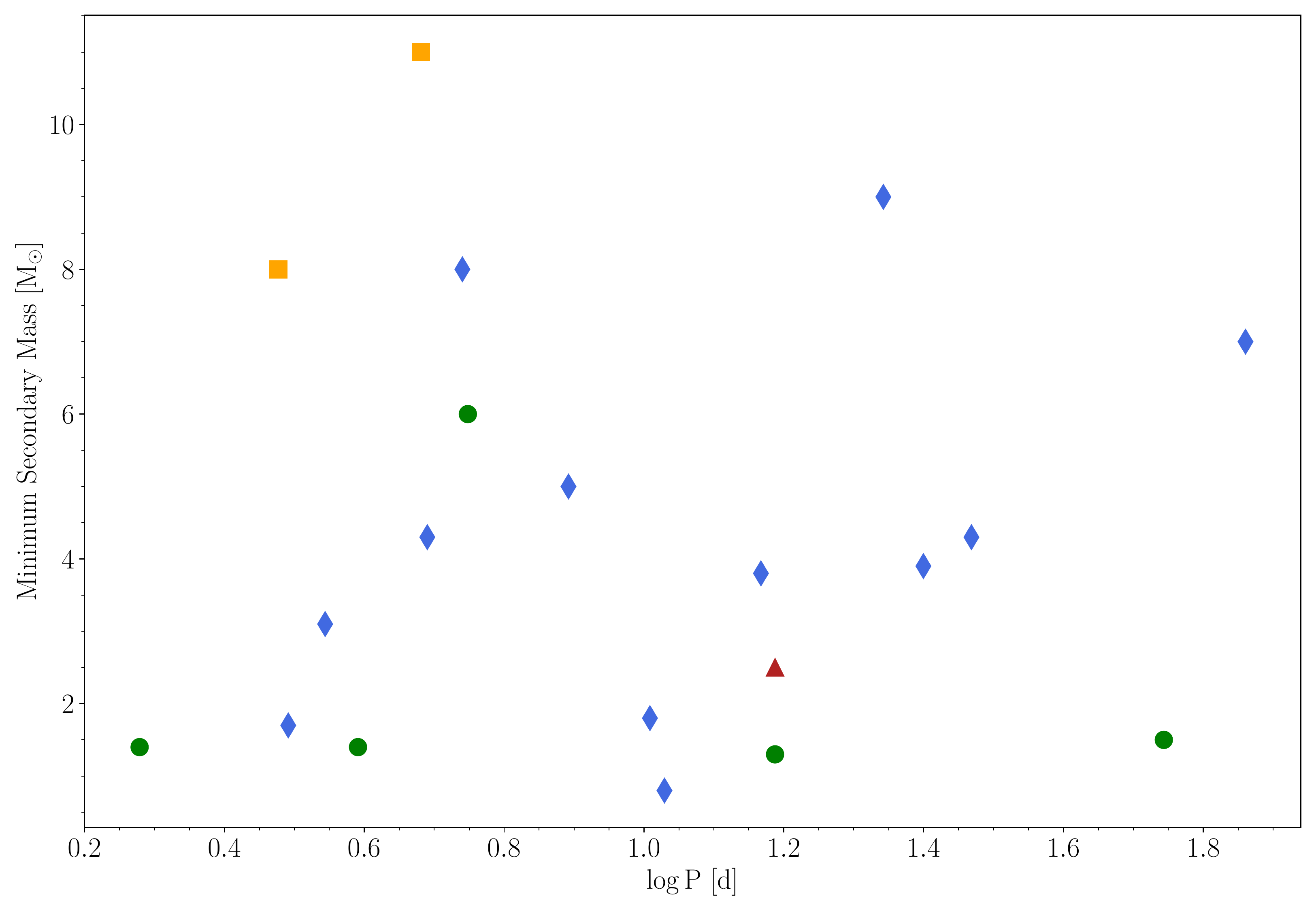}
    \caption{
    Distributions of the orbital parameter of the MONOS~II sample.
    Left panel: Eccentricity-period distribution of the MONOS~II SB1 sample. Systems in a pre-mass-transfer configuration are marked as blue diamonds, and those that are undergoing a mass-transfer event are marked as orange squares. Systems with a collapsed companion are marked as green circles. The systems marked as unc. in this work are also shown here as red stars.
    Right panel: Minimum secondary mass-period distribution of the MONOS~II SB1 sample. The colors are the same as for the upper panel. Cyg~OB2-A11 is shown here as a red triangle.}
    \label{fig:ecc-p}
\end{figure*}

\subsection{Orbital-parameter distributions and statistics}

$\,\!$\indent Short periods are favored within our sample. If we accept the published orbital solutions for six objects in the right column of Table~\ref{t-obj} (i.e.,\ those for which we have not improved existing solutions, with the exception of Cyg~OB2-22C), we find 11 systems out of a sample of 27  (41\%) that display periods shorter than $10$~d. 
Another 11 systems have intermediate periods, with the majority below 30~d. For the nearby systems $\theta^{1}$~Ori~CaCb and 15~Mon~Aa, the combination of astrometric and spectroscopic data allows the detection of very long periods  (several years). But the next longest period among the orbits confirmed corresponds to the supergiant Cyg~OB2-11, with 72.5~d. There are four other systems with periods longer than $P=100$~d within the group of unchecked orbits. They all come from the targeted search for binaries in Cygnus OB2  \citep{Kobuetal14} and are fainter than the other stars in the sample. This clearly illustrates the observational biases present in our sample and implies a need for caution when interpreting the period distribution. Although all other similar studies have found approximately the same distribution \citep{Sanaetal12a, Barbaetal17, Kobuetal14, Almeidaetal17}, it may be to a large degree dictated by an observational bias, as a star needs to have been detected as an SB1 to be included in our sample. Only the sample of \citet{Kobuetal14} is free from this initial bias, although it is still subject to those imposed by their observational strategy.

Figure~\ref{fig:ecc-p} shows secondary minimum mass-period and eccentricity-period distributions. If we analyze the eccentricity distribution, we find that the sample is roughly evenly distributed, with a slight preference for low values. The short-period systems have lower eccentricities, while longer-period systems have higher eccentricities; again, this trend has also been reported before by other authors. 
Our sample falls on the central region of the eccentricity-period plane.
The lack of eccentric systems with short periods is a fully expected physical effect due to the limitation imposed by the size of the stars themselves. 
There is also a significant lack of longer-period systems with small eccentricities. This could be an observational bias due to the difficulty in detecting low amplitude binaries with long periods. If we take into account the aforementioned findings of \citet{SimonDetal20a} regarding RV variability due to stellar pulsations and consider that they will necessarily be superimposed on putative orbital RVs, it becomes clear that detecting long-period systems with supergiant or fast-rotating primaries is a very complex task. Therefore, a meaningful interpretation of the implications of our findings will require a sophisticated analysis of all these biases to be conducted when the whole sample of binary systems has been revised.




\begin{acknowledgements}
      E.T.P., I.N.D. and S.S.D. acknowledge support from the Spanish Government Ministerio de Ciencia through grant AYA2015-68 012-C2-1/2-P. E.T.P. and J.M.A. acknowledge support from the Spanish Government Ministerio de Ciencia through grants AYA2016-75\,931-C2-2-P and PGC2018-095\,049-B-C22. 
      R.H.B. acknowledges support from the ESAC Faculty Council Visitor Program and ANID FONDECYT Project 1211903. 
      GH acknowledges support from the Spanish Government Ministerio de Ciencia through grant PGC2018-95049-B-CC22 and ESA Contract No. 4000126507/19/ES/CM. 
      This research has made use of the SIMBAD database, operated at CDS (Strasbourg, France), NASA's Astrophysics Data System Bibliographic Services and the Python programming language (Python Software Foundation), Lightkurve, a Python package for Kepler and TESS data analysis (Lightkurve Collaboration, 2018). 
\end{acknowledgements}

\bibliographystyle{aa} 
\interlinepenalty=10000
\bibliography{monos-II-ref} 

\begin{thebibliography}{185}
\expandafter\ifx\csname natexlab\endcsname\relax\def\natexlab#1{#1}\fi

\bibitem[{ESA(1997)}]{ESA97a}
 1997, {The HIPPARCOS and TYCHO catalogues. Astrometric and photometric star
  catalogues derived from the ESA HIPPARCOS Space Astrometry Mission}, Vol.
  1200

\bibitem[{{Abdo} {et~al.}(2009){Abdo}, {Ackermann}, {Ajello}, {Atwood},
  {Axelsson}, {Baldini}, {Ballet}, {Barbiellini}, {Bastieri}, {Baughman},
  {Bechtol}, {Bellazzini}, {Berenji}, {Bland ford}, {Bloom}, {Bonamente},
  {Borgland }, {Bregeon}, {Brez}, {Brigida}, {Bruel}, {Burnett}, {Buson},
  {Caliand ro}, {Cameron}, {Caraveo}, {Casand jian}, {Cavazzuti}, {Cecchi},
  {{\c{C}}elik}, {Chaty}, {Chekhtman}, {Cheung}, {Chiang}, {Ciprini}, {Claus},
  {Cohen-Tanugi}, {Cominsky}, {Conrad}, {Corbel}, {Corbet}, {Cutini}, {Dermer},
  {de Angelis}, {de Palma}, {Digel}, {Silva}, {Drell}, {Dubois}, {Dubus},
  {Dumora}, {Farnier}, {Favuzzi}, {Fegan}, {Focke}, {Fortin}, {Frailis},
  {Fukazawa}, {Funk}, {Fusco}, {Gargano}, {Gasparrini}, {Gehrels}, {Germani},
  {Giebels}, {Giglietto}, {Giordano}, {Glanzman}, {Godfrey}, {Grenier},
  {Grondin}, {Grove}, {Guillemot}, {Guiriec}, {Hanabata}, {Harding},
  {Hayashida}, {Hays}, {Hill}, {Horan}, {Hughes}, {Jackson}, {J{\'o}hannesson},
  {Johnson}, {Johnson}, {Johnson}, {Kamae}, {Katagiri}, {Kataoka}, {Kawai},
  {Kerr}, {Kn{\"o}dlseder}, {Kocian}, {Kuehn}, {Kuss}, {Lande}, {Larsson},
  {Latronico}, {Lemoine-Goumard}, {Longo}, {Loparco}, {Lott}, {Lovellette},
  {Lubrano}, {Madejski}, {Makeev}, {Marelli}, {Mazziotta}, {McEnery}, {Meurer},
  {Michelson}, {Mitthumsiri}, {Mizuno}, {Moiseev}, {Monte}, {Monzani},
  {Morselli}, {Moskalenko}, {Murgia}, {Nolan}, {Norris}, {Nuss}, {Ohsugi},
  {Omodei}, {Orlando}, {Ormes}, {Ozaki}, {Paneque}, {Panetta}, {Parent},
  {Pelassa}, {Pepe}, {Pesce-Rollins}, {Piron}, {Porter}, {Rain{\`o}}, {Rando},
  {Ray}, {Razzano}, {Rea}, {Reimer}, {Reimer}, {Reposeur}, {Ritz}, {Rochester},
  {Rodriguez}, {Romani}, {Roth}, {Ryde}, {Sadrozinski}, {Sanchez}, {Sander},
  {Saz Parkinson}, {Scargle}, {Sgr{\`o}}, {Sierpowska-Bartosik}, {Siskind},
  {Smith}, {Smith}, {Spandre}, {Spinelli}, {Strickman}, {Suson}, {Tajima},
  {Takahashi}, {Takahashi}, {Tanaka}, {Tanaka}, {Thayer}, {Thompson},
  {Tibaldo}, {Torres}, {Tosti}, {Tramacere}, {Uchiyama}, {Usher}, {Vasileiou},
  {Venter}, {Vilchez}, {Vitale}, {Waite}, {Wallace}, {Wang}, {Winer}, {Wood},
  {Ylinen}, \& {Ziegler}}]{Fermi09}
{Abdo}, A.~A., {Ackermann}, M., {Ajello}, M., {et~al.} 2009, \apjl, 706, L56

\bibitem[{{Aerts} \& {Rogers}(2015)}]{AaertsRogers15}
{Aerts}, C. \& {Rogers}, T.~M. 2015, \apjl, 806, L33

\bibitem[{{Aerts} {et~al.}(2017){Aerts}, {S{\'\i}mon-D{\'\i}az}, {Bloemen},
  {Debosscher}, {P{\'a}pics}, {Bryson}, {Still}, {Moravveji}, {Williamson},
  {Grundahl}, {Fredslund Andersen}, {Antoci}, {Pall{\'e}},
  {Christensen-Dalsgaard}, \& {Rogers}}]{Aertsetal17}
{Aerts}, C., {S{\'\i}mon-D{\'\i}az}, S., {Bloemen}, S., {et~al.} 2017, \aap,
  602, A32

\bibitem[{{Aharonian} {et~al.}(2006){Aharonian}, {Akhperjanian}, {Bazer-Bachi},
  {Beilicke}, {Benbow}, {Berge}, {Bernl{\"o}hr}, {Boisson}, {Bolz}, {Borrel},
  {Braun}, {Brown}, {B{\"u}hler}, {B{\"u}sching}, {Carrigan}, {Chadwick},
  {Chounet}, {Cornils}, {Costamante}, {Degrange}, {Dickinson},
  {Djannati-Ata{\"\i}}, {O'C. Drury}, {Dubus}, {Egberts}, {Emmanoulopoulos},
  {Espigat}, {Feinstein}, {Ferrero}, {Fiasson}, {Fontaine}, {Funk}, {Funk},
  {F{\"u}{\ss}ling}, {Gallant}, {Giebels}, {Glicenstein}, {Goret},
  {Hadjichristidis}, {Hauser}, {Hauser}, {Heinzelmann}, {Henri}, {Hermann},
  {Hinton}, {Hoffmann}, {Hofmann}, {Holleran}, {Horns}, {Jacholkowska}, {de
  Jager}, {Kendziorra}, {Kh{\'e}lifi}, {Komin}, {Konopelko}, {Kosack},
  {Latham}, {Le Gallou}, {Lemi{\`e}re}, {Lemoine-Goumard}, {Lohse}, {Martin},
  {Martineau-Huynh}, {Marcowith}, {Masterson}, {Maurin}, {McComb}, {Moulin},
  {de Naurois}, {Nedbal}, {Nolan}, {Noutsos}, {Orford}, {Osborne}, {Ouchrif},
  {Panter}, {Pelletier}, {Pita}, {P{\"u}hlhofer}, {Punch}, {Raubenheimer},
  {Raue}, {Rayner}, {Reimer}, {Reimer}, {Ripken}, {Rob}, {Rolland }, {Rowell},
  {Sahakian}, {Santangelo}, {Saug{\'e}}, {Schlenker}, {Schlickeiser},
  {Schr{\"o}der}, {Schwanke}, {Schwarzburg}, {Shalchi}, {Sol}, {Spangler},
  {Spanier}, {Steenkamp}, {Stegmann}, {Superina}, {Tavernet}, {Terrier},
  {Tluczykont}, {van Eldik}, {Vasileiadis}, {Venter}, {Vincent}, {V{\"o}lk},
  {Wagner}, \& {Ward}}]{HESS06b}
{Aharonian}, F., {Akhperjanian}, A.~G., {Bazer-Bachi}, A.~R., {et~al.} 2006,
  \aap, 460, 743

\bibitem[{Aldoretta {et~al.}(2015)Aldoretta, Caballero-Nieves, Gies, Nelan,
  Wallace, Hartkopf, Henry, Jao, Ma{\'\i}z~Apell{\'a}niz, Mason, Moffat,
  Norris, Richardson, \& Williams}]{Aldoretal15}
Aldoretta, E.~J., Caballero-Nieves, S.~M., Gies, D.~R., {et~al.} 2015, AJ, 149,
  26

\bibitem[{{Alduseva} {et~al.}(1982){Alduseva}, {Aslanov}, {Kolotilov}, \&
  {Cherepashchuk}}]{Aldusevaetal82}
{Alduseva}, V.~I., {Aslanov}, A.~A., {Kolotilov}, E.~A., \& {Cherepashchuk},
  A.~M. 1982, Soviet Astronomy Letters, 8, 386

\bibitem[{{Alexeeva} {et~al.}(2013){Alexeeva}, {Sobolev}, {Gorda}, {Yushkin},
  \& {McSwain}}]{Alexeeva13}
{Alexeeva}, S.~A., {Sobolev}, A.~M., {Gorda}, S.~Y., {Yushkin}, M.~V., \&
  {McSwain}, V. 2013, Astrophysical Bulletin, 68, 169

\bibitem[{{Almeida} {et~al.}(2017){Almeida}, {Sana}, {Taylor}, {Barb{\'a}},
  {Bonanos}, {Crowther}, {Damineli}, {de Koter}, {de Mink}, {Evans}, {Gieles},
  {Grin}, {H{\'e}nault-Brunet}, {Langer}, {Lennon}, {Lockwood}, {Ma{\'\i}z
  Apell{\'a}niz}, {Moffat}, {Neijssel}, {Norman}, {Ram{\'\i}rez-Agudelo},
  {Richardson}, {Schootemeijer}, {Shenar}, {Soszy{\'n}ski}, {Tramper}, \&
  {Vink}}]{Almeidaetal17}
{Almeida}, L.~A., {Sana}, H., {Taylor}, W., {et~al.} 2017, \aap, 598, A84

\bibitem[{{Aragona} {et~al.}(2009){Aragona}, {McSwain}, {Grundstrom}, {Marsh},
  {Roettenbacher}, {Hessler}, {Boyajian}, \& {Ray}}]{Aragona09}
{Aragona}, C., {McSwain}, M.~V., {Grundstrom}, E.~D., {et~al.} 2009, \apj, 698,
  514

\bibitem[{{Aslanov} \& {Barannikov}(1989)}]{Aslanovetal89}
{Aslanov}, A.~A. \& {Barannikov}, A.~A. 1989, Pisma v Astronomicheskii Zhurnal,
  15, 732

\bibitem[{{Aslanov} \& {Barannikov}(1992)}]{Aslanovetal92}
{Aslanov}, A.~A. \& {Barannikov}, A.~A. 1992, Soviet Astronomy Letters, 18, 58

\bibitem[{{Aslanov} {et~al.}(1984){Aslanov}, {Kornilova}, \&
  {Cherepashchuk}}]{Aslanovetal84}
{Aslanov}, A.~A., {Kornilova}, L.~N., \& {Cherepashchuk}, A.~M. 1984, Soviet
  Astronomy Letters, 10, 278

\bibitem[{{Balega} {et~al.}(2015){Balega}, {Chentsov}, {Rzaev}, \&
  {Weigelt}}]{Balegaetal15}
{Balega}, Y.~Y., {Chentsov}, E.~L., {Rzaev}, A.~K., \& {Weigelt}, G. 2015, in
  Astronomical Society of the Pacific Conference Series, Vol. 494, Physics and
  Evolution of Magnetic and Related Stars, ed. Y.~Y. {Balega}, I.~I.
  {Romanyuk}, \& D.~O. {Kudryavtsev}, 57

\bibitem[{Barannikov(1993)}]{Bara93}
Barannikov, A.~A. 1993, Astronomy Letters, 19, 420

\bibitem[{{Barannikov}(1999)}]{Barannikov99}
{Barannikov}, A.~A. 1999, Astronomy Letters, 25, 169

\bibitem[{{Barb{\'a}} {et~al.}(2017){Barb{\'a}}, {Gamen}, {Arias}, \&
  {Morrell}}]{Barbaetal17}
{Barb{\'a}}, R.~H., {Gamen}, R., {Arias}, J.~I., \& {Morrell}, N.~I. 2017, in
  IAU Symposium, Vol. 329, The Lives and Death-Throes of Massive Stars, ed.
  J.~J. {Eldridge}, J.~C. {Bray}, L.~A.~S. {McClelland}, \& L.~{Xiao}, 89--96

\bibitem[{Barb{\'a} {et~al.}(2010)Barb{\'a}, Gamen, Arias, Morrell,
  Ma{\'{\i}}z~Apell{\'a}niz, Alfaro, Walborn, \& Sota}]{barbetal10}
Barb{\'a}, R.~H., Gamen, R.~C., Arias, J.~I., {et~al.} 2010, in RMxAC, Vol.~38,
  30--32

\bibitem[{{Bareilles}(2017)}]{GBART}
{Bareilles}, F. 2017, {GBART: Determination of the orbital elements of
  spectroscopic binaries}

\bibitem[{{Bekenstein} \& {Bowers}(1974)}]{BekenBower74}
{Bekenstein}, J.~D. \& {Bowers}, R.~L. 1974, \apj, 190, 653

\bibitem[{{Berlanas} {et~al.}(2020){Berlanas}, {Herrero}, {Comer{\'o}n},
  {Sim{\'o}n-D{\'\i}az}, {Lennon}, {Pasquali}, {Ma{\'\i}z Apell{\'a}niz},
  {Sota}, \& {Peller{\'\i}n}}]{Sara20}
{Berlanas}, S.~R., {Herrero}, A., {Comer{\'o}n}, F., {et~al.} 2020, \aap, 642,
  A168

\bibitem[{{Boeche} {et~al.}(2004){Boeche}, {Munari}, {Tomasella}, \&
  {Barbon}}]{Boecheetal04}
{Boeche}, C., {Munari}, U., {Tomasella}, L., \& {Barbon}, R. 2004, \aap, 415,
  145

\bibitem[{{Bohannan} \& {Garmany}(1978)}]{Bohannanetal78}
{Bohannan}, B. \& {Garmany}, C.~D. 1978, \apj, 223, 908

\bibitem[{{Bolton} \& {Rogers}(1978)}]{BoltonRogers78}
{Bolton}, C.~T. \& {Rogers}, G.~L. 1978, \apj, 222, 234

\bibitem[{Bouret {et~al.}(2012)Bouret, Hillier, Lanz, \&
  Fullerton}]{Bouretal12}
Bouret, J.-C., Hillier, D.~J., Lanz, T., \& Fullerton, A.~W. 2012, A\&A, 544,
  A67

\bibitem[{{Bowman} {et~al.}(2020){Bowman}, {Burssens}, {Sim{\'o}n-D{\'\i}az},
  {Edelmann}, {Rogers}, {Horst}, {R{\"o}pke}, \& {Aerts}}]{Bowmanetal20}
{Bowman}, D.~M., {Burssens}, S., {Sim{\'o}n-D{\'\i}az}, S., {et~al.} 2020,
  \aap, 640, A36

\bibitem[{{Boyajian} {et~al.}(2005){Boyajian}, {Beaulieu}, {Gies},
  {Grundstrom}, {Huang}, {McSwain}, {Riddle}, {Wingert}, \& {De
  Becker}}]{Boyajianetal05}
{Boyajian}, T.~S., {Beaulieu}, T.~D., {Gies}, D.~R., {et~al.} 2005, \apj, 621,
  978

\bibitem[{{Brasseur} {et~al.}(2019){Brasseur}, {Phillip}, {Fleming},
  {Mullally}, \& {White}}]{Brasseur19}
{Brasseur}, C.~E., {Phillip}, C., {Fleming}, S.~W., {Mullally}, S.~E., \&
  {White}, R.~L. 2019, {Astrocut: Tools for creating cutouts of TESS images}

\bibitem[{{Brocksopp} {et~al.}(1999){Brocksopp}, {Tarasov}, {Lyuty}, \&
  {Roche}}]{Brocksoppetal99a}
{Brocksopp}, C., {Tarasov}, A.~E., {Lyuty}, V.~M., \& {Roche}, P. 1999, A\&A,
  343, 861

\bibitem[{{Burggraaff} {et~al.}(2018){Burggraaff}, {Talens}, {Spronck},
  {Lesage}, {Stuik}, {Otten}, {Van Eylen}, {Pollacco}, \&
  {Snellen}}]{Burggraaffetal18}
{Burggraaff}, O., {Talens}, G.~J.~J., {Spronck}, J., {et~al.} 2018, \aap, 617,
  A32

\bibitem[{{Burssens} {et~al.}(2020){Burssens}, {Sim{\'o}n-D{\'\i}az}, {Bowman},
  {Holgado}, {Michielsen}, {de Burgos}, {Castro}, {Barb{\'a}}, \&
  {Aerts}}]{Burssens20}
{Burssens}, S., {Sim{\'o}n-D{\'\i}az}, S., {Bowman}, D.~M., {et~al.} 2020,
  \aap, 639, A81

\bibitem[{Caballero-Nieves {et~al.}(2014)Caballero-Nieves, Nelan, Gies,
  Wallace, DeGioia-Eastwood, Herrero, Jao, Mason, Massey, Moffat, \&
  Walborn}]{CabNetal14}
Caballero-Nieves, S.~M., Nelan, E.~P., Gies, D.~R., {et~al.} 2014, AJ, 147, 40

\bibitem[{{Campillay} {et~al.}(2019){Campillay}, {Arias}, {Barb{\'a}},
  {Morrell}, {Gamen}, \& {Ma{\'\i}z Apell{\'a}niz}}]{Campillayetal19}
{Campillay}, A.~R., {Arias}, J.~I., {Barb{\'a}}, R.~H., {et~al.} 2019, \mnras,
  484, 2137

\bibitem[{{Casares} {et~al.}(2005){Casares}, {Rib{\'o}}, {Ribas}, {Paredes},
  {Mart{\'\i}}, \& {Herrero}}]{Casaresetal05}
{Casares}, J., {Rib{\'o}}, M., {Ribas}, I., {et~al.} 2005, \mnras, 364, 899

\bibitem[{{Cherepashchuk} \& {Aslanov}(1984)}]{1984Ap+SS.102...97C}
{Cherepashchuk}, A.~M. \& {Aslanov}, A.~A. 1984, \apss, 102, 97

\bibitem[{Chini {et~al.}(2012)Chini, Hoffmeister, Nasseri, Stahl, \&
  Zinnecker}]{Chinetal12}
Chini, R., Hoffmeister, V.~H., Nasseri, A., Stahl, O., \& Zinnecker, H. 2012,
  MNRAS, 424, 1925

\bibitem[{{Chlebowski} {et~al.}(1989){Chlebowski}, {Harnden}, \&
  {Sciortino}}]{Chlebowskietal89}
{Chlebowski}, T., {Harnden}, F.~R., J., \& {Sciortino}, S. 1989, \apj, 341, 427

\bibitem[{Conti \& Alschuler(1971)}]{ContAlsc71}
Conti, P.~S. \& Alschuler, W.~R. 1971, ApJ, 170, 325

\bibitem[{{Conti} {et~al.}(1977){Conti}, {Leep}, \& {Lorre}}]{Contietal77}
{Conti}, P.~S., {Leep}, E.~M., \& {Lorre}, J.~J. 1977, ApJ, 214, 759

\bibitem[{Cruz-Gonz\'alez {et~al.}(1974)Cruz-Gonz\'alez, Recillas-Cruz,
  Costero, Peimbert, \& Torres-Peimbert}]{1974rmxaa...1..211c}
Cruz-Gonz\'alez, C., Recillas-Cruz, E., Costero, R., Peimbert, M., \&
  Torres-Peimbert, S. 1974, \rmxaa, 1, 211

\bibitem[{Cvetkovic {et~al.}(2009)Cvetkovic, Vince, \& Ninkovic}]{Cvetetal09}
Cvetkovic, Z., Vince, I., \& Ninkovic, S. 2009, Publications of the
  Astronomical Observatory of Belgrade, 86, 331

\bibitem[{Cvetkovi{\'c} {et~al.}(2010)Cvetkovi{\'c}, Vince, \&
  Ninkovi{\'c}}]{Cvetetal10}
Cvetkovi{\'c}, Z., Vince, I., \& Ninkovi{\'c}, S. 2010, New Astronomy, 15, 302

\bibitem[{De~Becker {et~al.}(2008)De~Becker, Linder, \& Rauw}]{DeBeetal08}
De~Becker, M., Linder, N., \& Rauw, G. 2008, Information Bulletin on Variable
  Stars, 5841, 1

\bibitem[{De~Becker \& Rauw(2004)}]{DeBeRauw04}
De~Becker, M. \& Rauw, G. 2004, A\&A, 427, 995

\bibitem[{{de Jager} {et~al.}(1979){de Jager}, {Lamers}, {Macchetto}, \&
  {Snow}}]{de-Jageretal79}
{de Jager}, C., {Lamers}, H.~J.~G.~L.~M., {Macchetto}, F., \& {Snow}, T.~P.
  1979, \aap, 79, L28

\bibitem[{de~Wit {et~al.}(2005)de~Wit, Testi, Palla, \& Zinnecker}]{deWietal05}
de~Wit, W.~J., Testi, L., Palla, F., \& Zinnecker, H. 2005, A\&A, 437, 247

\bibitem[{{Dubus}(2013)}]{Dubus13}
{Dubus}, G. 2013, \aapr, 21, 64

\bibitem[{{Ebbets}(1980)}]{Ebbetsetal80}
{Ebbets}, D. 1980, \apj, 235, 97

\bibitem[{Etzel(2004)}]{SBOP}
Etzel, P. 2004, SBOP: Spectroscopic Binary Orbit Program

\bibitem[{{Ferrario} \& {Wickramasinghe}(2005)}]{FerrarioWick05}
{Ferrario}, L. \& {Wickramasinghe}, D.~T. 2005, \mnras, 356, 615

\bibitem[{{Frost} \& {Adams}(1904)}]{Frostetal04}
{Frost}, E.~B. \& {Adams}, W.~S. 1904, \apj, 19, 151

\bibitem[{{Fullerton} {et~al.}(1996){Fullerton}, {Gies}, \&
  {Bolton}}]{Fullertonetal96}
{Fullerton}, A.~W., {Gies}, D.~R., \& {Bolton}, C.~T. 1996, \apjs, 103, 475

\bibitem[{Gamen {et~al.}(2015)Gamen, Barb{\'a}, Walborn, Morrell, Arias,
  Ma{\'{\i}}z~Apell{\'a}niz, Sota, \& Alfaro}]{Gameetal15}
Gamen, R., Barb{\'a}, R.~H., Walborn, N.~R., {et~al.} 2015, A\&A, 583, L4

\bibitem[{{Garmany} {et~al.}(1980){Garmany}, {Conti}, \&
  {Massey}}]{Garmanyetal80}
{Garmany}, C.~D., {Conti}, P.~S., \& {Massey}, P. 1980, ApJ, 242, 1063

\bibitem[{{Gies} \& {Bolton}(1982)}]{GiesBolton82}
{Gies}, D.~R. \& {Bolton}, C.~T. 1982, \apj, 260, 240

\bibitem[{{Gies} \& {Bolton}(1986)}]{Giesetal86}
{Gies}, D.~R. \& {Bolton}, C.~T. 1986, The Astrophysical Journal Supplement
  Series, 61, 419

\bibitem[{Gies {et~al.}(2008)Gies, Bolton, Blake, Caballero-Nieves, Crenshaw,
  Hadrava, Herrero, Hillwig, Howell, Huang, Kaper, Koubsk{\'y}, \&
  McSwain}]{gies08}
Gies, D.~R., Bolton, C.~T., Blake, R.~M., {et~al.} 2008, ApJ, 678, 1237

\bibitem[{{Gies} {et~al.}(2003){Gies}, {Bolton}, {Thomson}, {Huang}, {McSwain},
  {Riddle}, {Wang}, {Wiita}, {Wingert}, {Cs{\'a}k}, \& {Kiss}}]{giesetal03}
{Gies}, D.~R., {Bolton}, C.~T., {Thomson}, J.~R., {et~al.} 2003, ApJ, 583, 424

\bibitem[{Gies {et~al.}(1997)Gies, Mason, Bagnuolo, Hahula, Hartkopf,
  McAlister, Thaller, McKibben, \& Penny}]{Giesetal97}
Gies, D.~R., Mason, B.~D., Bagnuolo, Jr., W.~G., {et~al.} 1997, ApJL, 475, L49

\bibitem[{Gies {et~al.}(1993)Gies, Mason, Hartkopf, McAlister, Frazin, Hahula,
  Penny, Thaller, Fullerton, \& Shara}]{Giesetal93}
Gies, D.~R., Mason, B.~D., Hartkopf, W.~I., {et~al.} 1993, AJ, 106, 2072

\bibitem[{{Grassitelli} {et~al.}(2015){Grassitelli}, {Fossati},
  {Sim{\'o}n-Di{\'a}z}, {Langer}, {Castro}, \& {Sanyal}}]{Grassitellietal15}
{Grassitelli}, L., {Fossati}, L., {Sim{\'o}n-Di{\'a}z}, S., {et~al.} 2015,
  \apjl, 808, L31

\bibitem[{{Gravity Collaboration} {et~al.}(2018){Gravity Collaboration},
  {Karl}, {Pfuhl}, {Eisenhauer}, {Genzel}, {Grellmann}, {Habibi}, {Abuter},
  {Accardo}, {Amorim}, {Anugu}, {{\'A}vila}, {Benisty}, {Berger}, {Blind},
  {Bonnet}, {Bourget}, {Brandner}, {Brast}, {Buron}, {Caratti O Garatti},
  {Chapron}, {Cl{\'e}net}, {Collin}, {Coud{\'e} Du Foresto}, {de Wit}, {de
  Zeeuw}, {Deen}, {Delplancke-Str{\"o}bele}, {Dembet}, {Derie}, {Dexter},
  {Duvert}, {Ebert}, {Eckart}, {Esselborn}, {F{\'e}dou}, {Finger}, {Garcia},
  {Garcia Dabo}, {Garcia Lopez}, {Gao}, {Gendron}, {Gillessen}, {Gont{\'e}},
  {Gordo}, {Gr{\"o}zinger}, {Guajardo}, {Guieu}, {Haguenauer}, {Hans},
  {Haubois}, {Haug}, {Hau{\ss}mann}, {Henning}, {Hippler}, {Horrobin}, {Huber},
  {Hubert}, {Hubin}, {Jakob}, {Jochum}, {Jocou}, {Kaufer}, {Kellner},
  {Kendrew}, {Kern}, {Kervella}, {Kiekebusch}, {Klein}, {K{\"o}hler}, {Kolb},
  {Kulas}, {Lacour}, {Lapeyr{\`e}re}, {Lazareff}, {Le Bouquin}, {L{\'e}na},
  {Lenzen}, {L{\'e}v{\^e}que}, {Lin}, {Lippa}, {Magnard}, {Mehrgan},
  {M{\'e}rand}, {Moulin}, {M{\"u}ller}, {M{\"u}ller}, {Neumann}, {Oberti},
  {Ott}, {Pallanca}, {Pand uro}, {Pasquini}, {Paumard}, {Percheron}, {Perraut},
  {Perrin}, {Pfl{\"u}ger}, {Duc}, {Plewa}, {Popovic}, {Rabien}, {Ram{\'\i}rez},
  {Ramos}, {Rau}, {Riquelme}, {Rodr{\'\i}guez-Coira}, {Rohloff}, {Rosales},
  {Rousset}, {Sanchez-Bermudez}, {Scheithauer}, {Sch{\"o}ller}, {Schuhler},
  {Spyromilio}, {Straub}, {Straubmeier}, {Sturm}, {Suarez}, {Tristram},
  {Ventura}, {Vincent}, {Waisberg}, {Wank}, {Widmann}, {Wieprecht}, {Wiest},
  {Wiezorrek}, {Wittkowski}, {Woillez}, {Wolff}, {Yazici}, {Ziegler}, \&
  {Zins}}]{Gravity18}
{Gravity Collaboration}, {Karl}, M., {Pfuhl}, O., {et~al.} 2018, \aap, 620,
  A116

\bibitem[{Hadasch {et~al.}(2012)Hadasch, Torres, Tanaka, Corbet, Hill, Dubois,
  Dubus, Glanzman, Corbel, Li, Chen, Zhang, Caliandro, Kerr, Richards,
  Max-Moerbeck, Readhead, \& Pooley}]{Hadaetal12}
Hadasch, D., Torres, D.~F., Tanaka, T., {et~al.} 2012, ApJ, 749, 54

\bibitem[{{Higgins} \& {Vink}(2019)}]{Higginsetal19}
{Higgins}, E.~R. \& {Vink}, J.~S. 2019, \aap, 622, A50

\bibitem[{{Hill}(1967)}]{Hilletal67}
{Hill}, G. 1967, \apjs, 14, 263

\bibitem[{{Holgado} {et~al.}(2018){Holgado}, {Sim{\'o}n-D{\'\i}az},
  {Barb{\'a}}, {Puls}, {Herrero}, {Castro}, {Garcia}, {Ma{\'\i}z
  Apell{\'a}niz}, {Negueruela}, \& {Sab{\'\i}n-Sanjuli{\'a}n}}]{Holgadoetal18}
{Holgado}, G., {Sim{\'o}n-D{\'\i}az}, S., {Barb{\'a}}, R.~H., {et~al.} 2018,
  \aap, 613, A65

\bibitem[{{Holgado} {et~al.}(2020){Holgado}, {Sim{\'o}n-D{\'\i}az},
  {Haemmerl{\'e}}, {Lennon}, {Barb{\'a}}, {Cervi{\~n}o}, {Castro}, {Herrero},
  {Meynet}, \& {Arias}}]{Holgadoetal20}
{Holgado}, G., {Sim{\'o}n-D{\'\i}az}, S., {Haemmerl{\'e}}, L., {et~al.} 2020,
  \aap, 638, A157

\bibitem[{Hoogerwerf {et~al.}(2000)Hoogerwerf, de~Bruijne, \&
  de~Zeeuw}]{Hoogetal00}
Hoogerwerf, R., de~Bruijne, J. H.~J., \& de~Zeeuw, P.~T. 2000, ApJ, 544, 133L

\bibitem[{{Hutchings}(1975)}]{Hutchings75}
{Hutchings}, J.~B. 1975, \apj, 200, 122

\bibitem[{{Kaper} {et~al.}(1996){Kaper}, {Henrichs}, {Nichols}, {Snoek},
  {Volten}, \& {Zwarthoed}}]{Kaperetal96}
{Kaper}, L., {Henrichs}, H.~F., {Nichols}, J.~S., {et~al.} 1996, \ aaps, 116,
  257

\bibitem[{Kendall {et~al.}(1995)Kendall, Lennon, Brown, \& Dufton}]{Kendetal95}
Kendall, T.~R., Lennon, D.~J., Brown, P.~J.~F., \& Dufton, P.~L. 1995, A\&A,
  298, 489

\bibitem[{{Kholtygin} {et~al.}(2007){Kholtygin}, {Fabrika}, {Chountonov},
  {Burlakova}, {Valyavin}, \& {Kang}}]{Kholtyginetal07}
{Kholtygin}, A.~F., {Fabrika}, S.~N., {Chountonov}, G.~A., {et~al.} 2007,
  Astronomische Nachrichten, 328, 1170

\bibitem[{Kiminki {et~al.}(2009)Kiminki, Kobulnicky, Gilbert, Bird, \&
  Chunev}]{Kimietal09}
Kiminki, D.~C., Kobulnicky, H.~A., Gilbert, I., Bird, S., \& Chunev, G. 2009,
  AJ, 137, 4608

\bibitem[{Kiminki {et~al.}(2008)Kiminki, McSwain, \& Kobulnicky}]{Kimietal08}
Kiminki, D.~C., McSwain, M.~V., \& Kobulnicky, H.~A. 2008, ApJ, 679, 1478

\bibitem[{Kobulnicky {et~al.}(2014)Kobulnicky, Kiminki, Lundquist, Burke,
  Chapman, Keller, Lester, Rolen, Topel, Bhattacharjee, Smullen,
  Vargas~{\'A}lvarez, Runnoe, Dale, \& Brotherton}]{Kobuetal14}
Kobulnicky, H.~A., Kiminki, D.~C., Lundquist, M.~J., {et~al.} 2014, ApJS, 213,
  34

\bibitem[{Kobulnicky {et~al.}(2012)Kobulnicky, Smullen, Kiminki, Runnoe, Wood,
  Long, Alexander, Lundquist, \& Vargas-Alvarez}]{Kobuetal12}
Kobulnicky, H.~A., Smullen, R.~A., Kiminki, D.~C., {et~al.} 2012, ApJ, 756, 50

\bibitem[{Kraus {et~al.}(2009)Kraus, Weigelt, Balega, Docobo, Hofmann,
  Preibisch, Schertl, Tamazian, Driebe, Ohnaka, Petrov, Sch{\"o}ller, \&
  Smith}]{Krauetal09b}
Kraus, S., Weigelt, G., Balega, Y.~Y., {et~al.} 2009, A\&A, 497, 195

\bibitem[{Lanz \& Hubeny(2007)}]{LanzHube07}
Lanz, T. \& Hubeny, I. 2007, ApJS, 169, 83

\bibitem[{Laur {et~al.}(2015)Laur, Tempel, Tuvikene, Eenm{\"a}e, \&
  Kolka}]{Lauretal15}
Laur, J., Tempel, E., Tuvikene, T., Eenm{\"a}e, T., \& Kolka, I. 2015, A\&A,
  581, A37

\bibitem[{{Lef{\`e}vre} {et~al.}(2009){Lef{\`e}vre}, {Marchenko}, {Moffat}, \&
  {Acker}}]{Lefevreetal09}
{Lef{\`e}vre}, L., {Marchenko}, S.~V., {Moffat}, A.~F.~J., \& {Acker}, A. 2009,
  \aap, 507, 1141

\bibitem[{Lehmann {et~al.}(2010)Lehmann, Vitrichenko, Bychkov, Bychkova, \&
  Klochkova}]{Lehmetal10}
Lehmann, H., Vitrichenko, E., Bychkov, V., Bychkova, L., \& Klochkova, V. 2010,
  A\&A, 514, A34

\bibitem[{{Lightkurve Collaboration} {et~al.}(2018){Lightkurve Collaboration},
  {Cardoso}, {Hedges}, {Gully-Santiago}, {Saunders}, {Cody}, {Barclay}, {Hall},
  {Sagear}, {Turtelboom}, {Zhang}, {Tzanidakis}, {Mighell}, {Coughlin}, {Bell},
  {Berta-Thompson}, {Williams}, {Dotson}, \& {Barentsen}}]{Lightkurve18}
{Lightkurve Collaboration}, {Cardoso}, J. V. d. M.~a., {Hedges}, C., {et~al.}
  2018, {Lightkurve: Kepler and TESS time series analysis in Python}

\bibitem[{Liu {et~al.}(1989)Liu, Janes, \& Bania}]{Liuetal89}
Liu, T., Janes, K.~A., \& Bania, T.~M. 1989, \aj, 98, 626

\bibitem[{{Lomb}(1976)}]{Lomb76}
{Lomb}, N.~R. 1976, \apss, 39, 447

\bibitem[{{Lorenz} {et~al.}(1994){Lorenz}, {Mayer}, \&
  {Drechsel}}]{Lorenzetal94}
{Lorenz}, R., {Mayer}, P., \& {Drechsel}, H. 1994, \aap, 291, 185

\bibitem[{Maehara(2014)}]{Maehara14}
Maehara, H. 2014, Journal of Space Science Informatics Japan, 3, 119

\bibitem[{Mahy {et~al.}(2009)Mahy, Naz{\'e}, Rauw, Gosset, De~Becker, Sana, \&
  Eenens}]{Mahyetal09}
Mahy, L., Naz{\'e}, Y., Rauw, G., {et~al.} 2009, A\&A, 502, 937

\bibitem[{Mahy {et~al.}(2013)Mahy, Rauw, De~Becker, Eenens, \&
  Flores}]{Mahyetal13}
Mahy, L., Rauw, G., De~Becker, M., Eenens, P., \& Flores, C.~A. 2013, A\&A,
  550, A27

\bibitem[{{Ma{\'\i}z Apell{\'a}niz}(2019)}]{Apellaniz19a}
{Ma{\'\i}z Apell{\'a}niz}, J. 2019, \aap, 630, A119

\bibitem[{Ma{\'{\i}}z~Apell{\'a}niz {et~al.}(2017)Ma{\'{\i}}z~Apell{\'a}niz,
  Alonso~Morag{\'o}n, Ortiz~de Z{\'a}rate~Alcarazo, \& {The GOSSS
  Team}}]{Maizetal17c}
Ma{\'{\i}}z~Apell{\'a}niz, J., Alonso~Morag{\'o}n, A., Ortiz~de
  Z{\'a}rate~Alcarazo, L., \& {The GOSSS Team}. 2017, in HSA9, 509--509

\bibitem[{{Ma{\'\i}z Apell{\'a}niz} \& {Barb{\'a}}(2020)}]{ApellanizBarba20}
{Ma{\'\i}z Apell{\'a}niz}, J. \& {Barb{\'a}}, R.~H. 2020, arXiv e-prints,
  arXiv:2002.12149

\bibitem[{Ma{\'\i}z~Apell{\'a}niz {et~al.}(2018)Ma{\'\i}z~Apell{\'a}niz,
  Pantaleoni~Gonz{\'a}lez, Barb{\'a}, Sim{\'o}n-D{\'{\i}}az, Negueruela,
  Lennon, Sota, \& Trigueros~P{\'a}ez}]{Maizetal18b}
Ma{\'\i}z~Apell{\'a}niz, J., Pantaleoni~Gonz{\'a}lez, M., Barb{\'a}, R.~H.,
  {et~al.} 2018, A\&A, 616, A149

\bibitem[{Ma{\'{\i}}z~Apell{\'a}niz {et~al.}(2012)Ma{\'{\i}}z~Apell{\'a}niz,
  Pellerin, Barb{\'a}, Sim{\'o}n-D{\'{\i}}az, Alfaro, Morrell, Sota,
  Penad{\'e}s~Ordaz, \& Gallego~Calvente}]{Maizetal12}
Ma{\'{\i}}z~Apell{\'a}niz, J., Pellerin, A., Barb{\'a}, R.~H., {et~al.} 2012,
  in Astronomical Society of the Pacific Conference Series, ed. L.~Drissen,
  C.~Robert, N.~St-Louis, \& A.~F.~J. Moffat, Vol. 465, 484

\bibitem[{Ma{\'\i}z~Apell{\'a}niz {et~al.}(2016)Ma{\'\i}z~Apell{\'a}niz, Sota,
  Arias, Barb{\'a}, Walborn, Sim{\'o}n-D{\'{\i}}az, Negueruela, Marco,
  Le{\~a}o, Herrero, Gamen, \& Alfaro}]{Maizetal16}
Ma{\'\i}z~Apell{\'a}niz, J., Sota, A., Arias, J.~I., {et~al.} 2016, ApJS, 224,
  4

\bibitem[{Ma{\'{\i}}z~Apell{\'a}niz {et~al.}(2011)Ma{\'{\i}}z~Apell{\'a}niz,
  Sota, Walborn, Alfaro, Barb{\'a}, Morrell, Gamen, \& Arias}]{Maizetal11}
Ma{\'{\i}}z~Apell{\'a}niz, J., Sota, A., Walborn, N.~R., {et~al.} 2011, in
  HSA6, ed. {M.~R.~Zapatero Osorio, J.~Gorgas, J.~Ma{\'{\i}}z Apell{\'a}niz,
  J.~R.~Pardo, \& A.~Gil de Paz}, 467--472

\bibitem[{Ma{\'{\i}}z~Apell{\'a}niz {et~al.}(2019)Ma{\'{\i}}z~Apell{\'a}niz,
  Trigueros~P{\'a}ez, Jim{\'e}nez~Mart{\'{\i}}nez, Barb{\'a},
  Sim{\'o}n-D{\'{\i}}az, Pellerin, Negueruela, \& Souza~Le{\~a}o}]{Maizetal19a}
Ma{\'{\i}}z~Apell{\'a}niz, J., Trigueros~P{\'a}ez, E.,
  Jim{\'e}nez~Mart{\'{\i}}nez, I., {et~al.} 2019, in HSA 10, 420

\bibitem[{Ma{\'\i}z~Apell{\'a}niz {et~al.}(2019)Ma{\'\i}z~Apell{\'a}niz,
  Trigueros~P{\'a}ez, Negueruela, Barb{\'a}, Sim{\'o}n-D{\'\i}az, Lorenzo,
  Sota, Gamen, Fari{\~n}a, \& Salas}]{Apellanizetal19}
Ma{\'\i}z~Apell{\'a}niz, J., Trigueros~P{\'a}ez, E., Negueruela, I., {et~al.}
  2019, \aap, 626, A20

\bibitem[{Ma{\'{\i}}z~Apell{\'a}niz {et~al.}(2004)Ma{\'{\i}}z~Apell{\'a}niz,
  Walborn, Galu{\'e}, \& Wei}]{Maizetal04b}
Ma{\'{\i}}z~Apell{\'a}niz, J., Walborn, N.~R., Galu{\'e}, H.~{\'A}., \& Wei,
  L.~H. 2004, ApJS, 151, 103

\bibitem[{Majaess {et~al.}(2008)Majaess, Turner, Lane, \&
  Moncrieff}]{Majaetal08}
Majaess, D.~J., Turner, D.~G., Lane, D.~J., \& Moncrieff, K.~E. 2008, JAVSO,
  36, 90

\bibitem[{{Markova}(2002)}]{Markova02}
{Markova}, N. 2002, \aap, 385, 479

\bibitem[{Martins {et~al.}(2010)Martins, Donati, Marcolino, Bouret, Wade,
  Escolano, \& Howarth}]{Martetal10a}
Martins, F., Donati, J., Marcolino, W.~L.~F., {et~al.} 2010, MNRAS, 407, 1423

\bibitem[{Martins {et~al.}(2015)Martins, Herv{\'e}, Bouret, Marcolino, Wade,
  Neiner, Alecian, Grunhut, \& Petit}]{Martetal15b}
Martins, F., Herv{\'e}, A., Bouret, J.-C., {et~al.} 2015, A\&A, 575, A34

\bibitem[{Martins {et~al.}(2005)Martins, Schaerer, \& Hillier}]{Martetal05a}
Martins, F., Schaerer, D., \& Hillier, D.~J. 2005, A\&A, 436, 1049

\bibitem[{Mason {et~al.}(1998)Mason, Gies, Hartkopf, Bagnuolo, Brummelaar, \&
  McAlister}]{Masoetal98}
Mason, B.~D., Gies, D.~R., Hartkopf, W.~I., {et~al.} 1998, AJ, 115, 821

\bibitem[{Mason {et~al.}(2009)Mason, Hartkopf, Gies, Henry, \&
  Helsel}]{Masoetal09}
Mason, B.~D., Hartkopf, W.~I., Gies, D.~R., Henry, T.~J., \& Helsel, J.~W.
  2009, AJ, 137, 3358

\bibitem[{{Mastroserio} {et~al.}(2019){Mastroserio}, {Ingram}, \& {van der
  Klis}}]{Mastroserioetal19}
{Mastroserio}, G., {Ingram}, A., \& {van der Klis}, M. 2019, \mnras, 488, 348

\bibitem[{Mayer {et~al.}(2017)Mayer, Harmanec, Chini, Nasseri, Nemravov{\'a},
  Drechsel, Catalan-Hurtado, Barlow, Fr{\'e}mat, \& Kotkov{\'a}}]{Mayeretal17}
Mayer, P., Harmanec, P., Chini, R., {et~al.} 2017, A\&A, 600

\bibitem[{McSwain(2003)}]{McSw03}
McSwain, M.~V. 2003, ApJ, 595, 1124

\bibitem[{{McSwain} {et~al.}(2007){McSwain}, {Boyajian}, {Grundstrom}, \&
  {Gies}}]{McSwetal07}
{McSwain}, M.~V., {Boyajian}, T.~S., {Grundstrom}, E.~D., \& {Gies}, D.~R.
  2007, \apj, 655, 473

\bibitem[{{McSwain} {et~al.}(2011){McSwain}, {De Becker}, {Roberts}, \&
  {Boyajian}}]{McSwainetal11}
{McSwain}, M.~V., {De Becker}, M., {Roberts}, M.~S.~E., \& {Boyajian}, T.~S.
  2011, Bulletin de la Societe Royale des Sciences de Liege, 80, 565

\bibitem[{McSwain {et~al.}(2010)McSwain, De~Becker, Roberts, Boyajian, Gies,
  Grundstrom, Aragona, Marsh, \& Roettenbacher}]{McSwetal10}
McSwain, M.~V., De~Becker, M., Roberts, M.~S.~E., {et~al.} 2010, AJ, 139, 857

\bibitem[{{McSwain} {et~al.}(2004){McSwain}, {Gies}, {Huang}, {Wiita},
  {Wingert}, \& {Kaper}}]{McSwainetal04}
{McSwain}, M.~V., {Gies}, D.~R., {Huang}, W., {et~al.} 2004, \apj, 600, 927

\bibitem[{{McSwain} {et~al.}(2001){McSwain}, {Gies}, {Riddle}, {Wang}, \&
  {Wingert}}]{McSwainetal01}
{McSwain}, M.~V., {Gies}, D.~R., {Riddle}, R.~L., {Wang}, Z., \& {Wingert},
  D.~W. 2001, \apj, 558, L43

\bibitem[{Mdzinarishvili(2004)}]{Mdzi04}
Mdzinarishvili, T.~G. 2004, Astrophysics, 47, 155

\bibitem[{{Miller-Jones} {et~al.}(2021){Miller-Jones}, {Bahramian}, {Orosz},
  {Mandel}, {Gou}, {Maccarone}, {Neijssel}, {Zhao}, {Zi{\'o}{\l}kowski},
  {Reid}, {Uttley}, {Zheng}, {Byun}, {Dodson}, {Grinberg}, {Jung}, {Kim},
  {Marcote}, {Markoff}, {Rioja}, {Rushton}, {Russell}, {Sivakoff}, {Tetarenko},
  {Tudose}, \& {Wilms}}]{MillerJetal21}
{Miller-Jones}, J. C.~A., {Bahramian}, A., {Orosz}, J.~A., {et~al.} 2021, arXiv
  e-prints, arXiv:2102.09091

\bibitem[{Motch {et~al.}(1997)Motch, Haberl, Dennerl, Pakull, \&
  Janot-Pacheco}]{Motcetal97}
Motch, C., Haberl, F., Dennerl, K., Pakull, M., \& Janot-Pacheco, E. 1997,
  A\&A, 323, 853

\bibitem[{{Musaev} \& {Chentsov}(1989)}]{Musaevetal89}
{Musaev}, F.~A. \& {Chentsov}, E.~L. 1989, Soviet Astronomy Letters, 15, 360

\bibitem[{Naz{\'e} {et~al.}(2004)Naz{\'e}, Rauw, Vreux, \&
  De~Becker}]{Nazeetal04b}
Naz{\'e}, Y., Rauw, G., Vreux, J.-M., \& De~Becker, M. 2004, A\&A, 417, 667

\bibitem[{{Naz{\'e}} {et~al.}(2010){Naz{\'e}}, {Ud-Doula}, {Spano}, {Rauw}, {De
  Becker}, \& {Walborn}}]{Nazeetal10a}
{Naz{\'e}}, Y., {Ud-Doula}, A., {Spano}, M., {et~al.} 2010, \aap, 520, A59

\bibitem[{Naz{\'e} {et~al.}(2008)Naz{\'e}, Walborn, \& Martins}]{Nazeetal08d}
Naz{\'e}, Y., Walborn, N.~R., \& Martins, F. 2008, RMxAA, 44, 331

\bibitem[{{Naze} {et~al.}(2001){Naze}, {Vreux}, \& {Rauw}}]{Nazeetal01b}
{Naze}, Y., Y., {Vreux}, J.~M., \& {Rauw}, G. 2001, \aap, 372, 195

\bibitem[{Negueruela {et~al.}(2015)Negueruela, Ma{\'{\i}}z-Apell{\'a}niz,
  Sim{\'o}n-D{\'{\i}}az, Alfaro, Herrero, Alonso, Barb{\'a}, Lorenzo, Marco,
  Mongui{\'o}, Morrell, Pellerin, Sota, \& Walborn}]{Neguetal15a}
Negueruela, I., Ma{\'{\i}}z-Apell{\'a}niz, J., Sim{\'o}n-D{\'{\i}}az, S.,
  {et~al.} 2015, in HSA8, ed. A.~J. Cenarro, F.~Figueras,
  C.~Hern{\'a}ndez-Monteagudo, J.~Trujillo~Bueno, \& L.~Valdivielso, 524--529

\bibitem[{{Ninkov} {et~al.}(1987){Ninkov}, {Walker}, \& {Yang}}]{Ninkovetal87}
{Ninkov}, Z., {Walker}, G.~A.~H., \& {Yang}, S. 1987, \apj, 321, 425

\bibitem[{Orosz {et~al.}(2011)Orosz, McClintock, Aufdenberg, Remillard, Reid,
  Narayan, \& Gou}]{Orosetal11}
Orosz, J.~A., McClintock, J.~E., Aufdenberg, J.~P., {et~al.} 2011, ApJ, 742, 84

\bibitem[{Patience {et~al.}(2008)Patience, Zavala, Prato, Franz, Wasserman,
  Tycner, Hutter, \& Hummel}]{Patietal08}
Patience, J., Zavala, R.~T., Prato, L., {et~al.} 2008, ApJL, 674, L97

\bibitem[{{Petrie} \& {Pearce}(1961)}]{PetriePearce61}
{Petrie}, R.~M. \& {Pearce}, J.~A. 1961, Publications of the Dominion
  Astrophysical Observatory Victoria, 12, 1

\bibitem[{{Pigulski} \& {Ko{\l}aczkowski}(1998)}]{Pigulskietal98}
{Pigulski}, A. \& {Ko{\l}aczkowski}, Z. 1998, \mnras, 298, 753

\bibitem[{Pourbaix {et~al.}(2004)Pourbaix, Tokovinin, Batten, Fekel, Hartkopf,
  Levato, Morrell, Torres, \& Udry}]{Pouretal04}
Pourbaix, D., Tokovinin, A.~A., Batten, A.~H., {et~al.} 2004, A\&A, 424, 727

\bibitem[{{Pozo Nu{\~n}ez} {et~al.}(2019){Pozo Nu{\~n}ez}, {Chini}, {Barr
  Dom{\'\i}nguez}, {Fein}, {Hackstein}, {Pietrzy{\'n}ski}, \&
  {Murphy}}]{PozoNetal19}
{Pozo Nu{\~n}ez}, F., {Chini}, R., {Barr Dom{\'\i}nguez}, A., {et~al.} 2019,
  \mnras, 490, 5147

\bibitem[{{Prinja} {et~al.}(2006){Prinja}, {Markova}, {Scuderi}, \&
  {Markov}}]{Prinjaetal06}
{Prinja}, R.~K., {Markova}, N., {Scuderi}, S., \& {Markov}, H. 2006, \aap, 457,
  987

\bibitem[{Puls {et~al.}(2005)Puls, Urbaneja, Venero, Repolust, Springmann,
  Jokuthy, \& Mokiem}]{Pulsetal05}
Puls, J., Urbaneja, M.~A., Venero, R., {et~al.} 2005, A\&A, 435, 669

\bibitem[{{Ramiaramanantsoa} {et~al.}(2018){Ramiaramanantsoa}, {Ratnasingam},
  {Shenar}, {Moffat}, {Rogers}, {Popowicz}, {Kuschnig}, {Pigulski}, {Hand ler},
  {Wade}, {Zwintz}, \& {Weiss}}]{Ramiaetal18}
{Ramiaramanantsoa}, T., {Ratnasingam}, R., {Shenar}, T., {et~al.} 2018, \mnras,
  480, 972

\bibitem[{{Rauw}(2011)}]{Rauw11}
{Rauw}, G. 2011, \aap, 536, A31

\bibitem[{{Rauw} {et~al.}(2015){Rauw}, {Naz{\'e}}, {Wright}, {Drake},
  {Guarcello}, {Prinja}, {Peck}, {Albacete Colombo}, {Herrero}, {Kobulnicky},
  {Sciortino}, \& {Vink}}]{Rawetal15}
{Rauw}, G., {Naz{\'e}}, Y., {Wright}, N.~J., {et~al.} 2015, \apjs, 221, 1

\bibitem[{Rib{\'o} {et~al.}(2002)Rib{\'o}, Paredes, Romero, Benaglia,
  Mart{\'{\i}}, Fors, \& Garc{\'{\i}}a-S{\'a}nchez}]{Riboetal02}
Rib{\'o}, M., Paredes, J.~M., Romero, G.~E., {et~al.} 2002, A\&A, 384, 954

\bibitem[{Rivero~Gonz{\'a}lez {et~al.}(2012)Rivero~Gonz{\'a}lez, Puls, Massey,
  \& Najarro}]{RivGetal12}
Rivero~Gonz{\'a}lez, J.~G., Puls, J., Massey, P., \& Najarro, F. 2012, A\&A,
  543, A95

\bibitem[{{Rogers}(1974)}]{Rogers74}
{Rogers}, G.~L. 1974, M.Sc. thesis

\bibitem[{Salas {et~al.}(2015)Salas, Ma{\'{\i}}z~Apell{\'a}niz, \&
  Barb{\'a}}]{Salaetal15}
Salas, J., Ma{\'{\i}}z~Apell{\'a}niz, J., \& Barb{\'a}, R.~H. 2015, in HSA8,
  ed. A.~J. Cenarro, F.~Figueras, C.~Hern{\'a}ndez-Monteagudo,
  J.~Trujillo~Bueno, \& L.~Valdivielso, 615--615

\bibitem[{Sana {et~al.}(2013)Sana, de~Koter, de~Mink, Dunstall, Evans,
  H{\'e}nault-Brunet, Ma{\'{\i}}z~Apell{\'a}niz, Ram{\'{\i}}rez-Agudelo,
  Taylor, Walborn, Clark, Crowther, Herrero, Gieles, Langer, Lennon, \&
  Vink}]{Sanaetal13b}
Sana, H., de~Koter, A., de~Mink, S.~E., {et~al.} 2013, A\&A, 550, A107

\bibitem[{Sana {et~al.}(2012)Sana, de~Mink, de~Koter, Langer, Evans, Gieles,
  Gosset, Izzard, Le~Bouquin, \& Schneider}]{Sanaetal12a}
Sana, H., de~Mink, S.~E., de~Koter, A., {et~al.} 2012, Science, 337, 444

\bibitem[{Sana \& Evans(2011)}]{SanaEvan11}
Sana, H. \& Evans, C.~J. 2011, in IAUS, Vol. 272, 474--485

\bibitem[{{Santolaya-Rey} {et~al.}(1997){Santolaya-Rey}, {Puls}, \&
  {Herrero}}]{SantolayaRetal97}
{Santolaya-Rey}, A.~E., {Puls}, J., \& {Herrero}, A. 1997, \aap, 323, 488

\bibitem[{Sarty {et~al.}(2011)Sarty, Szalai, Kiss, Matthews, Wu, Kuschnig,
  Guenther, Moffat, Rucinski, Sasselov, Weiss, Huziak, Johnston, Phillips, \&
  Ashley}]{Sartetal11}
Sarty, G.~E., Szalai, T., Kiss, L.~L., {et~al.} 2011, MNRAS, 411, 1293

\bibitem[{{Scargle}(1982)}]{Scargle82}
{Scargle}, J.~D. 1982, \apj, 263, 835

\bibitem[{Schilbach \& R{\"o}ser(2008)}]{SchiRose08}
Schilbach, E. \& R{\"o}ser, S. 2008, A\&A, 489, 105

\bibitem[{{Schneider} {et~al.}(2020){Schneider}, {Ohlmann}, {Podsiadlowski},
  {R{\"o}pke}, {Balbus}, \& {Pakmor}}]{Schneideretal20}
{Schneider}, F.~R.~N., {Ohlmann}, S.~T., {Podsiadlowski}, P., {et~al.} 2020,
  \mnras, 495, 2796

\bibitem[{{Schneider} {et~al.}(2019){Schneider}, {Ohlmann}, {Podsiadlowski},
  {R{\"o}pke}, {Balbus}, {Pakmor}, \& {Springel}}]{Schneideretal19}
{Schneider}, F. R.~N., {Ohlmann}, S.~T., {Podsiadlowski}, P., {et~al.} 2019,
  \nat, 574, 211

\bibitem[{{Shultz} \& {Wade}(2017)}]{ShultzWade17}
{Shultz}, M. \& {Wade}, G.~A. 2017, \mnras, 468, 3985

\bibitem[{{Sim{\'o}n-D{\'\i}az} {et~al.}(2018){Sim{\'o}n-D{\'\i}az}, {Aerts},
  {Urbaneja}, {Camacho}, {Antoci}, {Fredslund Andersen}, {Grundahl}, \&
  {Pall{\'e}}}]{Simonetal18}
{Sim{\'o}n-D{\'\i}az}, S., {Aerts}, C., {Urbaneja}, M.~A., {et~al.} 2018, \aap,
  612, A40

\bibitem[{{Simon-Diaz} {et~al.}(2021){Simon-Diaz}, {Britavskiy}, {Castro},
  {Holgado}, \& {de Burgos}}]{Simonetal21}
{Simon-Diaz}, S., {Britavskiy}, N., {Castro}, N., {Holgado}, G., \& {de
  Burgos}, A. 2021, A\&A, subm.

\bibitem[{{Sim{\'o}n-D{\'\i}az}
  {et~al.}(2020{\natexlab{a}}){Sim{\'o}n-D{\'\i}az}, {Britvaskiy}, {Castro}, \&
  {Holgado}}]{SimonDetal20a}
{Sim{\'o}n-D{\'\i}az}, S., {Britvaskiy}, N., {Castro}, N., \& {Holgado}, G.
  2020{\natexlab{a}}, in Contributions to the XIV.0 Scientific Meeting
  (virtual) of the Spanish Astronomical Society, 186

\bibitem[{Sim{\'o}n-D{\'{\i}}az {et~al.}(2011)Sim{\'o}n-D{\'{\i}}az, Castro,
  Herrero, Puls, Garc{\'\i}a, \& Sab{\'{\i}}n-Sanjuli{\'a}n}]{SimDetal11d}
Sim{\'o}n-D{\'{\i}}az, S., Castro, N., Herrero, A., {et~al.} 2011, Journal of
  Physics Conference Series, 328, 012021

\bibitem[{{Sim{\'o}n-D{\'\i}az} {et~al.}(2017){Sim{\'o}n-D{\'\i}az}, {Godart},
  {Castro}, {Herrero}, {Aerts}, {Puls}, {Telting}, \&
  {Grassitelli}}]{SimDetal17}
{Sim{\'o}n-D{\'\i}az}, S., {Godart}, M., {Castro}, N., {et~al.} 2017, \aap,
  597, A22

\bibitem[{Sim{\'o}n-D{\'{\i}}az \& Herrero(2014)}]{SimDHerr14}
Sim{\'o}n-D{\'{\i}}az, S. \& Herrero, A. 2014, A\&A, 562, A135

\bibitem[{Sim{\'o}n-D{\'\i}az {et~al.}(2006)Sim{\'o}n-D{\'\i}az, Herrero,
  Esteban, \& Najarro}]{SimDetal06}
Sim{\'o}n-D{\'\i}az, S., Herrero, A., Esteban, C., \& Najarro, F. 2006, A\&A,
  448, 351

\bibitem[{Sim{\'o}n-D{\'{\i}}az {et~al.}(2015)Sim{\'o}n-D{\'{\i}}az,
  Negueruela, Ma{\'{\i}}z~Apell{\'a}niz, Castro, Herrero, Garcia,
  P{\'e}rez-Prieto, Caon, Alacid, Camacho, Dorda, Godart,
  Gonz{\'a}lez-Fern{\'a}ndez, Holgado, \& R{\"u}bke}]{SimDetal15b}
Sim{\'o}n-D{\'{\i}}az, S., Negueruela, I., Ma{\'{\i}}z~Apell{\'a}niz, J.,
  {et~al.} 2015, in HSA8, ed. A.~J. Cenarro, F.~Figueras,
  C.~Hern{\'a}ndez-Monteagudo, J.~Trujillo~Bueno, \& L.~Valdivielso, 576--581

\bibitem[{{Sim{\'o}n-D{\'\i}az}
  {et~al.}(2020{\natexlab{b}}){Sim{\'o}n-D{\'\i}az}, {P{\'e}rez Prieto},
  {Holgado}, {de Burgos}, \& {Iacob Team}}]{SimonDetal20}
{Sim{\'o}n-D{\'\i}az}, S., {P{\'e}rez Prieto}, J.~A., {Holgado}, G., {de
  Burgos}, A., \& {Iacob Team}. 2020{\natexlab{b}}, in Contributions to the
  XIV.0 Scientific Meeting (virtual) of the Spanish Astronomical Society, 187

\bibitem[{{Sota} {et~al.}(2014){Sota}, {Ma{\'\i}z Apell{\'a}niz}, {Morrell},
  {Barb{\'a}}, {Walborn}, {Gamen}, {Arias}, \& {Alfaro}}]{Sotaetal14}
{Sota}, A., {Ma{\'\i}z Apell{\'a}niz}, J., {Morrell}, N.~I., {et~al.} 2014,
  \apjs, 211, 10

\bibitem[{Sota {et~al.}(2008)Sota, Ma{\'{\i}}z~Apell{\'a}niz, Walborn, \&
  Shida}]{Sotaetal08}
Sota, A., Ma{\'{\i}}z~Apell{\'a}niz, J., Walborn, N.~R., \& Shida, R.~Y. 2008,
  The Galactic O Star Catalog v.2.0

\bibitem[{Stahl {et~al.}(1996)Stahl, Kaufer, Rivinius, Szeifert, Wolf, Gaeng,
  Gummersbach, Jankovics, Kovacs, Mandel, Pakull, \& Peitz}]{Stahetal96}
Stahl, O., Kaufer, A., Rivinius, T., {et~al.} 1996, A\&A, 312, 539

\bibitem[{Stahl {et~al.}(2008)Stahl, Wade, Petit, Stober, \&
  Schanne}]{Stahetal08}
Stahl, O., Wade, G., Petit, V., Stober, B., \& Schanne, L. 2008, A\&A, 487, 323

\bibitem[{Stickland \& Lloyd(2001)}]{SticLloy01}
Stickland, D.~J. \& Lloyd, C. 2001, The Observatory, 121, 1

\bibitem[{{Stone}(1982)}]{Stoneetal82}
{Stone}, R.~C. 1982, \apj, 261, 208

\bibitem[{{Tanner}(1948)}]{Tanner48}
{Tanner}, R.~W. 1948, \jrasc, 42, 177

\bibitem[{Tetzlaff {et~al.}(2011)Tetzlaff, Neuh{\"a}user, \&
  Hohle}]{Tetzetal11}
Tetzlaff, N., Neuh{\"a}user, R., \& Hohle, M.~M. 2011, MNRAS, 410, 190

\bibitem[{{Thompson} {et~al.}(2012){Thompson}, {Everett}, {Mullally},
  {Barclay}, {Howell}, {Still}, {Rowe}, {Christiansen}, {Kurtz}, {Hambleton},
  {Twicken}, {Ibrahim}, \& {Clarke}}]{Thompsonetal12}
{Thompson}, S.~E., {Everett}, M., {Mullally}, F., {et~al.} 2012, \apj, 753, 86

\bibitem[{{Underhill}(1994{\natexlab{a}})}]{Underhill94b}
{Underhill}, A.~B. 1994{\natexlab{a}}, \apj, 420, 869

\bibitem[{{Underhill}(1994{\natexlab{b}})}]{Underhilletal94}
{Underhill}, A.~B. 1994{\natexlab{b}}, \apj, 420, 869

\bibitem[{{Underhill} \& {Matthews}(1995)}]{Underhilletal95}
{Underhill}, A.~B. \& {Matthews}, J.~M. 1995, Publications of the Astronomical
  Society of the Pacific, 107, 513

\bibitem[{{VanderPlas}(2018)}]{VanderPlas18}
{VanderPlas}, J.~T. 2018, \apjs, 236, 16

\bibitem[{{Vitrichenko}(2002)}]{Vitrichenko02}
{Vitrichenko}, E.~A. 2002, Astronomy Letters, 28, 324

\bibitem[{{Vreux} \& {Conti}(1979)}]{Vreuxetal79}
{Vreux}, J.~M. \& {Conti}, P.~S. 1979, \apj, 228, 220

\bibitem[{{Walborn}(1970)}]{Walborn70}
{Walborn}, N.~R. 1970, \apjl, 161, L149

\bibitem[{Walborn(1971)}]{Walb71c}
Walborn, N.~R. 1971, ApJL, 164, L67

\bibitem[{Walborn(1976)}]{Walb76}
Walborn, N.~R. 1976, ApJ, 205, 419

\bibitem[{{Walborn} {et~al.}(2016){Walborn}, {Morrell}, {Barba}, \&
  {Sota}}]{2016aj....151...91w}
{Walborn}, N.~R., {Morrell}, N.~I., {Barba}, R.~H., \& {Sota}, A. 2016, \aj,
  151, 91

\bibitem[{{Watson} {et~al.}(2006){Watson}, {Henden}, \& {Price}}]{Watson06}
{Watson}, C.~L., {Henden}, A.~A., \& {Price}, A. 2006, Society for Astronomical
  Sciences Annual Symposium, 25, 47

\bibitem[{{Webster} \& {Murdin}(1972)}]{WebsterMurdin72}
{Webster}, B.~L. \& {Murdin}, P. 1972, \nat, 235, 37

\bibitem[{Williams {et~al.}(2013)Williams, Gies, Hillwig, McSwain, \&
  Huang}]{Willetal13}
Williams, S.~J., Gies, D.~R., Hillwig, T.~C., McSwain, M.~V., \& Huang, W.
  2013, AJ, 145, 29

\bibitem[{Williams {et~al.}(2009)Williams, Gies, Matson, \&
  Huang}]{willetal09b}
Williams, S.~J., Gies, D.~R., Matson, R.~A., \& Huang, W. 2009, ApJL, 696, L137

\bibitem[{Yan {et~al.}(2008)Yan, Liu, \& Hadrava}]{Yanetal08}
Yan, J., Liu, Q., \& Hadrava, P. 2008, AJ, 136, 631

\bibitem[{{Zeinalov} \& {Musaev}(1986)}]{Zeinalov86}
{Zeinalov}, S.~K. \& {Musaev}, F.~A. 1986, Soviet Astronomy Letters, 12, 125

\bibitem[{Zinnecker \& Yorke(2007)}]{ZinnYork07}
Zinnecker, H. \& Yorke, H.~W. 2007, ARA\&A, 45, 481

\bibitem[{{Zi{\'o}{\l}kowski}(2005)}]{Ziolkowski05}
{Zi{\'o}{\l}kowski}, J. 2005, \mnras, 358, 851

\bibitem[{{Ziolkowski}(2014)}]{Ziolkowski14}
{Ziolkowski}, J. 2014, \mnras, 440, L61

\end{thebibliography}


\onecolumn
\begin{appendix} 
\section{Tables} \label{app:tables}
\vspace{0cm}


\begin{table}[!htp]
\small
\centering
\renewcommand{\arraystretch}{1.3}
\caption{Number of spectra of each object analyzed in this study. The code for the identification of each instrument is listed below; in the cases of both NoMaDS and CARMENES the extra suffix is added to identify the filter used for the observation. It should be noted that both NoMaDS and CARMENES spectra where taken in two bands simultaneously. In order to identify a spectrum, we use the notation YYMMDD\_Instrument. If more than one spectrum of the same object have been taken the same night, an additional counter \_\# is added. \label{t-numspc}}
\begin{tabular}{c c c c c c c c c c c}
\hline \hline
Name & FEROS & CAF\'E & STELLA & MERCATOR & FIES & NoMaDS & OHP & HARPS & CARMENES & Total \\
 & F & C & S & M & I & H\_V/G/R/B & P & T & C\_J/V &  \\
\hline
HD~\num{164438} & 5 & 4 & 8 & 1 & 2 & \ldots & \ldots & \ldots & 7 & 27 \\
V479~Sct & 8 & \ldots & \ldots & \ldots & 3 & \ldots & 7 & \ldots & \ldots & 18 \\
9~Sge & 5 & 3 & 1 & 78 & 15 & \ldots & 2 & \ldots & 4 & 108 \\
Cyg~X-1 & \ldots & 5 & \ldots & 1 & 9 & 14 & \ldots & \ldots & 2 & 31 \\
BD~$+$36~4063 & \ldots & \ldots & \ldots & \ldots & 9 & \ldots & \ldots & \ldots & \ldots & 9 \\
HDE~\num{229234} & \ldots & 4 & \ldots & \ldots & 6 & \ldots & \ldots & \ldots & \ldots & 10 \\
HD~\num{192281} & \ldots & 3 & \ldots & 1 & 4 & \ldots & 5 & \ldots & 10 & 23 \\
HDE~\num{229232}~AB & \ldots & 1 & \ldots & \ldots & 8 & \ldots & \ldots & \ldots & \ldots & 9 \\
ALS~\num{15133} & \ldots & \ldots & \ldots & \ldots & 1 & \ldots & \ldots & \ldots & \ldots & 1 \\
Cyg~OB2-A11 & \ldots & \ldots & \ldots & 2 & 3 & 8 & \ldots & \ldots & \ldots & 13 \\
Cyg~OB2-22~C & \ldots & \ldots & \ldots & \ldots & 1 & 6 & \ldots & \ldots & \ldots & 7 \\
Cyg~OB2-22~B & \ldots & \ldots & \ldots & \ldots & \ldots & 4 & \ldots & \ldots & \ldots & 4 \\
Cyg~OB2-41 & \ldots & \ldots & \ldots & \ldots & 2 & \ldots & \ldots & \ldots & \ldots & 2 \\
ALS~\num{15148} & \ldots & \ldots & \ldots & \ldots & 4 & 6 & \ldots & \ldots & \ldots & 10 \\
Cyg~OB2-1 & \ldots & \ldots & \ldots & \ldots & 2 & \ldots & \ldots & \ldots & \ldots & 2 \\
ALS~\num{15131} & \ldots & \ldots & \ldots & \ldots & 1 & \ldots & \ldots & \ldots & \ldots & 1 \\
Cyg~OB2-20 & \ldots & \ldots & \ldots & \ldots & 3 & \ldots & \ldots & \ldots & \ldots & 3 \\
Cyg~OB2-70 & \ldots & \ldots & \ldots & \ldots & \ldots & \ldots & \ldots & \ldots & \ldots & \ldots \\
Cyg~OB2-15 & \ldots & \ldots & \ldots & \ldots & 4 & \ldots & \ldots & \ldots & \ldots & 4 \\
ALS~\num{15115} & \ldots & \ldots & \ldots & \ldots & \ldots & \ldots & \ldots & \ldots & \ldots & \ldots \\
Cyg~OB2-29 & \ldots & \ldots & \ldots & \ldots & \ldots & \ldots & \ldots & \ldots & \ldots & \ldots \\
Cyg~OB2-11 & \ldots & \ldots & \ldots & \ldots & 2 & 4 & \ldots & \ldots & \ldots & 6 \\
68~Cyg & \ldots & 5 & \ldots & 22 & 5 & \ldots & 1 & \ldots & \ldots & 33 \\
HD~108 & 3 & 6 & 1 & 5 & 6 & 8 & 10 & \ldots & 2 & 41 \\
V747~Cep & \ldots & 8 & \ldots & 1 & 7 & 4 & \ldots & \ldots & \ldots & 20 \\
HD~\num{12323} & \ldots & \ldots & \ldots & 2 & 6 & \ldots & 6 & \ldots & 6 & 20 \\
HD~\num{16429}~Aa & \ldots & 12 & \ldots & 2 & 4 & 8 & 3 & \ldots & \ldots & 29 \\
HD~\num{15137} & \ldots & 6 & 4 & 3 & 4 & \ldots & 1 & \ldots & \ldots & 18 \\
HD~\num{14633}~AaAb & \ldots & 6 & 2 & 1 & 10 & \ldots & 3 & \ldots & \ldots & 22 \\
$\alpha$~Cam & \ldots & 3 & 1 & 29 & 9 & \ldots & 5 & \ldots & 8 & 55 \\
HD~\num{37737} & \ldots & \ldots & 1 & 2 & 13 & \ldots & 1 & \ldots & \ldots & 17 \\
15~Mon~Aa & 2 & 4 & \ldots & 6 & 28 & \ldots & 11 & 12 & 2 & 65 \\
HD~\num{46573} & 6 & \ldots & 1 & \ldots & 7 & \ldots & \ldots & \ldots & 2 & 16 \\
$\theta^{1}$~Ori~CaCb & 4 & 4 & \ldots & 31 & 22 & \ldots & 2 & \ldots & \ldots & 63 \\
HD~\num{52533}~A & 3 & 1 & 5 & \ldots & 6 & \ldots & \ldots & \ldots & \ldots & 15 \\
Total & 36 & 75 & 24 & 187 & 206 & 62 & 57 & 12 & 43 & 702 \\
\hline
\end{tabular}
\renewcommand{\arraystretch}{1.0}
\end{table}



\begin{landscape}
\begin{table}
\small
\centering
\renewcommand{\arraystretch}{1.3} \tabcolsep=0.14cm
\caption{Published and new spectroscopic solutions of the SB1 systems and candidates in the MONOS sample. T0 corresponds to the periastron passage for the eccentric orbits. The orbits derived in this work are marked as TP. 
Errors in parentheses corresponds to the last digit.
An ''f'' means that the parameter was fixed for that solution. 
An asterisk after the mass function indicates that the value was calculated by us with the parameters of the orbital solution.
We denote ellipsoidal systems with an ''$\dag$'' after the SBS classification.
For the rotational velocities ($ v \ sin i $) measured in this work, we used the IACOB-BROAD tool. \label{t-orbsol-all}}
\begin{tabular}{c c r@{.}l r@{.}l r@{.}l r@{.}l r@{.}l r@{.}l r@{.}l l c}
\hline \hline
\multicolumn{1}{c}{Name} & \multicolumn{1}{c}{SBS Class.} &  \multicolumn{2}{c}{Period} & \multicolumn{2}{c}{$T_{0}$} & \multicolumn{2}{c}{$e$} & \multicolumn{2}{c}{$\omega$} & \multicolumn{2}{c}{$K_1$} & \multicolumn{2}{c}{$\gamma_{1}$} & \multicolumn{2}{c}{f(m)} & \multicolumn{1}{c}{$v \sin i$} & \multicolumn{1}{c}{Ref} \\
\multicolumn{1}{c}{} & \multicolumn{1}{c}{MONOS~II} &  \multicolumn{2}{c}{d} & \multicolumn{2}{c}{JD$-2\,400\,000$} & \multicolumn{2}{c}{} & \multicolumn{2}{c}{degrees} & \multicolumn{2}{c}{km\,s$^{-1}$} & \multicolumn{2}{c}{km\,s$^{-1}$} & \multicolumn{2}{c}{M$_{\odot}$} & \multicolumn{1}{c}{km\,s$^{-1}$} & \multicolumn{1}{c}{} \\
\hline
HD~\num{164438} &  SB1 & 10&249\,74(17) & \num{55501}&114(74) & 0&296(13) & 220&7(26) & 27&78(40) & -14&4(25) & 0&019\,8(9) & 55(3) & TP. HeI \\
\ldots    & \ldots & 10&249\,73(26) & \num{55501}&148(114)& 0&302(20) & 222&3(41) & 27&11(57) & -15&95(39)& 0&018\,3(12)& 55(3) & TP. CC \\
\ldots    & \ldots & 10&250\,42(40) & \num{55501}&050(60) & 0&310(13) & 218&5(25) & 26&90(40) & -8&00(30) & 0&017\,8(8) & 89(2) & Ma17 \\

V479~Sct & SB1    & 3&906\,09(12)   & \num{55017}&468(240)  & 0&272(107) & 263&5(270) & 19&73(243) & 26&52(185)& 0&002\,77(105) & 157(8) & TP. OIII \\
\ldots   & \ldots & 3&906\,06(7)    & \num{55017}&133(208)  & 0&282(113) & 232&9(246) & 21&73(288) & 10&61(165)& 0&003\,67(151) & 157(8) & TP. HeII \\
\ldots   & \ldots & 3&906\,00(f)    & \num{55017}&080(60)   & 0&240(80)  & 237&3(218) & 23&60(400) & 3&90(130) & 0&004\,90(60)  & \ldots & Sa11 \\
\ldots   & \ldots & 3&906\,08(10)   & \num{52825}&985(53)   & 0&337(36)  & 236&0(58)  & 19&74(86)  & 4&01(31)  & 0&002\,61(36)  & \ldots & Ar09 \\
\ldots   & \ldots & 3&906\,03(17)   & \num{51943}&090(100)  & 0&350(40)  & 225&8(33)  & 25&20(140) & 17&20(70) & 0&005\,30(90)  & 113(8) & Ca05 \\
\ldots   & \ldots & 4&426\,70(50)   & \num{52756}&490(70)   & 0&480(60)  & 268&0(100) & 17&60(130) & 4&10(80)  & 0&001\,70(50)  & 140(8) & Mc04 \\
\ldots   & \ldots & 4&117\,00(1100) & \num{51822}&120(90)   & 0&410(50)  & 217&0(90)  & 14&70(90)  & 4&60(50)  & 0&001\,03(20)  & 131(6) & Mc01 \\

9~Sge   & Single  & 78&740(f)       & \num{44075}&770(1130) & 0&556(44) & 262&00(4000) & 11&45(112) & 17&0(4) & 0&007\,00  & \ldots & Un95 \\
\ldots  & \ldots  & 32&514          &  \num{4817}&201       & 0&600     & 356&22       & 9&00       & 11&0    & 0&001\,25  & \ldots & As92 \\
\ldots  & \ldots  & 78&300          & \num{44073}&630       & 0&380     & 164&00       & 9&00       & 13&6    & 0&005\,70  & \ldots & As84 \\

Cyg~X-1 & SB1E   & 5&599\,754(37)  & \num{56796}&219(12) & 0&0(f)   & .&.       & 83&79(130) & 2&84(125)    & 0&3408(161) & 95(5)   & TP. HeII \\
\ldots  & \ldots & 5&599\,742(35)  & \num{56796}&781(11) & 0&0(f)   & .&.       & 77&55(118) & 7&42(115)    & 0&2701(124) & 95(5)   & TP. OIII \\
\ldots  & \ldots & 5&599\,664(86)  & \num{56793}&040(29) & 0&0(f)   & .&.       & 74&25(271) & -14&70(252)  & 0&2371      & 95(5)   & TP. HeI \\
\ldots  & \ldots & 5&599\,829(f)   & \num{41163}&529(9)  & 0&018(3) & 307&6(53) & .&.        & .&.          & 0&2256 *    & 95(6)   & Or11 \\
\ldots  & \ldots & 5&599\,829(f)   & \num{52872}&788(9)  & 0&0(f)   & .&.       & 73&00(70)  & .&.          & 0&2257 *    & \ldots  & Gi08 \\
\ldots  & \ldots & 5&599\,829(f)   & \num{51730}&449(8)  & 0&0(f)   & .&.       & 75&60(70)  & .&.          & 0&2510(70)  & \ldots  & Gi03 \\
\ldots  & \ldots & 5&599\,829(16)  & \num{41874}&707(9)  & 0&0(f)   & .&.       & 74&93(56)  & .&.          & 0&2440      & \ldots  & Br99 \\
\ldots  & \ldots & 5&601\,700(100) &  \num{5869}&170(10) & 0&0(f)   & .&.       & 75&00(10)  & -4&44(50)    & 0&2248 *    & 94.3(5) & Ni87 \\
\ldots  & \ldots & 5&599\,740(80)  &  \num{5869}&110(14) & 0&0(f)   & .&.       & 74&60(130) & -0&90(100)   & 0&2420(90)  & \ldots  & Gi82 \\
\ldots  & \ldots & 5&607\,500      & \num{41166}&060     & 0&0(f)   & .&.       & 64&00      & .&.          & 0&1200      & 100(2)  & We72 \\

BD~$+$36~4063 & SB1E$^\dag$& 4&812\,17(3)  & \num{58381}&04(2) & 0&01(1) & 22&(83) & 160&(3) & 0&5(18)   & 2&06(10) & 96(5)   & TP. comb. HeII \\
\ldots       & \ldots & 4&812\,60(40) & \num{52448}&60(8) & 0&0(f)  & .&.     & 163&(3) & -17&0(30) & 2&18(12) & 126(15) & Wi09 \\

HDE~\num{229234} & SB1E$^\dag$& 3&510\,395(12)   & \num{56527}&114(13) & 0&04(1) & 56&0(20) & 49&9(6)  & -29&6(5) & 0&0451(16) & 95(5)  & TP. comb. HeI \\
\ldots       & \ldots & 3&510\,590(1750) & \num{54714}&326(11) & 0&0(f)  & .&.      & 48&5(11) & -16&3(8) & 0&0420(30) & \ldots     & Mh13 \\
\ldots       & \ldots & 3&510\,420(50)   & \num{50892}&400(10) & 0&0(f)  & 0&0(f)   & 50&9(8)  & -15&0(6) & 0&0480(20) & 103(2) & Bo04 \\

HD~\num{192281}  & Single & 5&480(6) & \num{45922}&469  & 0&19(11) & 26&9(400) & 16&8(24) & -54&5(27) & 0&002\,55  & 270  & Ba93 \\

HDE~\num{229232}~AB & Single & 6&2(2) & \num{54790}&6(1) & 0&0(f) & .&. & 15&6(12) & -46&6(8) & 0&002(1) & 273(19) & Wi13 \\

ALS~\num{15133} & SB1 unc. & 2\,259&0(460) & \num{53355}&0(1340) & 0&34(11) & 157&9(260) & 9&0(10) & -13&0(10) & 0&1419 * & \ldots & Ko12 \\

Cyg~OB2-A11 & SB1  & 15&4428(25)  & \num{55761}&47(99) & 0&136(59) & 47&0(240) & 23&4(14) & -18&58(99)  & 0&0199(37) & 100(7) & TP. comb. HeI \\
\ldots    & \ldots & 15&5110(560) & \num{55745}&70(80) & 0&210(70) & 37&0(190) & 24&0(20) & -13&00(100) & 0&0208 *   & \ldots      & Ko12 \\

\hline
\end{tabular}
\end{table}
\addtocounter{table}{-1}
\end{landscape}

\begin{landscape}
\begin{table}
\small
\centering
\renewcommand{\arraystretch}{1.3} \tabcolsep=0.14cm
\caption{Continued. It should be noted that the ephemerides of Cyg~OB2-22~C do not correspond to a spectroscopic orbit (see text).}
\begin{tabular}{c c r@{.}l r@{.}l r@{.}l r@{.}l r@{.}l r@{.}l r@{.}l l c}
\hline \hline
\multicolumn{1}{c}{Name} & \multicolumn{1}{c}{SBS Class.} &  \multicolumn{2}{c}{Period} & \multicolumn{2}{c}{$T_{0}$} & \multicolumn{2}{c}{$e$} & \multicolumn{2}{c}{$\omega$} & \multicolumn{2}{c}{$K_1$} & \multicolumn{2}{c}{$\gamma_{1}$} & \multicolumn{2}{c}{f(m)} & \multicolumn{1}{c}{$v \sin i$} & \multicolumn{1}{c}{Ref} \\
\multicolumn{1}{c}{} & \multicolumn{1}{c}{MONOS~II} &  \multicolumn{2}{c}{d} & \multicolumn{2}{c}{JD$-2\,400\,000$} & \multicolumn{2}{c}{} & \multicolumn{2}{c}{degrees} & \multicolumn{2}{c}{km\,s$^{-1}$} & \multicolumn{2}{c}{km\,s$^{-1}$} & \multicolumn{2}{c}{M$_{\odot}$} & \multicolumn{1}{c}{km\,s$^{-1}$} & \multicolumn{1}{c}{} \\
\hline
Cyg~OB2-22~C &   SB1 unc. E & 4&162\,083  & \num{56102}&8600  & 0&0(f) & .&. & .&. & .&. & .&. & 266(14)  & TP. HeI \\
\ldots       & \ldots       & 4&161\,100  & \num{50284}&2635  & 0&0(f) & .&. & .&. & .&. & .&. & \ldots         & Pi98 \\

Cyg~OB2-22~B & SB1 unc. & 38&0(2) & \num{56719}&9(16) & 0&21(20) & 244&0(310) & 9&5(17) & -20&4(12) & 0&0032 * & \ldots & Ko14 \\

Cyg~OB2-41 & SB1   & 29&3703(49)  & \num{56542}&66(91)  & 0&21(5) & 255&0(120) & 41&8(15) & -2&4(12) & 0&208(29) & \ldots & TP. comb. HeI \\
\ldots &    \ldots & 29&4100(300) & \num{56310}&98(116) & 0&23(5) & 299&0(140) & 36&3(19) & 0&4(14)  & 0&134 *   & \ldots & Ko14 \\

ALS~\num{15148} & SB1E$^\dag$& 3&170\,36(11) & \num{58679}&22(9)  & 0&06(5) & 71&0(490) & 28&3(13) & -6&9(9)   & 0&0074(11) & 177(9)  & TP. comb. HeI \\
\ldots &    \ldots & 3&170\,40(40) & \num{55971}&20(30) & 0&10(6) & 3&0(340)  & 27&7(17) & -10&6(12) & 0&0069 *   & \ldots        & Ko14 \\

Cyg~OB2-1 & SB1E+Ca & 4&85224(21) & \num{56338}&35(29) & 0&17(7) & 238&0(200) & 69&9(45) & -27&5(32) & 0&164(33) & 203(10)  & TP. comb. HeI \\
\ldots & \ldots     & 4&85230(30) & \num{56338}&20(30) & 0&14(7) & 225&0(220) & 71&0(41) & -30&1(29) & 0&175 *   & \ldots   & Ko14 \\
\ldots & \ldots     & 4&85270(20) & \num{53916}&72(22) & 0&11(4) & 212&0(170) & 72&3(23) & -24&2(6)  & .&.       & \ldots   & Ki08 \\

ALS~\num{15131} & SB1 unc. & 4&6252(12) & \num{55971}&80(67) & 0&15(15) & 240&0(520) & 5&4(9) & -17&4(6) & 0&0001 * & \ldots & Ko14 \\

Cyg~OB2-20 & SB1   & 25&1252(11) & \num{53922}&81(22) & 0&375(22) & 124&0(41) & 43&0(12) & -11&13(81) & 0&164(14) & 14(1)  & TP. comb. HeI \\
\ldots    & \ldots & 25&1261(12) & \num{53922}&94(17) & 0&370(20) & 125&0(30) & 42&8(8)  & -15&80(60) & 0&164 *   & \ldots & Ko14 \\
\ldots    & \ldots & 25&1400(80) & \num{51789}&08(4)  & 0&291(9)  & 158&3(6)  & 48&9(7)  & -19&30(40) & .&.       & \ldots & Ki09 \\

Cyg~OB2-70 & SB1 unc. & 245&1(3) & \num{56549}&1(84) & 0&51(17) & 242&0(160) & 14&5(29) & -17&8(14) & 0&049 * & \ldots & Ko14 \\

Cyg~OB2-15 & SB1   & 14&657\,47(38)  & \num{52162}&41(54)  & 0&138(36) & 53&0(140) & 40&0(13) & -10&2(10) & 0&095(11) & 163(8)  & TP. comb. HeI \\
\ldots    & \ldots & 14&658\,40(190) & \num{54287}&50(10)  & 0&150(70) & 43&0(280) & 40&0(32) & -14&4(21) & 0&094 *   & \ldots  & Ko14 \\
\ldots    & \ldots & 14&660\,00(200) & \num{53922}&03(287) & 0&030(50) & 68&0(710) & 41&2(17) & -20&8(12) & .&.       & \ldots  & Ki08 \\

ALS~\num{15115} & SB1 unc. & \num{4066}&0(450) & \num{52229}&0(290) & 0&75(f) & 352&0(90) & 15&0(23) & -12&1(11) & 0&412 * & \ldots & Ko14 \\

Cyg~OB2-29  & SB1 unc. & 151&2(8) & \num{56543}&7(38) & 0&60(20) & 15&0(100) & 17&5(23) & -10&7(8) & 0&043 * & \ldots & Ko14 \\

Cyg~OB2-11 & SB1.  & 72&488(50) & \num{54925}&4(11) & 0&374(58) & 269&4(72) & 23&5(11) & -27&1(8) & 0&078(16) & 75(3)  & TP. comb. HeI \\
\ldots    & \ldots & 72&430(70) & \num{55722}&0(10) & 0&500(60) & 262&6(80) & 26&0(10) & -31&0(10) & 0&086 *  & \ldots & Ko12 \\
 
68~Cyg & Single   & 3&1781(14) & \num{44577}&53(14) & 0&0(f) & .&. & 6&9(15) & 9&4(11) & 0&0001 * & \ldots & Gi86 \\
\ldots   & \ldots & 5&1000     & \num{44842}&10     & 0&0(f) & .&. & 30&0    & .&.     & 0&0140   & \ldots & Al82 \\

HD~108 & Single   & 1\,627&600  & \num{26260}&901      & 0&43   & 173&6  & 10&5    &  -69&7    & 0&144     & \ldots & Ba99 \\
\ldots   & \ldots & 4&612(1)    & \num{42296}&500(100) & 0&0(f) & .&.    & 9&3(13) & -70&0(10) & 0&0004(2) & \ldots & Hu75 \\

V747~Cep & SB1E   & 5&332\,37(15) & \num{54401}&172\,3(1367) & 0&37(3) & 180&2(65) & 99&6(36) & -10&7(25) & 0&436(49) & 220(11)  & TP. HeII \\
\ldots   & \ldots & 5&331\,46(20) & \num{54400}&532\,2(20)   & 0&30    & 180&0     & .&.      & .&.       & .&.       & \ldots   & Mj08 \\

HD~\num{12323} & SB1E$^\dag$& 1&925\,140(6)   & \num{56229}&34(2)  & 0&0(f)   & .&.        & 29&82(221) & -46&1(12) & 0&0053(12) & 121(6)  & TP. comb. CC \\
\ldots    & \ldots & 2&070\,974(20)  & \num{40001}&06(13) & 0&22(11) & 263&0(270) & 38&40(440) & -46&2(30) & 0&0113(40) & 124     & St01 \\
\ldots    & \ldots & 3&070\,500(700) & \num{41972}&26(11) & 0&21(3)  & 355&0(120) & 22&50(70)  & -41&5(6)  & 0&0033(3)  & \ldots  & Bo78 \\

HD~\num{16429}~Aa & SB1E+Cas & 3&054\,42(f) & 51\,893&22(10) & 0&17(4) & 169&0(120) & 136&0(60) & -50&0(30) & 0&76(1) & \ldots & Mc03 \\

HD~\num{15137} & SB1   & 55&3994(30)  & \num{40017}&483(819) & 0&585\,2(267) & 153&18(351)  & 13&92(57)  & -41&67(32) & 0&0082(16) & 270(14)  & TP. HeI+HeII \\
\ldots    & \ldots & 55&3957(38)  & \num{54421}&991(64)  & 0&623\,9(88)  & 152&17(86)   & 13&56(15)  & -42&06(11) & 0&0069(3)  & 234 & Mc10 \\
\ldots    & \ldots & 30&3500(f)   & \num{51879}&700(600) & 0&4800(700)   & 148&00(1000) & 13&00(100) & -48&40(60) & 0&0040(20) & 234 & Mc07 \\
\ldots    & \ldots & 28&6100(900) & \num{51904}&000(700) & 0&5200(700)   & 125&00(1100) & 12&90(130) & -49&00(70) & 0&0039(13) & 336 & Bo05 \\

\hline
\end{tabular}
\end{table}
\addtocounter{table}{-1}
\end{landscape}

\begin{landscape}
\begin{table}
\small
\centering
\renewcommand{\arraystretch}{1.3} \tabcolsep=0.14cm
\caption{Continued.}
\begin{tabular}{c c r@{.}l r@{.}l r@{.}l r@{.}l r@{.}l r@{.}l r@{.}l l c}
\hline \hline
\multicolumn{1}{c}{Name} & \multicolumn{1}{c}{SBS Class.} &  \multicolumn{2}{c}{Period} & \multicolumn{2}{c}{$T_{0}$} & \multicolumn{2}{c}{$e$} & \multicolumn{2}{c}{$\omega$} & \multicolumn{2}{c}{$K_1$} & \multicolumn{2}{c}{$\gamma_{1}$} & \multicolumn{2}{c}{f(m)} & \multicolumn{1}{c}{$v \sin i$} & \multicolumn{1}{c}{Ref} \\
\multicolumn{1}{c}{} & \multicolumn{1}{c}{MONOS~II} &  \multicolumn{2}{c}{d} & \multicolumn{2}{c}{JD$-2\,400\,000$} & \multicolumn{2}{c}{} & \multicolumn{2}{c}{degrees} & \multicolumn{2}{c}{km\,s$^{-1}$} & \multicolumn{2}{c}{km\,s$^{-1}$} & \multicolumn{2}{c}{M$_{\odot}$} & \multicolumn{1}{c}{km\,s$^{-1}$} & \multicolumn{1}{c}{} \\
\hline
HD~\num{14633}~AaAb & SB1+Ca & 15&409\,24(8)  & \num{51792}&635(32) & 0&695(13) & 139&6(14) & 19&1(3) & -41&55(16) & 0&0041(3) & 121(6) & TP. comb. HeI \\
\ldots & \ldots          & 15&408\,25(24) & \num{44227}&297(99) & 0&677(35) & 139&2(65) & 17&0(13)& -38&88(55) & 0&0031(8) & 138     & Mc11 \\
\ldots & \ldots          & 15&408\,20(40) & \num{51885}&070(10) & 0&700(3)  & 138&7(5)  & 19&0(1) & -38&17(8)  & 0&0040(1) & 138     & Mc07 \\
\ldots & \ldots          & 15&408\,30(40) & \num{51854}&280(50) & 0&698(10) & 140&3(22) & 19&0(4) & -37&90(3)  & 0&0040(3) & 134     & Bo05 \\
\ldots & \ldots          & 15&335\,0      & \num{42007}&300     & 0&680     & 166&3     & 31&3    & -46&00     & 0&0190    & \ldots  & Ro74 \\

$\alpha$~Cam & Single & 3&678\,4(1) & \num{43087}&57(10) & 0&49(8) & 148&0(60) & 8&7(10) & 13&0(5) & 0&000\,16(6) & \ldots & Ze86 \\

HD~\num{37737} & SB1E   & 7&847\,031(32)  & \num{56342}&540(63)  & 0&407(21) & 161&0(32) & 70&8(22) & -9&4(10) & 0&220(21) & 201(11)  & TP. comb. CC \\
\ldots     & \ldots & 7&847\,050      & \num{53690}&232      & 0&440(10) & 158&1(12) & 73&1(17) & -8&0(20) & 0&230(20) & 182     & Al13 \\
\ldots     & \ldots & 7&840\,000(200) & \num{53690}&200(100) & 0&430(20) & 158&0(50) & 72&0(20) &  8&0(20) & 0&220(20) & 182     & Mc07 \\
\ldots     & \ldots & 2&489\,222(70)  & \num{44921}&542(56)  & 0&210(58) & 154&6(86) & 75&9(43) & .&.      & 0&105(18) & 187     & St01 \\
\ldots     & \ldots & 2&489\,880(390) & \num{44921}&340(190) & 0&132(83) & 130&0(27) & 74&1(52) & .&.      & 0&103(22) & \ldots  & Gi86 \\

15~Mon~Aa & SB2a  & \num{27112}&2(14819) & \num{50105}&48(151941) & 0&716(98) & 69&2(111) & 13&0(10) & 33&6(10) & 2&12     & \ldots & Cv10 \\
\ldots & \ldots   & \num{27027}&9(1059)  & \num{50109}&12(10592)  & 0&760(17) & 82&6(15)  & .&.      & 28&4(f)  & 3&81     & \ldots & Cv09 \\
\ldots & \ldots   &  \num{8633}&0(230)   & \num{24508}&00(5700)   & 0&780(20) & 348&0(30) & 15&1(20) & 23&3(2)  & 0&77(32) & 62     & Gi97 \\
\ldots & \ldots   &  \num{9247}&0(640)   & \num{23371}&00(8400)   & 0&670(50) & 349&0(50) & 9&4(9)   & 22&0(3)  & 0&32(11) & 63     & Gi93 \\

HD~\num{46573} & SB1   & 10&6539(f) & \num{53732}&53(14) & 0&63(3) & 265&04(720)  & 11&2(5) & 50&1(4) & 0&0007(1) & 77(3)  & TP. comb. HeI \\
\ldots & \ldots    & 10&6700(f) & \num{53731}&68(48) & 0&47(13)& 254&63(1420) & 8&5(11) & 50&9(8) & 0&0005(2) & 110(18) & Mh09 \\
\ldots & \ldots    & 10&6700(f) & \num{53732}&24(26) & 0&0(f)  & .&.          & 6&9(9)  & 49&7(7) & 0&0003(1) & 110(18) & Mh09 \\

$\theta^{1}$~Ori~CaCb & SB1+Sa  & 4201&0(2000)  & \num{56546}&0(1070)  & 0&546(90)  & 85&9(146) & 15&2(19) & 29&3(9)  & 0&9(4)   & 23(2)  & TP. OIII \\
\ldots & \ldots                 & 4164&0(730)   & \num{52348}&0(730)   & 0&590(40)  & 283&0(20) & .&.      & .&.      & .&.      & \ldots & Gr18 \\
\ldots & \ldots                 & 4120&0(70)    & \num{56610}&0(70)    & 0&590(10)  & 286&1(2)  & 14&7(11) & 30&2(6)  & .&.      & 25(1)  & Ba15 \\
\ldots & \ldots                 & 4112&0(1820)  & \num{52483}&0(1820)  & 0&592(7)   & 285&8(85) & .&.      & 23&6     & .&.      & \ldots & Kr09 \\
\ldots & \ldots                 & 9880&0        & \num{24932}&0        & 0&142      & 99&3      & 24&4     & 13&0     & 14&4 *   & \ldots & St08 \\
\ldots & \ldots                 & 9497&0(14610) & \num{54101}&0(21540) & 0&160(140) & 96&9(118) & 19&9     & 13&0(30) & 7&5 *    & \ldots & Pa08 \\
\ldots & \ldots                 & 61&49(2)      & \num{52550}&0(10)    & 0&490(50)  & 198&0(60) & 6&1(4).  & 25&0(40) & 0&001(3) & 36     & Lh10 \\

HD~52\,533~A & SB1E & 21&9652(14) & \num{44615}&340(760) & 0&364(55) & 337&8(82) & 82&2(44)  & 39&25(324) & 1&02(17)   & 299(15)  & TP. comb. CC \\
\ldots & \ldots     & 22&1861(2)  & \num{44635}&010(30)  & 0&300(10) & 330&3(5)  & 105&0(10) & 76&70(60)  & 2&30(7)    & 270      & Mc07 \\
\ldots & \ldots     &  3&2951(3)  & \num{44988}&973(20)  & 0&234(35) & 4&8(24)   & 34&6(12)  & 41&76(92)  & 0&0131(14) & \ldots   & Gi86 \\

\hline
\end{tabular}
\renewcommand{\arraystretch}{1.0}

\tablebib{
(Al13) \citet{Alexeeva13}    ; (Al82) \citet{Aldusevaetal82}  ; (Ar09) \citet{Aragona09}       ; (As84) \citet{Aslanovetal84}   ; (As92) \citet{Aslanovetal92}   ; (Ba15) \citet{Balegaetal15}         ; (Ba93) \citet{Bara93}          ; (Ba99) \citet{Barannikov99}    ; (Bo04) \citet{Boecheetal04}    ; (Bo05) \citet{Boyajianetal05}  ; (Bo78) \citet{BoltonRogers78}       ; (Br99) \citet{Brocksoppetal99a};  (Ca05) \citet{Casaresetal05}  ; (Cv09) \citet{Cvetetal09}      ; (Cv10) \citet{Cvetetal10}      ; (Gi03) \citet{giesetal03}           ; (Gi08) \citet{gies08}          ; (Gi82) \citet{GiesBolton82}    ; (Gi86) \citet{Giesetal86}      ; (Gi93) \citet{Giesetal93}      ; (Gi97) \citet{Giesetal97}           ; (Gr18) \citet{Gravity18}       ; (Hu75) \citet{Hutchings75}     ; (Ki08) \citet{Kimietal08}      ; (Ki09) \citet{Kimietal09}      ; (Ko12) \citet{Kobuetal12}           ; (Ko14) \citet{Kobuetal14}      ; (Kr09) \citet{Krauetal09b}     ; (Lh10) \citet{Lehmetal10}      ; (Ma17) \citet{Mayeretal17}     ; (Mc01) \citet{McSwainetal01}        ; (Mc03) \citet{McSw03}          ; (Mc04) \citet{McSwainetal04}   ; (Mc07) \citet{McSwetal07}      ; (Mc10) \citet{McSwetal10}      ; (Mc11) \citet{McSwainetal11}        ; (Mh09) \citet{Mahyetal09}      ; (Mh13) \citet{Mahyetal13}      ; (Mj08) \citet{Majaetal08}      ; (Ni87) \citet{Ninkovetal87}    ; (Or11) \citet{Orosetal11}           ; (Pa08) \citet{Patietal08}      ; (Pi98) \citet{Pigulskietal98}  ; (Ro74) \citet{Rogers74}        ; (Sa11) \citet{Sartetal11}      ; (St01) \citet{SticLloy01}           ; (St08) \citet{Stahetal08}      ; (Un95) \citet{Underhilletal95} ; (We72) \citet{WebsterMurdin72} ; (Wi09) \citet{willetal09b}     ; (Wi13) \citet{Willetal13}           ; (Ze86) \citet{Zeinalov86}       
}
\end{table}

\end{landscape}


\section{RV and LC curves} \label{app:curves}

\subsection{RV curves}

\begin{figure*}[!htp]
    \includegraphics[width=.5\textwidth]{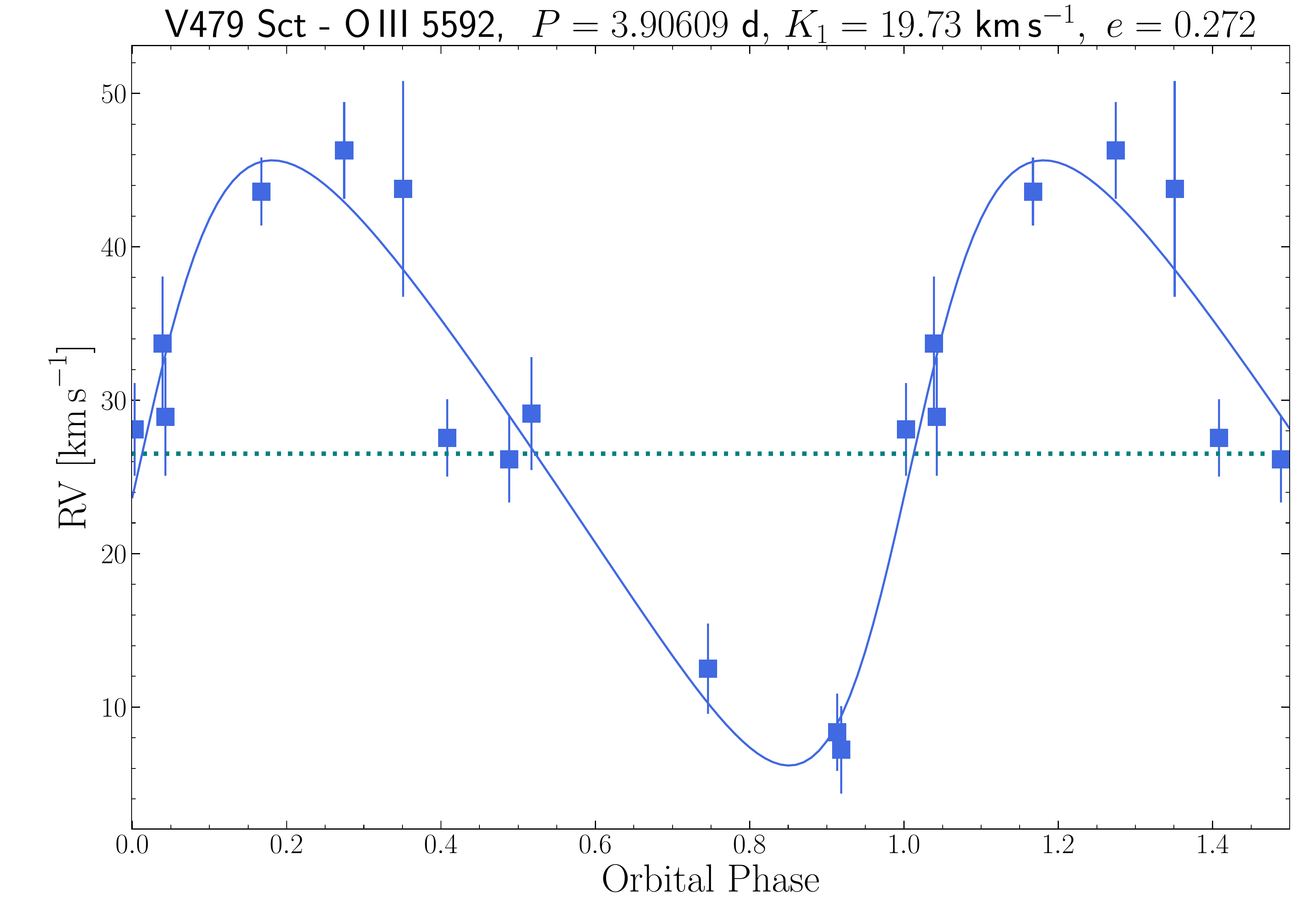}
    \includegraphics[width=.5\textwidth]{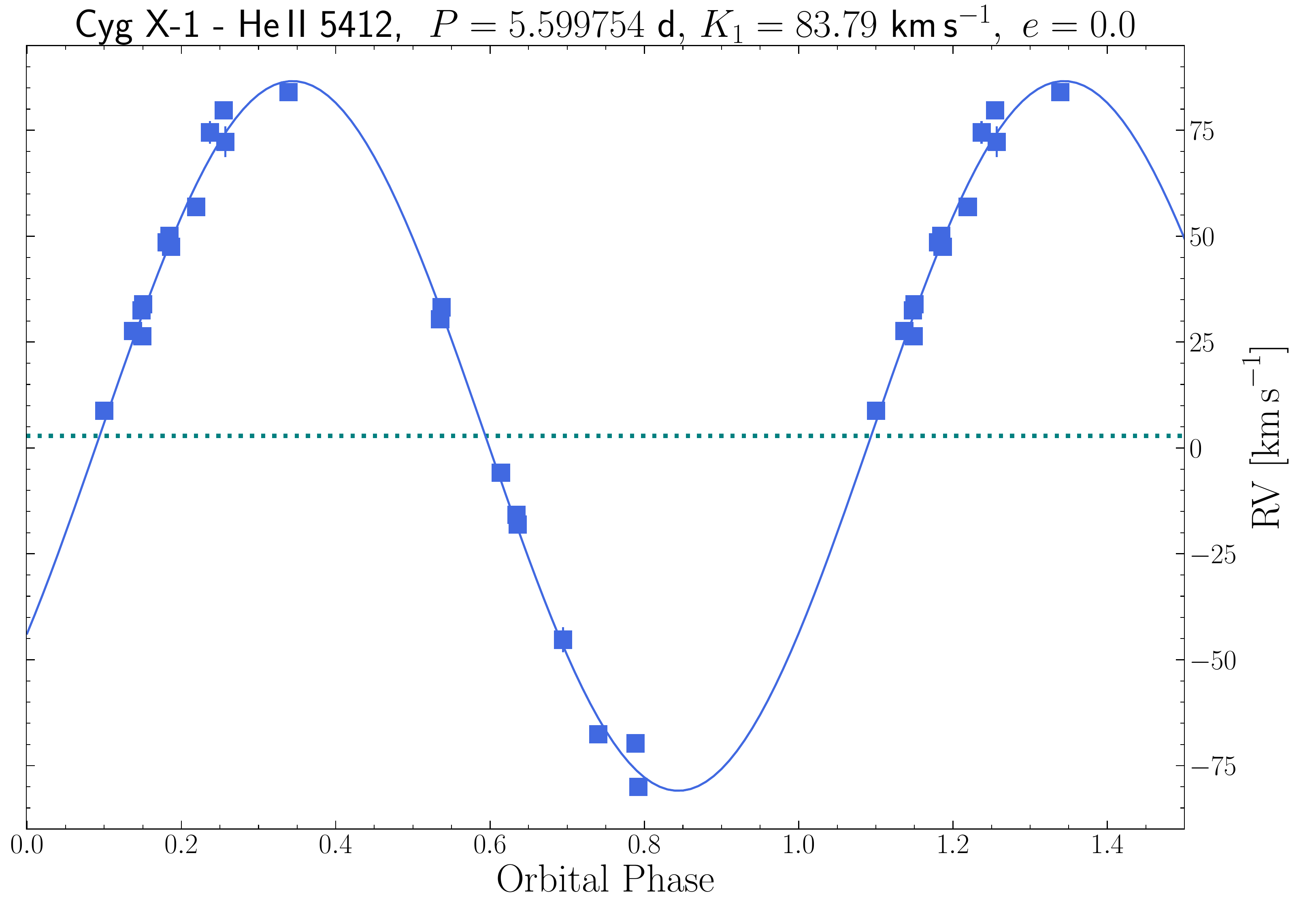} 
    
    \includegraphics[width=.5\textwidth]{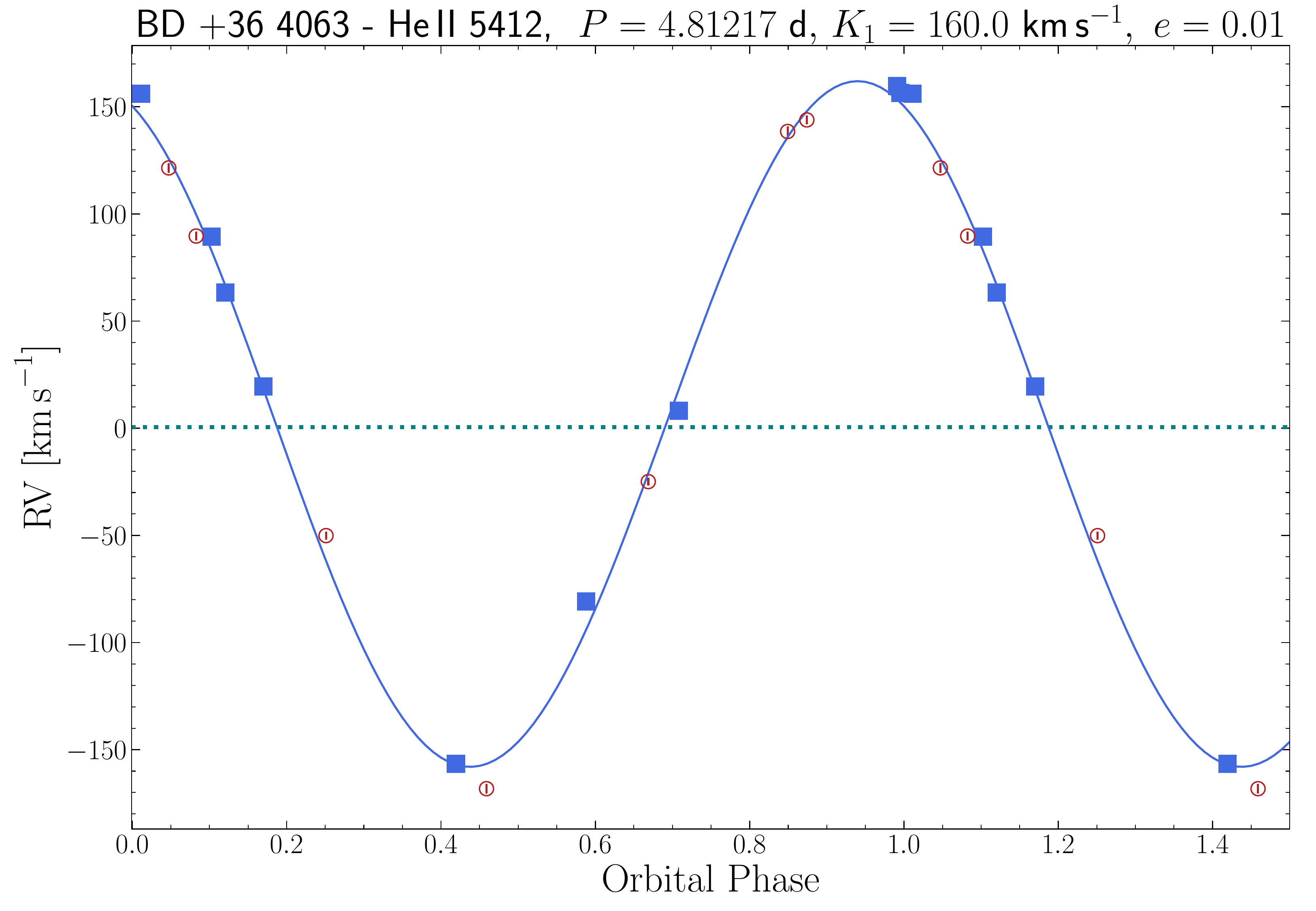} 
    \includegraphics[width=.5\textwidth]{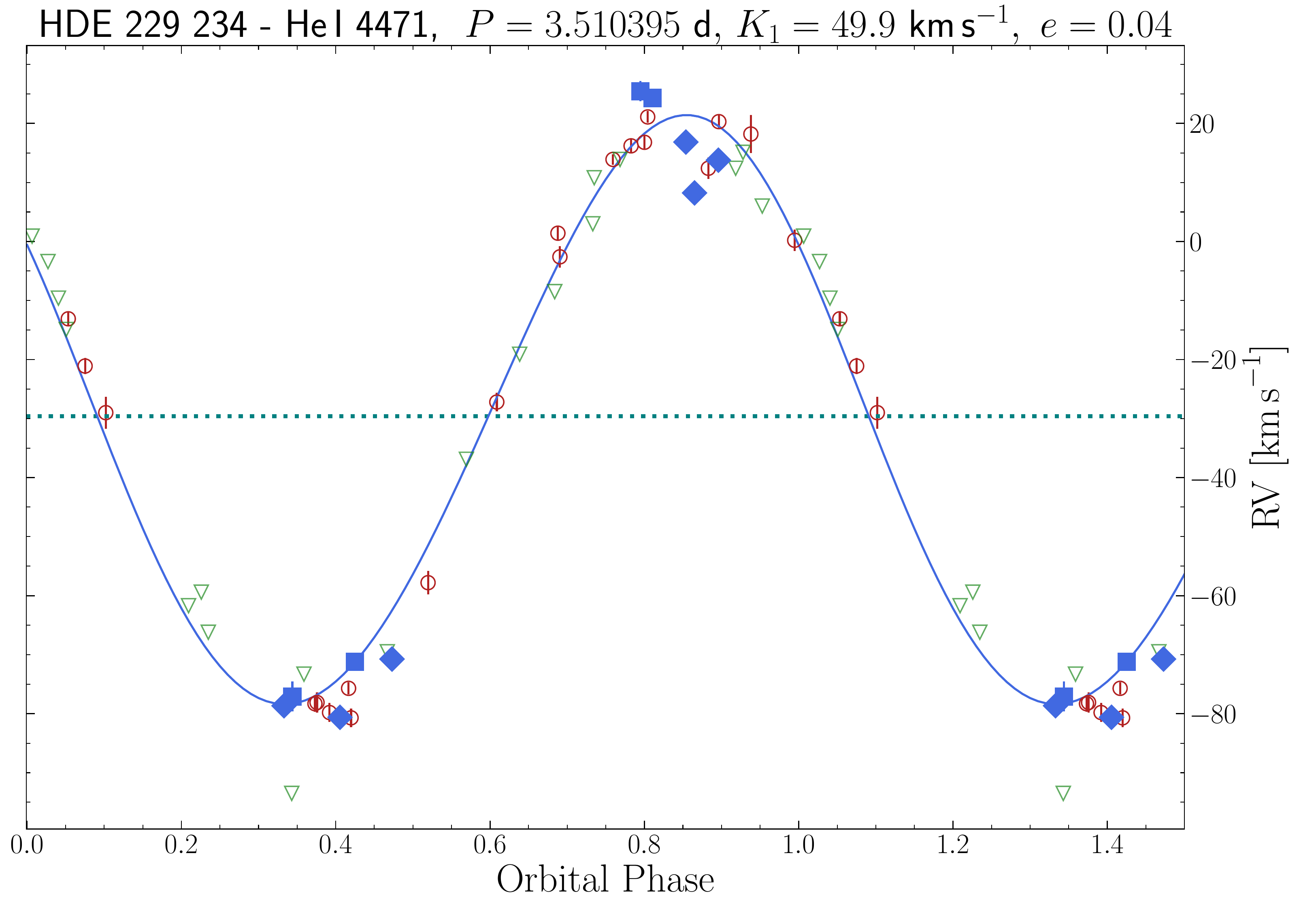}
    
    \includegraphics[width=.5\textwidth]{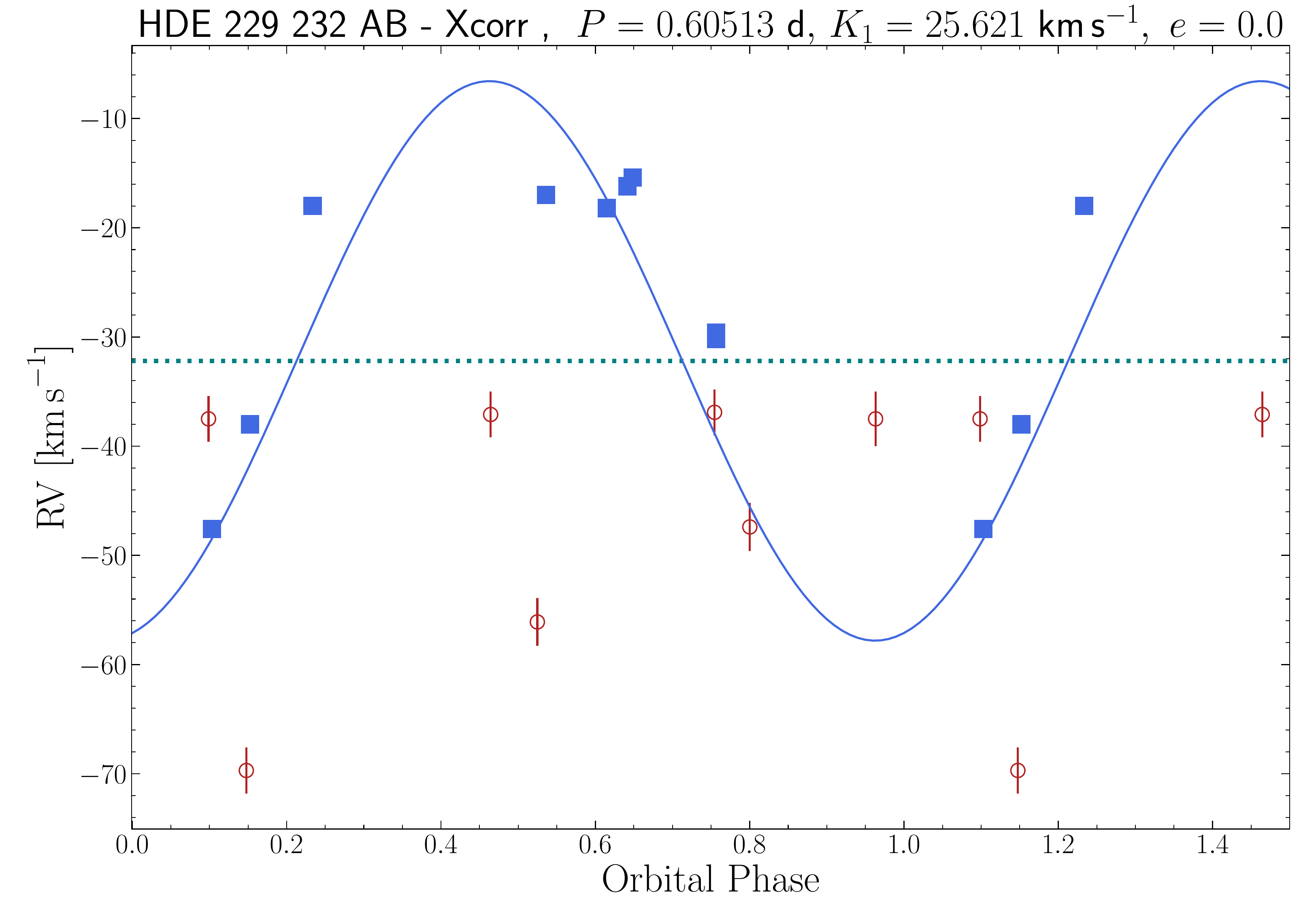} 
    \includegraphics[width=.5\textwidth]{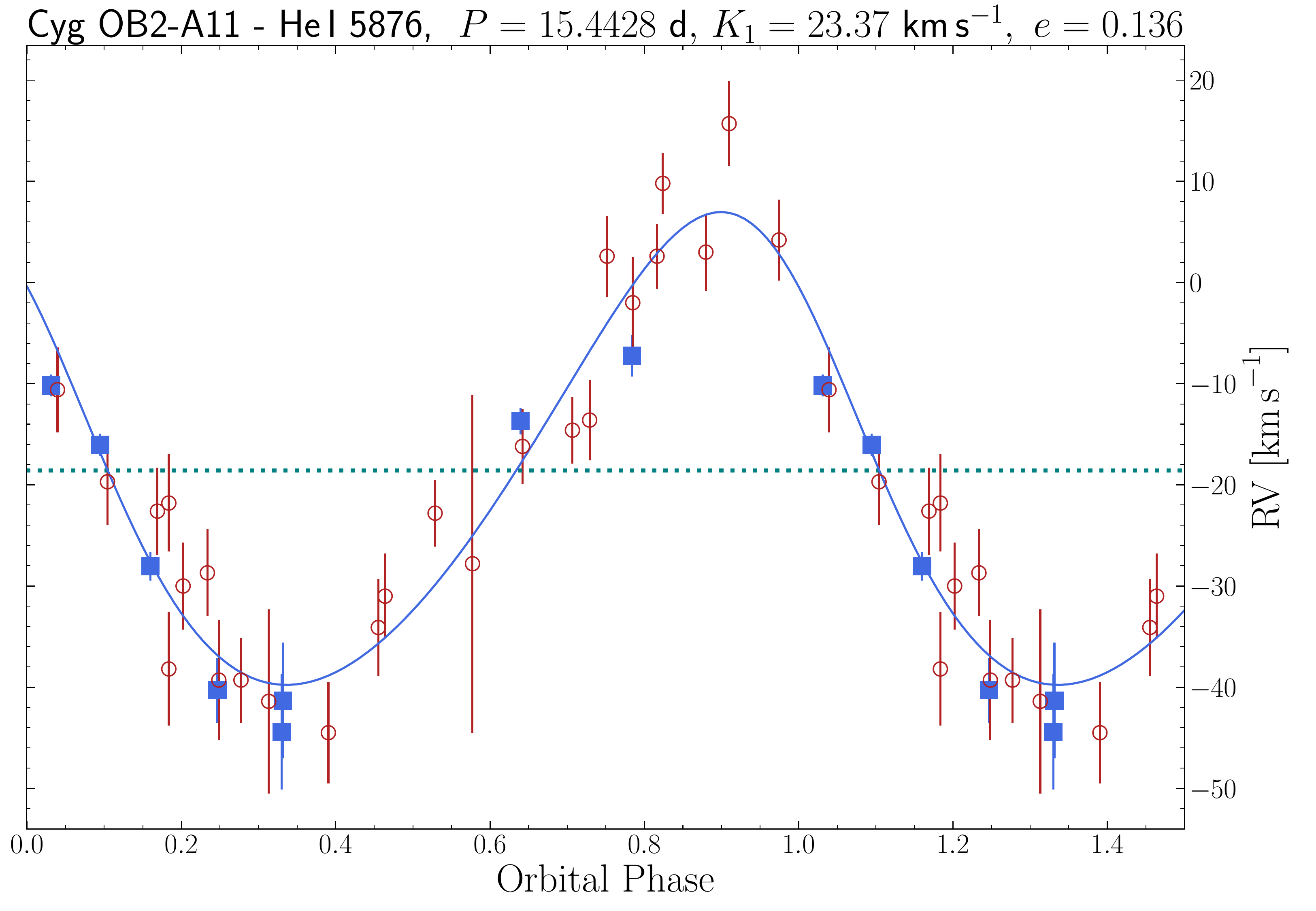}
    \caption{New orbital solutions of several systems. MONOS RVs of each diagnostic line are shown as navy squares. The dotted blue line shows the $\gamma$ of the system. 
        Upper left panel: New orbital solution for V479~Sct.
        Upper right panel: New orbital solution for Cyg X-1. 
        Middle left panel: New combined orbital solution for BD~$+$36~4063. Red circles are the RVs from \cite{willetal09b}.
        Middle right panel: New combined orbital solution for HDE~\num{229234}. Also shown are RVs from \cite{Boecheetal04} (open red circles), \cite{Mahyetal13} (green triangles), and MONOS (navy squares and diamonds). The MONOS RVs were split into two sets in order to highlight the RV scatter in different observing runs.
        Lower left panel: New combined orbital solution for HDE~\num{229232}~AB. Red circles are the RVs from \cite{Willetal13}.
        Lower right panel: New combined orbital solution for Cyg~OB2-A11. Red circles are the RVs from \cite{Kobuetal12}. 
     }\label{orb-fig:1}
\end{figure*}

\begin{figure*}[!htp]
    \includegraphics[width=.5\textwidth]{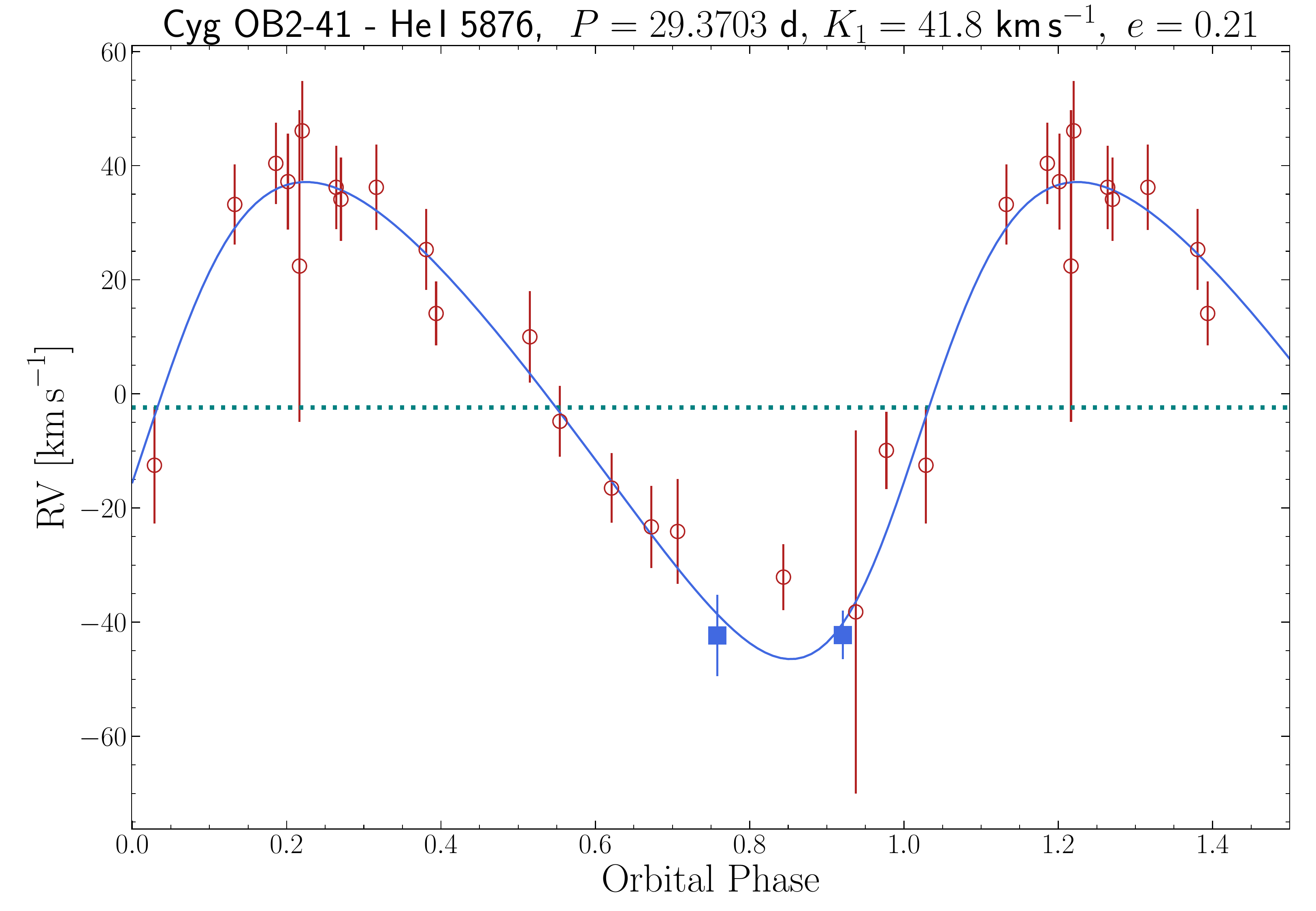} 
    \includegraphics[width=.5\textwidth]{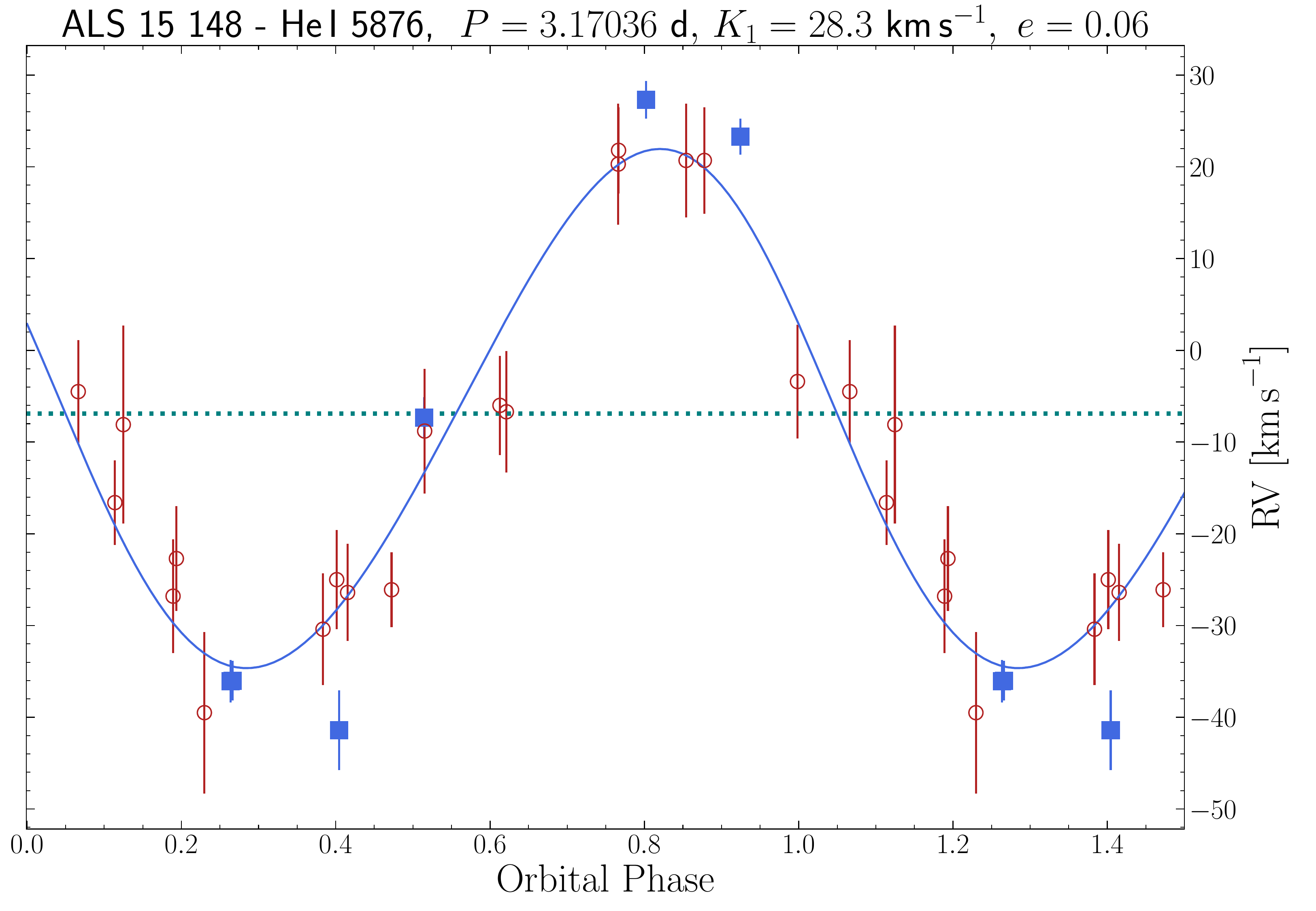} 
    
    \includegraphics[width=.5\textwidth]{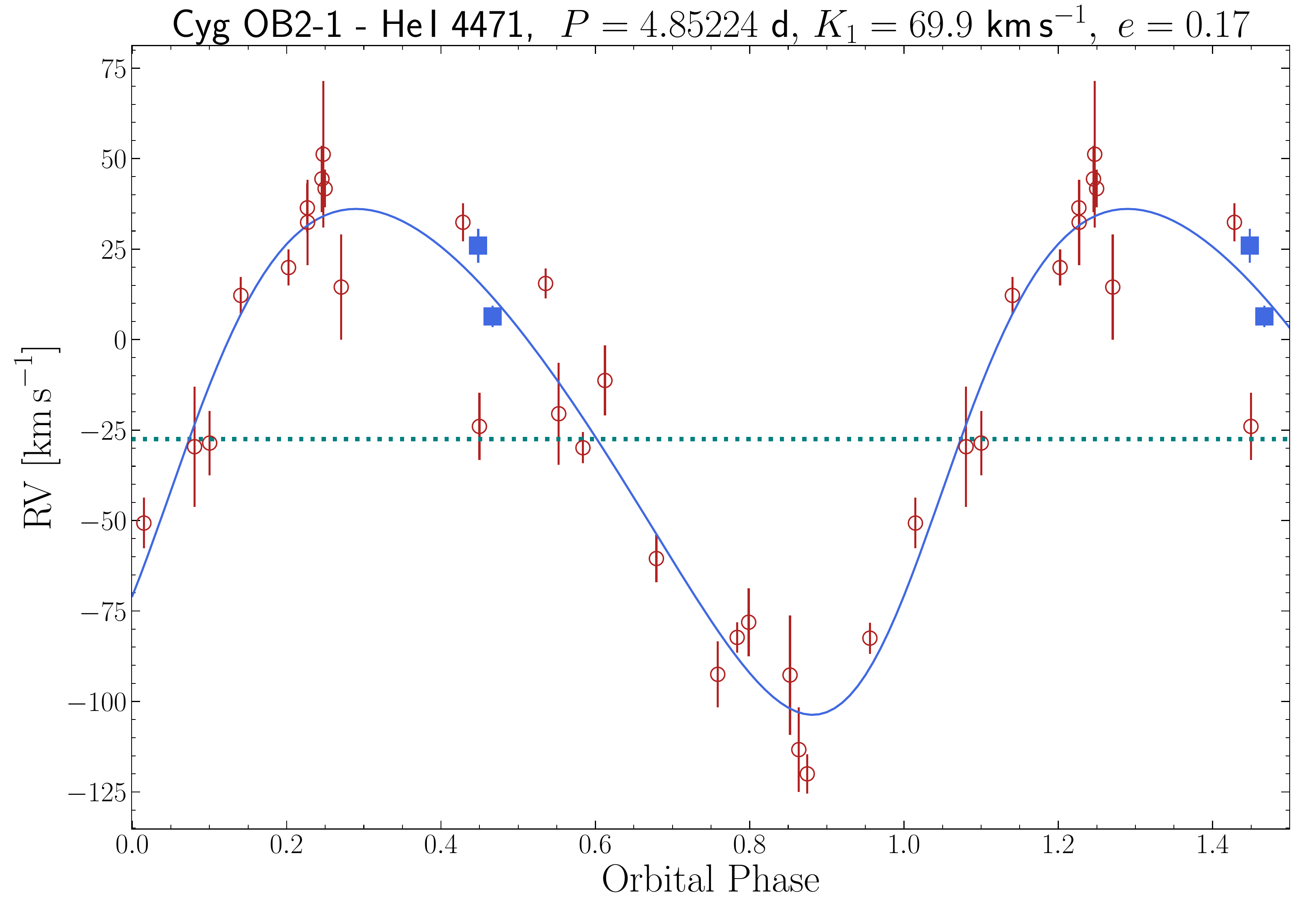} 
    \includegraphics[width=.5\textwidth]{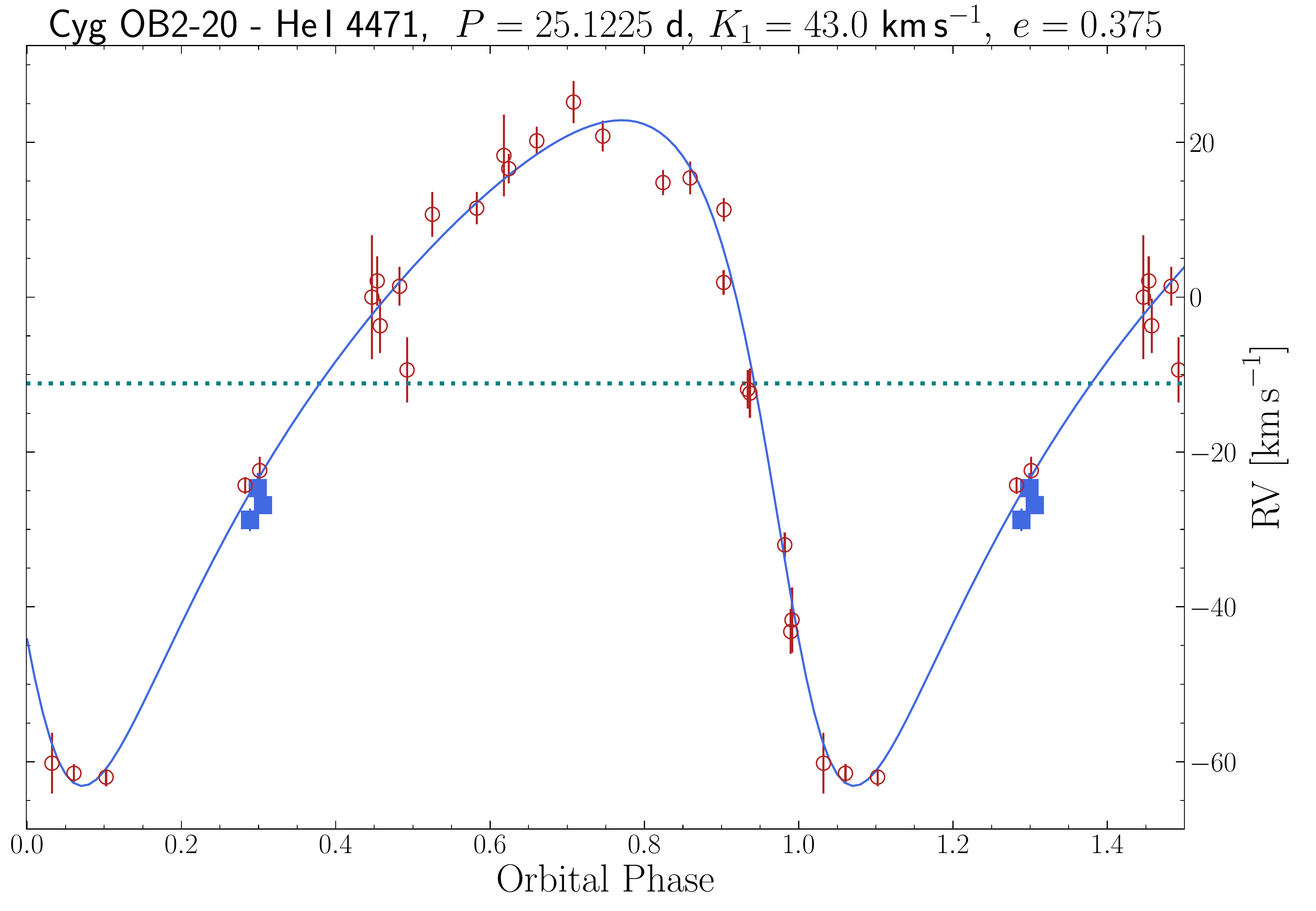} 
    
    \includegraphics[width=.5\textwidth]{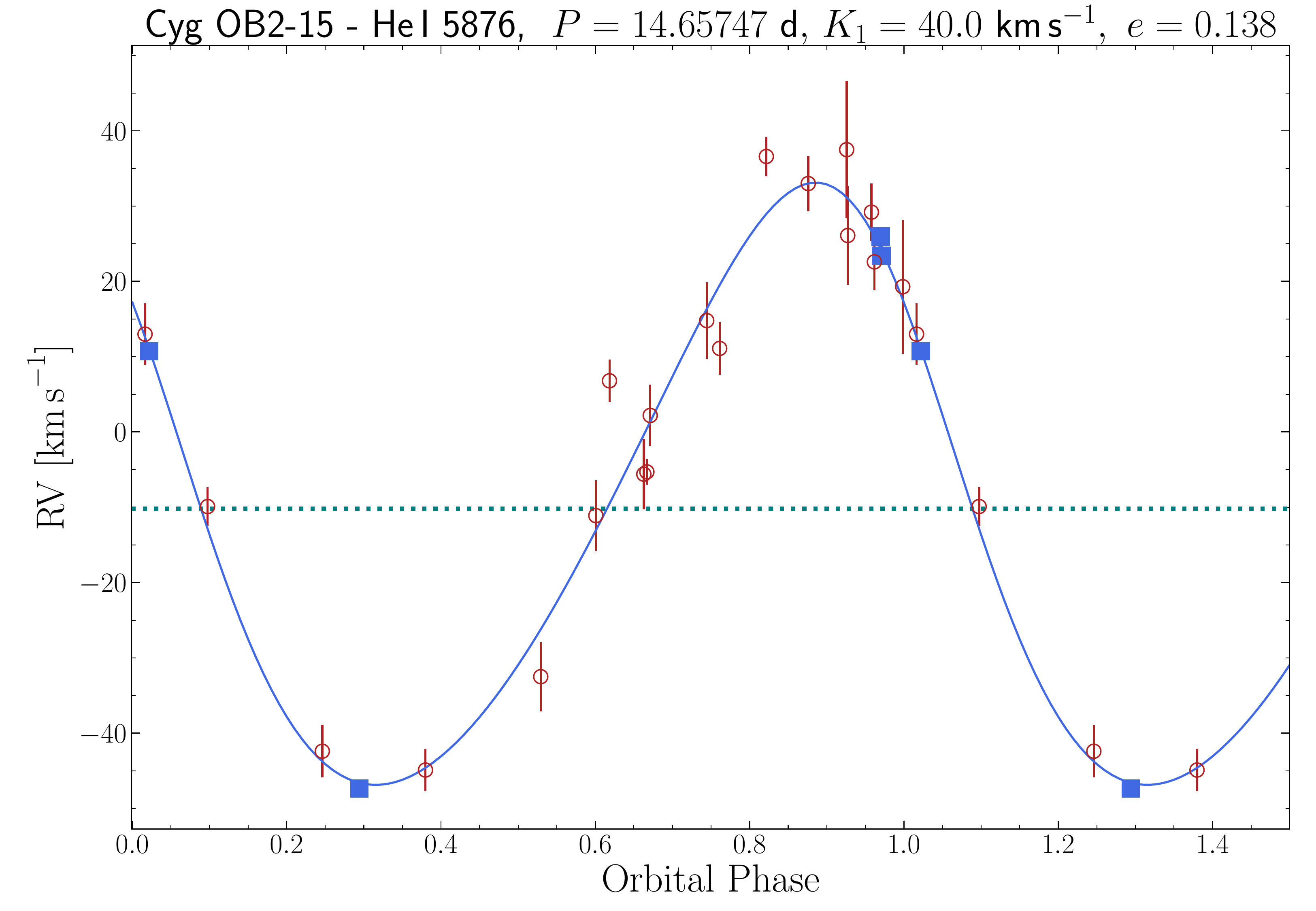} 
    \includegraphics[width=.5\textwidth]{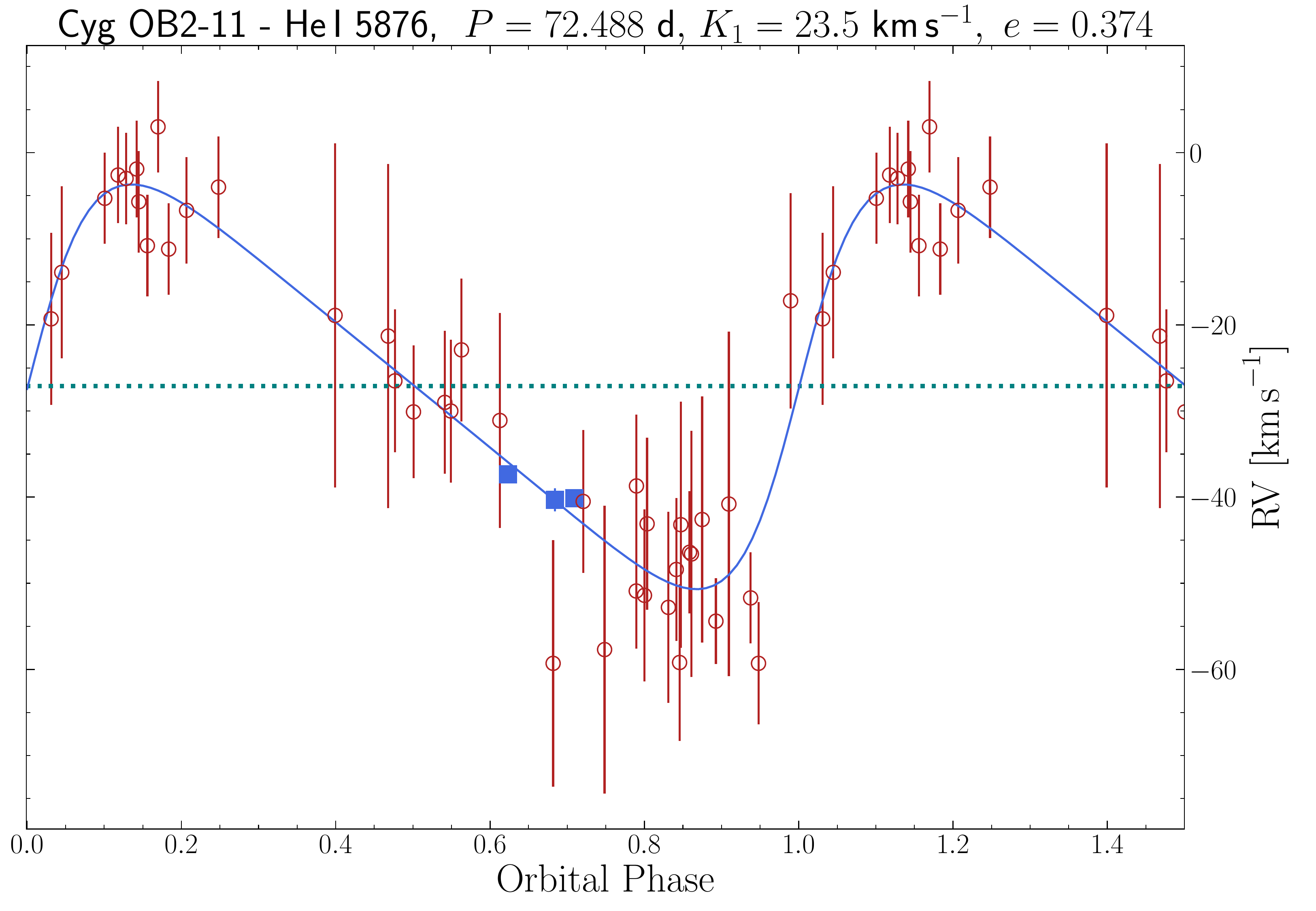} 
    \caption{New orbital solutions of several systems. MONOS RVs of each diagnostic line are shown as navy squares. The dotted blue line shows the $\gamma$ of the system.
        Upper left panel: New combined orbital solution for Cyg~OB2-41. Red circles are the RVs from \cite{Kobuetal14}.
        Upper right panel: New combined orbital solution for ALS~\num{15148}. Red circles are the RVs from \cite{Kobuetal14}.
        Middle left panel: New combined orbital solution for Cyg~OB2-1. Red circles are the RVs from \cite{Kobuetal14}.
        Middle right panel: New combined orbital solution for Cyg~OB2-20. Red circles are the RVs from \cite{Kobuetal14}.
        Lower left panel: New combined orbital solution for Cyg~OB2-15. Red circles are the RVs from \cite{Kobuetal14}.
        Lower right panel: New combined orbital solution for Cyg~OB2-11. Red circles are the RVs from \cite{Kobuetal14}.
    }\label{orb-fig:2}
\end{figure*}

\begin{figure*}[!htp]
    \includegraphics[width=.5\textwidth]{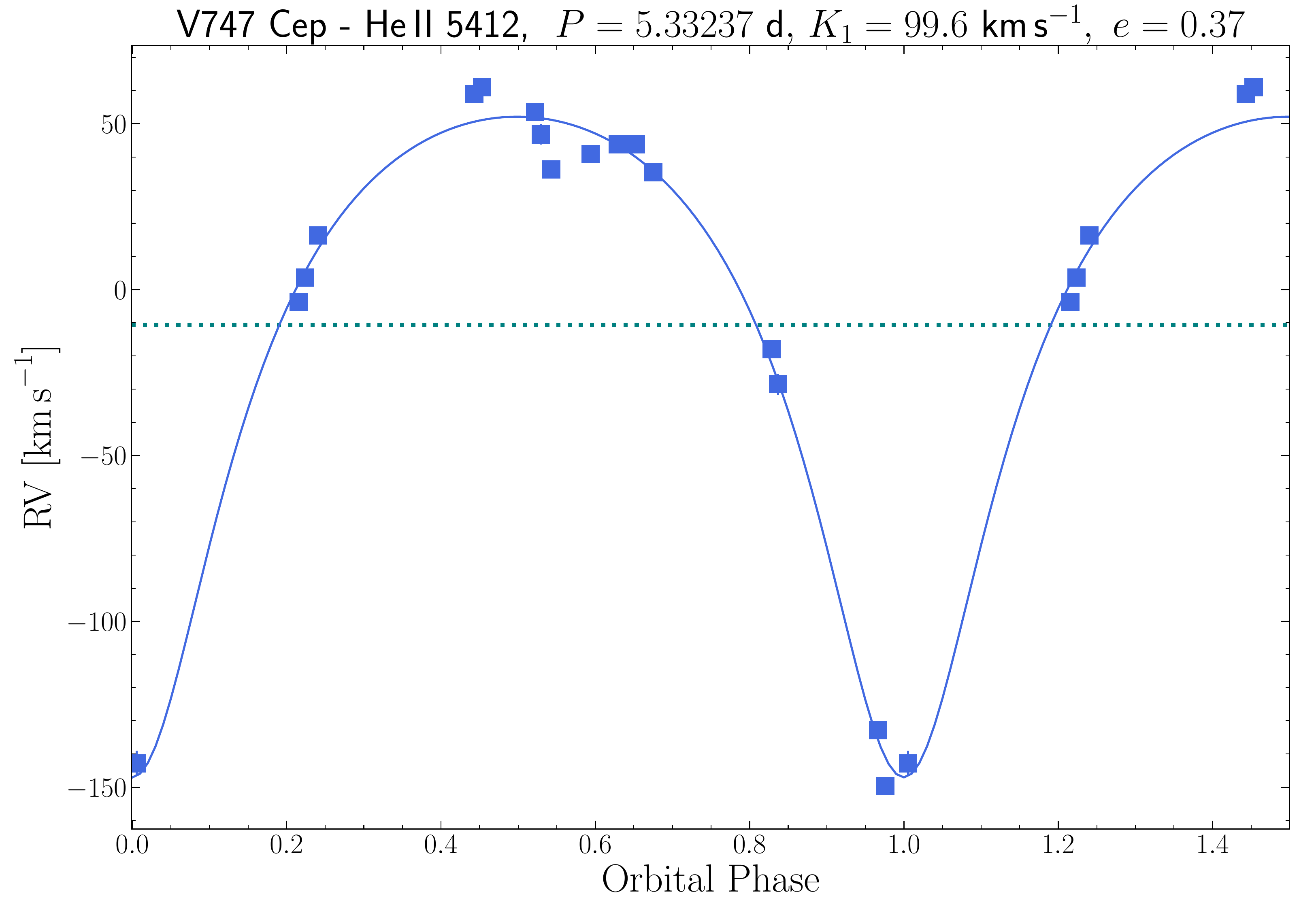}
    \includegraphics[width=.5\textwidth]{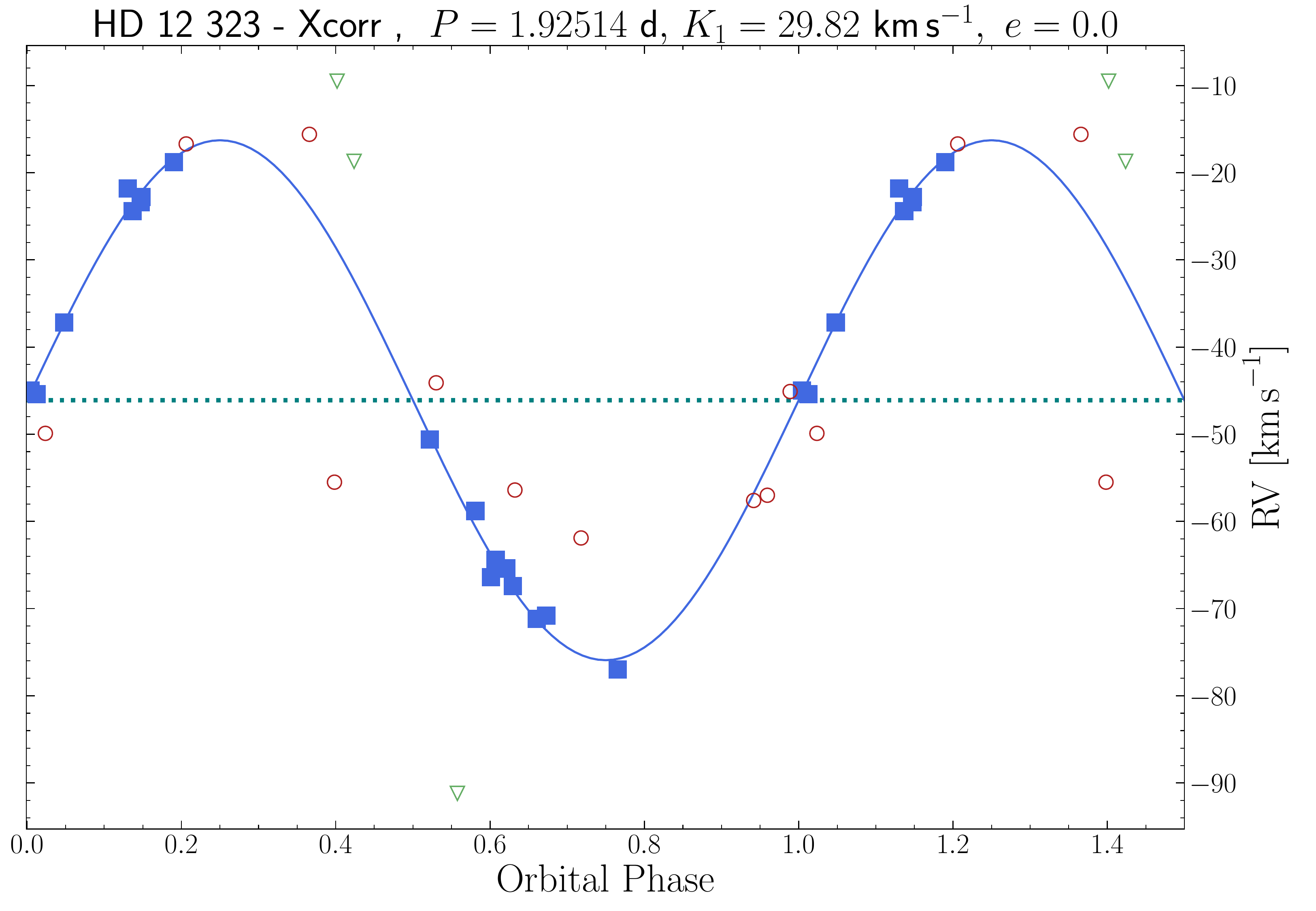} 
    
    \includegraphics[width=.5\textwidth]{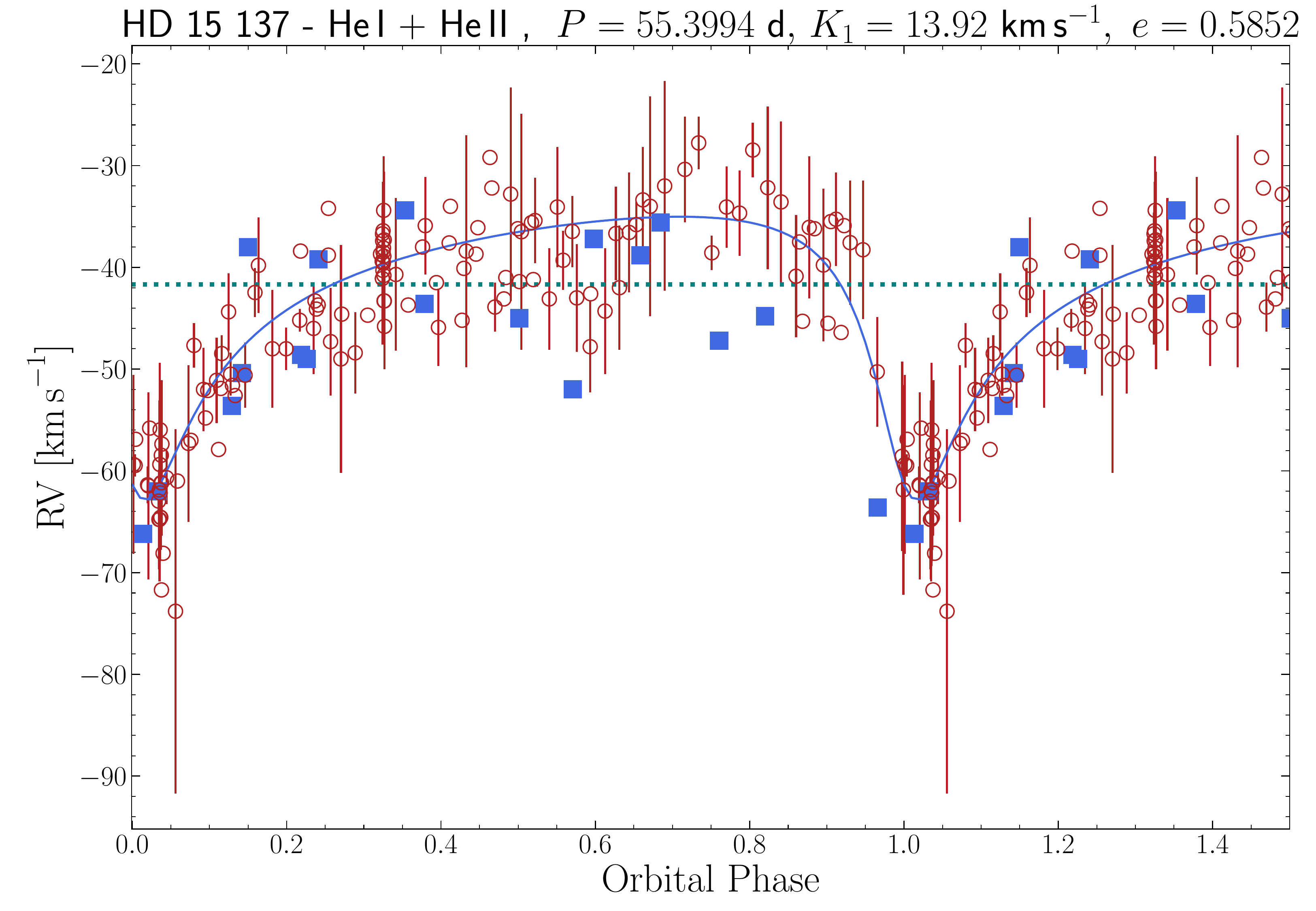}
    \includegraphics[width=.5\textwidth]{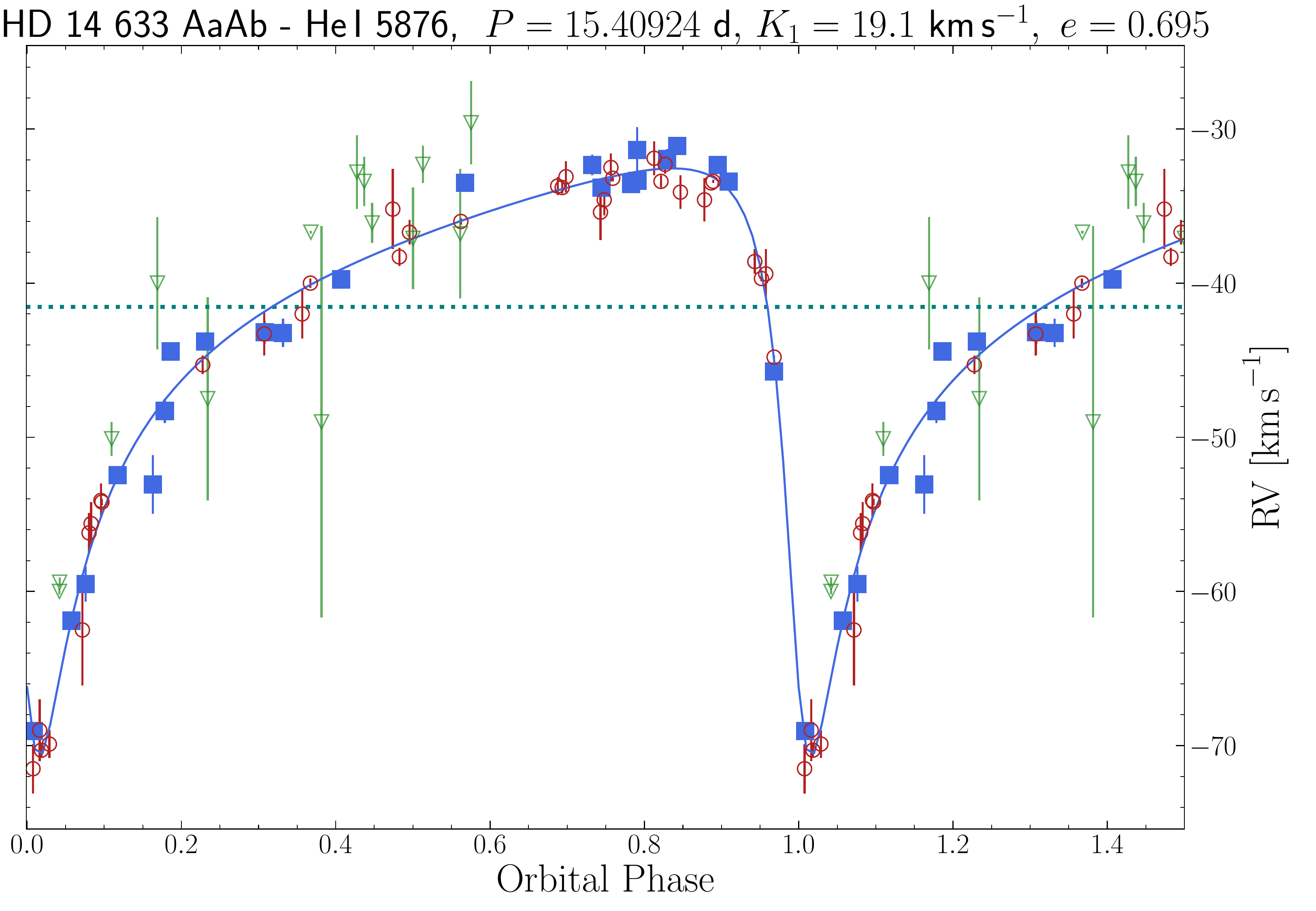} 
    
    \includegraphics[width=.5\textwidth]{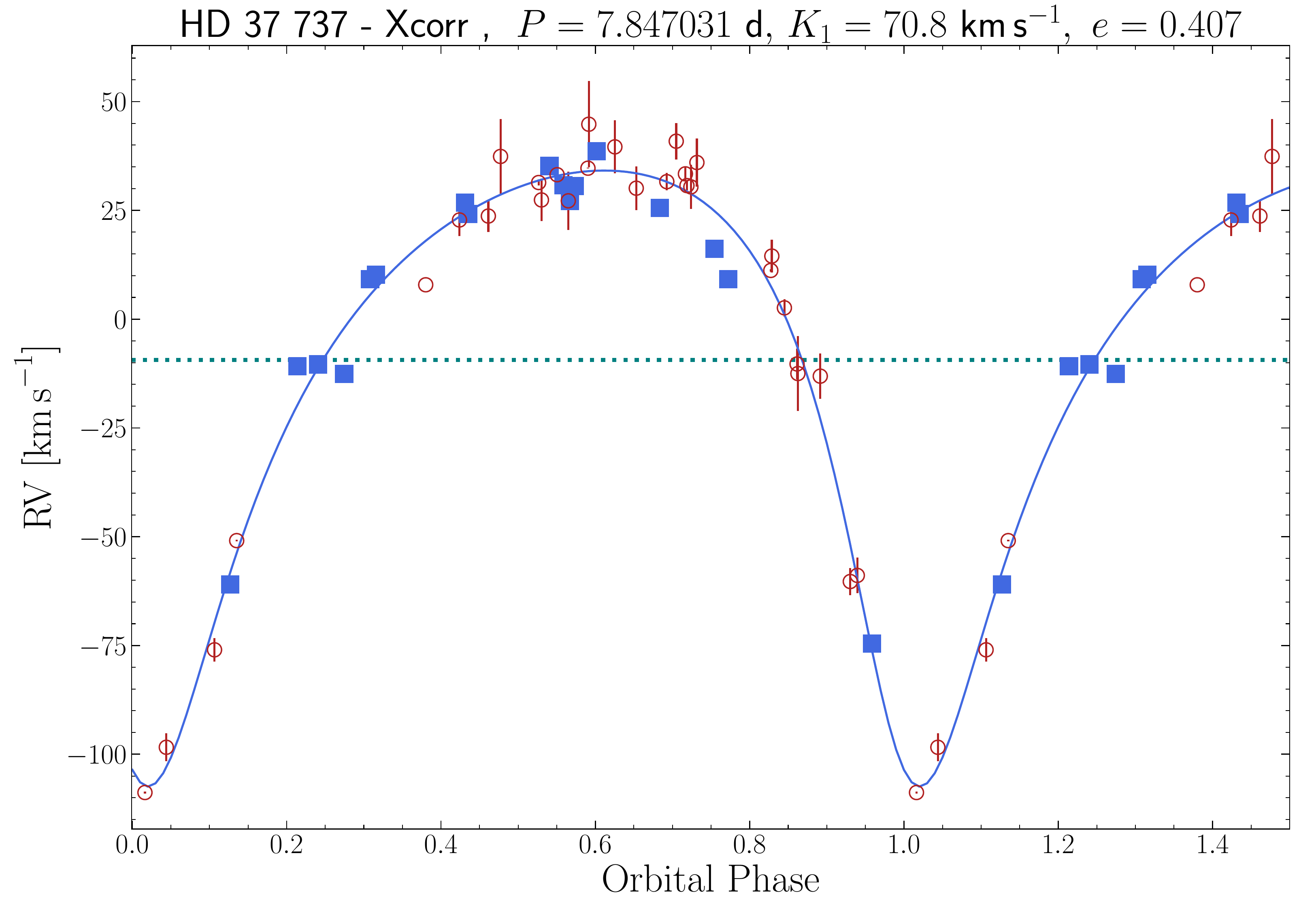} 
    \includegraphics[width=.5\textwidth]{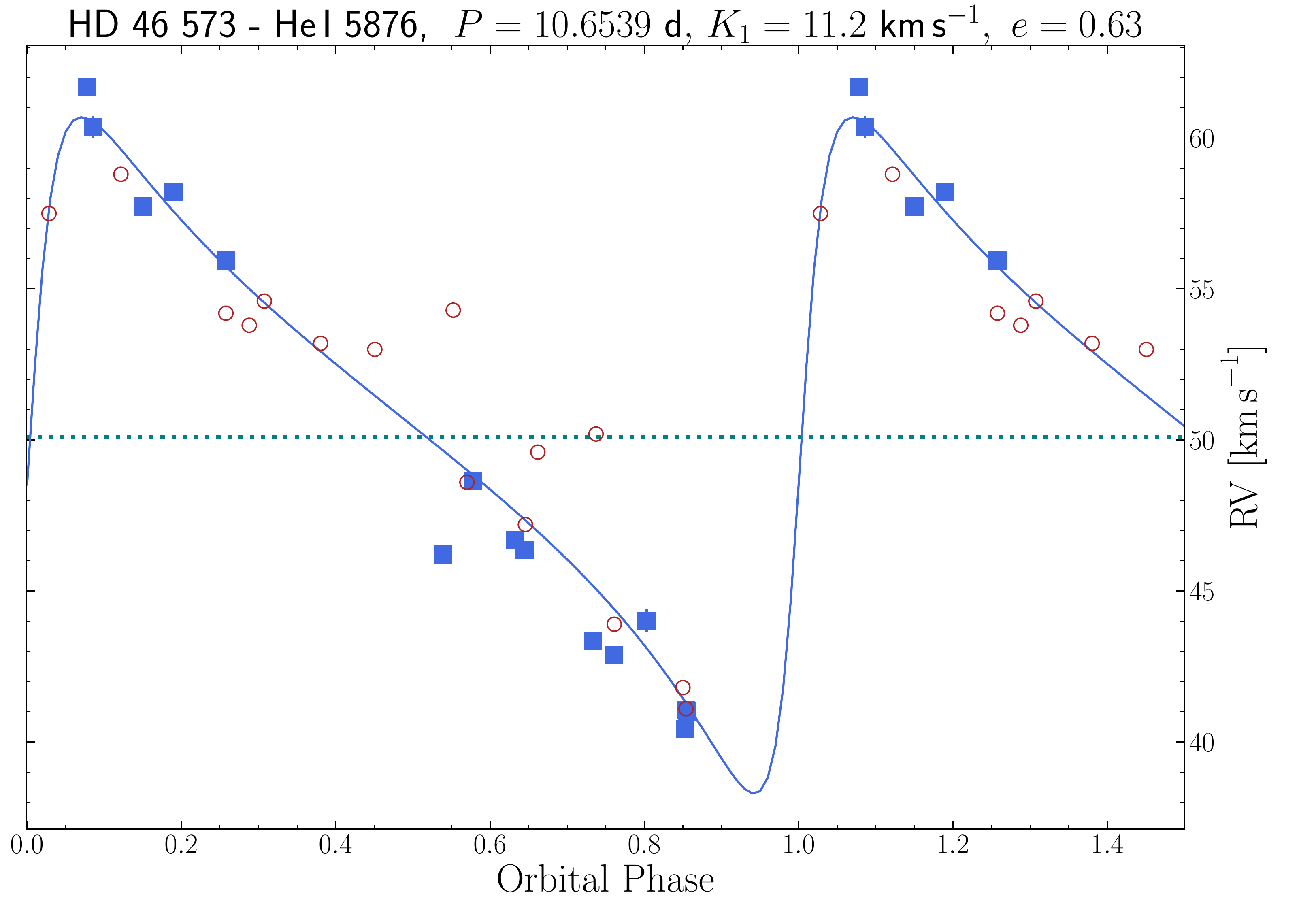} 
    \caption{New orbital solutions of several systems. MONOS RVs of each diagnostic line are shown as navy squares. The dotted blue line shows the $\gamma$ of the system.
        Upper left panel: Orbital solutions of V747~Cep. 
        Upper right panel: Orbital solution for HD~\num{12323}. Red circles are the RVs from \citet{BoltonRogers78}, and green triangles are from \citet{SticLloy01}.
        Middle left panel: New combined orbital solution for HD~\num{15137}.  Red circles are the RVs from \cite{McSwetal10} and \cite{Boyajianetal05}.
        Middle right panel: New combined orbital solution for HD~\num{14633}~AaAb. Red circles are the RVs from \cite{Boyajianetal05}. RVs from \cite{McSwetal07} (green triangles) are also shown but were not used to calculate the orbital solution.
        Lower left panel: New combined orbital solution for HD~\num{37737}. Red circles are the RVs from \cite{McSwetal07}, \cite{SticLloy01}, and \cite{Giesetal86}.
        Lower right panel: New combined orbital solution for HD~\num{46573}. Red circles are the RVs from \cite{Mahyetal09}.
    }\label{orb-fig:3}
\end{figure*}

\newpage
\newpage
\quad

\begin{figure*}[!htp]
\centering
    \includegraphics[width=.495\textwidth]{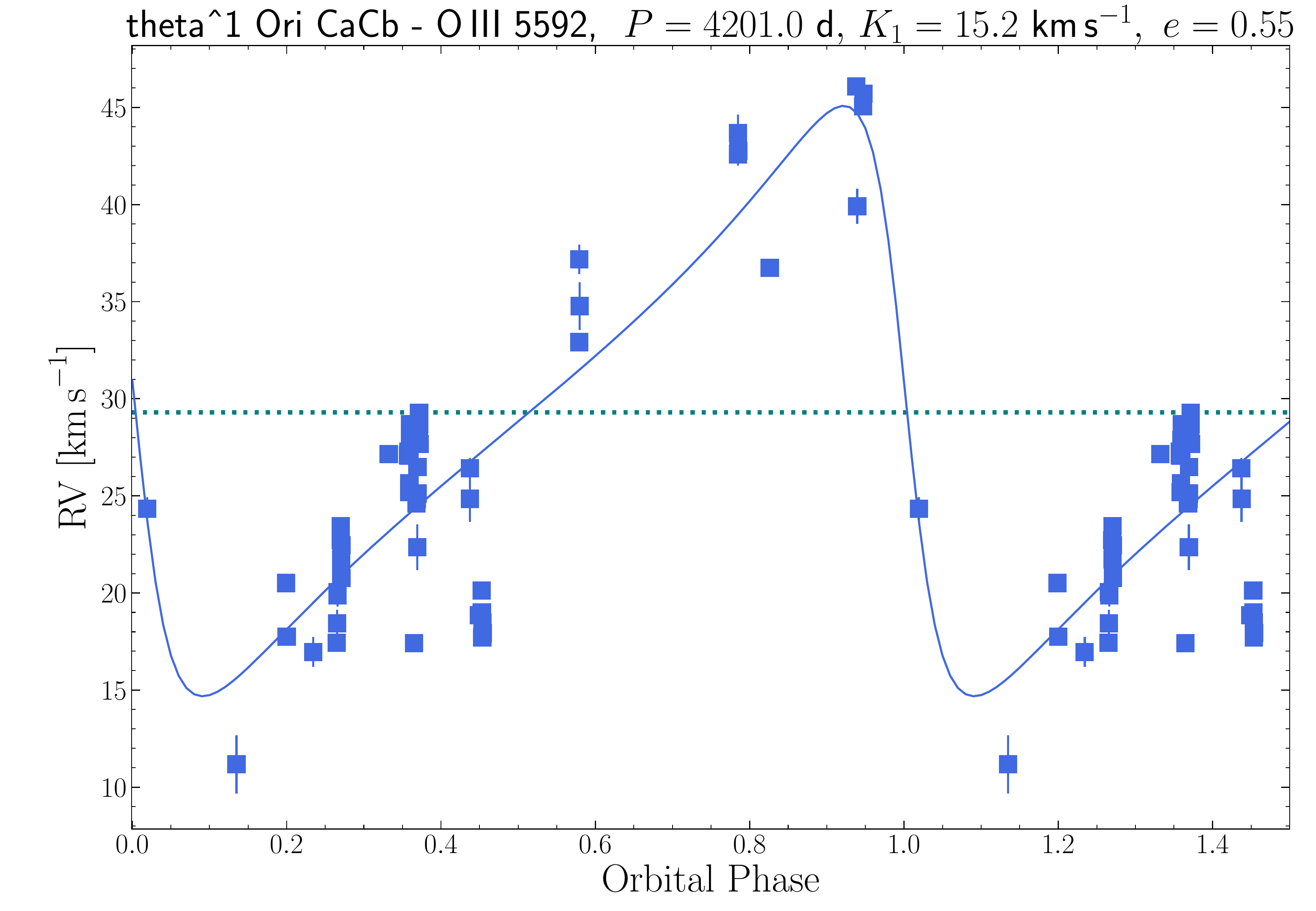} 
    \includegraphics[width=.495\textwidth]{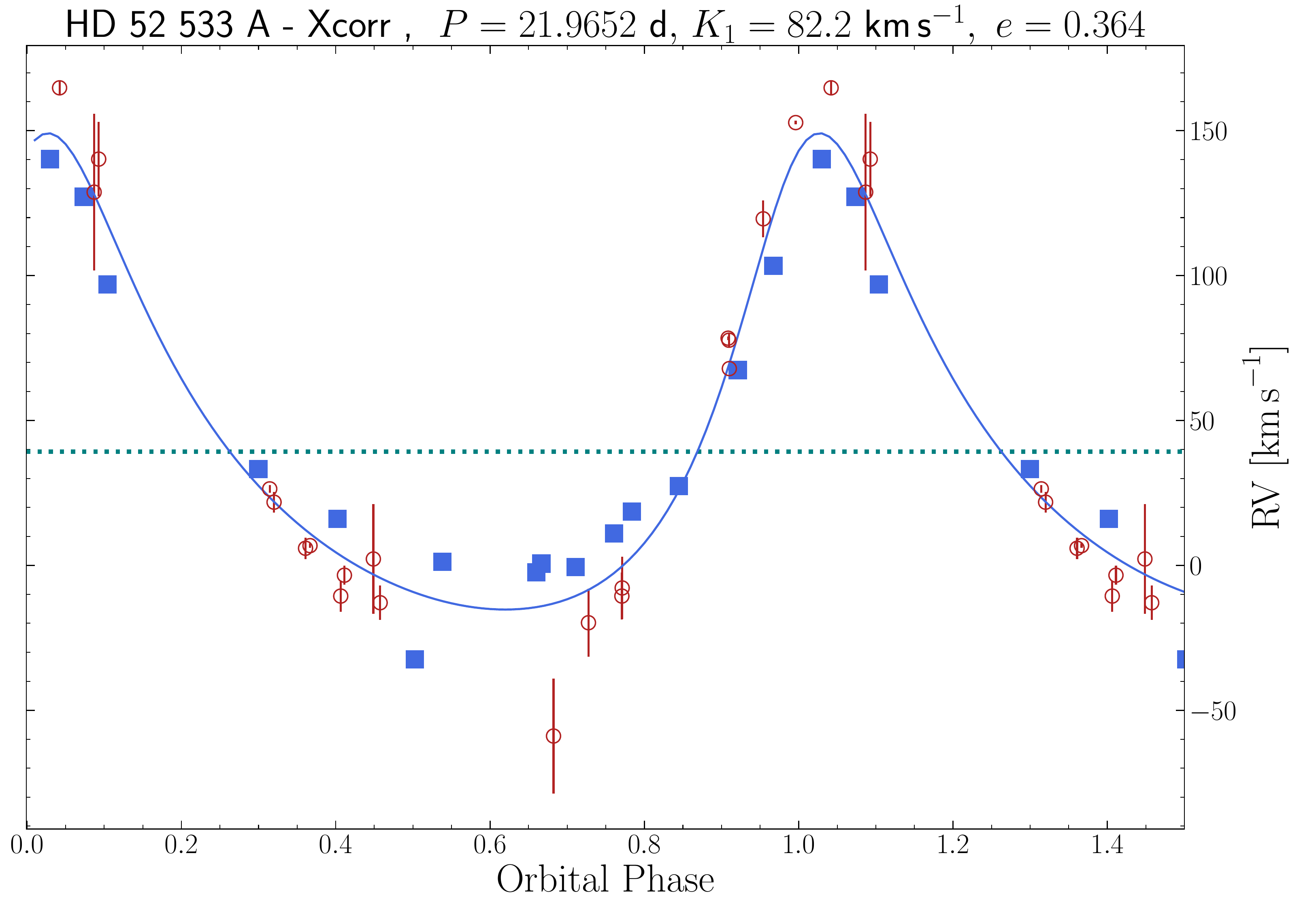}     
    \caption{New orbital solutions of several systems. MONOS RVs of each diagnostic line are shown as navy squares. The dotted blue line shows the $\gamma$ of the system.
        Left panel:New orbital solution for $\theta ^1$~Ori~CaCb.  
        Right panel: New combined orbital solution for HD~\num{52533}~A. Red circles are the RVs from \cite{McSwetal07}.
        }\label{orb-fig:4}
\end{figure*}

\subsection{LC}

\begin{figure*}[!htp]
    \includegraphics[width=.5\textwidth]{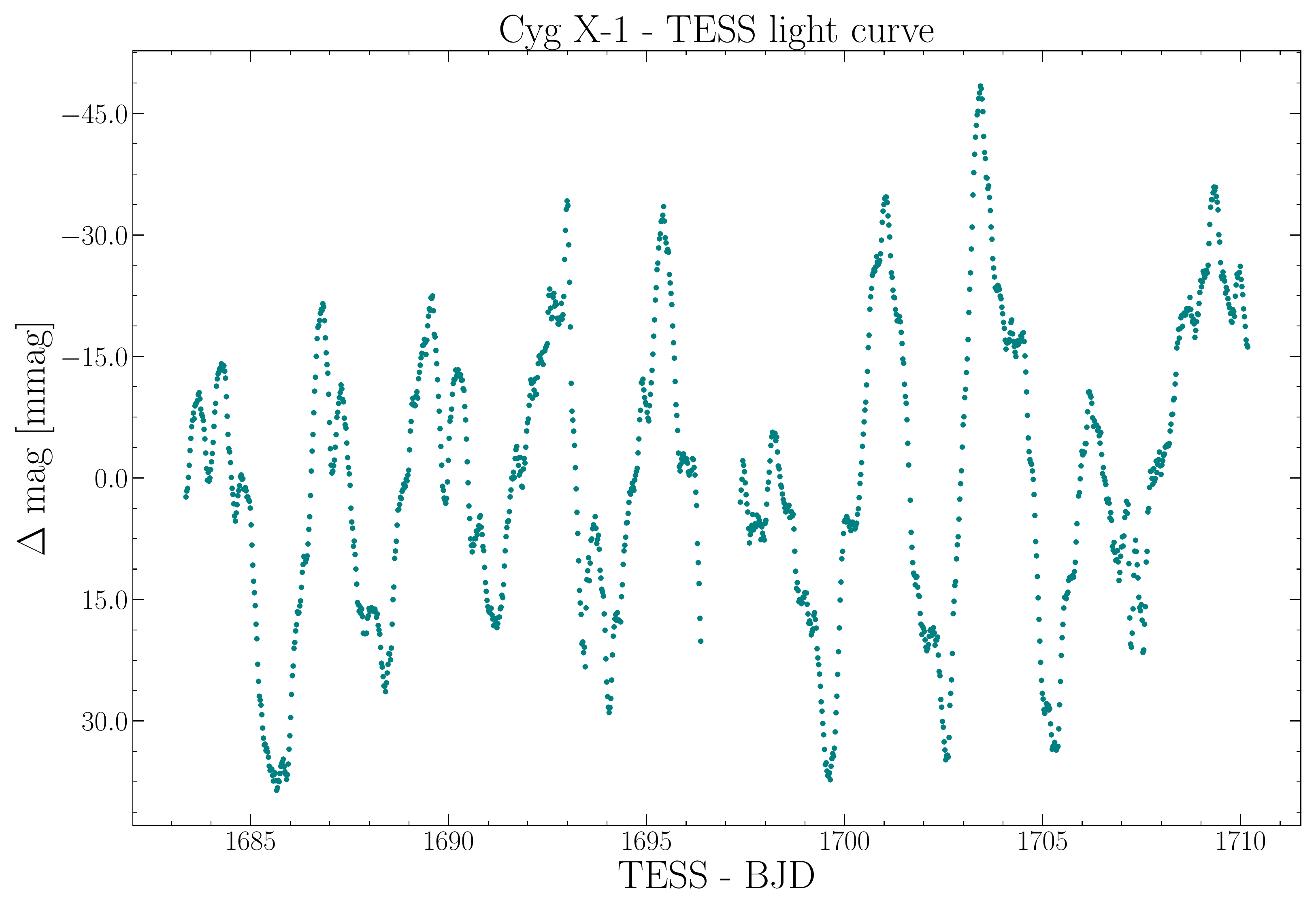}
    \includegraphics[width=.5\textwidth]{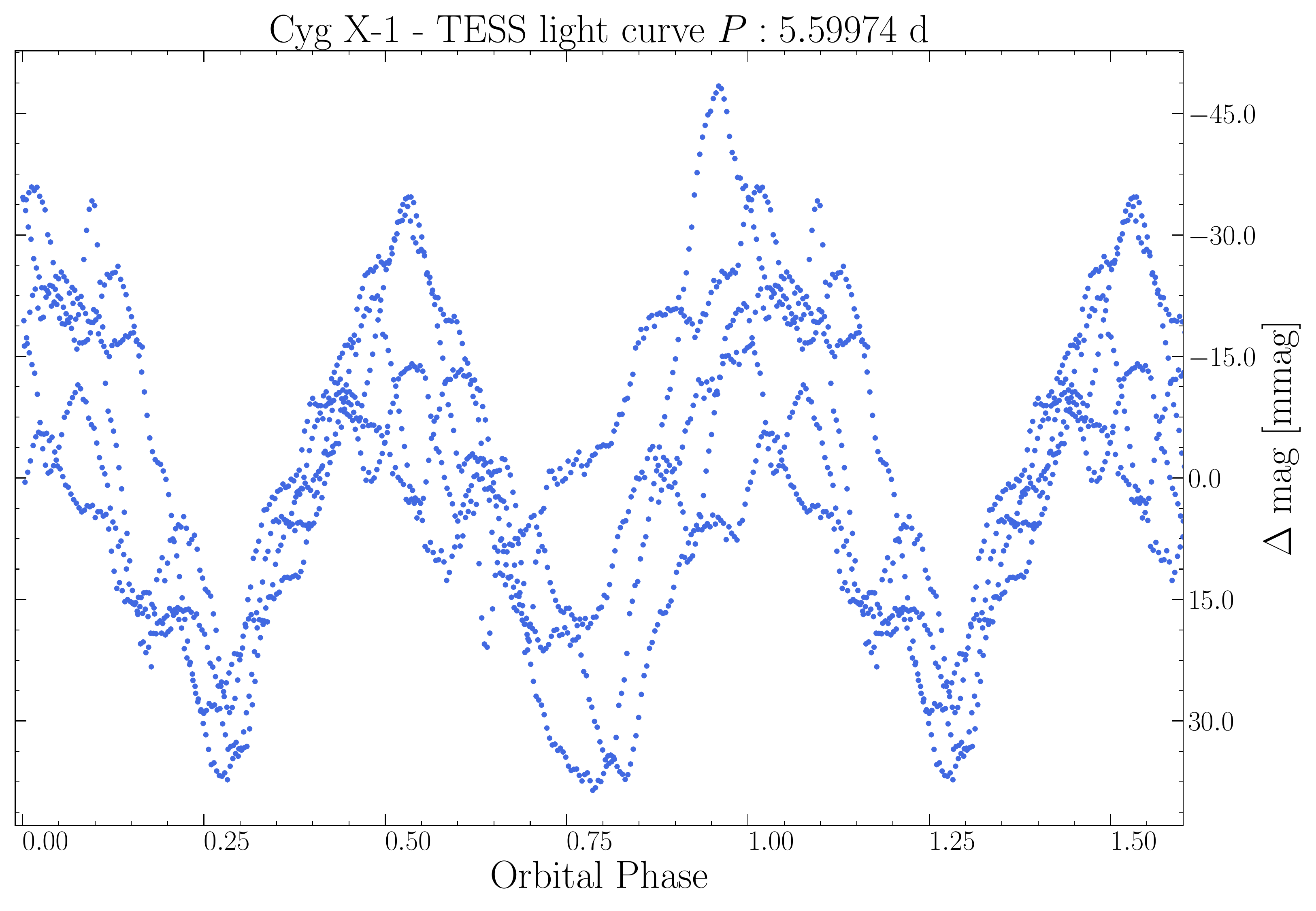} 
    \includegraphics[width=.5\textwidth]{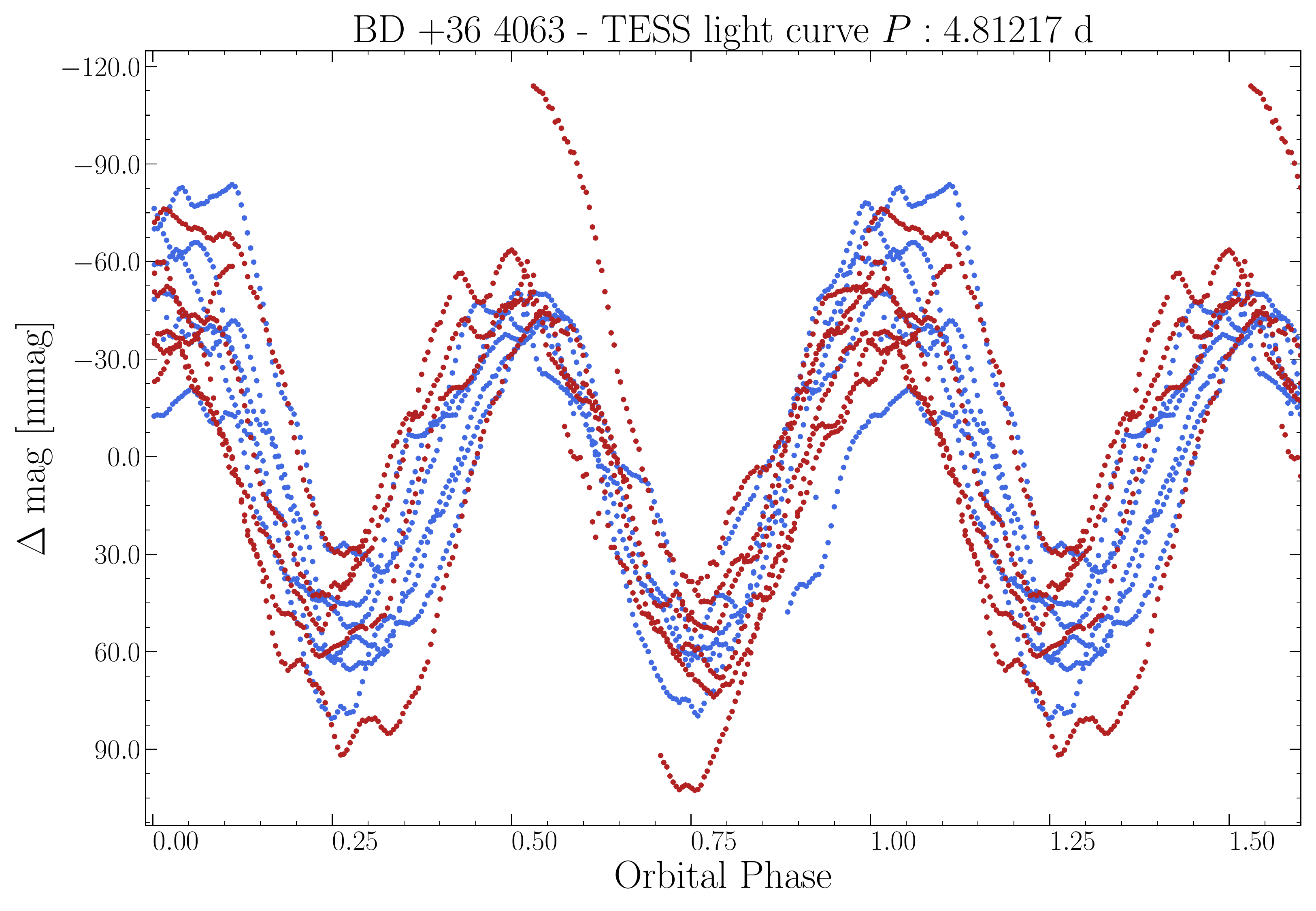}
    \includegraphics[width=.5\textwidth]{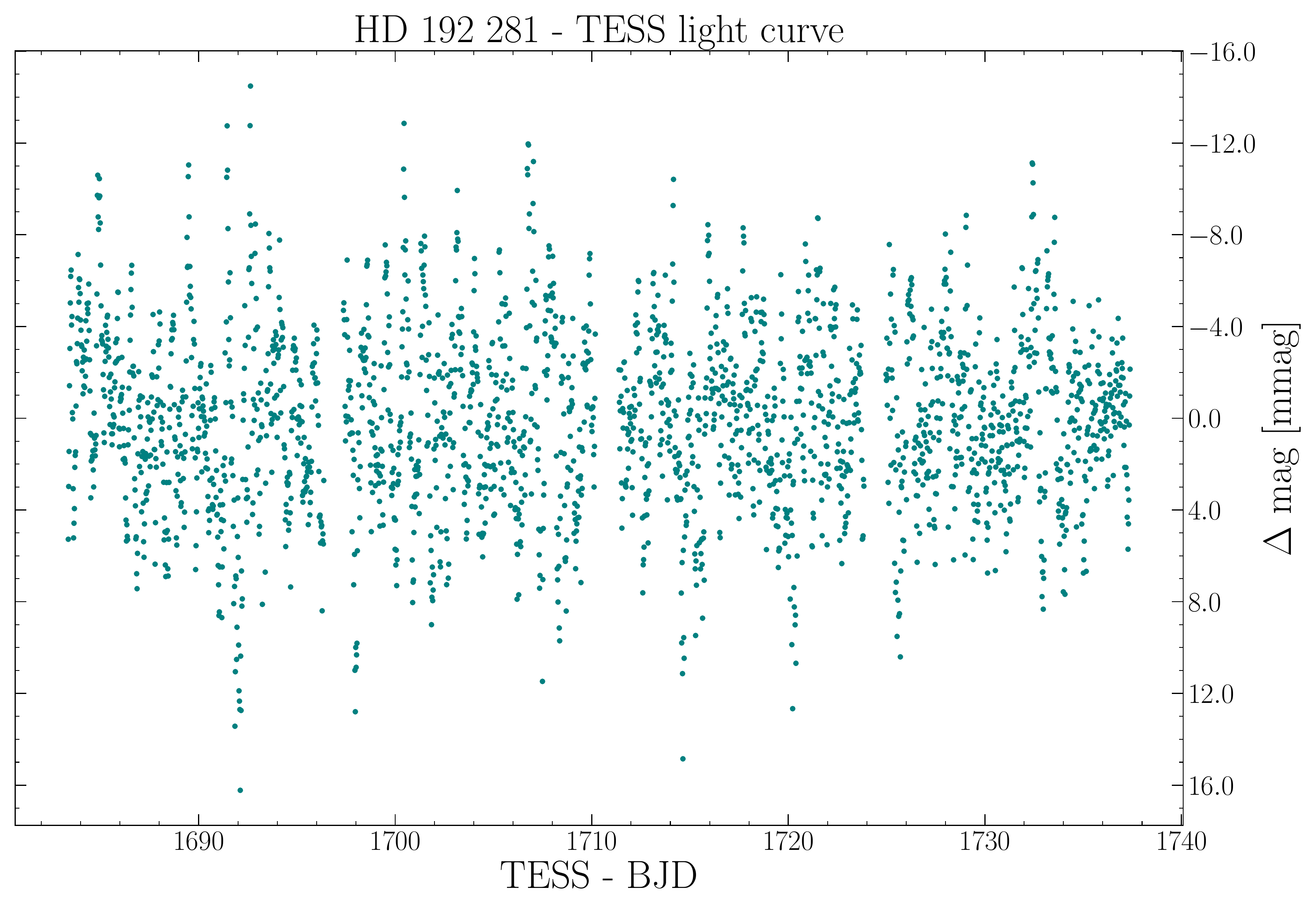} 
    
    \includegraphics[width=.5\textwidth]{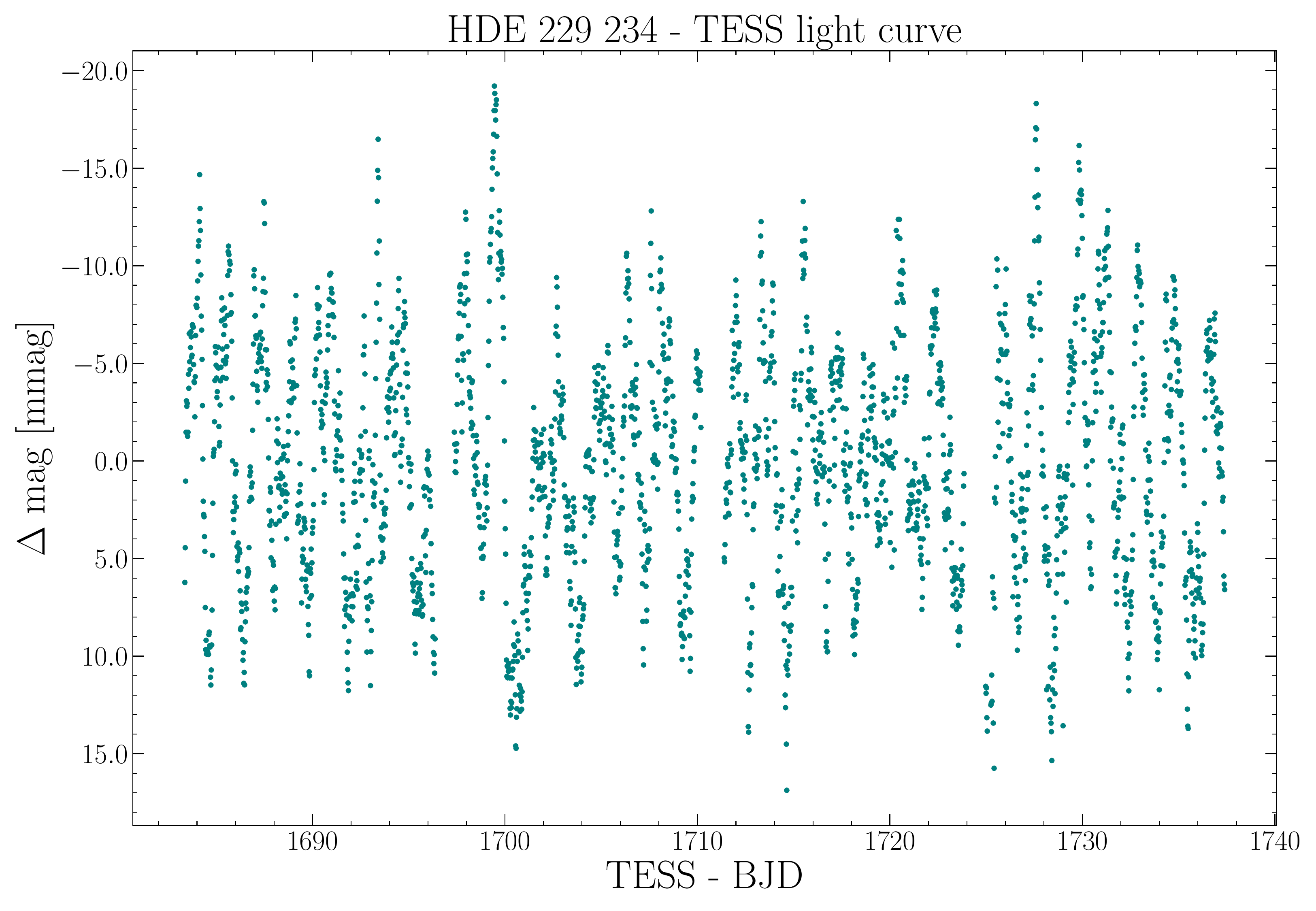}
    \includegraphics[width=.5\textwidth]{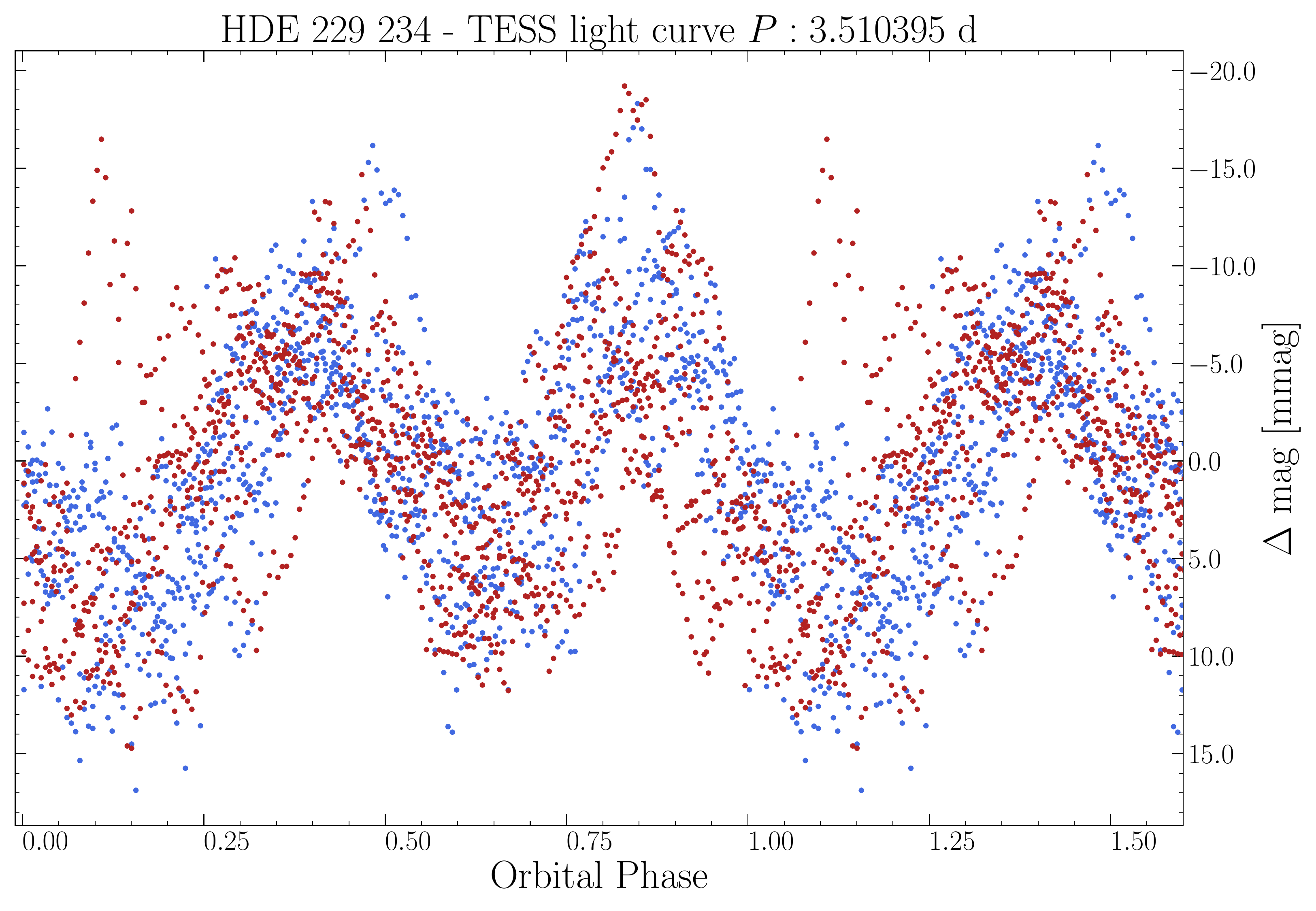} 
    \caption{Normalized and folded TESS LCs of the following systems. 
        Upper left panel: LC for Cyg~X-1 in sector 14. 
        Upper right panel: Folded  LC for Cyg~X-1 using the orbital period. 
        Middle left panel: Folded  LC for BD~$+$36~4063 using the orbital period. 
        Middle right panel: LC for HD~\num{192281} in sectors 14 and 15. 
        Lower left panel: LC for HDE~\num{229234} observations in sectors 14 (blue) and 15 (red). 
        Lower right panel: Folded LC using the orbital period for HDE~\num{229234}. 
        }\label{tess-fig:1}
\end{figure*}

\begin{figure*}[!htp]
    \includegraphics[width=.5\textwidth]{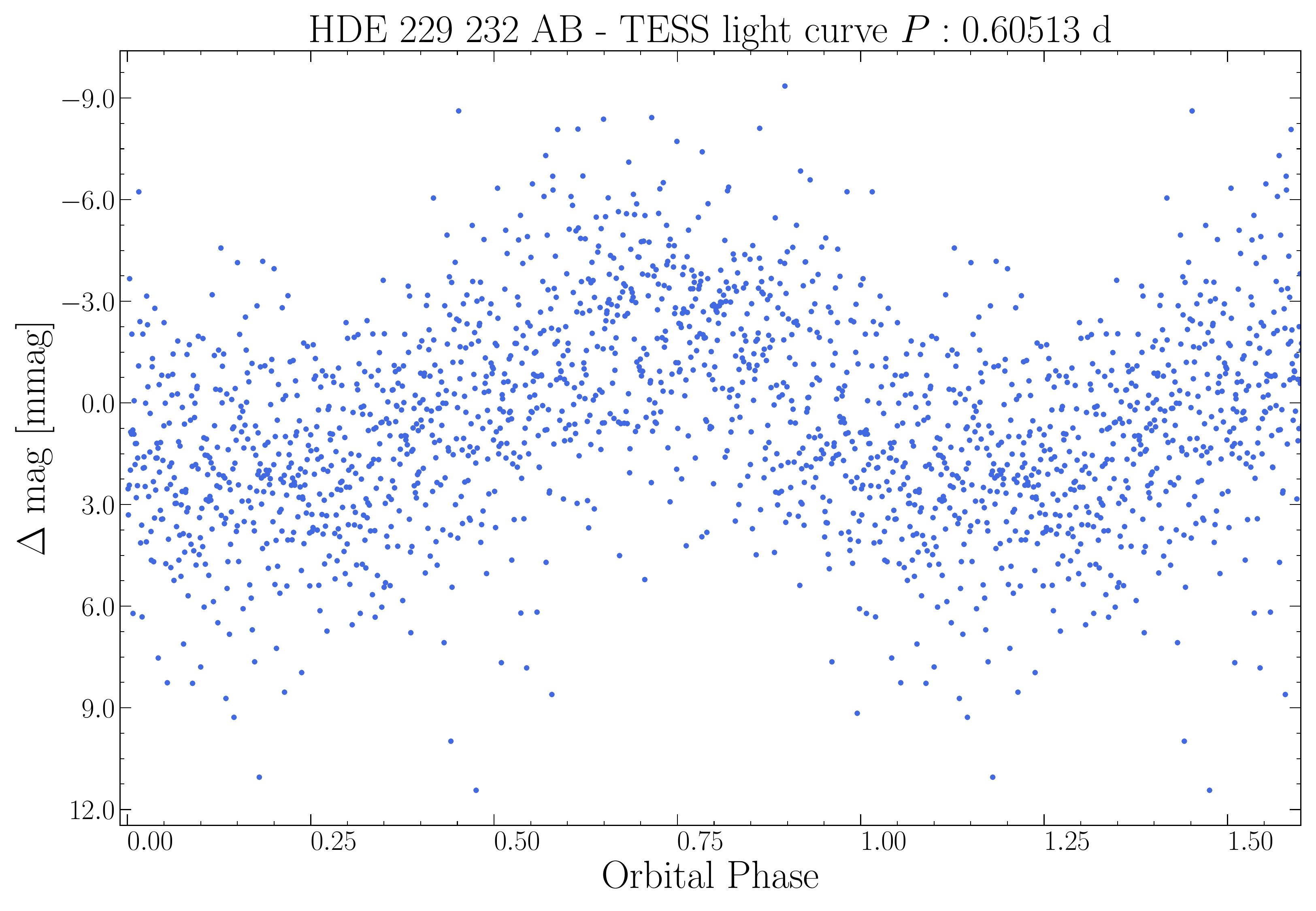} 
    \includegraphics[width=.5\textwidth]{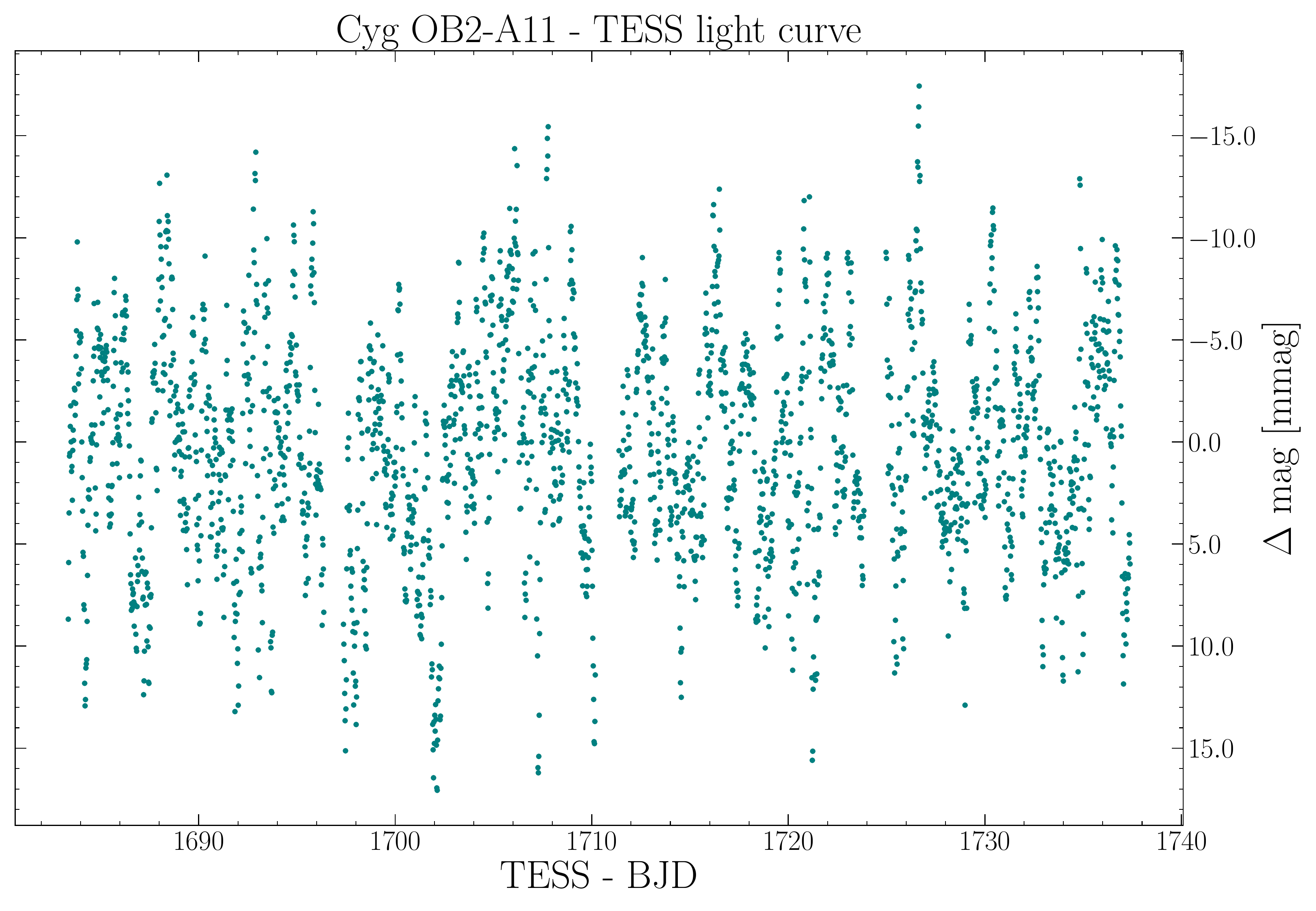} 
    
    \includegraphics[width=.5\textwidth]{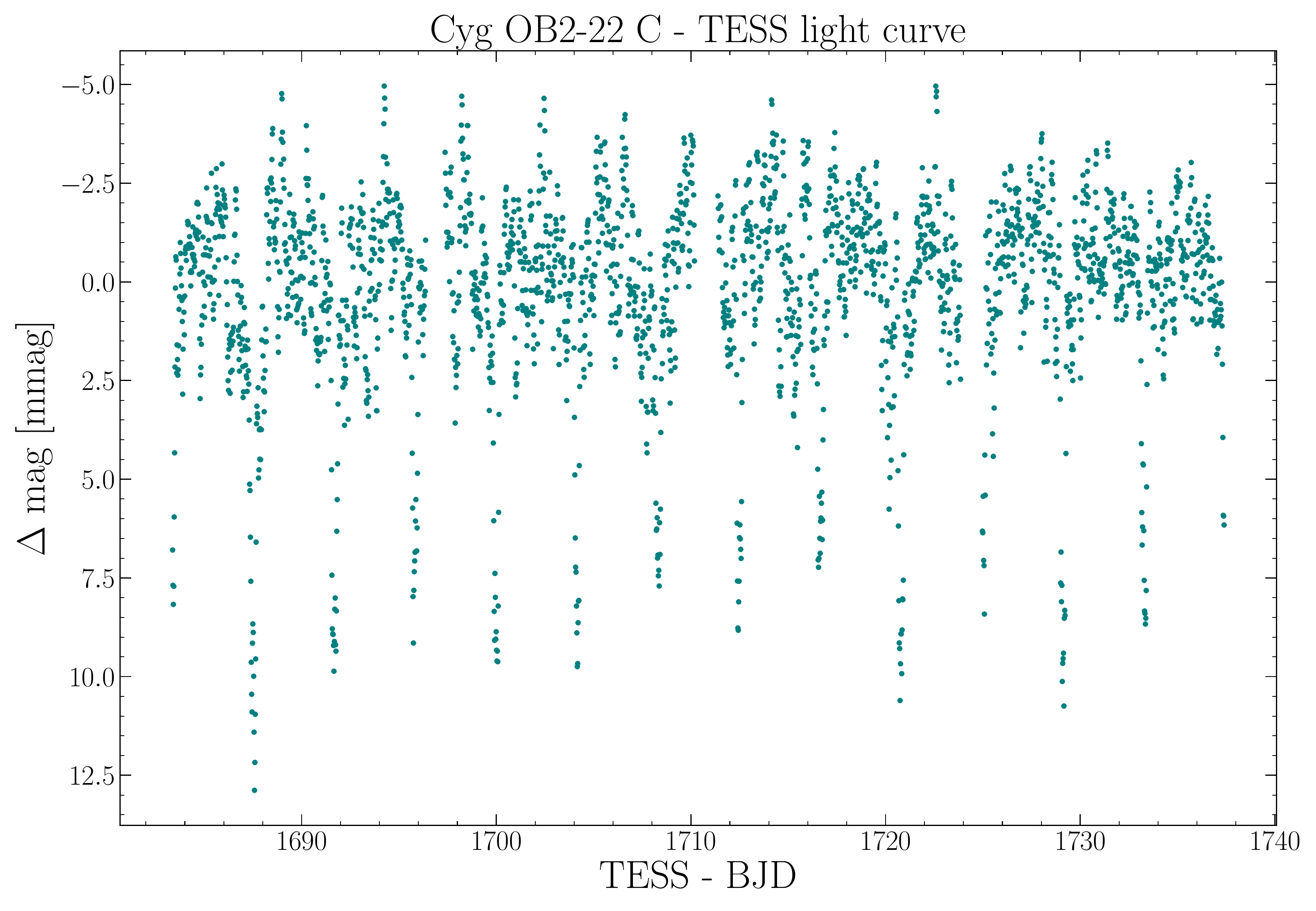}     
    \includegraphics[width=.5\textwidth]{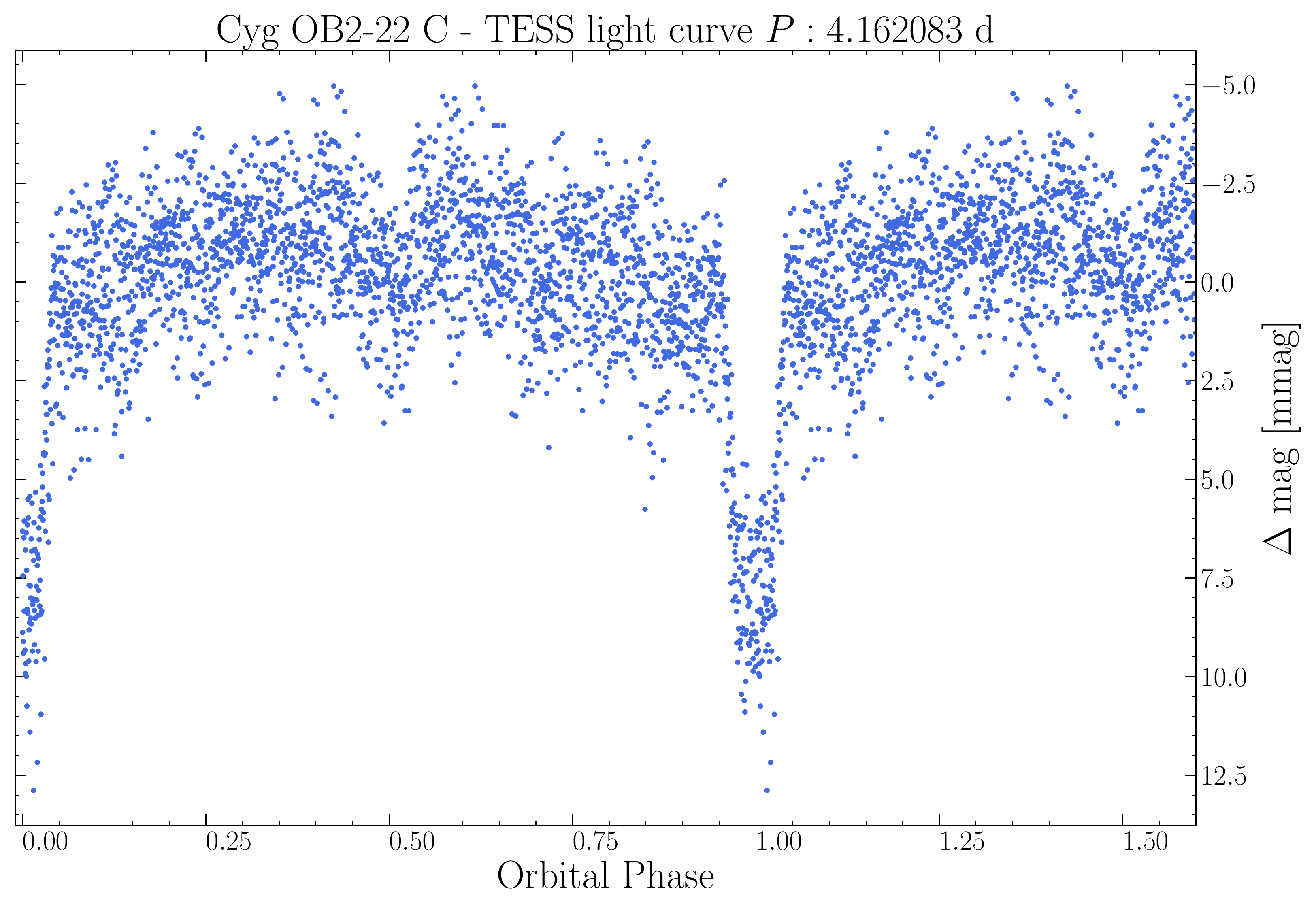}  
    
    \includegraphics[width=.5\textwidth]{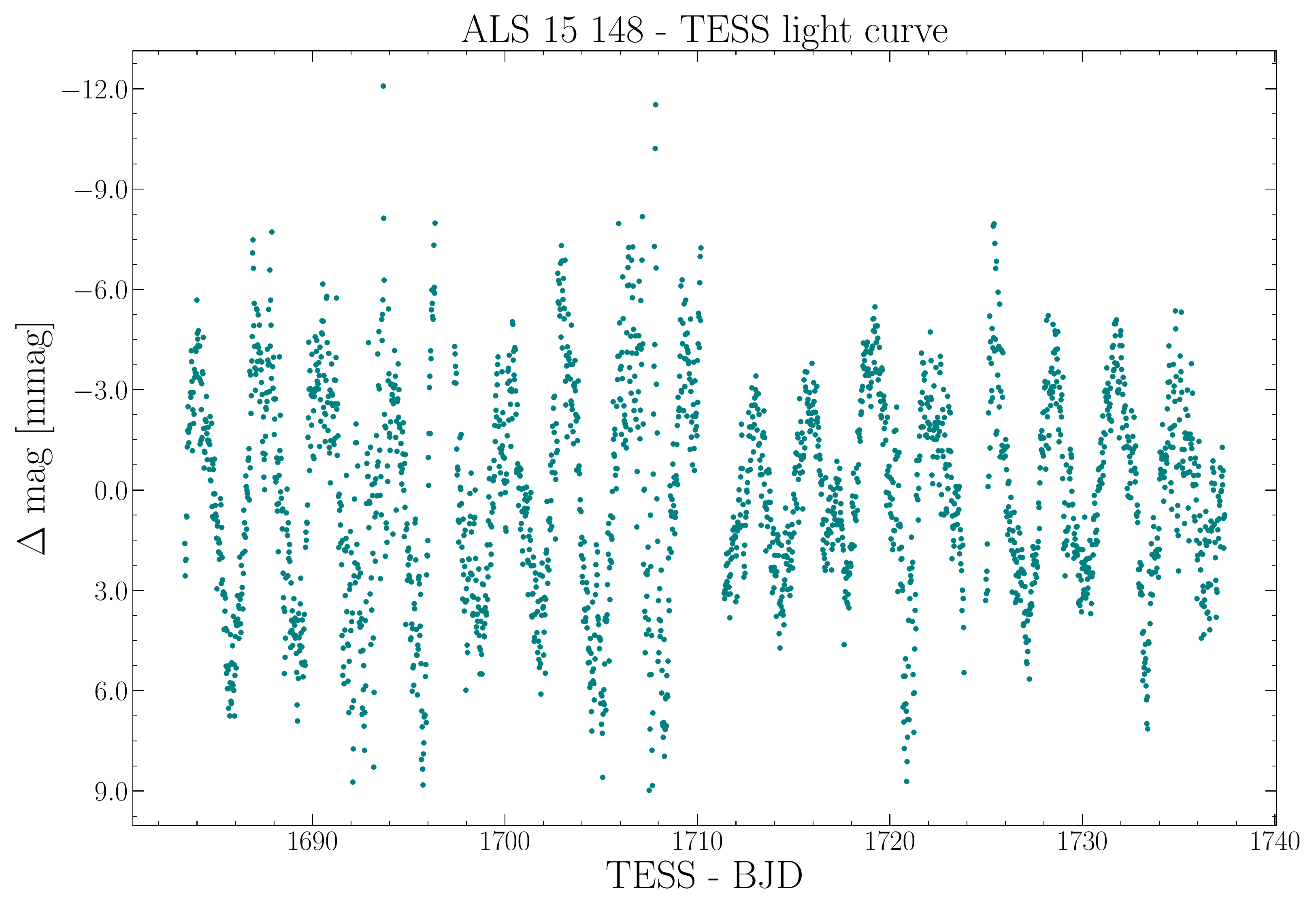}                              
    \includegraphics[width=.5\textwidth]{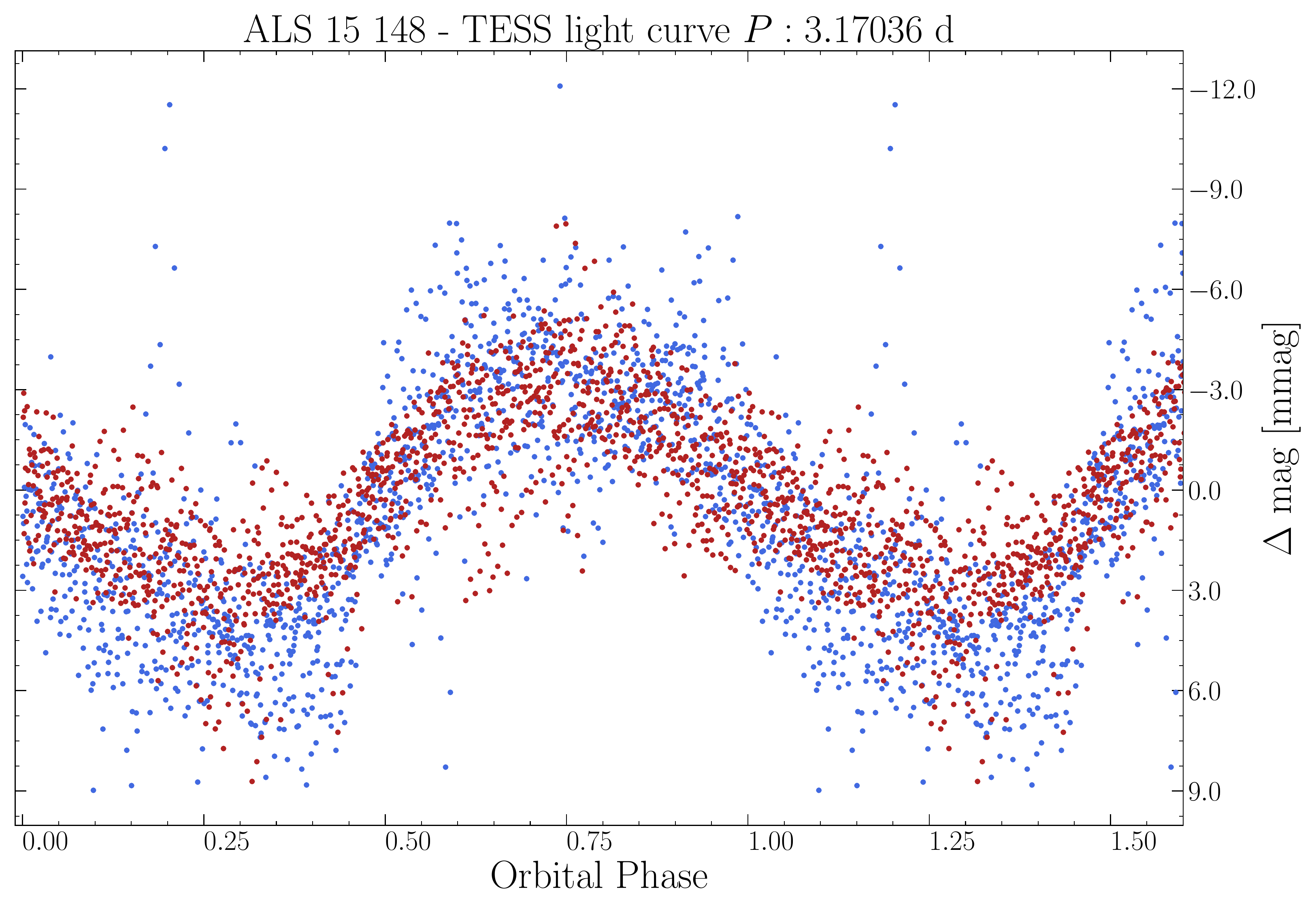} 
    \caption{Normalized  and folded TESS LCs of the following systems. 
        Upper left panel: LC for HDE~\num{229232}~AB in sector 14. 
        Upper right panel: LC for Cyg~OB2-A11 in sectors 14 and 15. 
        Middle left panel: LC for Cyg~OB2-22~C in sectors 10 and 11.
        Middle right panel: Folded LC using a period $P=\num{4.16083}$~d for Cyg~OB2-22~C. 
        Lower left panel: LC for ALS~\num{15148} in sectors 14 (blue) and 15 (red). The LC is diluted due to the contamination of neighbor stars. 
        Lower right panel: Folded LC for ALS~\num{15148} using the orbital period.
        }\label{tess-fig:2}
\end{figure*}

\begin{figure*}[!htp]
    \includegraphics[width=.5\textwidth]{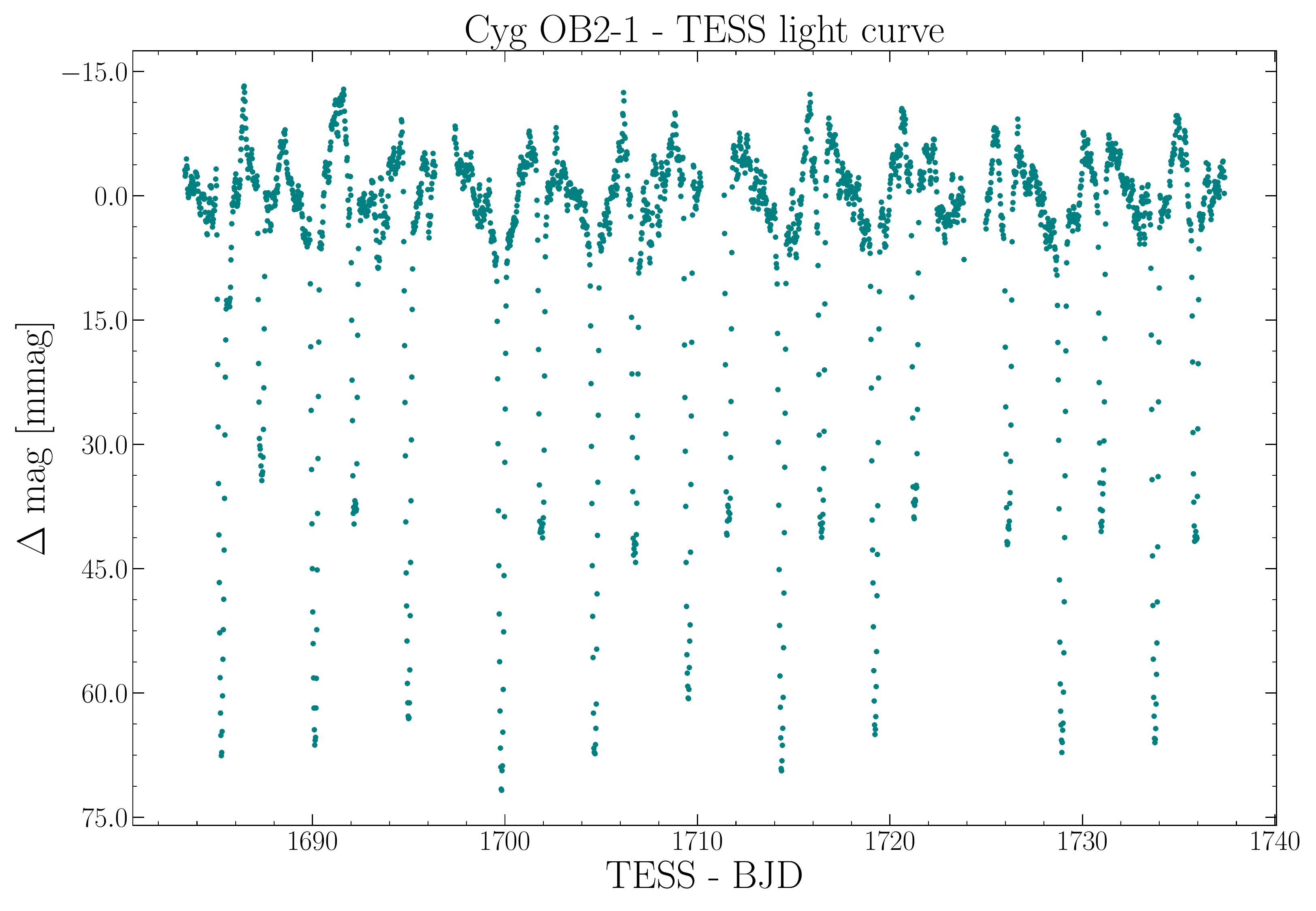}                             
    \includegraphics[width=.5\textwidth]{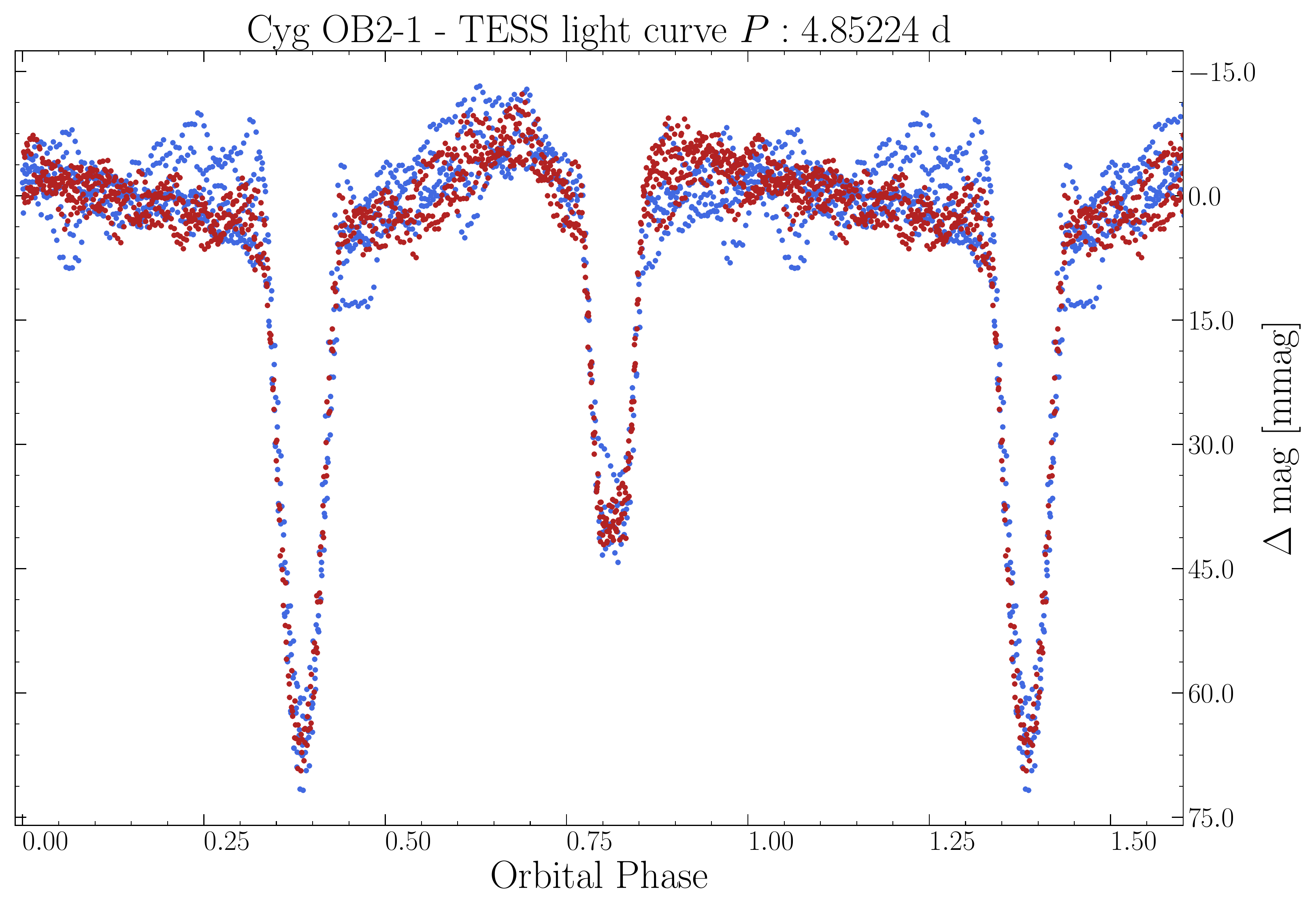} 
    
    \includegraphics[width=.5\textwidth]{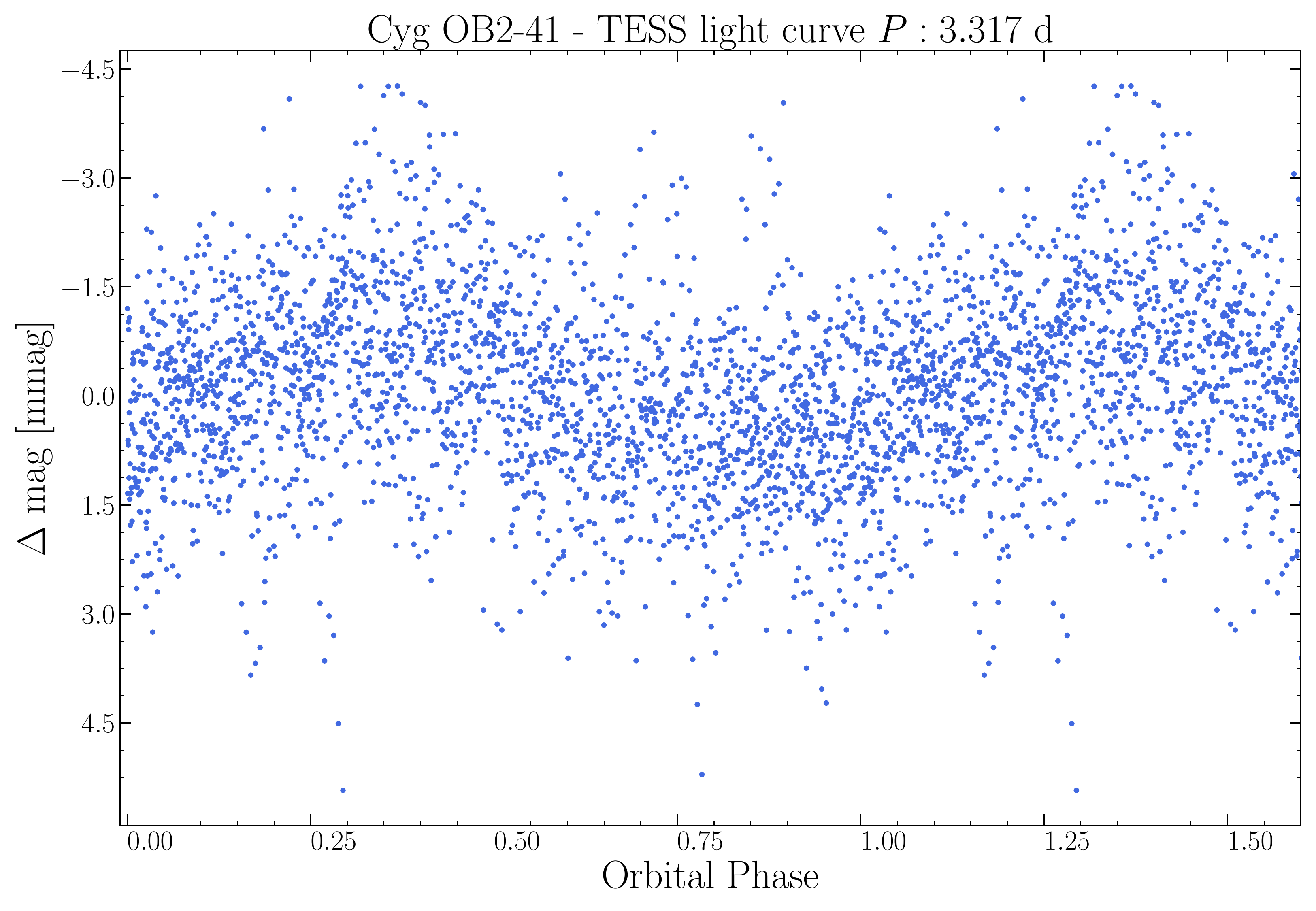} 
    \includegraphics[width=.5\textwidth]{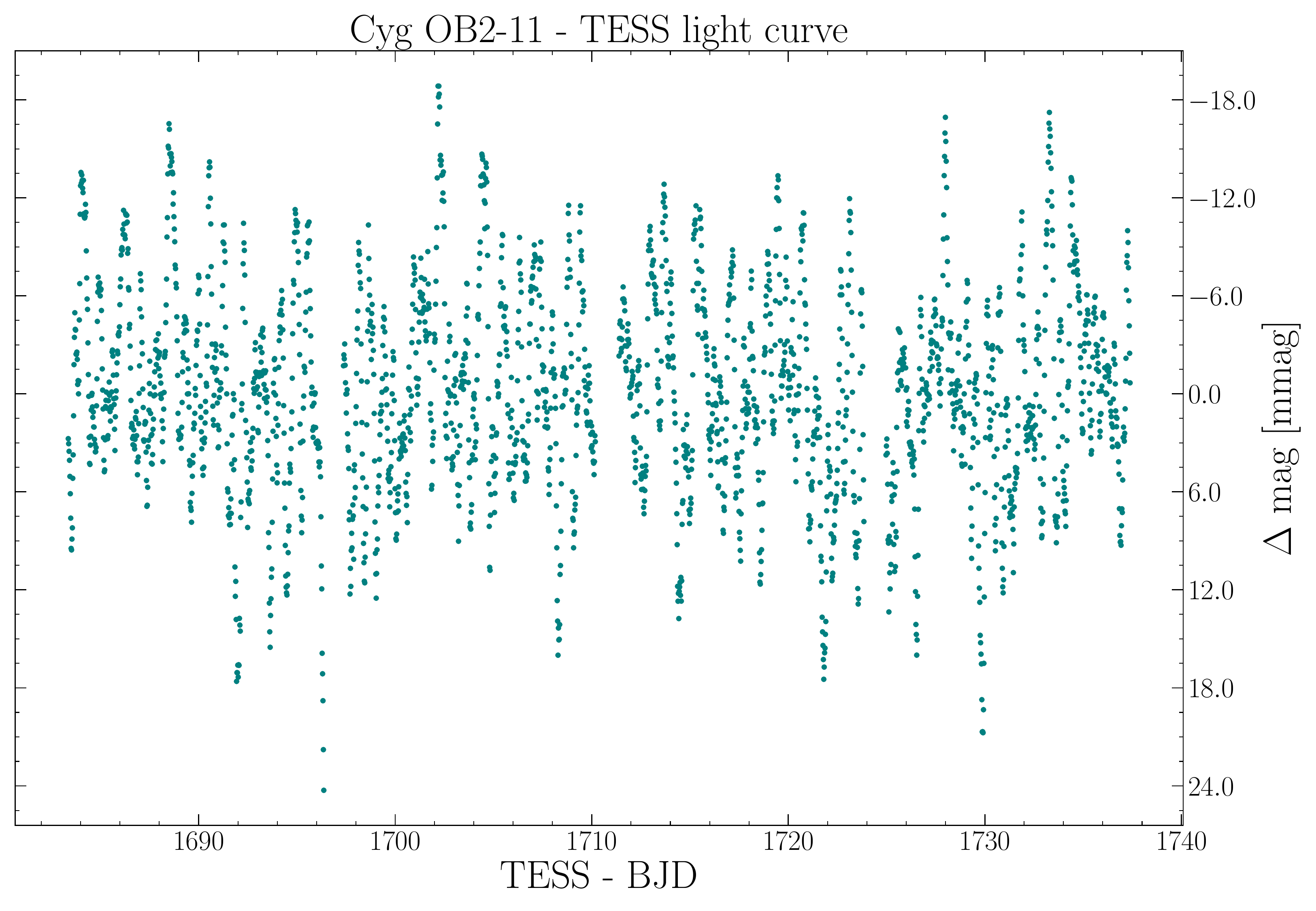} 
    
    \includegraphics[width=.5\textwidth]{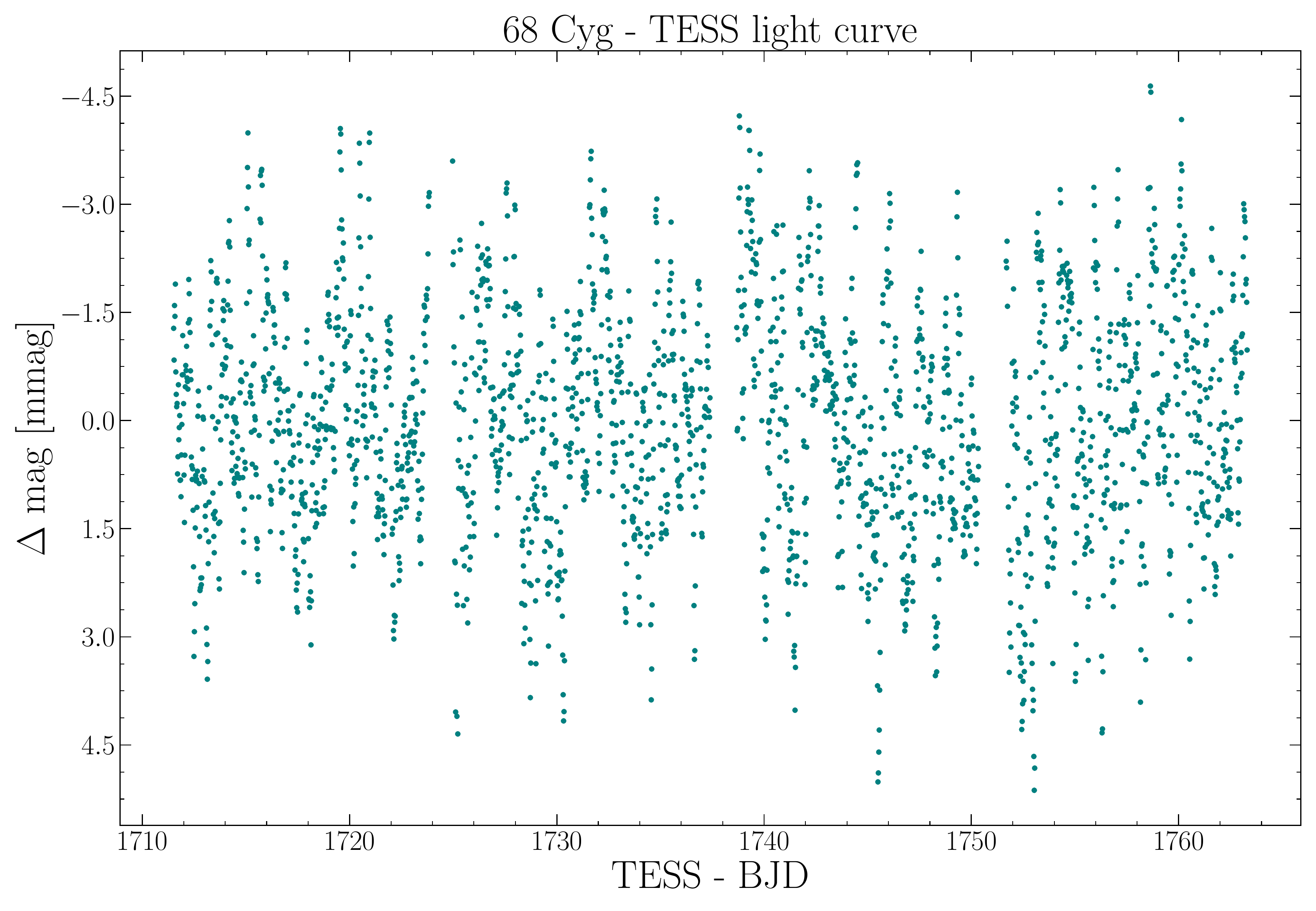}                            
    \includegraphics[width=.5\textwidth]{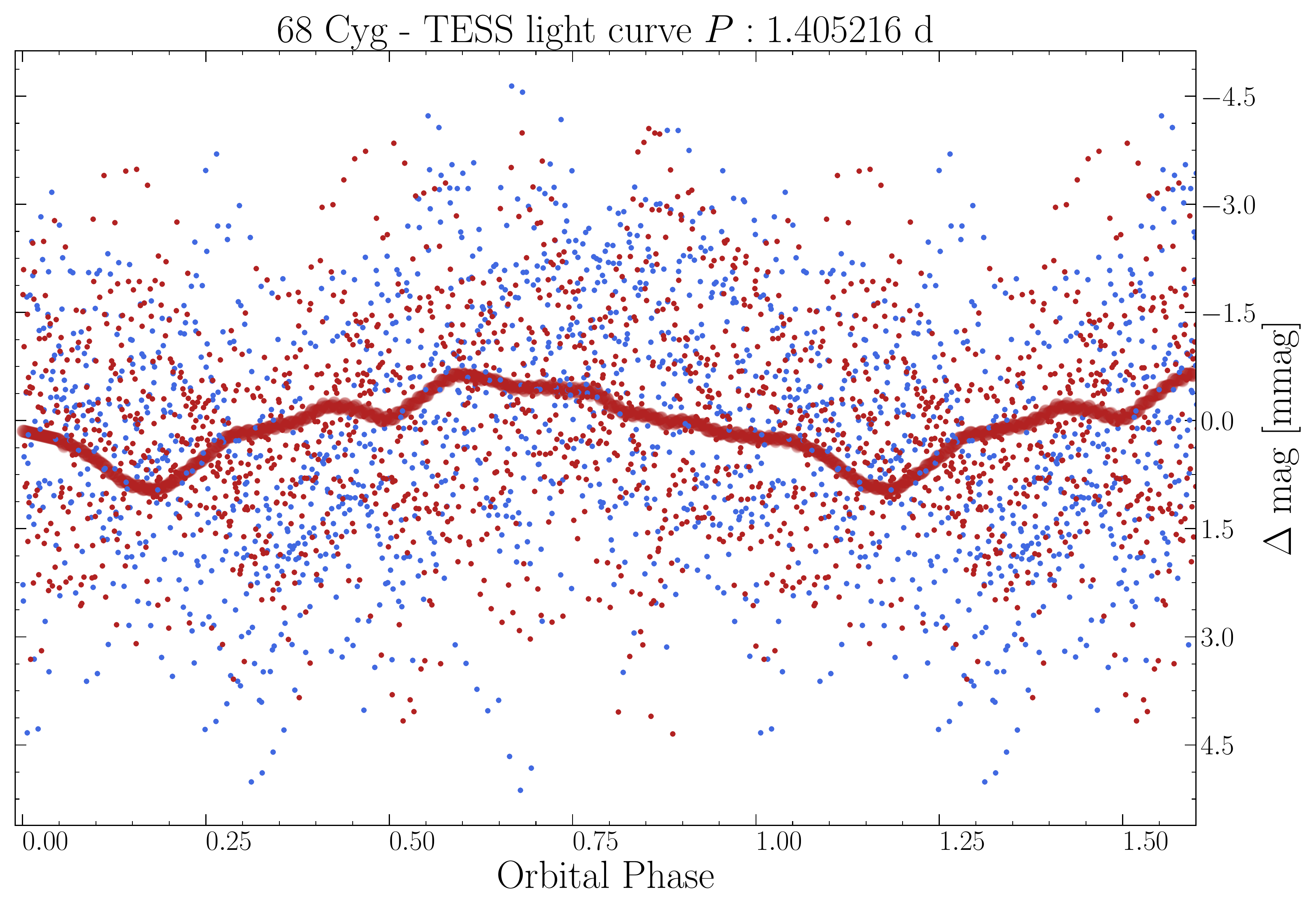} 
    \caption{Normalized and folded TESS LCs of the following systems.  
        Upper left panel: LC for Cyg~OB2-1 in sector 14 (blue) and 15 (red).  
        Upper right panel: LC folded for Cyg~OB2-1 using the orbital period. 
        Middle left panel: Folded LC using a period $P=\num{3.317}$~d for Cyg~OB2-41.
        Middle right panel: LC for Cyg OB2-11 in sectors 14 and 15.  
        Lower left panel: LC for 68~Cyg observed in sectors 15 and 16 (about 54 days).
        Lower right panel: Folded LC for 68~Cyg using the most significant period, $P=\num{1.405216}$~d. Red dots represent a smoothing done to enhance the perception of the periodicity in the data. 
        }\label{tess-fig:3}
\end{figure*}

\begin{figure*}[!htp]
    \includegraphics[width=.5\textwidth]{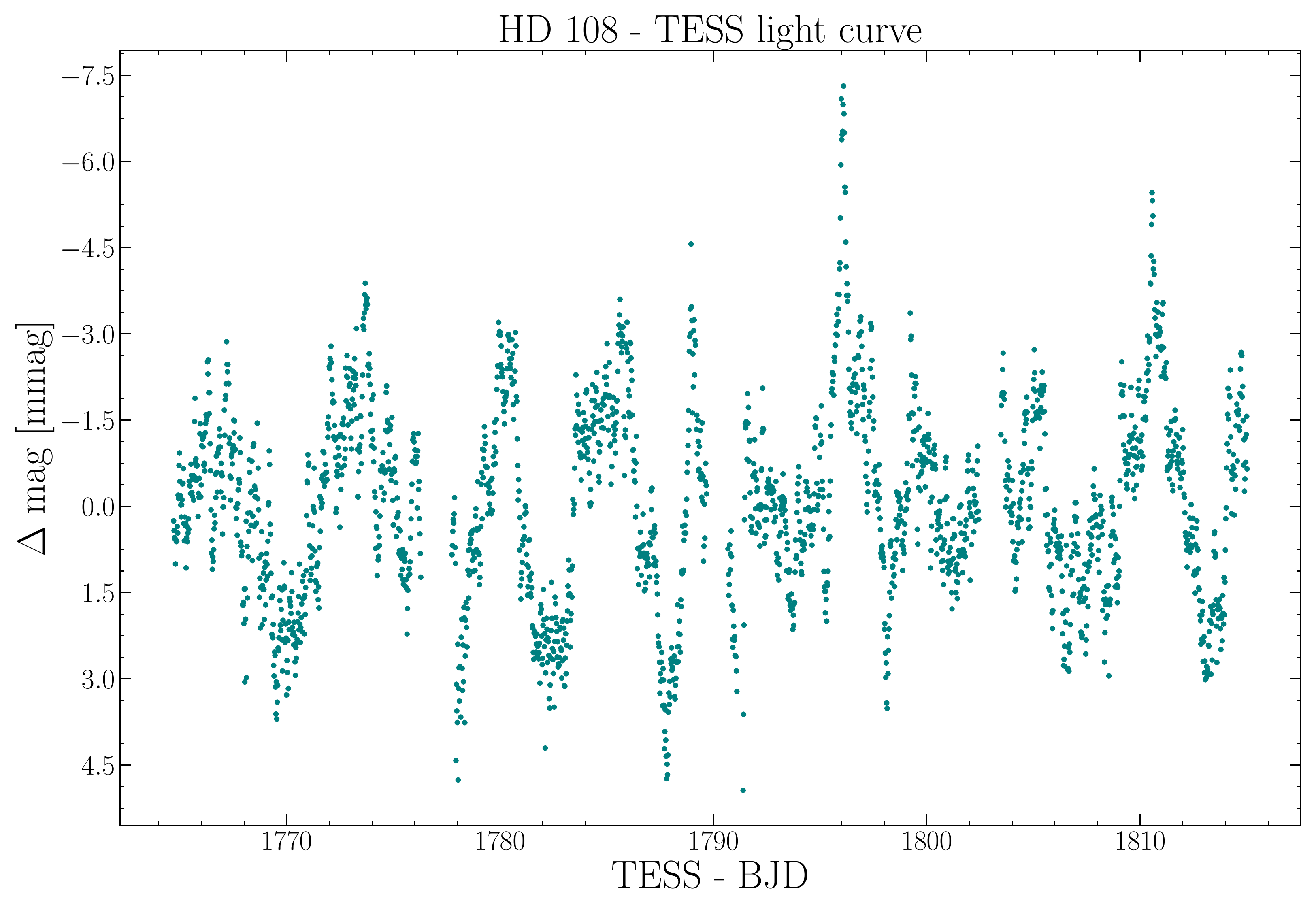}        
    \includegraphics[width=.5\textwidth]{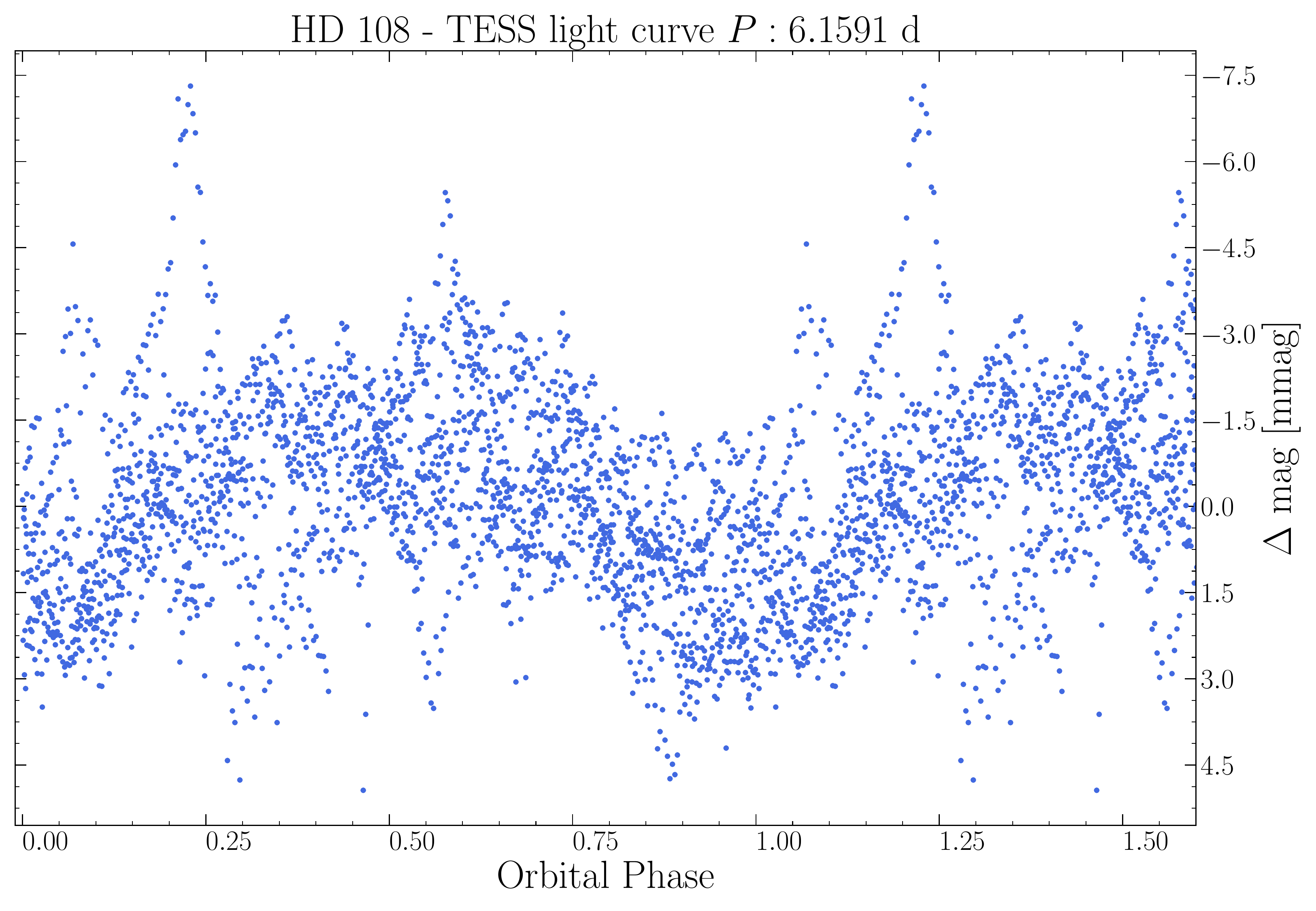}        
    
    \includegraphics[width=.5\textwidth]{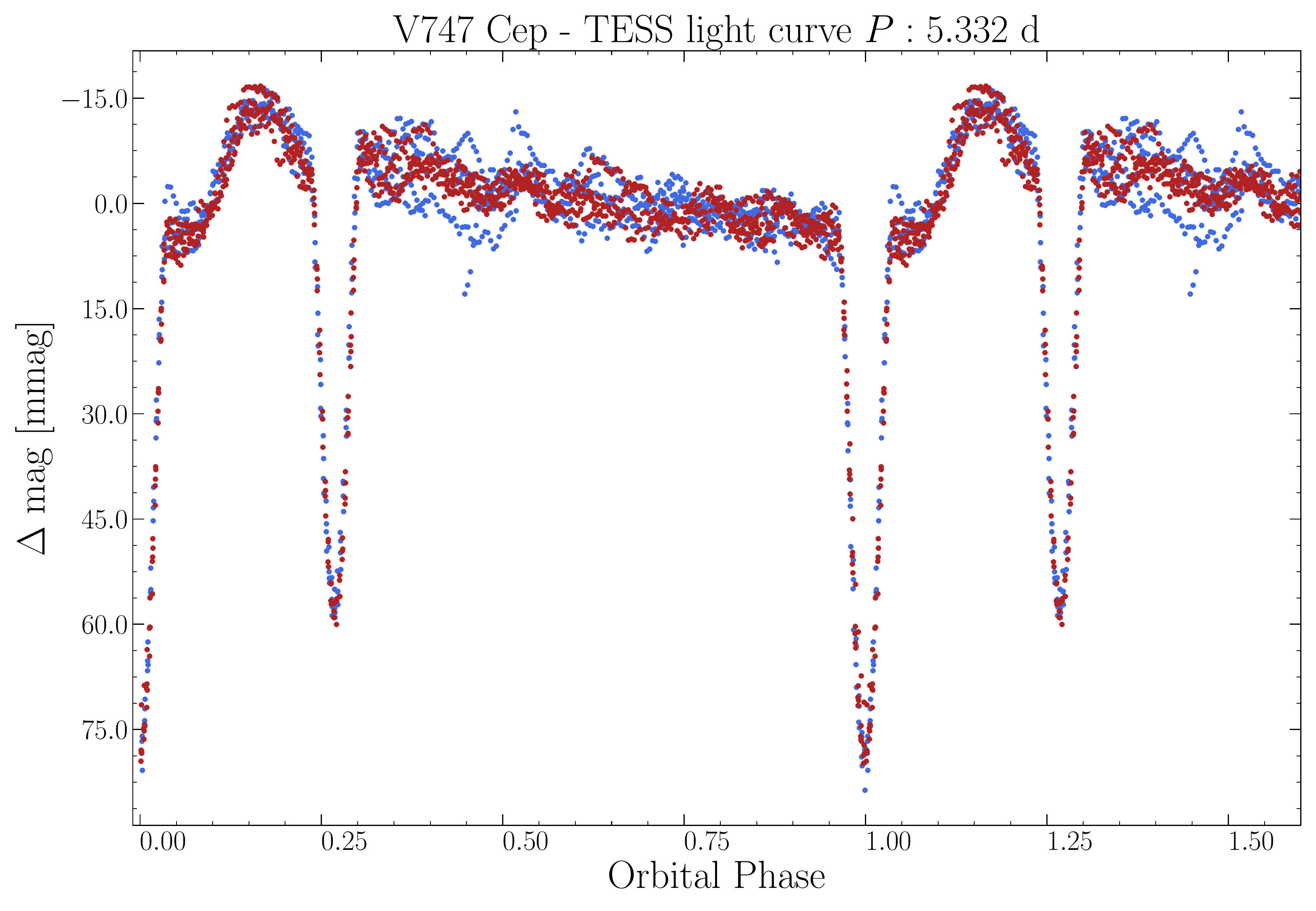} 
    \includegraphics[width=.5\textwidth]{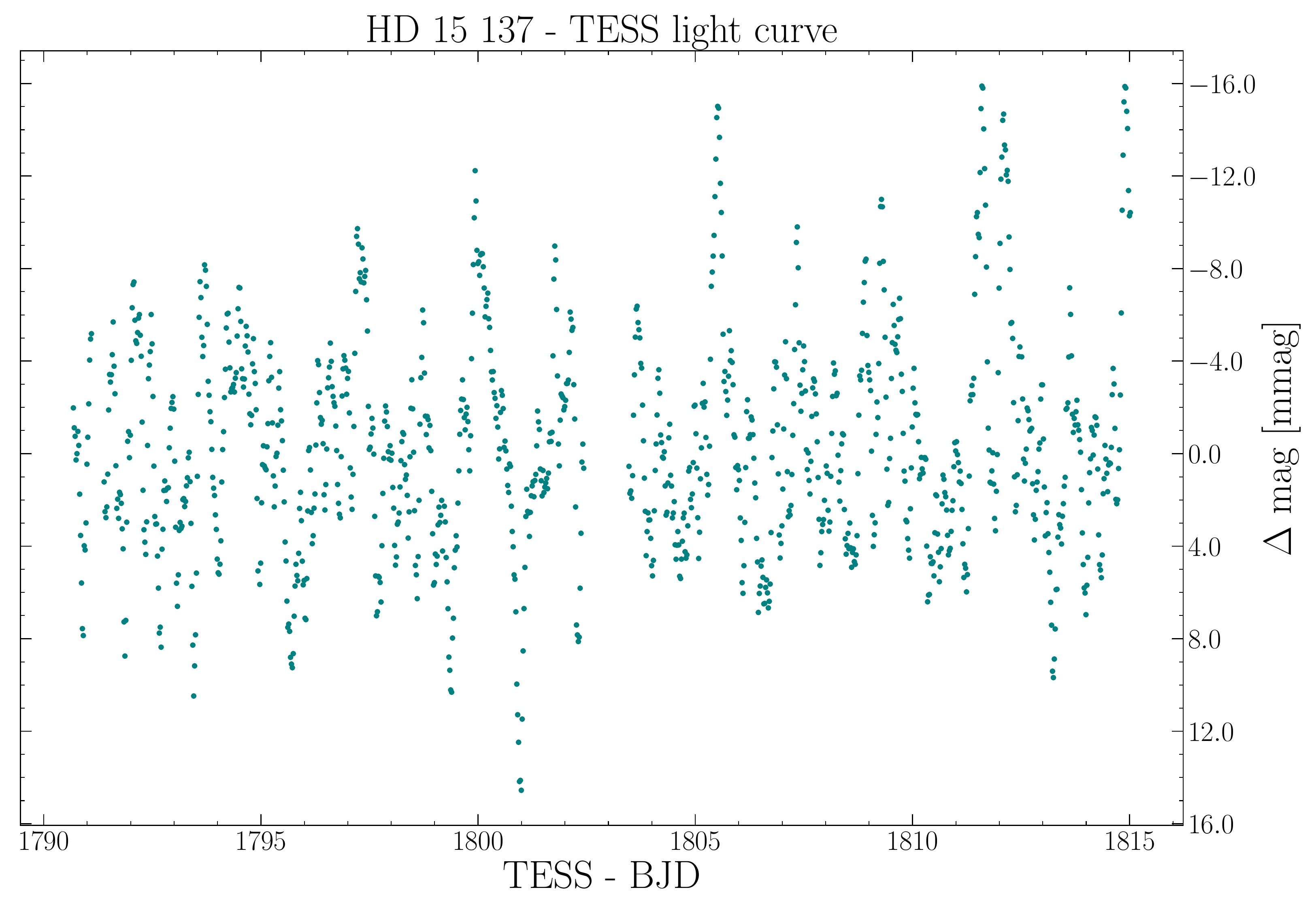} 
    
    \includegraphics[width=.5\textwidth]{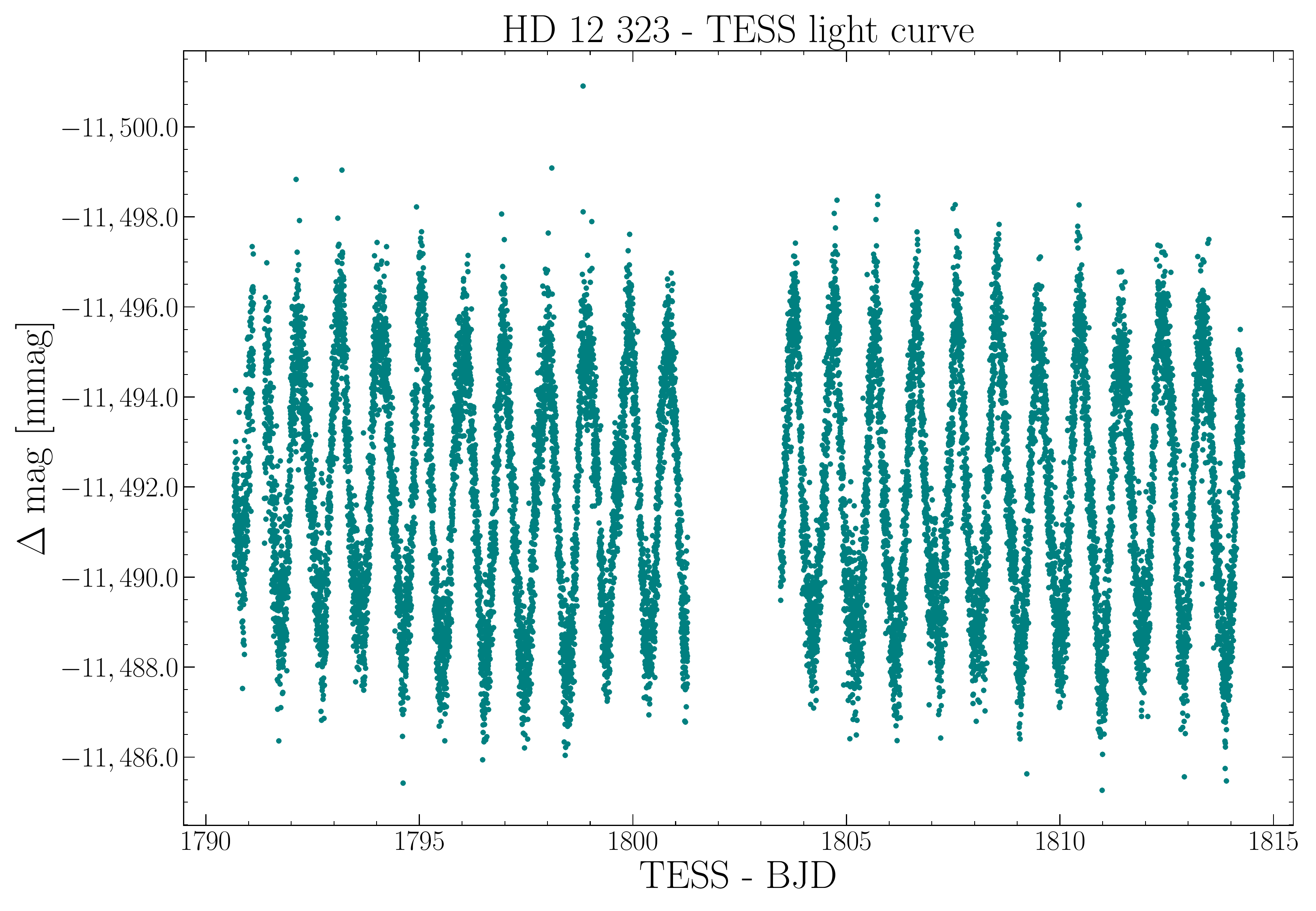}                             
    \includegraphics[width=.5\textwidth]{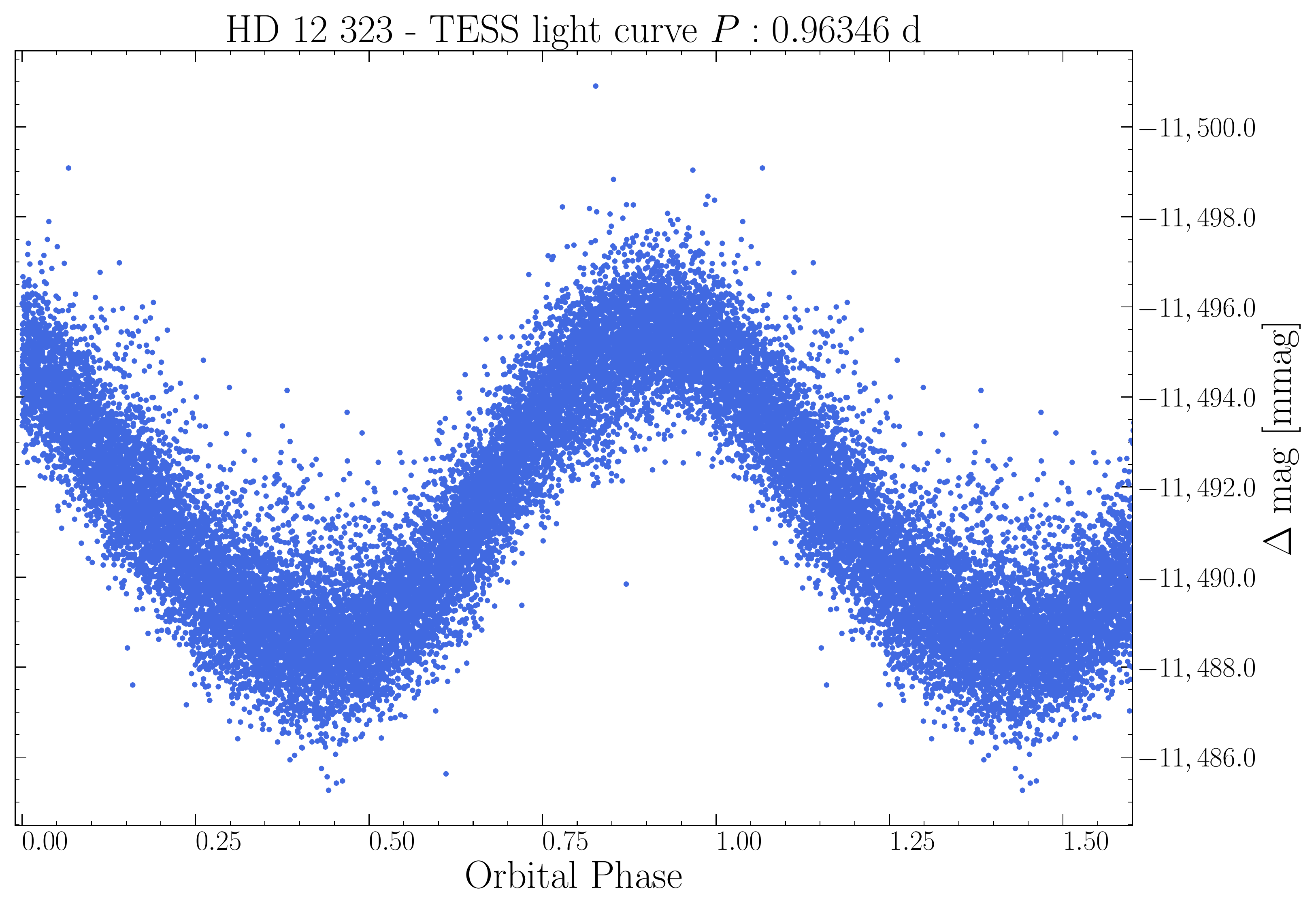} 
    \caption{Normalized and folded TESS LCs of the following systems. 
        Upper left panel: LC for HD~108 in sectors 14 and 15.
        Upper right panel: Folded LC using a period $P=\num{6.1769}$~d for HD~108. 
        Middle left panel: Folded LC for V747~Cep using the $P=5.332$~d~period.
        Middle right panel: LC for HD~\num{15137} in sector 18.
        Lower left panel: LC for HD~\num{12323} in sector 18. 
        Lower right panel: Folded LC for HD~\num{12323} using the period $P=\num{0.96346}$~d. 
        }\label{tess-fig:4}
\end{figure*}

\begin{figure*}[!htp]
\centering
    \includegraphics[width=.495\textwidth]{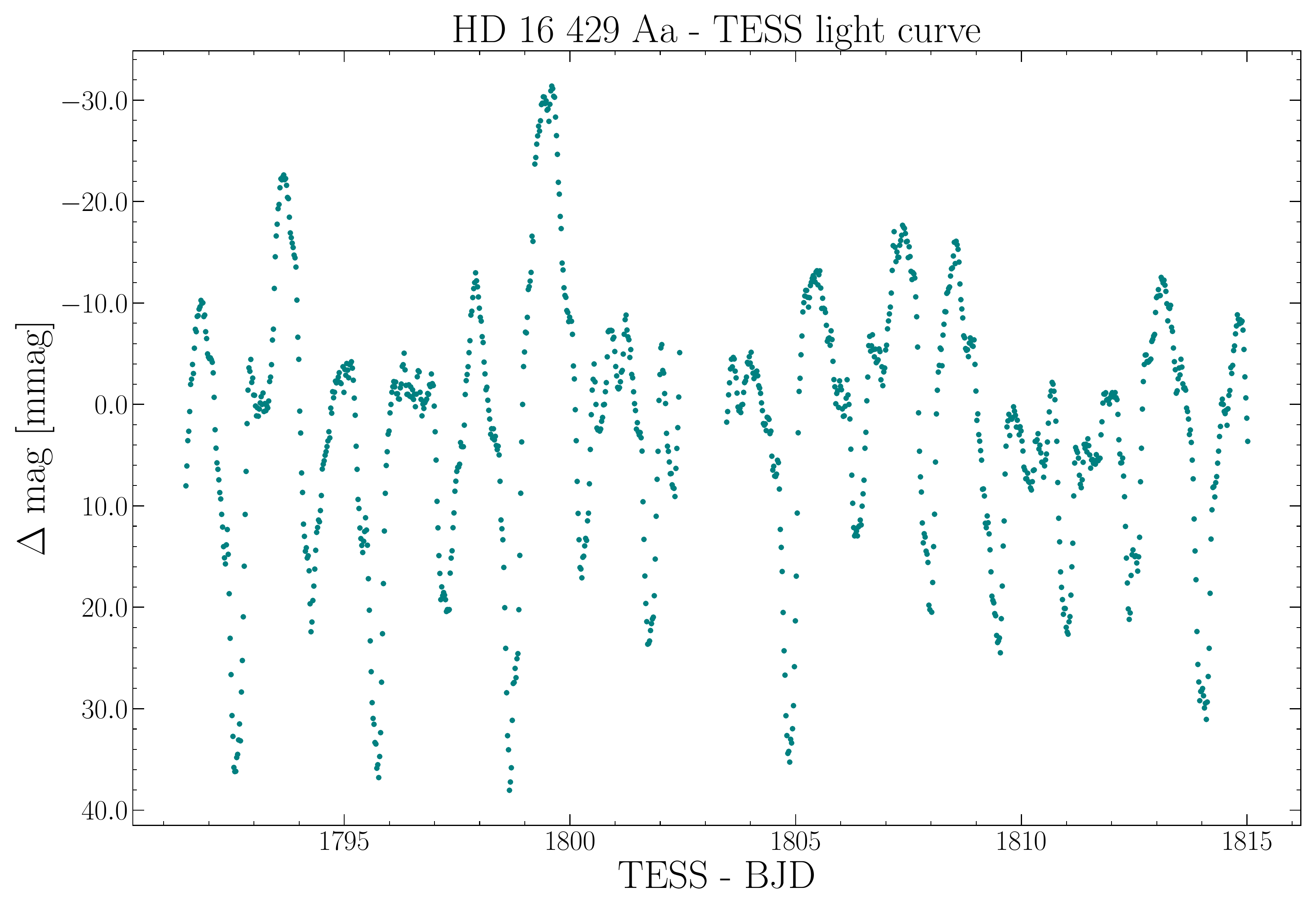} 
    \includegraphics[width=.495\textwidth]{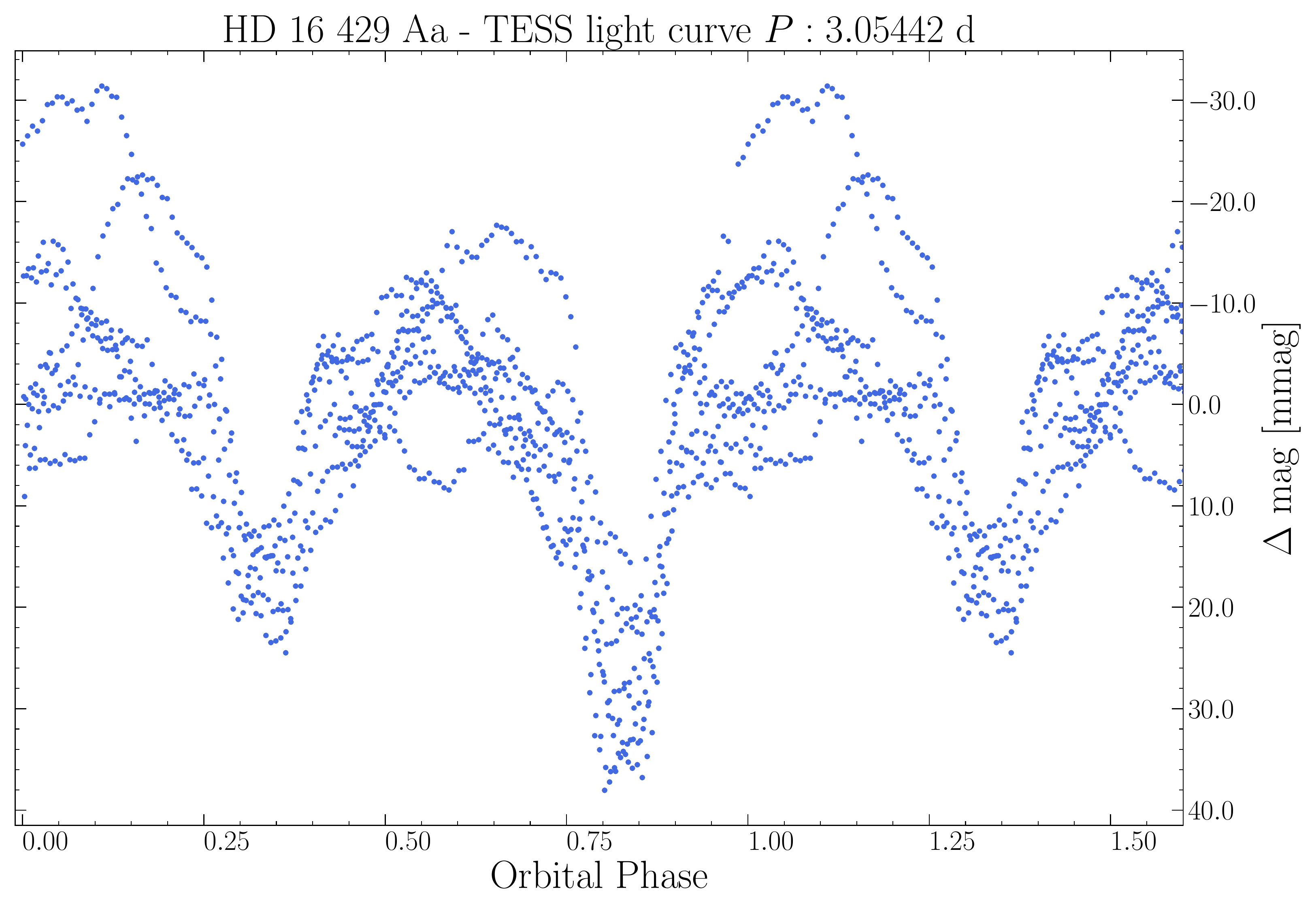} 
    
    \includegraphics[width=.495\textwidth]{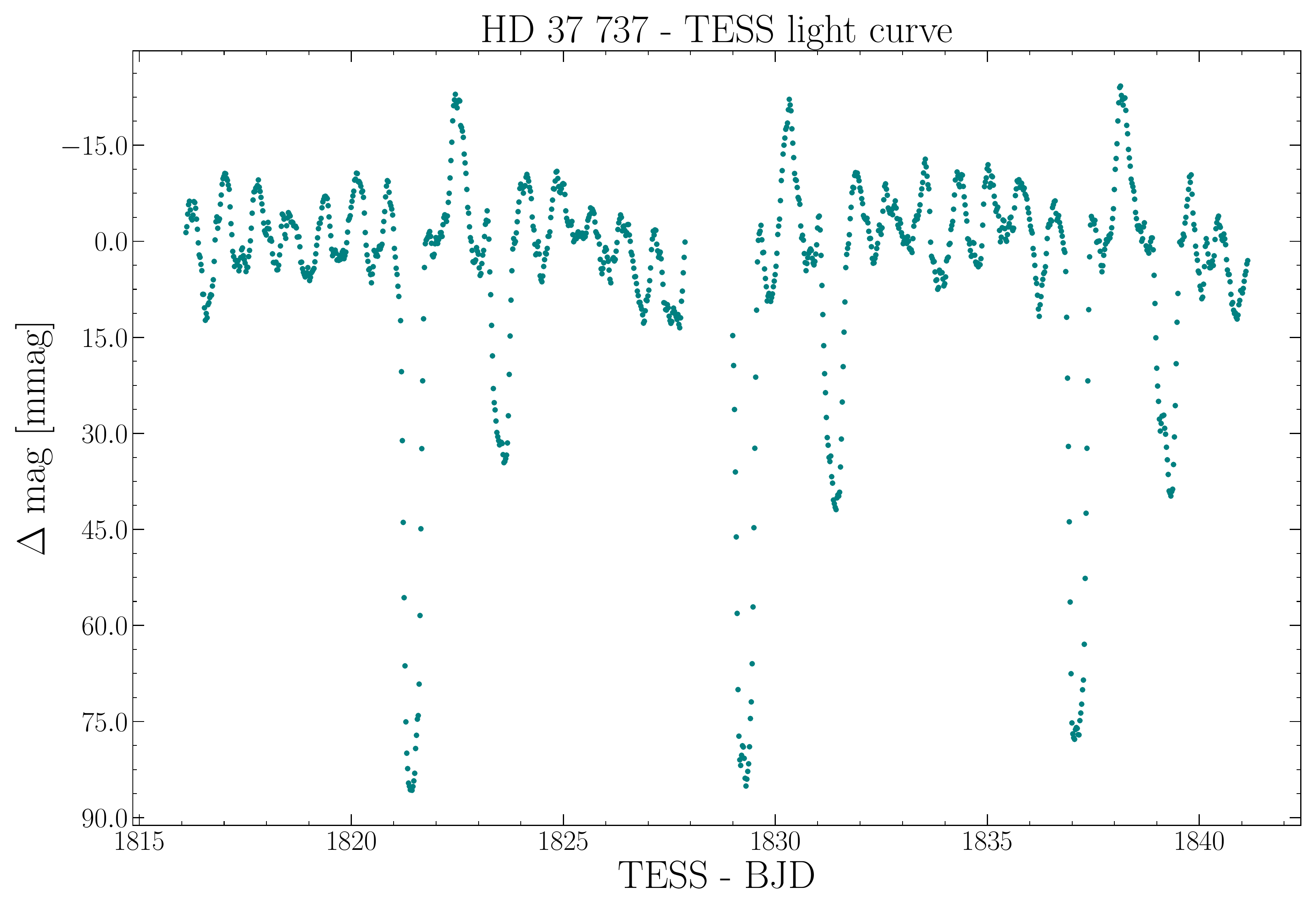}                          
    \includegraphics[width=.495\textwidth]{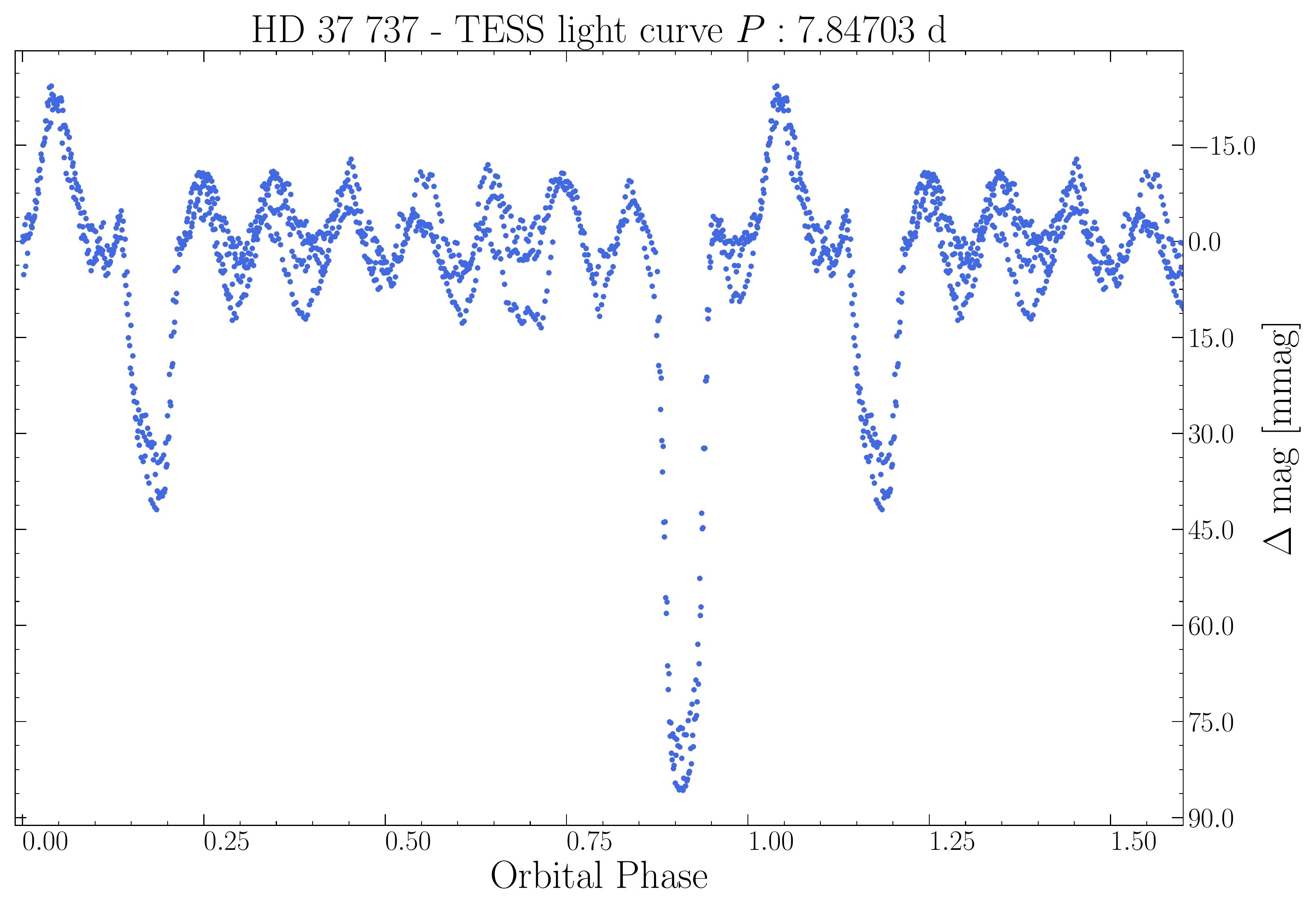} 
    
    \includegraphics[width=.495\textwidth]{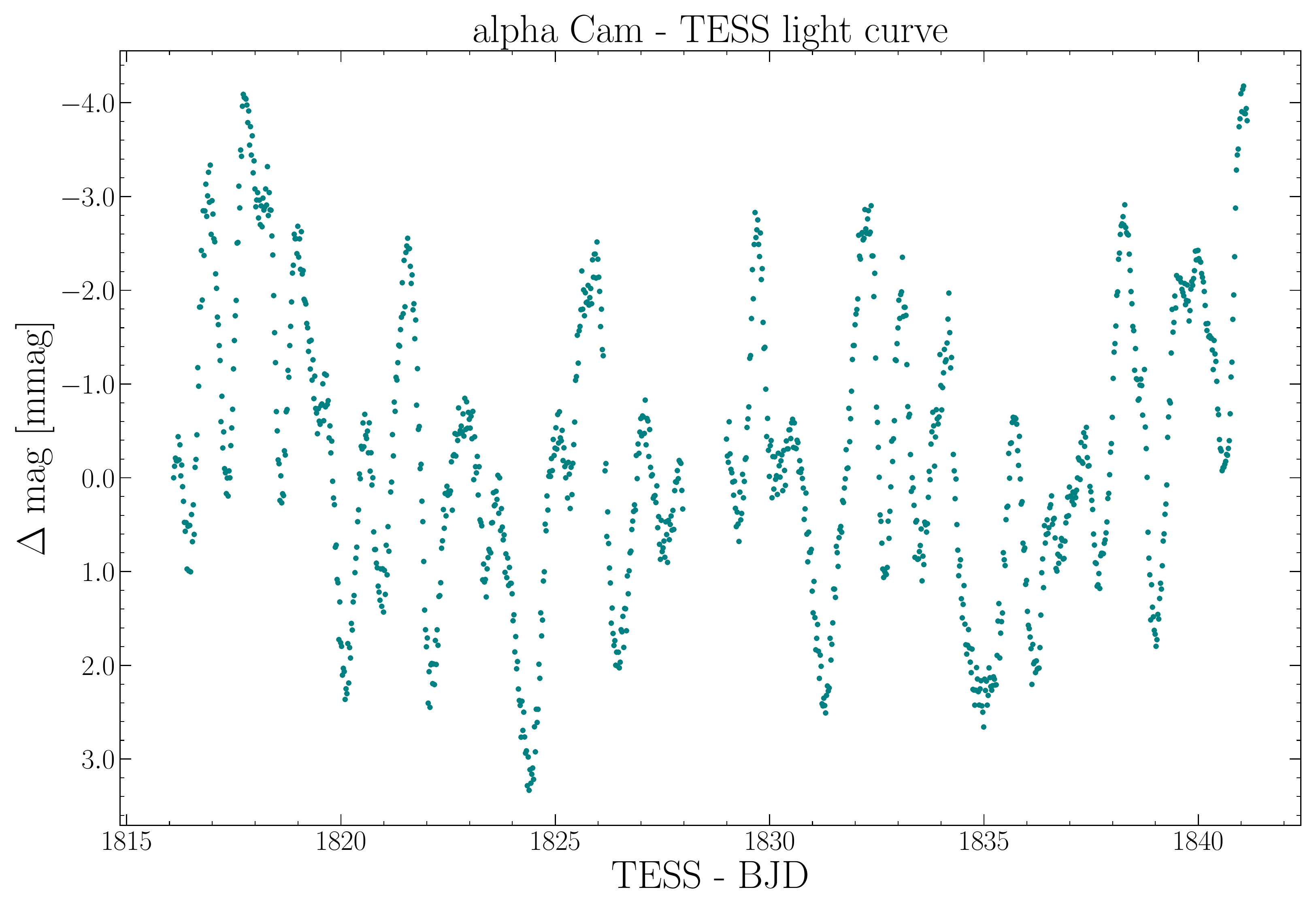} 
    \includegraphics[width=.495\textwidth]{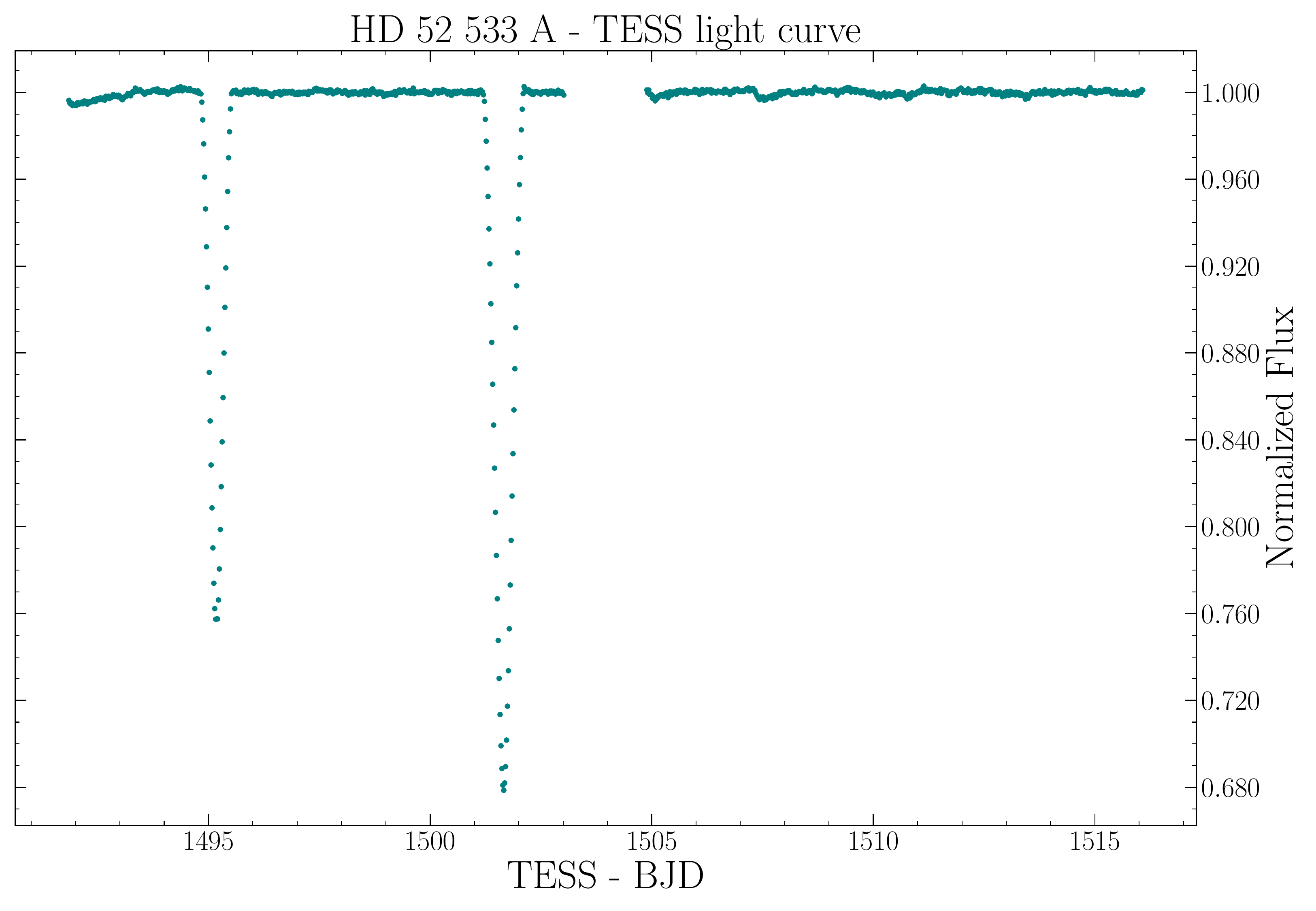} 
    
    \caption{Normalized and folded TESS LCs of the following systems. 
        Upper left panel: LC for HD~\num{16429}~AaAb in sector 18, which displays clear, unevenly spaced eclipses. Also, the LC shows apparent stochastic variations that should be analyzed in the future to unravel their nature. 
        Upper right panel: Folded LC with the spectroscopic period for HD~\num{16429}~AaAb. 
        Middle left panel: LC for HD~\num{37737} in sector 19. 
        Middle right panel: Folded LC for HD~\num{37737} using a period $P=\num{7.84703}$~d. 
        Lower left panel: LC for $\alpha$ Cam. 
        Lower right panel: LC for HD~\num{52533}~A in sector 7. 
        }\label{tess-fig:5}
\end{figure*}

\quad
\section{Radial velocity measurements}\label{app:rv}
In this appendix we present all the RVs measured in this work, including those that have not been used to analyze the orbital solutions of the systems. This Appendix is available in electronic form at the CDS via anonymous ftp to \rurl{cdsarc.u-strasbg.fr} (130.79.128.5) or via \url{http://cdsweb.u-strasbg.fr/cgi-bin/qcat?J/A+A/}.

\end{appendix}

\end{document}